%% file: thesis_final_clean.tex
\documentclass[11pt]{book}	%11pt

\usepackage[left=3.2cm,top=3.5cm,right=2.3cm,bottom=3.0cm,a4paper,twoside]{geometry}

\usepackage{setspace}
\usepackage[charter]{mathdesign}

\RequirePackage[avantgarde]{quotchap}

\usepackage{natbib}
\usepackage{graphicx}
\usepackage{xspace}
\usepackage{lscape}
\usepackage{longtable}
\usepackage{amsmath}
\usepackage{fancyhdr}
\usepackage{titlesec}
\usepackage[breaklinks]{hyperref}
\hypersetup{colorlinks=true,linkcolor=blue,citecolor=blue,filecolor=blue,urlcolor=blue}
\usepackage{afterpage}
\usepackage[hang,flushmargin]{footmisc}

\usepackage{threeparttable}
\usepackage{arydshln}
\usepackage{textcomp} 
\usepackage{multirow}
\usepackage{textcomp}
\usepackage{upgreek}
\usepackage{lscape}
\usepackage[counterclockwise]{rotating}
\usepackage[small,bf]{caption}

\include{definitions}

\usepackage{subcaption} 
\usepackage{breqn}

\titleformat{\paragraph}[runin]{}{}{}{\bfseries\em}[:]

\titlespacing*{\section}{0pt}{0.1in}{0.0in}
\titlespacing*{\subsection}{0pt}{0.1in}{0.0in}
\titlespacing*{\subsubsection}{0pt}{0.1in}{0.0in}
\titlespacing*{\paragraph}{0pt}{0.0in}{0.1in}

\newlength{\figwidth}
\newlength{\figmargin}
\newlength{\initialjump}
\renewcommand{\v}[1]{\ensuremath{\mathbf{#1}}}
\newcommand*{\dittoclosing}{--- \raisebox{-0.5ex}{''} ---}

\begin{document}

% Make title page

\pagestyle{empty}

\begin{titlepage}

\ \\
\vspace{0.35in}
\begin{center}

\noindent{\huge \bf Lightning on exoplanets and brown dwarfs}

\vspace{-0.1in}

\noindent{\large \bf Modelling and detection of lightning signatures throughout the electromagnetic spectrum}

\vspace{0.25in}

\textit{by}

\vspace{0.25in}

\noindent{\large \bf Gabriella Hodos\'an}

\vspace{0.9in}

\includegraphics[scale=0.24]{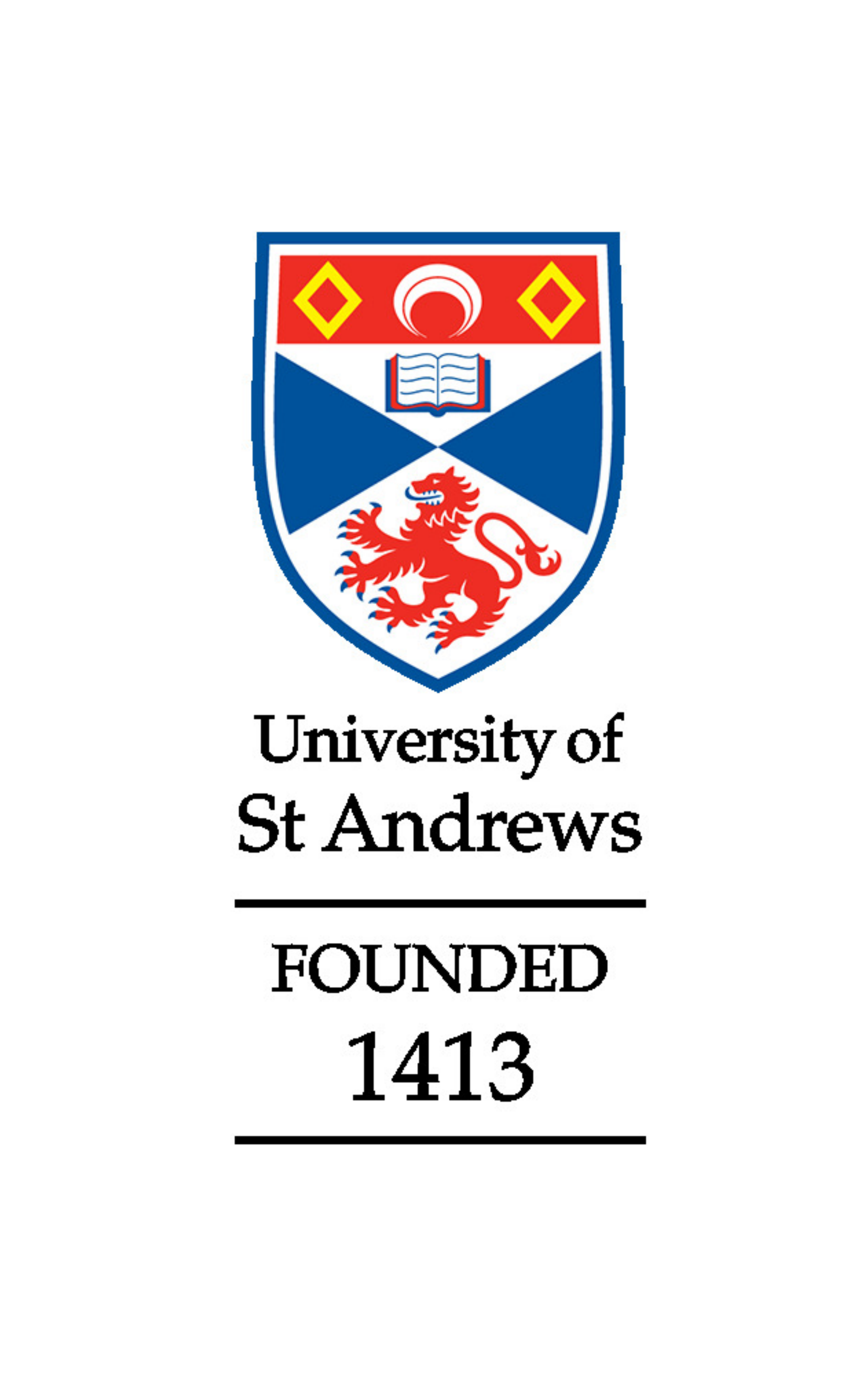}

\vspace{0.5in}

{\large \it Submitted for the degree of Doctor of Philosophy in Astrophysics}

\vspace{0.2in}

{\large September 2017}

\end{center}
\begin{center}

\vspace{0.0in}

\end{center}

\begin{flushleft}
Any\'anak \'es Ap\'anak	
\end{flushleft}

\end{titlepage}

\frontmatter

\chapter{Declaration}

I, Gabriella Hodos\'an, hereby certify that this thesis, which is approximately 50000 words in length, has been written by me, and that it is the record of work carried out by me, or principally by myself in collaboration with others as acknowledged, and that it has not been submitted in any previous application for a higher degree. \\

\noindent I was admitted as a research student in September, 2013 and as a candidate for the degree of PhD in Astrophysics in September, 2013; the higher study for which this is a record was carried out in the University of St Andrews between 2013 and 2017. \\

Date  \hspace{1.8in} Signature of candidate \\

\noindent I hereby certify that the candidate has fulfilled the conditions of the Resolution and Regulations appropriate for the degree of PhD in Astrophysics in the University of St Andrews and that the candidate is qualified to submit this thesis in application for that degree. \\

Date  \hspace{1.8in} Signature of supervisor \\

\chapter{Copyright Agreement}

In submitting this thesis to the University of St Andrews I understand that I am giving permission for it to be made available for use in accordance with the regulations of the University Library for the time being in force, subject to any copyright vested in the work not being affected thereby. I also understand that the title and the abstract will be published, and that a copy of the work may be made and supplied to any bona fide library or research worker, that my thesis will be electronically accessible for personal or research use unless exempt by award of an embargo as requested below, and that the library has the right to migrate my thesis into new electronic forms as required to ensure continued access to the thesis. I have obtained any third-party copyright permissions that may be required in order to allow such access and migration, or have requested the appropriate embargo below. \\

\noindent The following is an agreed request by candidate and supervisor regarding the publication of this thesis: No embargo on print or electronic copy of this thesis. I agree to the title and abstract being published. \\

Date  \hspace{1.8in} Signature of candidate \\

Date  \hspace{1.8in} Signature of supervisor \\

\setstretch{1.3}

\vspace{10cm}

\begin{flushright}
\small "All states of being are determined by mind. \\
It is mind that leads the way." \\
$-$ Ajahn Munindo, A Dhammapada for Contmeplation
\end{flushright}

\vspace{6cm}

\begin{flushright}
\small "Az eszesnek elm\'eje tudom\'anyt szerez, \\
\'es a b\"olcseknek f\"ule tudom\'anyt keres." \\
$-$ P\'eldabesz\'edek, 18:15 (K\'aroli G\'asp\'ar ford\'it\'asa)

\vspace{1cm}
	
\small "An intelligent mind acquires knowledge, \\
and the ear of the wise seeks knowledge" \\
$-$ Proverbs 18:15 \\
\end{flushright}

\chapter{Collaboration Statement}

This thesis is the result of my own work carried out at the University of St Andrews between September 2013 and March 2017. Parts of the work presented in this thesis have been published, or are submitted for publication in refereed scientific journals. In all cases, the text in the thesis has been written by me and the figures presented in the chapters have been produced by me, unless explicitly stated here, or in the figure's caption.

\begin{enumerate}
\item Chapter \ref{chap:stat} is based on "Lightning climatology of exoplanets and brown dwarfs guided by Solar System data", {\bf Hodos\'an, G.}; Helling, Ch.; Asensio-Torres, R.; Vorgul, I. \& Rimmer, P. B., 2016, Monthly Notices of the Royal Astronomical Society, 461, 3927. Data were obtained by means of Richard A. Hart from \textit{Vanus Express}, Carlos Augusto Morales Rodrigues from STARNET, Daniel J. Cecil from LIS/OTD, and Robert H. Holzworth from WWLLN. World Wide Lightning Location Network is a collaboration among over 50 universities and institutions. It provided the lightning location data used in this paper. Rub\'en Asensio-Torres produced the lightning maps for Earth (Figs \ref{fig:1}-\ref{fig:4b}), as part of a summer internship at the University of St Andrews. Paul B. Rimmer calculated the density curves presented in Fig. \ref{fig:mr}. All co-authors of this paper provided comments on the final manuscript.

\item Chapter \ref{chap:hatp11b} is based on "Is lightning a possible source of the radio emission on HAT-P-11b?", {\bf Hodos\'an, G.}; Rimmer, P. B. \& Helling, Ch., 2016, Monthly Notices of the Royal Astronomical Society, 461, 1222; and on "Exo-lightning radio emission: the case study of HAT-P-11b", {\bf Hodos\'an, G.}; Rimmer, P. B. \& Helling, Ch., 2017, submitted to the Conference proceedings of the 8$^{\rm th}$ International Workshop on Planetary, Solar and Heliospheric Radio Emissions (PRE 8), held in Seggauberg near Leibnitz/Graz, Austria, October 25$-$27, 2016. Paul B. Rimmer provided the chemistry estimates for this paper, and Sect. \ref{sec:chem} in Chapter \ref{chap:hatp11b} is based on his work. This piece of work carries relevant and important information regarding exoplanetary lightning, and is closely related to my work, therefore, I included it in this thesis. All co-authors of this paper provided comments on the final manuscript.

\item Chapter \ref{chap:model} is based on "Lightning radio emission on exoplanets and brown dwarfs I: modelling approach and testing for Solar System planets", {\bf Hodos\'an, G.}, Helling, Ch., \& Vorgul, I. This manuscript has been submitted to the Monthly Notices of the Royal Astronomical Society for peer review (Feb. 2017). All co-authors of this paper provided comments on the submitted manuscript.
\end{enumerate}

\chapter{Abstract}

Lightning is an important electrical phenomenon, known to exist in several Solar System planets. Amongst others, it carries information on convection and cloud formation, and may be important for pre-biotic chemistry. Exoplanets and brown dwarfs have been shown to host environments appropriate for the initiation of lightning discharges. In this PhD project, I aim to determine if lightning on exoplanets and brown dwarfs can be more energetic than it is known from Solar System planets, what are the most promising signatures to look for, and if these "exo-lightning" signatures can be detected from Earth.

This thesis focuses on three major topics. First I discuss a lightning climatology study of Earth, Jupiter, Saturn, and Venus. I apply the obtained lightning statistics to extrasolar planets in order to give a first estimate on lightning occurrence on exoplanets and brown dwarfs. Next, I introduce a short study of potential lightning activity on the exoplanet HAT-P-11b, based on previous radio observations. Related to this, I discuss a first estimate of observability of lightning from close brown dwarfs, with the optical Danish Telescope. The final part of my project focuses on a lightning radio model, which is applied to study the energy and radio power released from lightning discharges in hot giant gas planetary and brown dwarf atmospheres. The released energy determines the observability of signatures, and the effect lightning has on the local atmosphere of the object.

This work combines knowledge obtained from planetary and earth sciences and uses that to learn more about extrasolar systems. My main results show that lightning on exoplanets may be more energetic than in the Solar System, supporting the possibility of future observations and detection of lightning activity on an extrasolar body. My work provides the base for future radio, optical, and infrared search for "exo-lightning". 

\vspace{10cm}

\begin{flushright}
\small "My life amounts to no more than one drop in a limitless ocean. \\
Yet what is any ocean, but a multitude of drops?" \\
$-$ David Mitchell, Cloud Atlas
\end{flushright}

\vspace{6cm}

\begin{flushright}
\small "Fairy tales are more than true: not because they tell us that dragons exist, \\
but because they tell us that dragons can be beaten." \\
$-$ Neil Gaiman, Coraline
\end{flushright}

\chapter{Acknowledgements}

K\"osz\"on\"om sz\"uleimnek, hogy felneveltek \'es j\'o p\'eld\'at \'all\'itottak el\'em az \'elet minden ter\'en. K\"osz\"on\"om, hogy megteremtett\'ek sz\'amomra a biztos h\'atteret, \'es seg\'itenek c\'eljaim el\'er\'es\'eben. T\'amogat\'asuk \'es szeretet\"uk n\'elk\"ul ez a dolgozat soha nem sz\"ulethetett volna meg.
Hasonl\'ok\'eppen k\"osz\"on\"om testv\'ereimnek, Zsoltinak \'es Juditnak, \'es a sz\H{u}k csal\'adomnak, hogy mindig mellettem \'allnak.

First and foremost, I would like to thank my parents, Judit and Csaba, for their love and support, and my siblings, Zsolti and Judit, and my close family for being there for me. You will understand my opening words and how grateful I am to you.

I would like to thank my supervisor, Christiane, for guiding my career. In addition to shaping my academic work, she introduced me to a lot of important people, and helped me with conference talks. I am thankful for the staff members, postdocs, especially Aleks, Paul, Craig, Irena, and Claudia, whom I could always bother if I had questions or doubts about my work. Funding for the PhD was provided by the European Research Council (ERC) under the project 'Lightning', grant number 257431.

These 3.5 years would not have been as great and as fun as they actually were without the PhD students I started with, and the close friends I "picked up" on the road. Without your welcoming I could not have gotten through the first months of my life in a brand new country as easily as I did. Thank you, David and Josh for the game and movie nights, and Sunday lunches, and thank you, Tim and Alistair for making those parties even better. I am especially grateful to Victor, Alasdair, Inna, and Graham, who were always there when I needed someone to put my mind back on the right track. You guys will always have a special place in my heart. Thank you, Kirstin, Lisa, Annelies, Isabel, and Maya, for being amazing and fun to be around. Thank you, Milena, for being an awesome flatmate and friend. I will always remember the laughs and chats, and sometimes long discussions we had in the kitchen of 15 Shields Ave. I would like to thank Rub\'en for his friendship and help with my PhD work. 

The road to St Andrews was made interesting and exciting by my Masters' professors and supervisors, who thought me and showed me the beauty of astronomy. I would like to thank them for their help, especially to L\'aszl\'o Kiss, Gyula Szab\'o, Attila Simon, Peter McCullough, and B\'alint \'Erdi. 

Finally, I would like to thank my loving and fantastic partner, Zach, for his constant, never ending support and love. He always believed in me even when I did not. He encouraged me when I was down, listened to my rants, and complaints; laughed with me and was happy for my successes. I could not have gone through the PhD without his help. See, dear? I did it!

\newpage

\setstretch{1.0}

\tableofcontents

\listoffigures

\listoftables

\setstretch{1.5}

\mainmatter

\pagestyle{fancy}

% NOTE: Here come the chapters - you can have one file per chapter. When you are writing a given chapter, you can always comment out the others to speed up the typesetting.

\include{./chapters/1_introduction_v3}	% \ref{chap:intro} 

\include{./chapters/2a_lightphysics}	% \ref{chap:ligform}

\include{./chapters/2b_signatures}	% \ref{chap:ligsig}

\include{./chapters/2c_solsyslightning}  % \ref{chap:liginout}

\include{./chapters/3_lightningstat} 	% \ref{chap:stat}

\include{./chapters/4_hatp11b}		% \ref{chap:hatp11b}

\include{./chapters/4b_danishtel}	% \ref{chap:danish}

\include{./chapters/5_radiomodel}	% \ref{chap:model}

\include{./chapters/6_conclusions}	% \ref{chap:concl}

% NOTE: The \appendix command means that chapters will now change from numbers to letters

%\appendix

\backmatter

% NOTE: you can also add pages like the following. Using the empty pagestyle means that it won't count as a chapter (like the declaration, copyright, abstract, ...)

\pagestyle{empty}

%\chapter{Online resources}

%\begin{footnotesize}

%\noindent [1] - \href{http://circumstellardisks.org/}{http://circumstellardisks.org/}

%\noindent [2] - \href{http://stardust.jpl.nasa.gov/}{http://stardust.jpl.nasa.gov/}

%\noindent [3] - \href{http://gemelli.colorado.edu/~bwhitney/}{http://gemelli.colorado.edu/~bwhitney/}

%\end{footnotesize}

%\setstretch{1.1}

%\newpage

%\ \\

\newpage

% NOTE: finally, the bibliography. I used BibTex, hence all references were in references.bib.

\addcontentsline{toc}{chapter}{Bibliography} % Add bibliography to table of contents
%ez a style nem tetszik... kell masik
\bibliographystyle{apj_tr9} % limits to 8 authors, longer author lists go to et al.
\bibliography{references}   % main bib file

\end{document}

%% file: definitions.tex
\usepackage{xspace}

\defcitealias{lecav2013}{L13}

%% file: chapters/1_introduction_v3.tex
\chapter{Introduction} \label{chap:intro}

%__________________________________________________________________
%__________________________________________________________________
\section{Motivation}

\begin{figure}
\begin{center}
\includegraphics[scale=0.33]{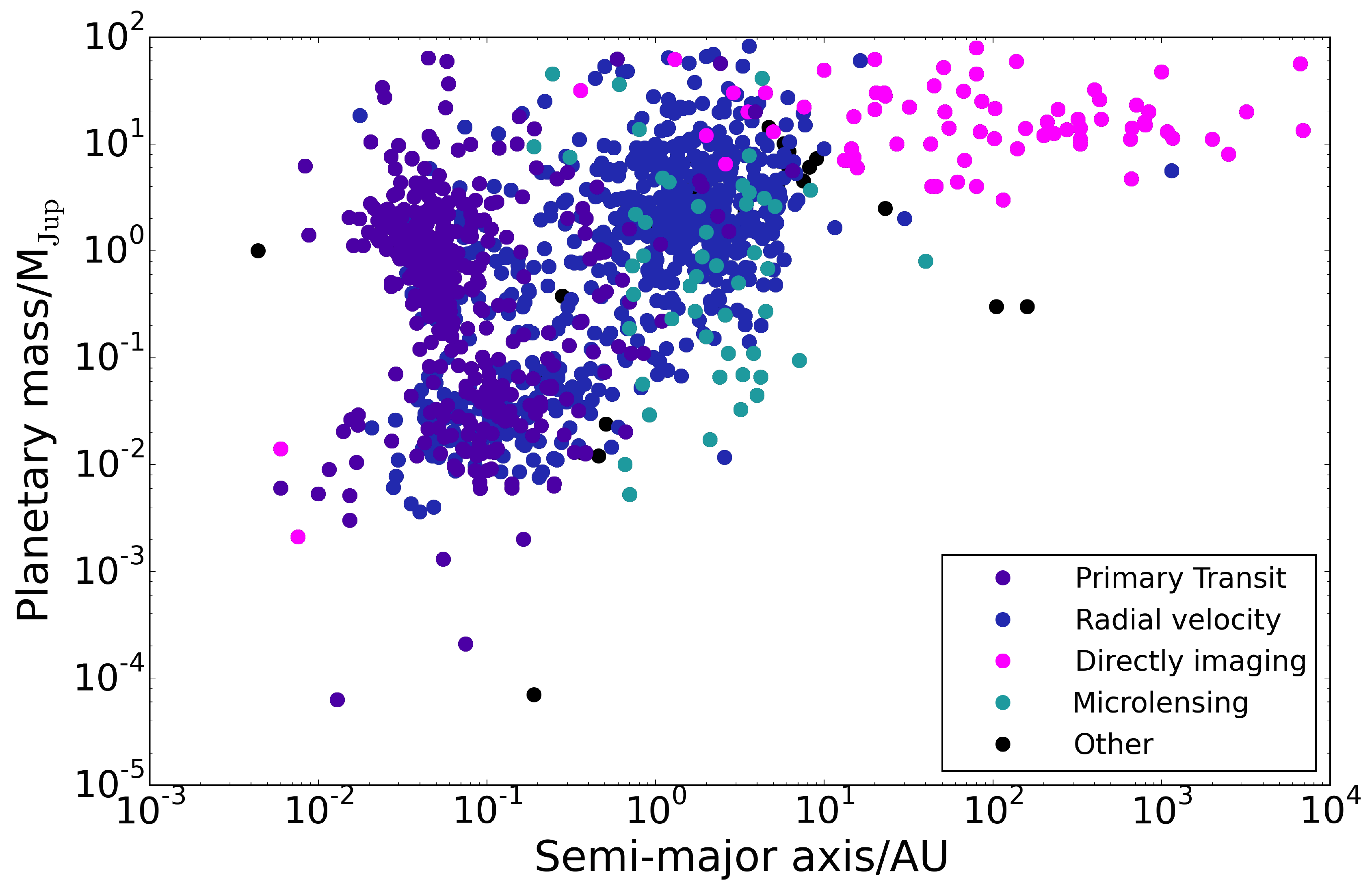}
\end{center}
\caption{Diversity of exoplanets. 
The different colours indicate the method the planet was detected with. Purple: transit method, blue: radial velocity, magenta: direct imaging, cyan: microlensing, black: other methods such as pulsar timing and astrometry. The large variety of exoplanet-types and environments suggest that lightning may be a common phenomenon in the universe. Data are from exoplanet.eu, downloaded on 2017/01/11.}
\label{fig:intro_1}
\vspace{0.5cm}
\end{figure}

The discovery of the first exoplanet around a neutron star \citep{wolszczan1992} and then around a Sun-like star \citep{mayor1995} opened the gates to a new astronomical field concerning extrasolar planetary systems. By March 2017, there have been $\sim3600$ exoplanets discovered\footnote{http://exoplanet.eu/; 2017 March 23.}. This large number allows us to focus on the more detailed characterization of the different types of planets, including atmospheric chemistry and internal composition of the planetary bodies.
Figure \ref{fig:intro_1} illustrates the large variety of exoplanets that have been observed, including hot Jupiters (e.g. HD 189733b), mini-Neptunes (e.g. Kepler-11c), super-Earths (e.g. 55 Cnc e) and even planets smaller than Earth (e.g. Kepler-70c). The different techniques used for detection allow the exploration of these extrasolar objects from different points of view. Radial-velocity measurements and transit observations together give a constraint on the radius and mass of the planet. Transmission spectroscopy reveals information regarding the planetary atmosphere. The orbit of the planet can be mapped with direct imaging, and microlensing could map the frequency of different sized planets around different stars in the Galaxy disc, since it is not biased towards certain stellar or planetary types. The diversity of exoplanets in mass and radius results in diverse bulk compositions, which, together with a wide range of distances from the host star, will inevitably suggest that these objects host a large variety of surface and atmospheric environments. As I will show in the following chapters, lightning occurs in various environments in the Solar System. Therefore, when looking for lightning signatures, we are not limited only to a certain type of exoplanet, but can probe a large variety of them.

\begin{figure}
\begin{center}
\includegraphics[scale=0.33]{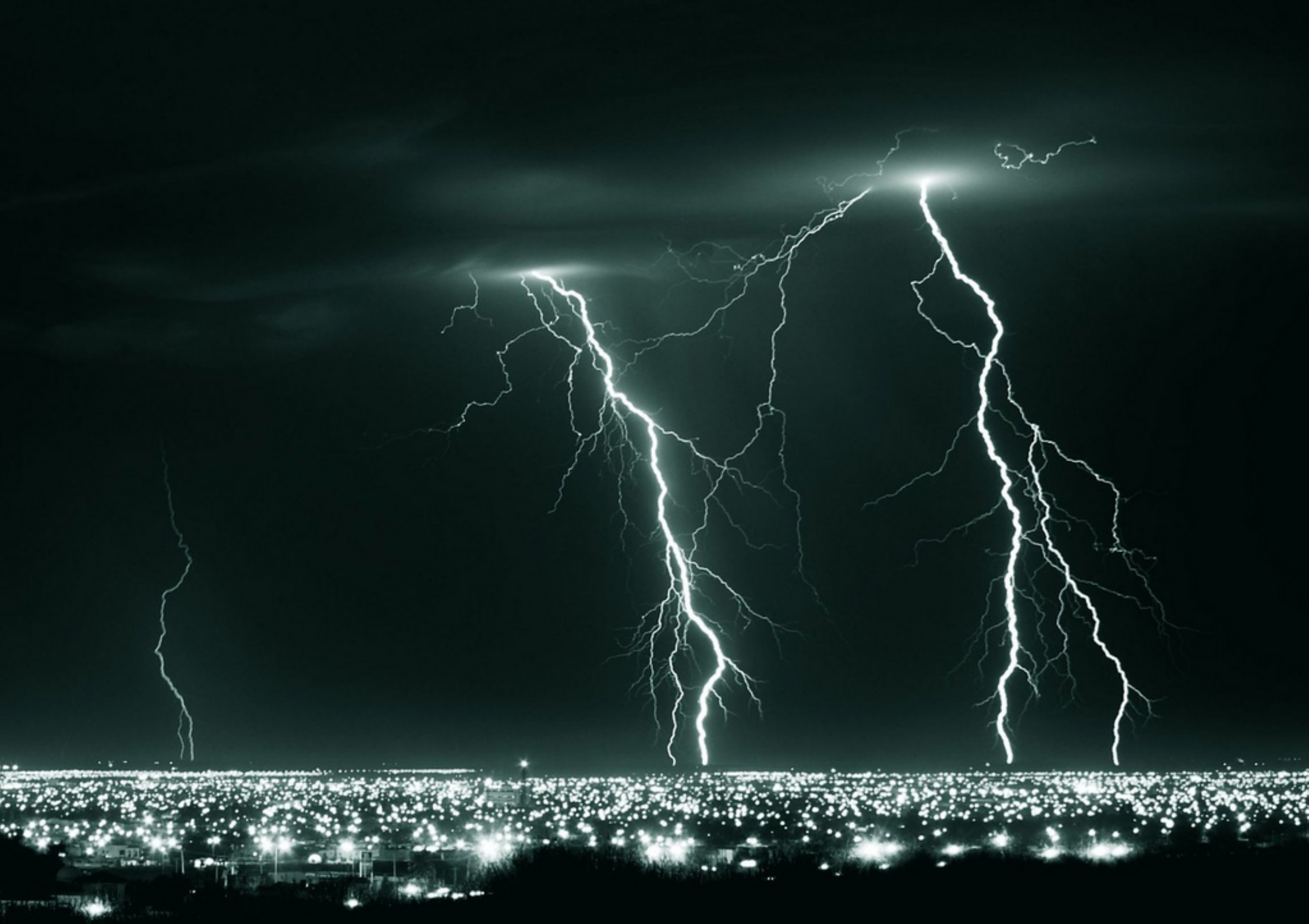}
\end{center}
\caption{Lightning flashes over Hermosillo, Sonora, Mexico, 2008. Credit and courtesy: Jos\'e Eugenio G\'omez Rodr\'iguez}
\label{fig:intro_2_lightn}
\vspace{0.5cm}
\end{figure}

Lightning (Fig. \ref{fig:intro_2_lightn}) and lightning related phenomena have been studied for decades \citep[selected reviews on the subject:][]{rakov2003,yair2008,yair2012,pasko2010,pasko2012,dwyer2012,dwyer2014,siingh2015}. It indicates cloud formation and large-scale convection in atmospheres of planetary objects. Lightning supports the global electric circuit \citep[GEC,][]{wilson1921}, a continuously present electric current between the earth-ionosphere cavity. The GEC is important, e.g., because it affects life on Earth. Life, as we know it, evolved around the environments found on Earth, which the GEC is part of, therefore, it is expected that organisms evolved in an Earth environment will need the presence of such a current. For example, sharks are known to be sensitive to the changes in oceanic electric currents caused by the changes in the Earth's magnetosphere \citep{meyer2005}. Lightning indicates the presence of GEC in an extrasolar atmosphere, therefore can be a potential habitability indicator combined with biosignatures in exoplanets similar to Earth. Lightning has been suggested as a tool to study, for example, earthquake occurrence, and a relation with global warming was indicated. \citet{mullayarov2007} investigated the relation between lightning radio signatures originating from thunderstorms passing over earthquake regions and earthquake activity. They observed a decrease in the lightning radio signal amplitude a few days before an earthquake occurred, and found the signal going back to the original amplitude on the day of the event. \citet{romps2014} suggested a link between global warming over the United States and flash rate variability. Their results showed an increase of flash numbers due to an increase of global precipitation rate and of the convective available potential energy (CAPE), a proxy of lightning activity. 

Furthermore, the Miller-Urey experiment showed that in highly reductive environments (composed of CH$_4$, NH$_3$, H$_2$, and H$_2$O) lightning discharges produce prebiotic molecules \citep{miller1953,miller1959}, which are important for the formation of life. The experiment was set up so that it would represent an "Early-Earth" atmosphere as it was thought to be in the 1950s. Though, now it is debatable whether the early atmosphere of the Earth was such a reductive one,  
the large variety of extrasolar planets (Fig. \ref{fig:intro_1}), suggest that such environments could exist outside the Solar System. \citet{kasting1993b} hypothesized a rather weakly reducing (made of mostly CO$_2$ and N$_2$) early atmosphere for Earth, while \citet{tian2005} suggested that a hydrogen-rich prebiotic atmosphere is also possible. \citet{cleaves2008} demonstrated that lightning supports the formation of amino acids even in weakly reducing environments. It has also been shown that for a large variety of prebiotic scenarios \citep[e.g.][]{patel2015}, hydrogen cyanide (HCN) is the key simple precursor molecular species for prebiotic chemistry. \citet{ardaseva2017} found that lightning is very effective at generating HCN in the atmosphere of the Early Earth, further supporting the theory that lightning is an important element for prebiotic chemistry. These studies indicate that lightning is an important electrical phenomenon on Earth. This suggests that lightning, if exists, will be important for numerous, previously listed reasons (e.g. indicator of cloud dynamics; it changes the chemical composition of an atmosphere; etc.), on exoplanets, and possibly on low-mass stellar objects (brown dwarfs) as well.

Studies have revealed that both exoplanets and brown dwarfs host environments with the necessary ingredients (i.e. charged particles, seed electrons, charge separation) for lightning to initiate: Both observations \citep[e.g.][]{kreidberg2014, sing2009, sing2013, sing2015} and kinetic cloud models \citep[e.g.][]{helling2008, helling2008b, helling2011, helling2011b} confirmed that clouds form in extrasolar planetary atmospheres. Cloud formation involves convection and gravitational settling, the main processes separating two oppositely charged regions from each other. The result of charge separation is the build-up of a large electric potential and therefore a large electric field that is necessary for the initiation of lightning discharges \citep{rakov2003,aplin2013,helling2013,helling2016b}. Works like \citet{helling2011b, rimmer2013, rodriguezbarrera2015} %stark2013, 
showed that various ionization processes occur in extrasolar atmospheres, which provide the necessary positive and negative charged particles. \citet{helling2013} and \citet{bailey2014} suggested that the processes building up the electric field necessary for lightning, are indeed capable of producing lightning discharges in extrasolar atmospheres. 

Lightning induced radio and optical emission has been observed in the Solar System on several planets. Earth lightning shows a large variety of signatures discussed in Chapter \ref{chap:ligsig}. Radio emission has been observed most probably on Venus \citep[e.g.][]{russell2008,russell2011}, Uranus \citep{zarka1986} and Neptune \citep{gurnett1990,kaiser1991}. Lightning on the giant planets Jupiter \citep[e.g.][]{gurnett1979} and Saturn \citep[e.g.][]{fischer2006,fischer2007} has been observed both in the optical and the radio bands (Chapter \ref{chap:liginout}). 

\citet{vorgul2016} suggested that present day radio observations of brown dwarfs may contain hints to the presence of lightning in these atmospheres. They found that discharges may affect the radio signatures of electron cyclotron maser emission, the hypothetic source of radio emission on brown dwarfs. \citet{zarka2012} analysed the possibility of detecting radio emission from extrasolar planets. They concluded that emission $10^5$ times stronger than radio emission observed on Jupiter and Saturn from a distance of $10$ pc would be possible to detect, though propagation effects will affect the radio signal below a few MHz. In their paper, \citet{zarka2012} scaled up the same radio emission that was observed from Jupiter and Saturn and used that for their estimates.

This thesis covers a new field of comparative planetology studies, and explores how lightning discharges occur in exoplanetary and brown dwarf atmospheres. It looks for the answers to the questions, whether lightning can be observed from outside the Solar System, what would be its signatures, and how much energy could be released from lightning that affects the local atmosphere. 

%__________________________________________________________________
\section{Thesis outline} \label{sec:thesis}

In this thesis, I give an overview of what has been known of lightning inside and outside the Solar System. Then, I will focus on my PhD work, including a statistical study and a model of lightning radio emission, in order to answer the questions raised above.

Chapter \ref{chap:ligform} focuses on the basic principle of lightning formation, discusses the different lightning-forming environments and the scientific nomenclature.
Chapter \ref{chap:ligsig} describes the signatures of lightning in more detail. The signatures are separated into three main categories based on the mechanism of origin: direct lightning emission, emission caused by secondary events, and signatures caused by the effects of lightning on the local chemistry. I discuss previous observations of lightning in the Solar System and overview past studies of lightning discharges on exoplanets and brown dwarfs in Chapter \ref{chap:liginout}. 

In Chapter \ref{chap:stat}, I study lightning climatology for the different atmospheric environments of Earth, Venus, Jupiter and Saturn. I present lightning distribution maps for Earth, Jupiter and Saturn, and flash densities [flashes km$^{-2}$ h$^{-1}$] for these planets and Venus, based on optical and/or radio data of spacecraft and radio networks. I also calculate flash densities for several phases of two volcanic eruptions, Eyjafjallaj\"okull's (2010) and Mt Redoubt's (2009). I apply these findings to estimate lightning occurrence on certain exoplanets and brown dwarfs, which I collect in 6 categories according to their characteristics. I briefly discuss the effects of stellar activity on lightning occurrence, and examine lightning energy distributions for Earth, Jupiter and Saturn. In an example, I apply the obtained flash densities to transiting exoplanets to provide a lower limit of the total number of flashes that might occur during their full transit.

In Chapter \ref{chap:hatp11b}, I analyse the possibility of detecting lightning radio emission from the exoplanet HAT-P-11b. \citet{lecav2013} carried out radio transit observations of this exoplanet, and suggested that a small part of the radio flux can be attributed to the planet. Here, I assume that this signal is real, and study if this radio emission could be caused by lightning with similar energetic properties like in the Solar System. I also estimate the optical emission of the hypothetical thunderstorm on HAT-P-11b, and present a short study of its chemical effects on the local atmosphere. I carry out a parameter study to explore the range of flash densities necessary to produce the observed radio flux (assuming the flux is lightning-induced).

In Chapter \ref{chap:danish}, I present lightning optical emission estimates for close brown dwarf systems: Luhman-16, $\epsilon$ Indi, SCR 1845-6357. The targets are chosen to be observable from La Silla, Chile, with the Danish 1.54-m telescope. I use this telescope to plan the observations for. I estimate optical fluxes and apparent magnitudes in I, V, and U bands.

In Chapter \ref{chap:model}, I estimate the energy dissipated from lightning discharges and the total power emitted at radio frequencies, in order to study possible differences and similarities of extrasolar and Solar System lightning. I construct a model of lightning radio emission based on models developed for Earth lightning and used for other Solar System planets, such as Jupiter and Saturn. I apply my model to extrasolar atmospheres based on characteristics of lightning discharges on exoplanets and brown dwarfs published in \citet{bailey2014}. I present energy and radio power estimates for extrasolar objects with the following properties: log($g$)=3.0 and 5.0; T$_{\rm eff}$ = 1500 \dots 2000; [M/H] = 0.0 and -0.3.

In Chapter \ref{chap:concl}, I summarize my work and present a lookout for future research possibilities.

%% file: chapters/2a_lightphysics.tex
\chapter{The theory of lightning formation} \label{chap:ligform}

The most common theory of lightning formation is as follows: Charges, produced by, e.g., different ionization processes, accumulate on cloud particles, which are separated by e.g. gravitational settling or convection. As the effectively more negatively charged particles settle towards the bottom of the cloud, a large electric potential is built up, creating an electric field that accelerates electrons to high energies producing an electron avalanche and leading to the lightning discharge process \citep{rakov2003,aplin2013,helling2013}. In this chapter, I introduce the processes involved with lightning formation and show that these processes occur in extrasolar atmospheres as well (Sect. \ref{sec:ion}-\ref{sec:propag}). I briefly discuss lightning forming environments and why they are interesting in terms of extrasolar lightning occurrence (Sect. \ref{sec:envir}). There are several reviews on lightning theory. Here, I use mostly the following ones: \citet{rakov2003,dwyer2014,james2008,rousseldupre2008}. For further details, I direct the reader to these reviews and references therein. At the end of each section, I summarize the current theory of extrasolar lightning. The theory of lightning formation on exoplanets and brown dwarfs has not yet been comprehensively explored, as the pioneers of the field \citep[e.g.][]{helling2013,helling2013b,helling2016b,rimmer2013,rimmer2016} have only started to study the subject in depths in the last few years. My work presented in this thesis \citep[and in][]{hodosan2016,hodosan2016b,hodosan2017} further explores the characteristics and properties of "exo-lightning", in connection with these previous studies.

%__________________________________________________________________
%__________________________________________________________________
\section{Ionization processes and charge generation} \label{sec:ion}

To initiate a lightning discharge, seed charged particles have to be present in lightning-producing environment. These ions are produced by different ionization processes discussed in this section.  
On Earth, the two major processes that produce the free electrons necessary for lightning initiation in thunderclouds, are cosmic ray ionization and the natural radioactivity of the Earth \citep{rakov2003,stozhkov2003}. Above about 1-3 km the major provider of electrons is cosmic rays \citep{rakov2003,ermakov2003}. \citet{hess1912} discovered cosmic ray ionization during balloon experiments above 1100 m altitude. The two major sources of cosmic rays are the Galaxy and the Sun, therefore the rate of ionization will depend on solar activity and the latitude representing the incident angle of solar particles \citep{rakov2003,ermakov2003}. Though the flux of cosmic rays is much smaller than the solar electromagnetic radiation, they are the major source of free electrons below $\sim$ 35 km altitude \citep{ermakov2003}. Apart from being important for initiating discharge events, cosmic rays may affect the Earth climate by intensifying aerosol formation \citep{pudovkin1995,shumilov1996}.
 
Lightning occurs in volcanic plumes and possibly dust devils. Triboelectrification or tribocharging (frictional charging) is thought to be one of the processes active in both of these environments. It is a charge transfer mechanism based on surface interactions, when materials are rubbed together and collide \citep{gilbert1991,zheng2013,yair2016}. Electrification of volcanic plumes is thought to correlate with strong water boiling (lava entering liquid salty water) and extensive magma fragmentation during explosive eruptions \citep{mather2006}. Plume particles can be electrified either as a result of their formation process or subsequently by radioactive decay, particle interaction with existing charges, or particle-particle interactions \citep{james2008}. Charge generation in plumes is not well understood, however there are many theories that could work under different conditions. Apart from triboelectrification, fractoemission is thought to be a major source of ions in the volcano plume. Fractoemission is the ejection of ions, neutral atoms and electromagnetic radiation from the fresh crack surfaces during fracture events \citep{dickinson1988,james2000}. The above mentioned mechanisms are based on the interaction of solid particles, while there are mechanisms, which involve liquid water in the charging process. In plumes that are related to lava entering the sea, boiling may have a role in charge generation \citep{pounder1980}. At the interface of water and another medium (such as air), a double electric layer will form because of the polar nature of water molecules. This way negative charges will align outwards while positive charges align inwards \citep{pounder1980}. These charged layers then can be separated by boiling of the water, if the electric layers are sheared more quickly than they can rearrange to maintain the charge balance \citep{james2008}. 

Ionization processes have been studied on extrasolar objects as well. \citet{helling2011b} studied the efficiency of dust collisional ionization similar to triboelectrification in volcano plumes, in dusty clouds of brown dwarfs. They found that turbulence-induced dust-dust collision is the most efficient kinetic process in ionizing the local atmosphere. They showed that this mechanism is the most efficient in the inner cloud, where the dust-to-gas ratio is high because of the quick growth of dust cloud particles. \citet{helling2011b} concluded that dust collisional ionization is efficient enough to lead to electron avalanches producing lightning discharges in brown dwarf dust clouds. \citet{rimmer2013} analysed the effects of cosmic ray ionization in extrasolar atmospheres. Their results showed that both for brown dwarfs and free-floating giant gas planets, the top of the atmosphere becomes partially ionized, reaching levels of a weakly interacting plasma, only due to cosmic ray ionization. In these objects, cosmic rays produce the majority of free electrons in the upper atmosphere \citep{rimmer2013}. \citet{rodriguezbarrera2015} considered thermal ionization of ultra-cool objects. They found that in L-dwarfs the ionization rate due to thermal ionization is high enough to seed lightning processes, and to reach levels, which may lead to the build-up of an ionosphere.

%__________________________________________________________________
%__________________________________________________________________
\section{Charge accumulation and separation} \label{sec:separ}

After charges are generated they have to be separated to create a potential difference to maintain a high electric field for the discharge process to occur. Here I discuss two main processes that can separate charged regions:

\noindent \textit{Convection in water clouds \citep{dwyer2014}:}
In water clouds on Earth, charges primarily accumulate on either soft hail (graupel) particles or small ice crystals. Charge transfer will occur between the two types of particles by collision in the presence of super-cooled (colder than 0 C$^{\circ}$, but not frozen) water droplets, in an environment with temperatures between $-10$ C$^{\circ}$ and $-20$ C$^{\circ}$. Due to the polar nature of the water molecules, an interface is formed on the ice particles, resulting in negative charges lining up outwards, while positive charges face inwards. When the larger graupel particles collide with the smaller ice particles, they break off the outer negative layer of the ice, accumulating that mass from the collision and becoming negatively charged, while the smaller ice particle remains positively charged \citep{saunders1993}. Due to its mass, the heavier, now negatively charged, graupel particle will either sink or remain stationary in the cloud. The positively charged ice crystals, being light, will be carried by the updraft to higher altitudes, hence the charge separation. The above described charge transfer may be reversed in the presence of different ambient temperatures or water content \citep[e.g.][]{saunders1991}. This procedure, is very specific to water clouds on Earth and maybe on Jupiter \citep{yair1995}, where the main charge separation process is convection. 

\noindent \textit{Gravitational settling in dusty environments:} The most accepted charge separation mechanism in volcanic plumes is gravitational settling. Due to different fall-velocities, lower regions in the plume will have larger ash particles while upper regions will contain more aerosols and gas particles \citep{james2008}. According to experiments and observations the net negative charge will be on the solid silicate particles in the lower region, and the net positive charge will be on volcanic gases and aerosols in the upper region \citep[e.g.][]{lane1992}. \citet{cimarelli2014} performed laboratory experiments where they created a volcanic plume with lightning activity. In their experiments when monodisperse coarse beads were used, the charged particles did not separate in two layers but clusterised, some clusters became more positively while others more negatively charged.

In dusty extrasolar atmospheres, gravitational settling has been studied as the main mechanism for separating different-sized cloud particles. Works like \citet{woitke2003,helling2008} modelled cloud formation in quasi-static atmospheres of brown dwarfs, studying processes like grain formation, gravitational settling, and grain evaporation, and found that gravitational settling is efficient in creating cloud layers with different sized particles. \citet{helling2013b} and \citet{helling2013} suggested that gravitational settling is the mechanism for large-scale charge separation in extrasolar mineral clouds.

%__________________________________________________________________
%__________________________________________________________________
\section{Breakdown field and lightning propagation} \label{sec:propag}

\begin{figure}
\begin{center}
\includegraphics[scale=0.6]{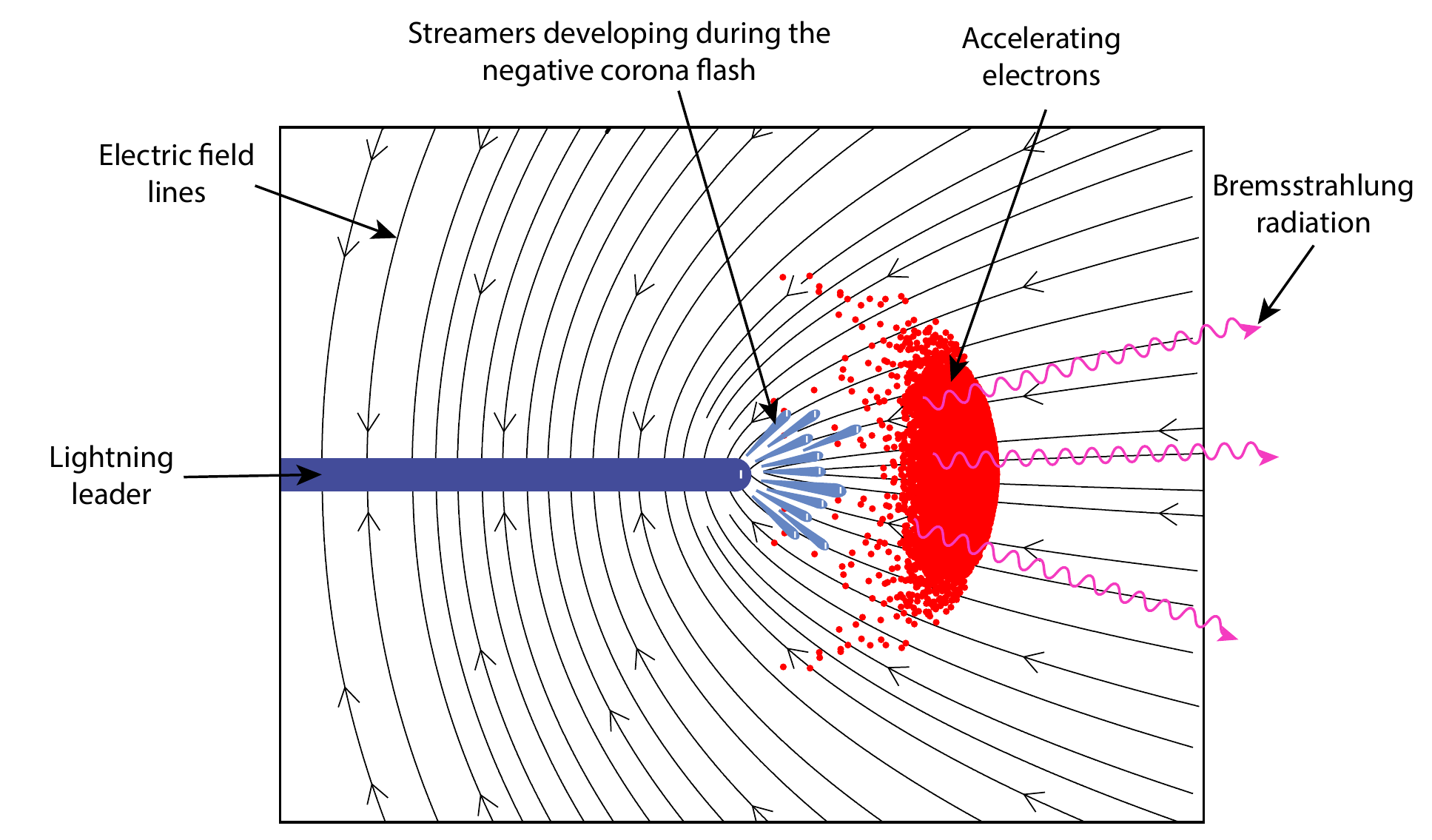}
\end{center}
\caption{Acceleration of runaway electrons in a lightning leader and corona streamer. Reproduced with permission from \citet{celestin2015}.} 
\label{fig:phys_1}
\vspace{0.8cm}
\end{figure}

If the charge separation is efficient enough, a large electric potential difference is built up between the oppositely charged regions. The electric field created this way, will accelerate the free electrons present in the medium, that have been produced by ionization processes (Sect. \ref{sec:ion}). These electrons will collide with other electrons or molecules either creating new free electrons further ionizing the medium, or attaching themselves to ions neutralizing the medium \citep{dwyer2014,helling2013}. For lightning to occur, the electric field has to overcome a breakdown threshold. Above this threshold, the rate with which the electrons will ionize the medium will be larger than the recombination and attachment rate, therefore supplying the process with further free electrons and producing an electron avalanche \citep{dwyer2014,rousseldupre2008}. The avalanche of free electrons will turn into the ionization front that propagates through the air, eventually resulting in the discharge process that is the lightning flash (Fig. \ref{fig:phys_1}). The breakdown field is lower at lower atmospheric pressures at the top of the cloud, than at the bottom of the cloud at higher pressures \citep{dwyer2014}. 
We distinguish between conventional and runaway breakdown when discussing lightning discharges. The main difference between them is the energy of the accelerated free electrons involved. For a conventional breakdown the electrons do not exceed a few eV energy level, therefore, to reach the breakdown threshold, a larger electric field is needed to accelerate these seed electrons \citep[][and references therein]{dwyer2014,rousseldupre2008}. Runaway breakdown involves free electrons, supplied mostly by cosmic rays, with energies in the keV$-$MeV range. During runaway breakdown electrons are accelerated to relativistic energies resulting in an electron beam propagating through the medium \citep[Fig. \ref{fig:phys_1};][]{rousseldupre2008,milikh2010}. The larger initial energy of electrons enables a lower electric field to accelerate the electrons to the breakdown threshold \citep{milikh2010}, therefore, runaway breakdown was suggested to play a big role in lightning initiation, since the threshold to overcome is much lower for that than for conventional breakdown \citep[e.g.][]{rousseldupre2008,helling2013}.

\citet{helling2013} carried out an extensive modelling study of breakdown fields and discharge behaviour in mineral clouds of exoplanetary and brown dwarf atmospheres. They investigated how the cloud properties such as grain size, cloud size, local chemistry, affect the breakdown conditions like the required minimum voltage. The results of \citet{helling2013} suggested that in high-pressure regions small-scale discharges will develop, while large-scale lightning discharges will occur in the higher atmosphere at lower pressures. They found their results of minimum voltage ($10^{-7}-10^7$ V cm$^{-1}$) and critical number of charges ($\sim 10^5-10^{23} e$ cm$^{-3}$) necessary for breakdown to be consistent with experimental results at 1 bar pressure. \citet{helling2013} also noted that the chemical composition of the atmosphere does not significantly influence the breakdown field in the studied extrasolar objects.

\begin{figure}
\begin{center}
\includegraphics[scale=0.9]{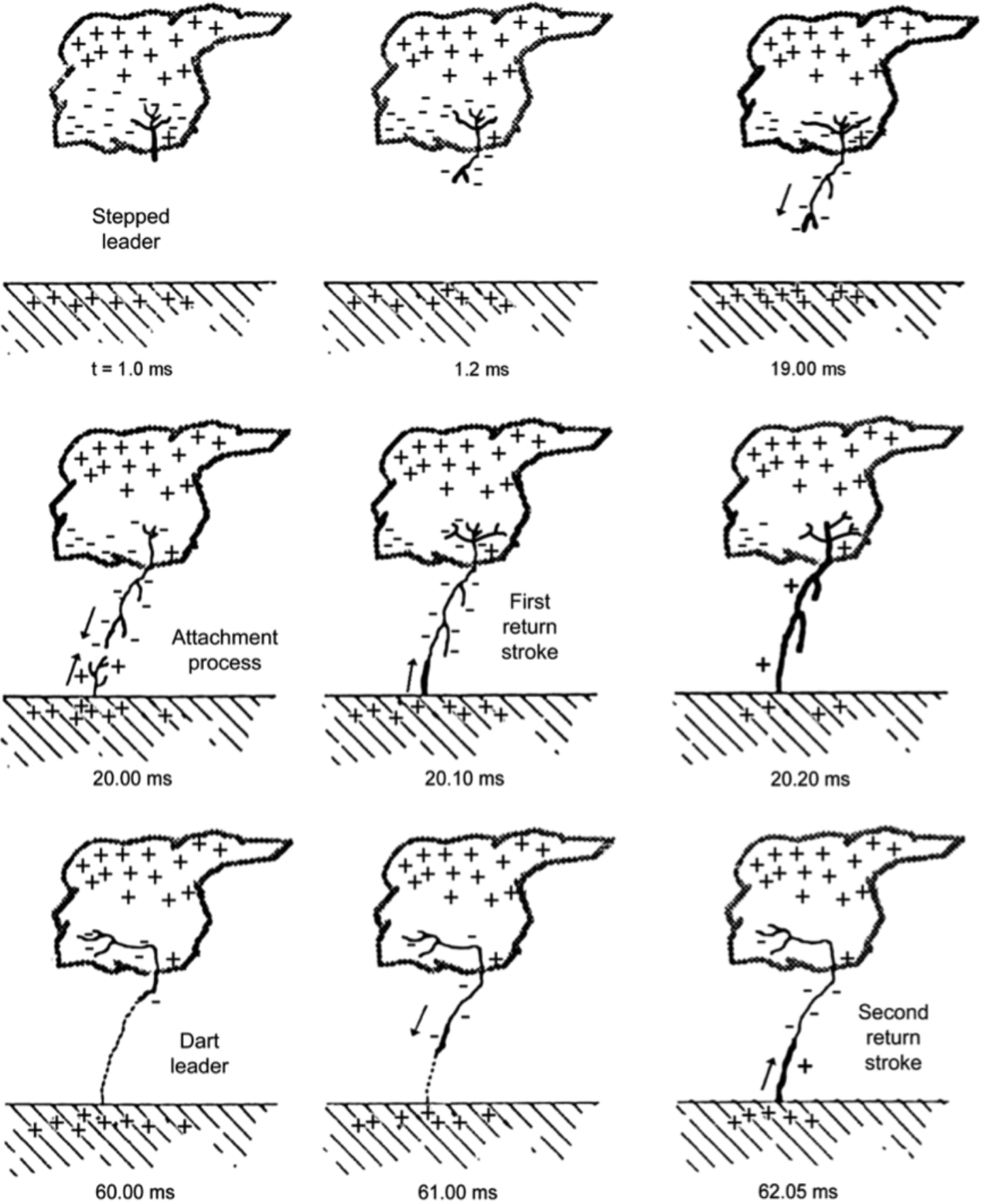}
\end{center}
\caption{A sketch of a negative cloud-to-ground flash, showing the different phases of the lightning discharge. The timescale [ms] starts from the first electrical breakdown processes in the cloud. Note that in the Global Electric Circuit, during fair weather the ground is negative. However, the negatively charged bottom of the thundercloud will produce a "mirror-charge" on the ground, resulting in a more positively charged surface of the Earth as is shown on the figure. Reproduced with permission from \citet{dwyer2014} \citep[original appeared in][]{uman1987}.}
\label{fig:phys_2}
\vspace{0.8cm}
\end{figure}

\subsubsection{Cloud-to-ground discharge}

Once the breakdown threshold is reached and overcome, a lightning discharge develops. In the second part of this section I discuss the propagation and the main parts of a \textit{\textbf{cloud-to-ground (CG)}} lightning discharge, following \citet{dwyer2014}. The steps are illustrated in Fig. \ref{fig:phys_2}:

\begin{itemize}
\item[1.] \textbf{stepped leader}: an electrical discharge propagating in discrete steps and moving the negative charges from the negative charged region to the positive one. During each step, the leader produces optical, radio and $X$-ray emission. Much rarer and less studied are positive "stepped" leaders, which are less luminous from the negative ones and emit less very low frequency radiation. The second type of leader may propagate not just in steps but in a more continuous way as well (Fig. \ref{fig:phys_1}). 
\item[2.] \textbf{corona streamer}: in large numbers, it surrounds the conducting core, which is made of the stepped leaders. It is a low-level, non-thermalized discharge, which makes the leader look much extended on photographs than the actual conducting part is (Fig. \ref{fig:phys_1}).
\item[3.] \textbf{first return stroke}: after the attachment occurs (see below), the negative charges move into the positive region, or in case of a cloud-to-ground discharge, into the Earth. At the meantime, the channel built up by the leaders becomes very luminous as a strong current propagates upwards and down on the branches. This is called the first return stroke, the part of the lightning flash that the human eye usually sees. After the first stroke, the lightning flash may end, in which case it is called a single-stroke flash.
\item[4.] \textbf{dart leader}: if more negative charges are available in the negative region where the original channel started from, a continuously propagating leader, called the dart leader, will move these charges towards the positive region depositing them along the previously built-up channel. It deposits fewer charges in the channel than the stepped leader. It is expected to follow the same channel as was created by the stepped leader and the first return stroke, since that channel is warmer and has a lower air-density resulting in a lower breakdown field \citep{tran2015}.
\item[5.] \textbf{subsequent return stroke}: it occurs after the dart leader attaches to the positive region, in a similar process as the first return stroke occur. However, because the dart leader deposits fewer charges, the subsequent return stroke is much fainter and less energetic than the first return stroke.
\end{itemize}

One of the biggest questions of lightning formation is how the \textbf{"attachment process"} works. As the stepped leader moves from the negative region towards the positive one, surrounded by corona streamers, it builds up the conducting channel between regions \citep{dwyer2014}. As it gets close to the ground (or the positive region), the relatively large number of negative charges on the leader will attract the relatively more positive charges from the other region. When the electric field becomes large enough between the leader and the ground, several upward propagating discharges will initiate from the positively charged region \citep{lu2012}. One of these upward propagating discharges will connect to the downward propagating negative leader through an attachment process \citep{dwyer2014}. Following the attachment, two return stroke currents are initiated, one moving upwards, towards the negative region, and a shorter one moving towards the ground, or positive region \citep{tran2015b}. The place where the attachment occurs will determine the "strike-point" and the primary current channel between the two charged regions. \citet{wang2015} noticed that the height at which the attachment occurred for three natural lightning flashes, was dependent on the peak current of the channel, with larger peak currents resulting in higher initiation points. Understanding the attachment process is especially important in terms of lightning protection, since this process will determine where lightning will strike on the ground \citep{dwyer2014}.

\subsubsection{Intra-cloud discharge}

\begin{figure}
\begin{center}
\includegraphics[scale=1.0]{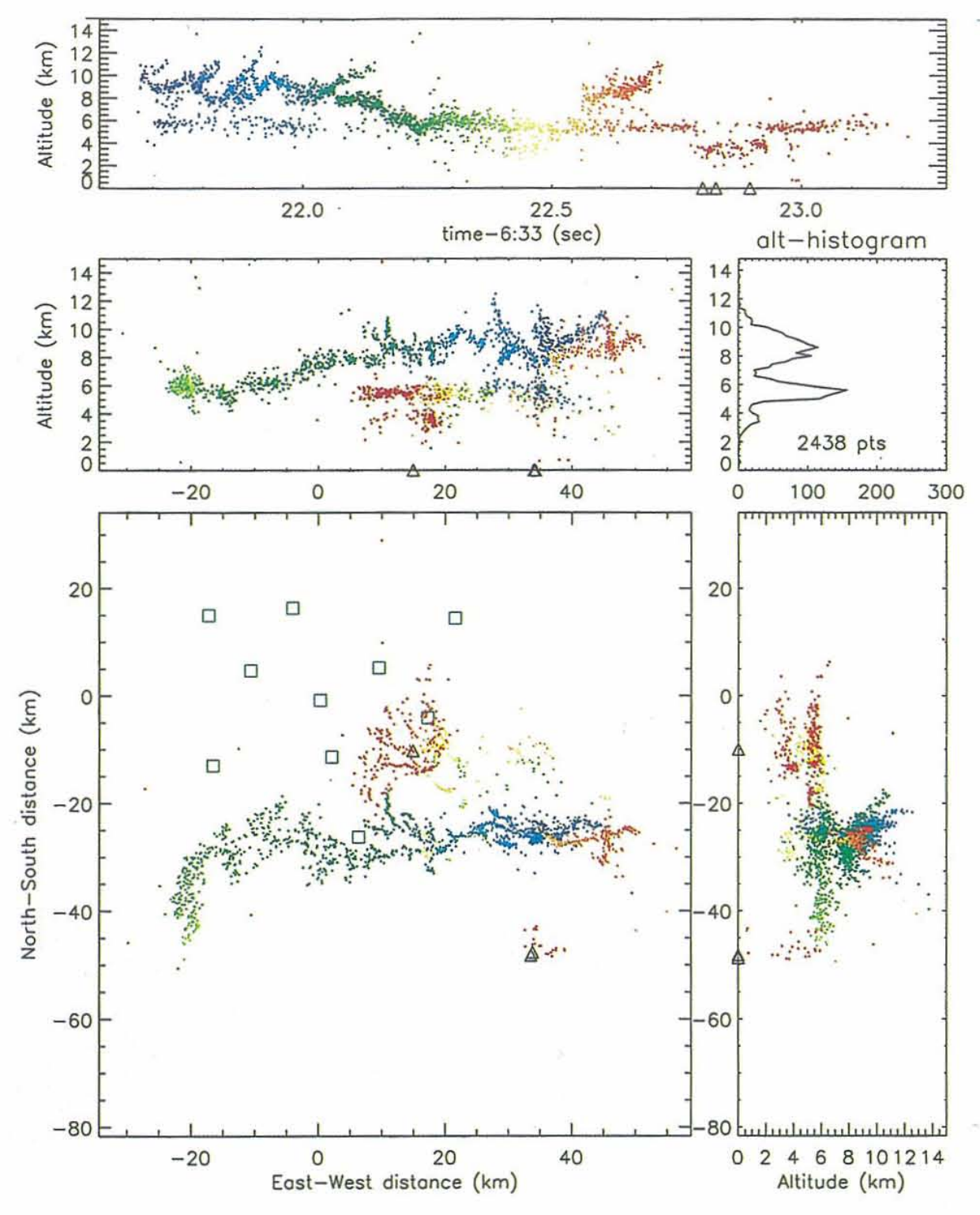}
\end{center}
\caption{A horizontally extensive (75 km in east-west direction) intra-cloud (IC) lightning flash and cloud-to-ground (CG) flashes measured by The New Mexico Tech 3D Lightning Mapping Array in  Oklahoma, June 10-11,1998. Small squares indicate the measurement stations, while the triangles show the locations of CG flashes. The duration of the IC discharges was 1.5 s (top panel). The colour-scale represents the propagation of the discharge in time as shown on the top panel. Reproduced with permission from \citet{krehbiel2000}.}
\label{fig:lma_ic}
\vspace{0.8cm}
\end{figure}

Lightning occurs not just between the ground and a cloud, but inside clouds as well. \textit{\textbf{Intra-cloud (IC) discharges}} form part of the most numerous types of lightning discharges, the cloud flashes \citep{rakov2003}. Here, I discuss the main similarities and differences between the development if an IC and a CG discharged following \citet[section 9]{rakov2003}. In principle, the main phases of an IC discharges are like the ones discussed for the CG discharges: in the initial stage of an IC flash the discharge channel extends with intermittent steps, similar to stepped leaders, from the negatively charged region to the positively charged one, inside a cloud. In the late stage of the process, further negative charges are transported to the origin of the discharge within the negative region from more remote areas of the cloud. This extends the discharge further in both time and space. IC discharges do not tend to host a return stroke-like phase.

The research of IC lightning compared to CG lightning was very scarce compared to CG lightning research. This was mostly because both visual and electrical measurements of IC flashes is much harder than CG flashes. Since the development of Lightning Mapping Arrays (LMA), however, a new gate to IC research has opened. LMA measures the time of arrival of intensive very high radio frequency (VHF) radiation of lightning discharges with the help of GPS at 6 or more stations. Using the measurements, lightning activity is mapped in three spatial dimensions and in time \citep{rison1999}. The detailed measurements make it possible to form an accurate picture of lightning flashes, even if they occur inside a cloud. Figure \ref{fig:lma_ic} shows the structure of an IC discharge mapped by the New Mexico Tech 3D in 1998 \citep{krehbiel2000}. The time evolution (colours as shown on the top panel) and altitude range the discharge spans shows a horizontally very extended event that occurred in 1.5 seconds. LMA observations contribute to the better understanding of IC discharges and the development of discharge models.

%__________________________________________________________________
%__________________________________________________________________
\section{Basic physical properties of a cloud-to-ground discharge}

In this section, I briefly present various physical properties of a lightning flash. Most of these properties are discussed in more details in other chapters, which are mentioned below. Here, I only consider Earth lightning features. Where possible, I use the symbols used in the following chapters for each property. Further references information can be found in, e.g., \citet{volland1984,rakov2003,dwyer2014}.

\begin{itemize}
\item \textit{discharge duration, $\tau$:} According to \citet[][table 1.1]{rakov2003}, a single stroke in a lightning flash usually lasts between 70 and 100 $\mu$s. In case of a multiple-stroke flash, the time between each stroke is several tens of ms, resulting in a total flash duration of a few hundred ms. I discuss and analyse the effects of $\tau$ on the energy and power release of lightning in Chapter \ref{chap:model}.

\item \textit{peak current, $i_0$:} The largest peak currents occur during the return stroke phase of a lightning flash, hence this is the most energetic, most luminous part of the flash. $i_0$ of a return stroke ranges between a few 100 A and a few hundred kA \citep{rakov2003,dwyer2014}. On average, a return stroke carries 30 kA current \citep{farrell1999}. The peak current determines the energy and power content of a lightning discharge, as described in Chapter \ref{chap:model}.

\item \textit{total charge, $Q$:} The total charge transfer during a lightning flash on Earth is on average $20-30$ C \citep{rakov2003,dwyer2014}. The amount of charge moved in the lightning channel will determine the current, and the resulting energy release of the discharge, as described in Chapter \ref{chap:model}.

\item \textit{energy (total, optical, radio, acoustic), $W$:} As we will see in Chapter \ref{chap:ligsig}, lightning radiates in the whole electromagnetic spectrum. However, this radiation consumes only a small part of the total energy dissipated from a discharge ($W_d$). On average, lightning produces $10^9-10^{10}$ J energy, of which $\sim 1$\% is radiated into the optical \citep[$W_{\rm opt}$;][]{borucki1987}, and $\sim 1$\% is radiated into the radio wavelengths \citep[$W_{\rm rad}$;][]{volland1984}. The thunder of lightning is produced by the shockwaves that are the consequence of the gas expanding and the clouds contracting due to the high temperature changes during a lightning discharge \citep{rakov2003}. \citet[][sect. 11.2.4]{rakov2003} suggested that the acoustic efficiency of lightning can be between 2 and 20\%. The rest of lightning energy manifests as thermal energy and contributes to the chemical changes in the atmosphere. (The radio energy released from lightning is the main topic of Chapter \ref{chap:model}.)

\item \textit{flash rate and flash density, $R_{\rm fl}, \rho_{\rm fl}$:} Lightning flash rate describes lightning distribution in time, while the flash density includes information on spatial distribution as well. Hence, the units are the following: $[R_{\rm fl}] = $flashes h$^{-1}$; $[\rho_{\rm fl}] = $ flashes km$^{-2}$ h$^{-1}$. Naturally, the units can be in the required quantities. On average, on Earth, there are $\sim 2 \times 10^{-4}$ flashes occurring every hour in every km$^{2}$. Lightning occurs more frequently over continents than over oceans ($\sim$ 2 orders of magnitude more), and more over lower latitude regions than higher latitude regions. (A more detailed discussion of these values and their comparison to the literature is found in Chapter \ref{chap:stat}.)

\item \textit{channel tortuosity and branches:} A lightning channel is not a strait line. Due to the inhomogeneities in the charge distribution, various branches and channel segments are built up during the leader process. The segmentation of the channel is called tortuosity. The length of the various segments ranges between a few cm and several m \citep[][p. 161-162, and references therein]{rakov2003}. \citet{levine1978} found in simulations that the tortuosity introduces a fine structure in the time domain radiation field waveform, and it increases the high frequency energy content of lightning. \citet{levine1995} confirmed the effects of tortuosity on the fine structure of the measured electric field based on natural and triggered lightning experiments.

\item \textit{multiplicity of flashes:} A lightning flash may contain more than one stroke. The number of strokes it contains is called flash multiplicity. According to \citet[][table 1.1, and references therein]{rakov2003}, about 15-20\% of lightning flashes contain a single stroke, and on average they contain 3 to 5 strokes. (The multiplicity of flashes is considered in Chapters \ref{chap:stat} and \ref{chap:model}.)
\end{itemize}

There are several other properties of lightning discharges (e.g. peak frequency, $f_0$; current velocity, $\v v$), most of which depend on the above discussed ones. In Chapter \ref{chap:stat} I will study the occurrence of lightning in the Solar System (flash rates and densities), while in Chapter \ref{chap:model}, I will further discuss the properties of the electric field, and frequency and power spectra of a return stroke.

%__________________________________________________________________
%__________________________________________________________________
\section{Types of lightning discharges and lightning hosting environments} \label{sec:envir}

\begin{figure}
\begin{center}
\includegraphics[scale=0.4]{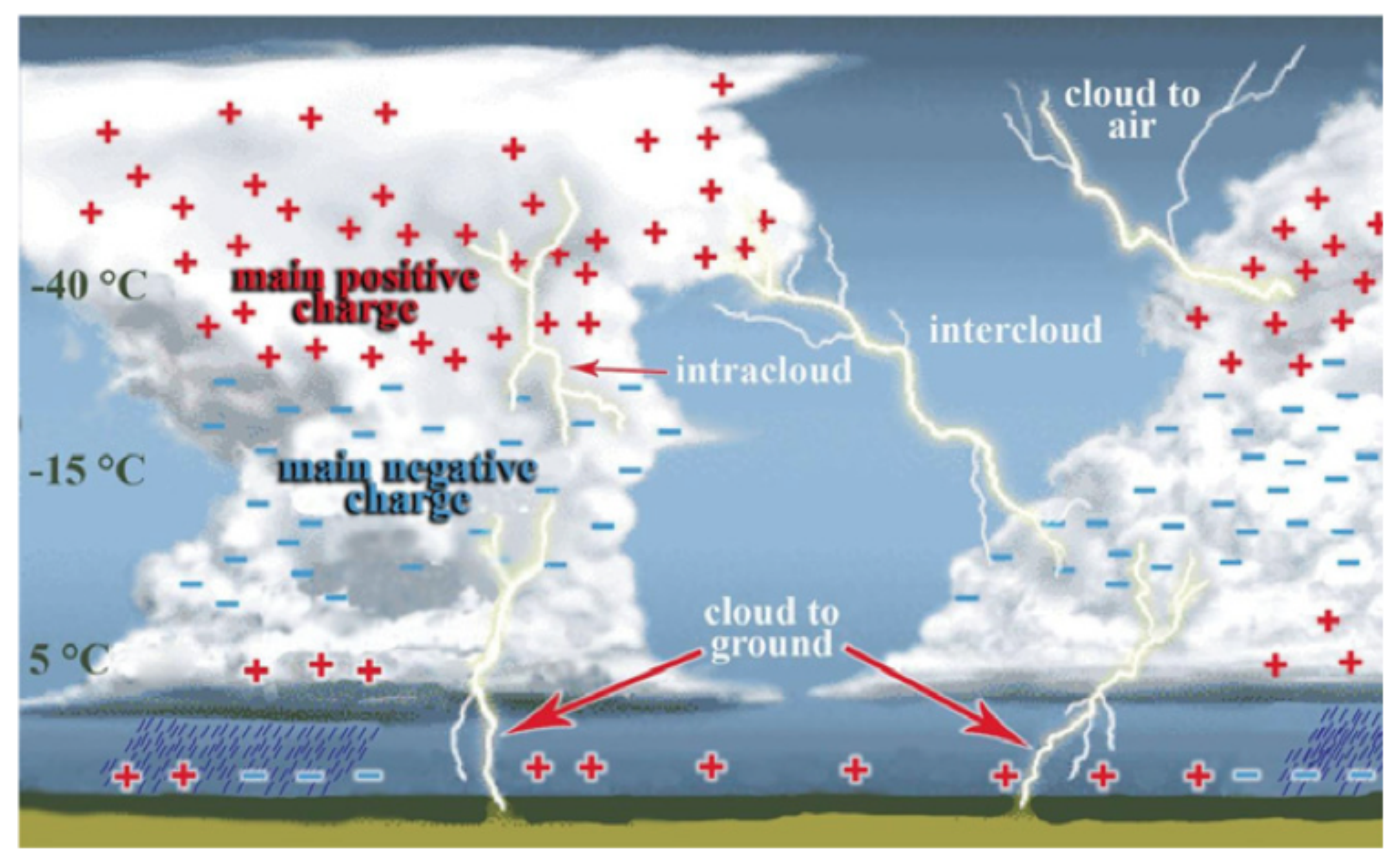}
\end{center}
\caption[Simple charge structure of thunderclouds, and types of lightning discharges based on the location of the charged regions involved with the process. Reproduced with permission from \citet{dwyer2014} who adopted the figure from Encyclopedia Britannica.]{Simple charge structure of thunderclouds, and types of lightning discharges based on the location of the charged regions involved with the process. Reproduced with permission from \citet{dwyer2014} who adopted the figure from Encyclopedia Britannica\protect\footnotemark.}
\label{fig:phys_3}
\vspace{0.8cm}
\end{figure}

It is common to use the term \textit{lightning flash} when talking about the lightning discharge process. A \textit{lightning stroke} represents only part of the lightning flash as defined in Sect. \ref{sec:propag}. Though it is very Earth specific, to complement the nomenclature, I also mention here that when a lightning flash hits an object on the ground or in the atmosphere, it is called a \textit{lightning strike} \citep{rakov2003}. I have already discussed CG and IC discharges, however in terms of where are the two charged regions in between which the lightning flash occurs, we distinguish between further types (Fig. \ref{fig:phys_3}): cloud-to-ground (CG), intra-cloud (IC), cloud-to-cloud (CC, or inter-cloud), and cloud-to-air (CA) discharge. CG lightning is the most well studied type of lightning flashes, therefore, in the literature they distinguish between further types based on the polarity of charges transported in the stepped leader and streamer (-CG, +CG), and whether it is an upward, or a downward propagating flash \citep{rakov2003}. 

\footnotetext{\url{https://www.britannica.com/science/lightning-meteorology}}

Other, less studied and rare type of lightning discharges, is called ball lightning. Ball lightning is usually seen after CG lightning strike and nearby lightning activity; it occurs close to the ground as a sphere or ellipsoid with diameter between 1 and 100 cm \citep{cen2014}. It usually lasts for 1 $-$ 10 s, and appears in various colours \citep{rakov2003}. Models explaining the origin and physics of ball lightning can be divided into two main categories: the proposed source of energy is either internal \citep[e.g.][]{abrahamson2000} or external \citep[e.g.][]{lowke2012}. \citet{cen2014} observed a ball lightning that occurred after a CG discharge, with slit-less spectrograph, and found that its spectrum contained lines produced by material present in the soil. They concluded that this example supports the theory of \citet{abrahamson2000} of internal energy source, however they emphasize that this may be only an example of different types of ball lightning produced by different mechanisms. Based on witness reports and laboratory experiments, the total energy of ball lightning has been found not to exceed the total energy of a lightning flash, in fact, in many occasions, it is way below that \citep[$10^{-2}-10^9$ J, e.g.][]{barry1980,paiva2010}. \citep[For further, less known types of lightning discharges see][section 20, p. 656.]{rakov2003}

Dedicated observational campaigns have revealed that lightning occurs in very diverse environments in the Solar System. On Earth, lightning is frequently produced in thunderclouds that are made of water and ice particles. Thunderstorms also occur in clouds of ice and snow particles, producing "winter lightning" \citep{brook1982, wu2013}. Modern volcano lightning observations showed that lightning occurs in volcano plumes, which are primarily composed of mineral dust particles. Lightning has been observed mostly after explosive eruptions, which is the result of volatiles stuck in the magma chamber due to the high viscosity of the magma. As the volatiles cannot escape or expand under the volcano, the internal pressure becomes so high that the magma will explosively fragment \citep{james2008}. Volcanic lightning has been detected by standard detection systems operating in the radio wavelengths, which indicate that it has similar radio signatures to thundercloud lightning. Optical observations of volcanic lightning have also been conducted \citep{mather2006}. Sand storms, dust storms and dust devils can also be electrified \citep[e.g.][]{yair2016} and may host lightning activity, where charging is the consequence of triboelectrification \citep[Sect. \ref{sec:ion};][and references therein]{harrison2016}. 

On Jupiter and Saturn, lightning is produced in dense, vertically extended, convective clouds \citep{cook1979, dyudina2001, dyudina2004, dyudina2013, read2011}. Lightning on Venus may appear as IC or CC discharge due to the high atmospheric pressure, which would not allow CG discharges to occur unless the electric field becomes extremely high \citep{yair2009}. Lightning on Mars was suggested to occur in electrified dust storms \citep[e.g.][]{farrell2006, yair2012}. I discuss planetary lightning in more details in Chapter \ref{chap:liginout}. The large variety of lightning hosting environments and extrasolar bodies (Chapter \ref{chap:intro}) suggests that lightning occurs outside the Solar System.

%% file: chapters/2b_signatures.tex
\chapter{Signatures of lightning} \label{chap:ligsig}

\begin{singlespace}
\begin{table*}
\scriptsize
\renewcommand{\arraystretch}{1.7}
\resizebox{\columnwidth}{!}{
\begin{threeparttable}
\caption{Lightning signatures observed in the Solar System. The right column lists instruments that are potentially useful, based on their wavelength of operation, for observing lightning on exoplanets and brown dwarfs. Note that instrumental sensitivity is not taken into account here. \citep[Adopted and updated from][]{bailey2014}.}
  \begin{tabular}{llllll}
	\hline
	Process & Signature & Wavelength & Planet & References & Instrument \\ %\tnote{(*)} \\
	\hline

	\vtop{\hbox{\strut Direct lightning}\hbox{\strut emission}} & \vtop{\hbox{\strut $\gamma$ - ray}\hbox{\strut (TGF)}} & 20 eV - 40 MeV & Earth & \vtop{\hbox{\strut \citet{lu2011, yair2012}}\hbox{\strut \citet{marisaldi2010}}} & \vtop{\hbox{\strut Fermi GBM, \citet{meegan2009}}\hbox{\strut AGILE, \citet{tavani2006}}}\\  \cdashline{2-6} 
	  
	  & X - ray & $30-250$ keV & Earth & \vtop{\hbox{\strut \citet{dwyer2004}}\hbox{\strut \citet{dwyer2012}}}  &  \vtop{\hbox{\strut AGILE}\hbox{\strut Astrosat - SXT\tnote{(1)}}\hbox{\strut Astrosat - LAXPC\tnote{(2)}}} \\ \cdashline{2-6}
	 
	  & He & 588 nm & Jupiter & \vtop{\hbox{\strut \citet{borucki1996}}\hbox{\strut \citet{aplin2013}}} & \vtop{\hbox{\strut VLT - X$-$SHOOTER,}\hbox{\strut \citet{vernet2011}}\hbox{\strut VLT - VIMOS,  \citet{lefevre2003}}} \\ \cdashline{2-6} 

	  & \vtop{\hbox{\strut NUV to NIR}\hbox{\strut many lines of}\hbox{\strut N$_2$, N(II),}\hbox{\strut O(I), O(II)}\hbox{\strut}\hbox{\strut}} & 310-980 nm & Earth & \citet{wallace1964} & \multirow{2}{*}{\vtop{\hbox{\strut Astrosat - UVIT, \citet{kumar2012}}\hbox{\strut Swift - UVOT, \citet{roming2005}}\hbox{\strut VLT - X$-$SHOOTER}\hbox{\strut VLT - VIMOS}\hbox{\strut HARPS, \citet{mayor2003}}\hbox{\strut HST - NICMOS, \citet{viana2009}}\hbox{\strut IRTF - TEXES, \citet{lacy2002}}\hbox{\strut Spitzer IRS, \citet{houck2004}}\hbox{\strut JWST - NIRCam and NIRSpec\tnote{(4)},}\hbox{\strut \citet{gardner2006}}}} \\ \cdashline{2-5}

	  & NIR emission & 0.35-0.85 $\mu$m & Jupiter & \vtop{\hbox{\strut \citet{baines2007}}\hbox{\strut}\hbox{\strut}} & \\ \cdashline{2-6}
	  
	  & \vtop{\hbox{\strut Schumann-}\hbox{\strut resonance (SR)}} & few Hz & Earth\tnote{(5)} & \citet{simoes2012} & \vtop{\hbox{\strut Not observable}\hbox{\strut outside the ionosphere}} \\ \cdashline{2-6}

	  & whistlers & tens of Hz - kHz & \vtop{\hbox{\strut Earth}\hbox{\strut Jupiter}\hbox{\strut Saturn}\hbox{\strut Neptune}} & \vtop{\hbox{\strut \citet{desch2002}}\hbox{\strut \citet{yair2008, yair2012}}\hbox{\strut \citet{akalin2006}}\hbox{\strut \citet{fischer2008}}\hbox{\strut \citet{gurnett1990}}} & \vtop{\hbox{\strut LOFAR, \citet{vanhaarlem2013}}\hbox{\strut UTR 2, \citet{braude1978}}\hbox{\strut LWA, \citet{kassim2005}}} \\ \cdashline{2-6}
	  
	  & sferics & 1 kHz - 100 MHz & \vtop{\hbox{\strut Earth}\hbox{\strut Jupiter}\hbox{\strut Saturn}\hbox{\strut Uranus}} & \vtop{\hbox{\strut \citet{desch2002}}\hbox{\strut \citet{yair2008}}\hbox{\strut \citet{fischer2008}}\hbox{\strut \citet{zarka1986}}} & \vtop{\hbox{\strut LOFAR}\hbox{\strut UTR 2}\hbox{\strut LWA}} \\
	  
	  \hline

	\vtop{\hbox{\strut Effect on}\hbox{\strut local}\hbox{\strut chemistry}} & NO$_x$ & \vtop{\hbox{\strut 439 nm (NO$_2$)}\hbox{\strut 445 nm (NO$_2$)}\hbox{\strut 5.3 $\mu$m (NO)}} & \vtop{\hbox{\strut Earth}\hbox{\strut Venus}} & \vtop{\hbox{\strut \citet{noxon1976}}\hbox{\strut \citet{lorenz2008}}\hbox{\strut \citet{krasnopolsky2006}}\hbox{\strut}} & \multirow{2}{*}{\vtop{\hbox{\strut HST - STIS,}\hbox{\strut \citet{hernandez2012}}\hbox{\strut VLT - X$-$SHOOTER}\hbox{\strut VLT - VIMOS}\hbox{\strut HARPS}\hbox{\strut HST - NICMOS}\hbox{\strut IRTF - TEXES}\hbox{\strut Spitzer IRS}\hbox{\strut JWST instruments\tnote{(4)}}}} \\ \cdashline{2-5}
	
 	  & O$_3$ & \vtop{\hbox{\strut $200-350$ nm}\hbox{\strut $420-830$ nm}\hbox{\strut $9.6$ $\mu$m}\hbox{\strut $14.3$ $\mu$m}} & Earth & \vtop{\hbox{\strut \citet{zhang2003}}\hbox{\strut \citet{lorenz2008}}\hbox{\strut \citet{tessenyi2013}}\hbox{\strut \citet{ehrenreich2006}}} & \\ \cdashline{2-6}	  
	 
	 & HCN & \vtop{\hbox{\strut 2.97525 $\mu$m}\hbox{\strut 3.00155 $\mu$m}} & Jupiter & \vtop{\hbox{\strut \citet{barnun1975}}\hbox{\strut \citet{barnun1985}}} & \multirow{3}{*}{\vtop{\hbox{\strut VLT - CRIRES,}\hbox{\strut \citet{kaeufl2004}}\hbox{\strut Keck - NIRSPEC,}\hbox{\strut \citet{mclean1998}}\hbox{\strut JWST - NIRCam, NIRISS, NIRSpec\tnote{(4)}}}} \\ \cdashline{2-5}
	  
	 & C$_2$H$_2$ & \vtop{\hbox{\strut 2.998 $\mu$m}\hbox{\strut 3.0137 $\mu$m}} & Jupiter & \vtop{\hbox{\strut \citet{barnun1975}}\hbox{\strut \citet{barnun1985}}} & \\ 
	\hline 
	  
	\multirow{3}{*}{\vtop{\hbox{\strut Emission caused}\hbox{\strut by secondary}\hbox{\strut events}\hbox{\strut (e.g. sprites)\tnote{(3)}}}} & 1PN$_2$ & $609-753$ nm & Earth & \citet{pasko2007} & \multirow{3}{*}{\vtop{\hbox{\strut HST - STIS}\hbox{\strut VLT - X$-$SHOOTER}\hbox{\strut VLT - VIMOS}\hbox{\strut HARPS}}} \\ 
	  
	  & 1NN$_2^+$ & $391.4$ nm & & & \\ 
	  
	  & 2PN$_2$ & $337$ nm & & & \\ \cdashline{2-6}
	  
	  & LBH N$_2$ & $150-280$ nm & Earth & \citet{pasko2007} & \vtop{\hbox{\strut HST - COS, \citet{green2012} }\hbox{\strut HST - STIS}} \\
	\hline
\label{tab:gab}
\end{tabular}
  \begin{tablenotes}
	\item[(1)] http://astrosat.iucaa.in/?q=node/14       
	\item[(2)] http://astrosat.iucaa.in/?q=node/12	
	\item[(3)] 1PN$_2$ is the first, 2PN$_2$ is the second positive,  LBH N$_2$ is the Lyman-Birge-Hopfield N$_2$ band system. 1NN$_2^+$ is the first negative band system of N$_2^+$. 
	\item[(4)] For further information on the James Webb Space Telescope and its instruments see: http://www.stsci.edu/jwst/instruments/
	\item[(5)] SR have only been observed on Earth, but in principle can be produced in any object with an ionosphere.
  \end{tablenotes}
\end{threeparttable}
}
\end{table*}
\end{singlespace}

Lightning signatures span the whole electromagnetic spectrum, from extremely low frequency (ELF) radio emission to high energetic X-rays. Table \ref{tab:gab} groups the main signatures of lightning into three categories: (i) emission coming from lightning directly, (ii) the effects of lightning on the local chemistry, and (iii) emission caused by secondary events, such as transient luminous events. Some of these signatures, like line emission and absorption in the spectrum, depend on the composition of the local atmosphere. Others, like e.g. radio or $X$-ray emission are independent of the chemistry. In the following sections, I discuss the three categories individually, based on Earth lightning observations. I also include detections of exoplanets and brown dwarfs in the appropriate wavelengths, if they exist, though the origin of such emission might not be lightning in the case of these objects.

%__________________________________________________________________
%__________________________________________________________________
\section{Direct lightning emission}

The photons emitted directly from the lightning channel have a broadband spectrum. The emission can be the result of either accelerated electrons (resulting in radio, X-ray or $\gamma$-ray emission) or the excitation of atomic states \citep[e.g. IR and optical emission;][]{bailey2014,fullekrug2013}. In this section I summarize signatures emitted by a lightning channel, from short (high-energy) to long (radio) wavelengths.

%__________________________________________________________________
\subsection{High-energy emission from lightning} \label{sec:highen}

In theory, we call an $X$-ray any energetic photon that is generated by an electron, and $\gamma$-ray an energetic photon that is generated by any other mechanism, e.g. by nuclear processes \citep{dwyer2012}. The most important mechanism producing high energy photons from lightning flashes is the Bremsstrahlung interaction of electrons with the air molecules \citep{dwyer2012}, which means, technically almost all high energy photons produced by lightning are $X$-rays. Another way to distinguish between $\gamma$-rays and $X$-rays is based on their energies: photons with energies greater than 1 MeV are called $\gamma$-rays, and below that they are called $X$-rays, regardless how they were formed \citep{dwyer2012}. Despite these conventions, the categorization of high-energy radiation from lightning is usually based on historical views and detection methods. For example, TGFs are called $\gamma$-rays because they were originally detected by a $\gamma$-ray detector, even though their energies may be lower than 1 MeV and they are produced by Bremsstrahlung \citep{dwyer2012}. Therefore, in this thesis, I discuss the two types of emission based on the referred literature.

\subsubsection{Terrestrial Gamma-Ray flashes}

\begin{figure}
\begin{center}
\includegraphics[scale=0.7]{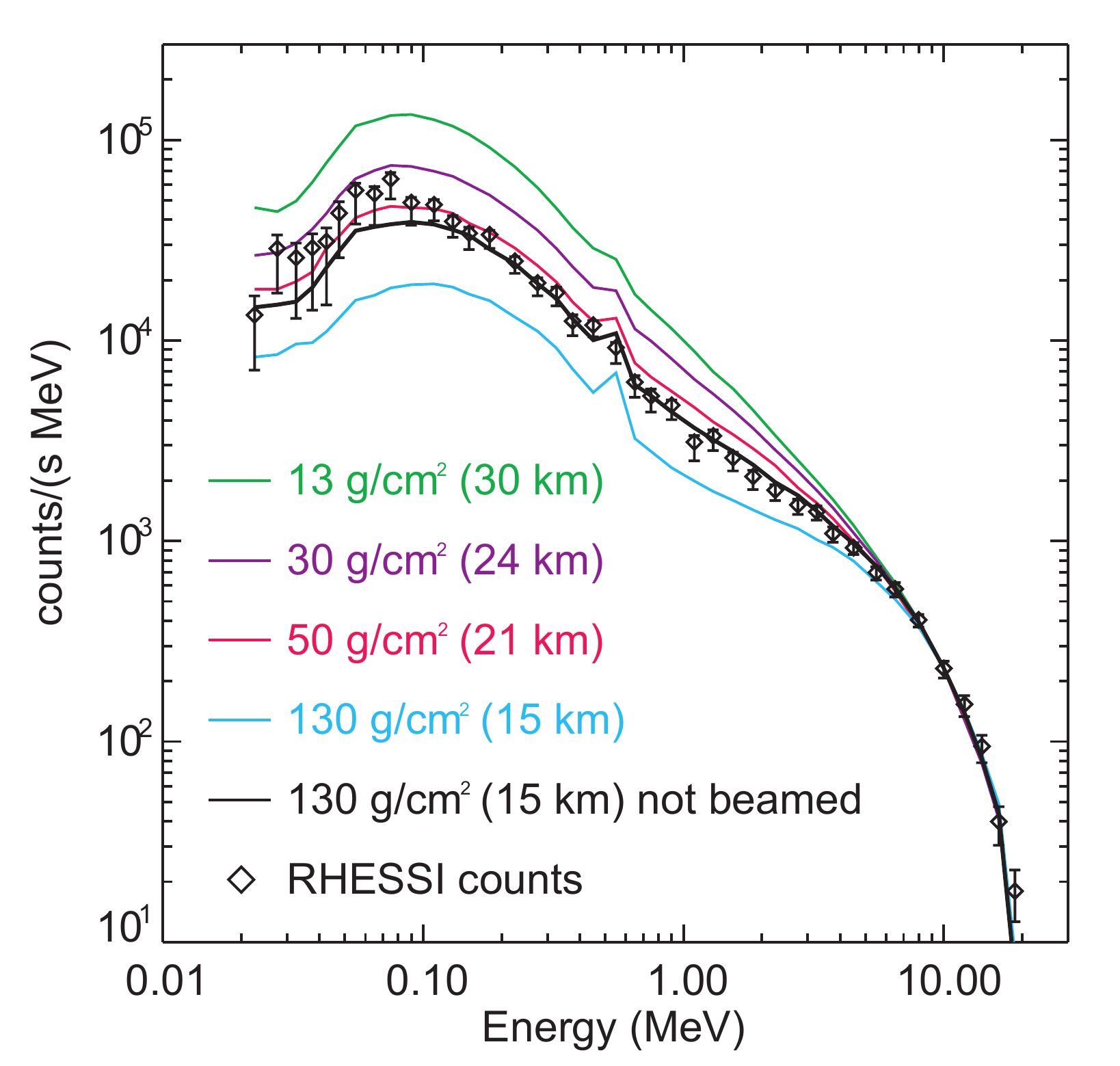} 
\end{center}
\caption{Summed spectrum of 289 TGFs measured by the RHESSI satellite (rhombus symbol) compared to relativistic runaway electron model spectra at different atmospheric depths (in g cm$^{-2}$ corresponding to altitudes in km). The slope of the TGF spectrum between $\sim 0.5$ and 4 MeV indicates the source altitude of the event, which for these TGFs was determined to be between 15 and 21 km. Reproduced with permission from \citet{dwyer2005b}. Courtesy for the original figure to Joseph Dwyer.}
\label{fig:gammar}
\vspace{0.8cm}
\end{figure}

Terrestrial Gamma-Ray flashes (TGFs) are short ($<$1 ms) bursts of $\gamma$-ray photons with a broadband spectrum between 20 kev and 40 MeV \citep[e.g.][]{lu2011, yair2012}. They are produced by relativistic runaway electrons with energies of $\sim 20-40$ MeV (via Bremsstrahlung), as they are accelerated in a strong electric field \citep{lu2011}. Early models attributed TGFs to cloud-to-ground discharges \citep{lehtinen1996}, however the altitudes of TGF-associated lightning flashes (10 to 17 km) indicate that the runaway electrons necessary for TGF events are formed inside the thundercloud attributing TGFs to intra-cloud discharges \citep{stanley2006, lu2011}. \citet{dwyer2005b} ran a Monte Carlo simulation to find the origin of TGFs detected by the RHESSI (Reuven Ramaty High Energy Solar Spectroscopic Imager) satellite and BATSE (Burst and Transient Source Experiment), and noted that while RHESSI observations are consistent with a source at 15$-$21 km altitude, the BATSE TGFs are most probably produced higher in the atmosphere around 30 km, suggesting two different sources for TGFs (Fig. \ref{fig:gammar}). The Monte Carlo model of the air breakdown included all relevant physical processes of electron interactions with the air \citep[e.g. ionization, atomic excitation;][]{dwyer2005b}. The simulation produced TGF spectra for different electric field strength, source altitude and source geometry, by propagating electrons in a pre-defined electric field ($E$). The simulation was terminated when the particles exited the avalanche region \citep[where $E \ne 0$;][]{dwyer2005b}. TGFs have been associated with runaway electron production during the leader or streamer phase of lightning discharges \citep{dwyer2008}. The total energy released during a TGF event is $\sim$ 1-10 kJ with a mean energy at $\sim$ 2 MeV \citep{lu2011, yair2012}.

TGFs have been only observed on Earth to date. They were first discovered by \citet{fishman1994} via the BATSE detector on board NASA's Compton Gamma Ray Observatory. Since then, detections were made, amongst others, by the RHESSI satellite \citep{smith2005}, the Fermi GBM \citep[Gamma-Ray Burst Monitor; e.g.][]{briggs2010}, and the AGILE \citep[Astrorivelatore Gamma a Immagini Leggero;][]{marisaldi2010, marisaldi2015} satellite. 

\citet{inan1996} associated TGFs with lightning events for the first time, when they found a direct relation between lightning generated radio emission and $\gamma$-ray flashes. \citet{lu2011} investigated the direct connection between TGFs and two, previously identified, lightning signals, i.e. the very low frequency (3-30 kHz) impulse from fast discharges \citep[$\le 100  \mu s$;][]{stanley2006}, and the ultra low frequency (ULF, 300-3000 Hz) pulse from slow discharges \citep{cummer2005, lu2010}. They realised that all TGF-associated radio emission occurs in the presence of an ULF pulse. \citet{smith2011} analysed the data produced by the ADELE $\gamma$-ray detector (Airborne Detector for Energetic Lightning Emissions) and estimated the TGF-to-lightning flash ratio to be of the order of $10^{-2}$ to $10^{-3}$ and concluded that TGFs are not a primary triggering mechanism for lightning. \citet{briggs2011} discovered strong positron annihilations lines in TGF spectra observed by Fermi GBM. Their results indicate that a considerable fraction of TGFs is made of positrons alongside with electrons. \citet{ostgaard2015} studied the data of the RHESSI satellite and reported the detection of a new set of weak TGFs. Though the new population did not increase the TGF/flash ratio significantly \citep{ostgaard2015}, they concluded that this detection suggest the existence of a population of TGFs not detected before because their emission is lower than the detection threshold of current instruments. \citet{smith2016} examined data from the RHESSI satellite focusing on the cumulative detection of $\gamma$-rays rather then individual events as \citet{ostgaard2015} did, and noticed a very low TGF count per lightning flash from the stacked TGF detection throughout a 9-year period. 

A new ESA project, ASIM (Atmosphere-Space Interaction Monitor) will be placed on the International Space Station (ISS) in 2018 to study TGFs and TLEs (Transient Luminous Events) and their relation to thunderstorms\footnote{http://asim.dk/}. Scheduled to launch in 2018\footnote{https://taranis.cnes.fr/en/TARANIS/index.htm}, TARANIS (Tool for the Analysis of RAdiations from lightNIng and Sprites) is a microsatellite dedicated to the study of, amongst others, characterization of TGFs and TLEs as well as the lightning event producing these phenomena \citep{blanc2007}.

$\gamma$-rays have been observed from the Moon by the \textit{Compton Gamma-Ray Observatory} \citep{thompson1997} and Fermi GBM \citep{giglietto2009,abdo2012}. The emission was attributed to the interaction of galactic cosmic rays with the surface of the Moon \citep{giglietto2009}. No $\gamma$-rays have been detected from Solar System objects other than the Earth, the Moon and the Sun \citep{dermer2013}, neither from exoplanets nor from brown dwarfs.

\subsubsection{$X$-rays}

\begin{figure}
\begin{center}
\includegraphics[scale=0.7]{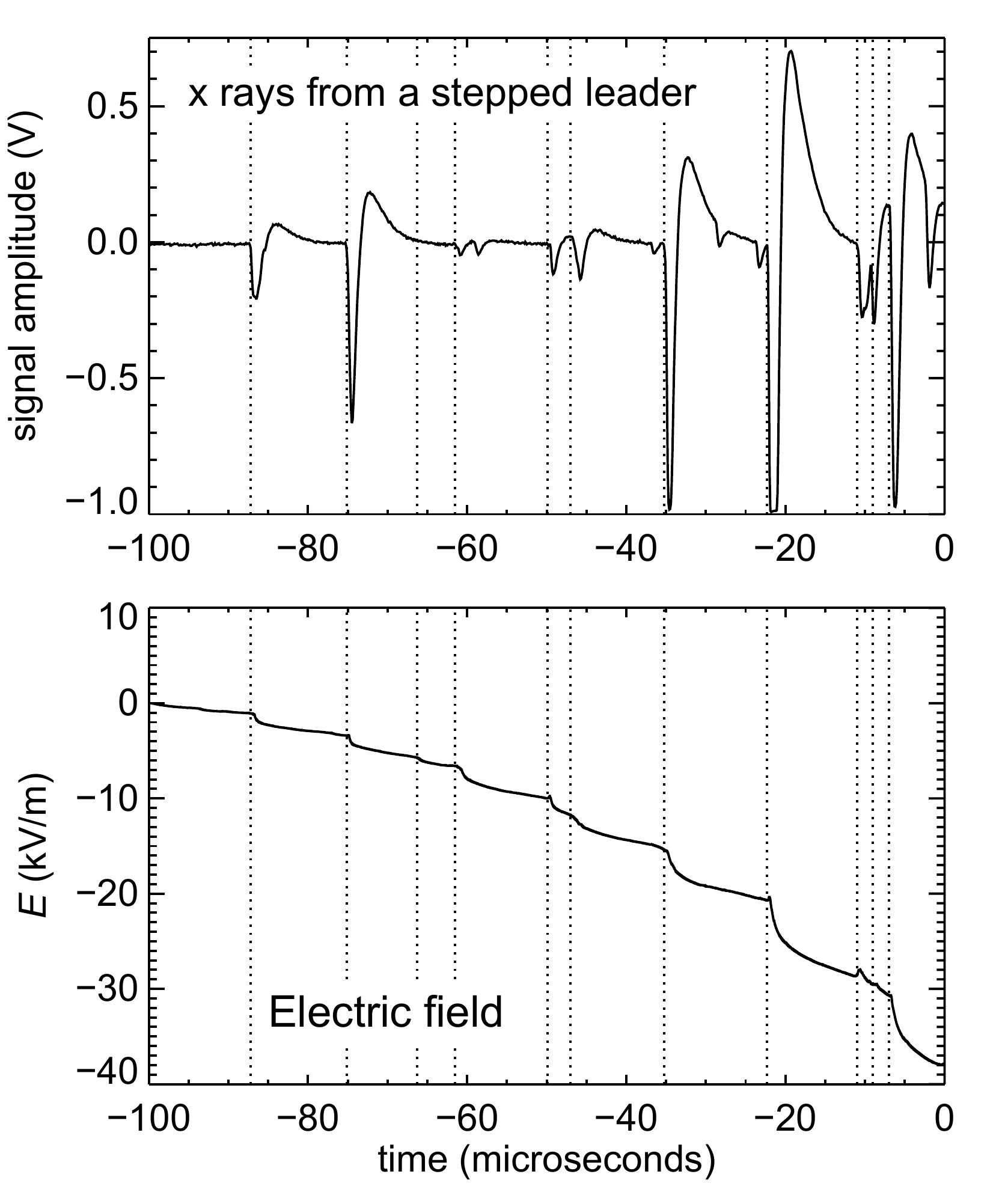} 
\end{center}
\caption{\textbf{Top:} X-ray emission from lightning stepped leader. \textbf{Bottom:} measured electric field waveform for a cloud-to-ground lightning flash. Time 0 corresponds to the beginning of the return stroke. Vertical dotted lines indicate the step formation times, which clearly show that the X-ray pulses are associated with the leader steps. Reproduced with permission from \citet{dwyer2005}. Courtesy for the original figure to Joseph Dwyer.}
\label{fig:xrays}
\vspace{0.8cm}
\end{figure}

The fact that lightning produces $X$-rays was not widely accepted before the 2000s, though \citet{wilson1925} predicted their existence suggesting that $X$-rays and $\gamma$-rays are produced when accelerated electrons interact with the nuclei of atoms in a discharge process. However, since 2001 the evidence for the existence of such emission has increased. \citet{moore2001} conducted experiments in New Mexico and detected energetic photons just before three lightning strikes. They suggested that the particles came from the stepped leader phase of lightning discharges (see Chapter \ref{chap:ligform}, Sect. \ref{sec:propag}), however they also concluded that it was not clear whether the photons were emitted by the stepped leader or electrons accelerated by cosmic rays. Later, rocket-triggered\footnote{Very similar in properties to natural subsequent lightning strokes \citep{dwyer2014}.} lightning tests confirmed the existence of high-energy $X$-rays emitted by lightning discharges \citep{dwyer2003,dwyer2004}.

\citet{dwyer2005} carried out $X$-ray and electric field measurements from natural lightning strikes and learnt that $X$-rays are emitted during the stepped-leader phase of a lightning flash (Fig. \ref{fig:xrays}), $\sim 1$ ms before the return stroke, and that they occurred in discrete, short ($< 1$ $\mu$s) bursts with energies of a few hundred keV. \citet{moss2006} developed a Monte Carlo model to study the acceleration of low energy electrons emitted from lightning streamers to high energies in a strong electric field, and found that part of these electrons can seed the Bremsstrahlung process producing $X$-ray bursts. In their model, \citet{moss2006} used several different cross-sections to describe the electrons' motion through air and the collision between particles. \citet{schaal2012} analysed data from the Thunderstorm Energetic Radiation Array (TERA) and obtained properties of both triggered and natural lightning $X$-rays. They inferred electron luminosities (e$^{-}$ s$^{-1}$) from observed $X$-rays and noted that the measured values are less than what had been suggested by theoretical calculations. They also found that the electron luminosity increases with return stroke currents up to 10 kA, then forms a plateau at large currents. \citet{xu2014} studied $X$-ray production through Bremsstrahlung via simulations of negative CG discharges, and predicted the observable energy spectrum from the ground. They concluded that their model results are comparable with measurements of the TERA instrument in terms of general shape and hardness of the spectrum. \citet{kochkin2015} coordinated in-flight measurements of lightning-emitted $X$-rays, and found that they are emitted from the leader phase of a lightning flash, and when the lightning strike attaches to the aircraft. In summary, both natural and rocket-triggered lightning measurements confirmed that all known leader types produce high energy $X$-rays \citep{dwyer2012,dwyer2014}.

$X$-ray emission has been detected from brown dwarfs (BDs), however the underlying processes are most likely not lightning related.
First \citet{neuhauser1998} detected $X$-rays from BDs, who analysed ROSAT (R\"OntgenSATellit) data of the Chamaeleon I star-forming cloud, and noticed that the object called Cha H$_{\alpha}$ 1, an M$7.5$ to M8 spectral type BD, has a typical $X$-ray luminosity that of an M-type star. \citet{grosso2007} reported the detection of $X$-rays from nine M type, young BDs out of which seven was a new detection. From their sample, \citet{grosso2007} found that BDs are less efficient $X$-ray emitters than low-mass stars, and concluded that there is an indication of decrease in coronal activity (indicated by $X$-ray emission) with spectral type. L-type BDs have also been searched for $X$-ray emission \citep[][and references therein]{helling2014}, however only one has been successfully detected so far \citep{audard2007}, which supports the findings of \citet{grosso2007}. \citet{audard2007} reported the first detection of an L dwarf, Kelu-1 AB, in $X$-rays with \textit{Chandra}. No T dwarfs or lower mass objects have been detected in $X$-rays yet. So far, $X$-ray emission from BDs have been attributed to the magnetic and coronal activity of the star \citep{helling2014}. 

%__________________________________________________________________
\subsection{Optical emission and, visible, and IR spectral signatures} \label{sec:optem}

\begin{figure}
\begin{center}
\includegraphics[scale=0.4]{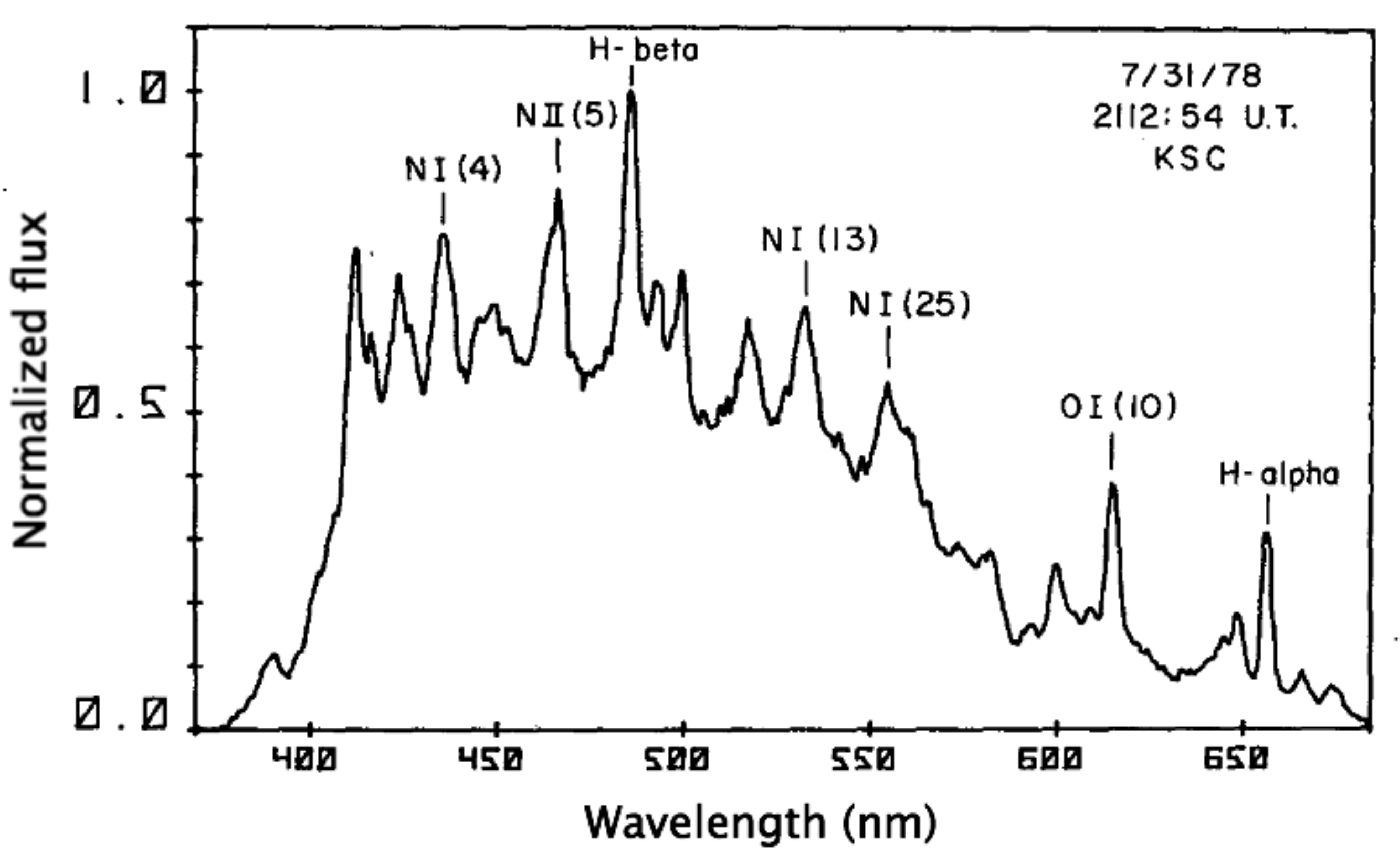} 
\end{center}
\caption{Earth lightning spectrum taken on 31/Jul/1978 at the Kennedy Space Center. It is the difference spectrum of a composite lightning$-$scattered daylight spectrum and a scattered daylight spectrum taken 5 s after the lightning discharge. The spectrum shows N and O lines, and the H Balmer series, as it is expected for a N-O dominated atmosphere. Figure reproduced with permission from \citet{orville1980}.}
\label{fig:emission_e}
\vspace{0.8cm}
\end{figure}

\begin{figure}
\begin{center}
\includegraphics[scale=0.6]{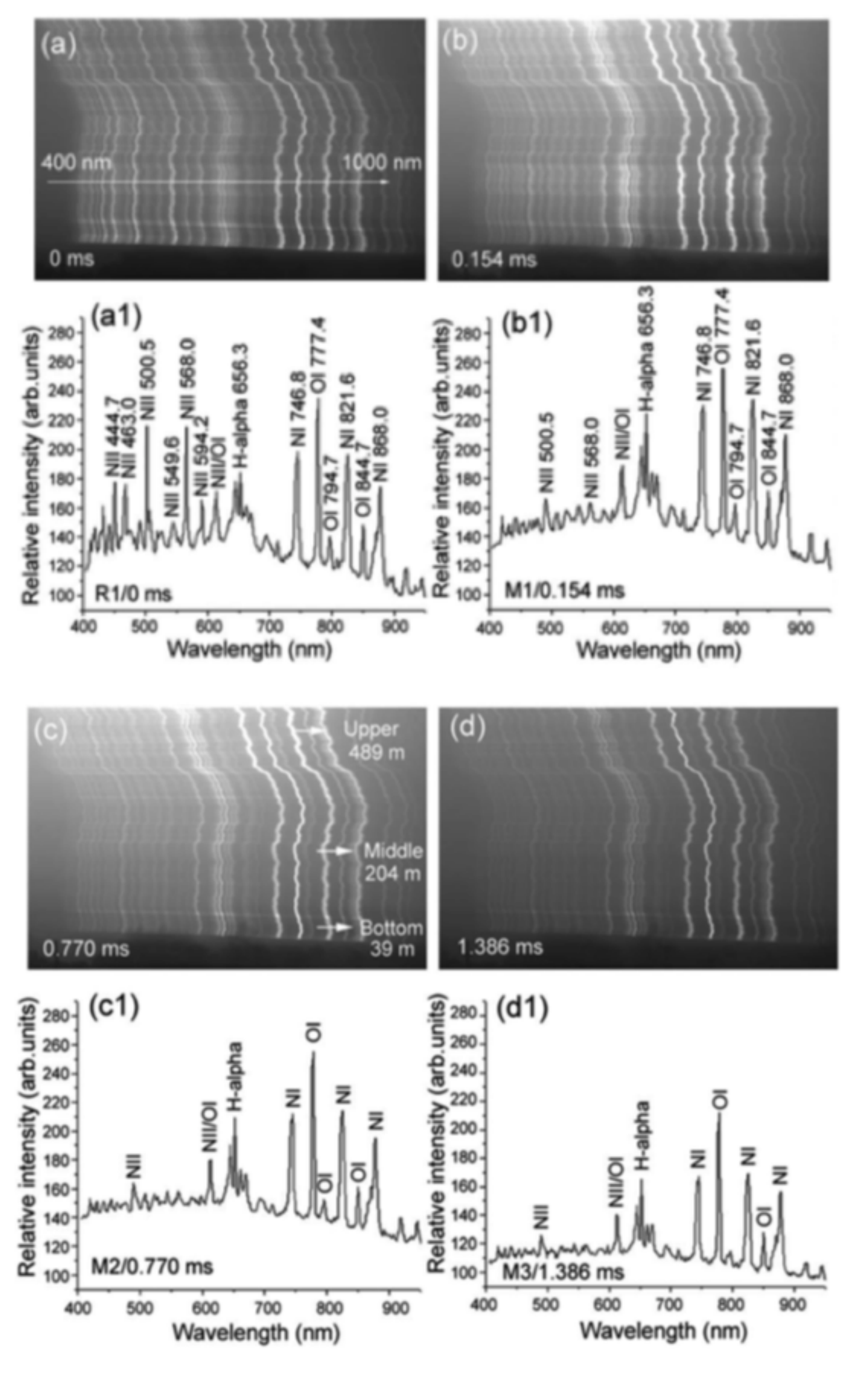} 
\end{center}
\caption{Earth lightning discharge spectra and corresponding spectral graphs of the first $\sim 1.4$ ms of the event. The most dominant lines are from O and N, with H$_\alpha$ appearing in the spectrum as well. Reproduced with permission from \citet{xue2015}.}
\label{fig:2_opt_e}
\vspace{0.8cm}
\end{figure}

With the advances of the nineteenth century, spectroscopy became one of the first tools used for scientific studies of lightning. In 1868, John Herschel was the first to note that nitrogen had the most luminous lines in the spectrum of lightning, and that the relative intensity of lines is different for different spectra \citep{herschel1868, rakov2003}. Spectral lines of lightning emission were systematically identified by \citet{schuster1880}. Further studies of lightning spectra from the nineteenth century were carried out by e.g. \citet{joule1872, procter1872, gledhill1881, konkoly1883, fowler1892}. \citet{slipher1917} took one of the first spectra of a lightning flash with a slit spectrograph in order to give better wavelength estimates for the observed emission lines. He observed several O and N lines and bands and compared his results with previous slitless observations. From the second half of the twentieth century, lightning spectra were examined in order to identify the physical processes acting during a lightning discharge \citep[][and references therein]{rakov2003}. In the meantime, further details of the chemical composition represented in a lightning flash were revealed \citep[e.g.][Fig. \ref{fig:emission_e}]{wallace1964,orville1980}. The first infrared spectrum of lightning was taken by \citet{jose1950}, who measured the emission between $\sim$ 6400$-$8800 \AA, and identified both N and O lines in it. An early review of lightning spectroscopy can be found in \citet{salanave1961}, and an extended review up to the twenty-first century in \citet{rakov2003}.

\begin{figure}
\begin{center}
\includegraphics[scale=0.38]{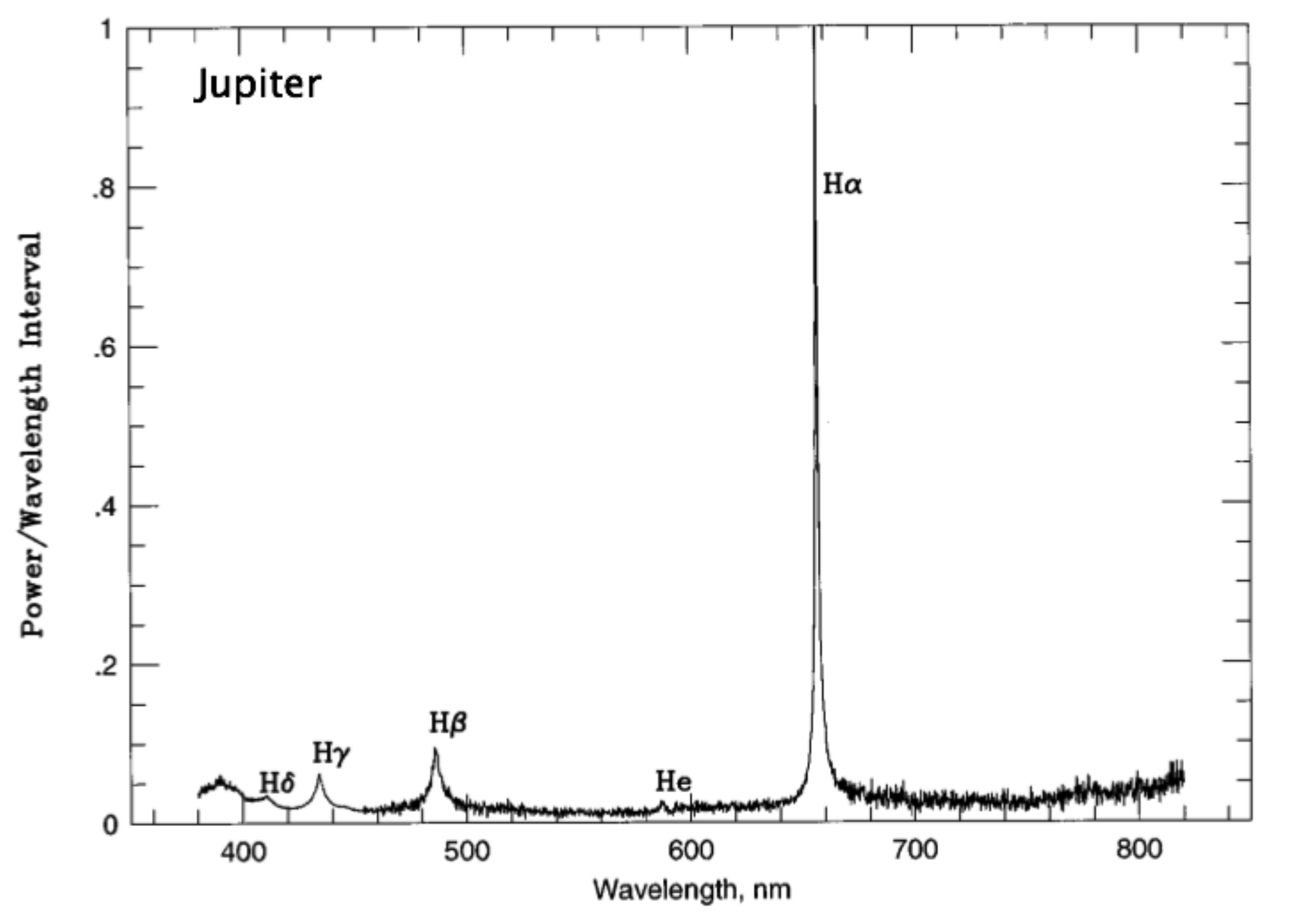} 
\includegraphics[scale=0.38]{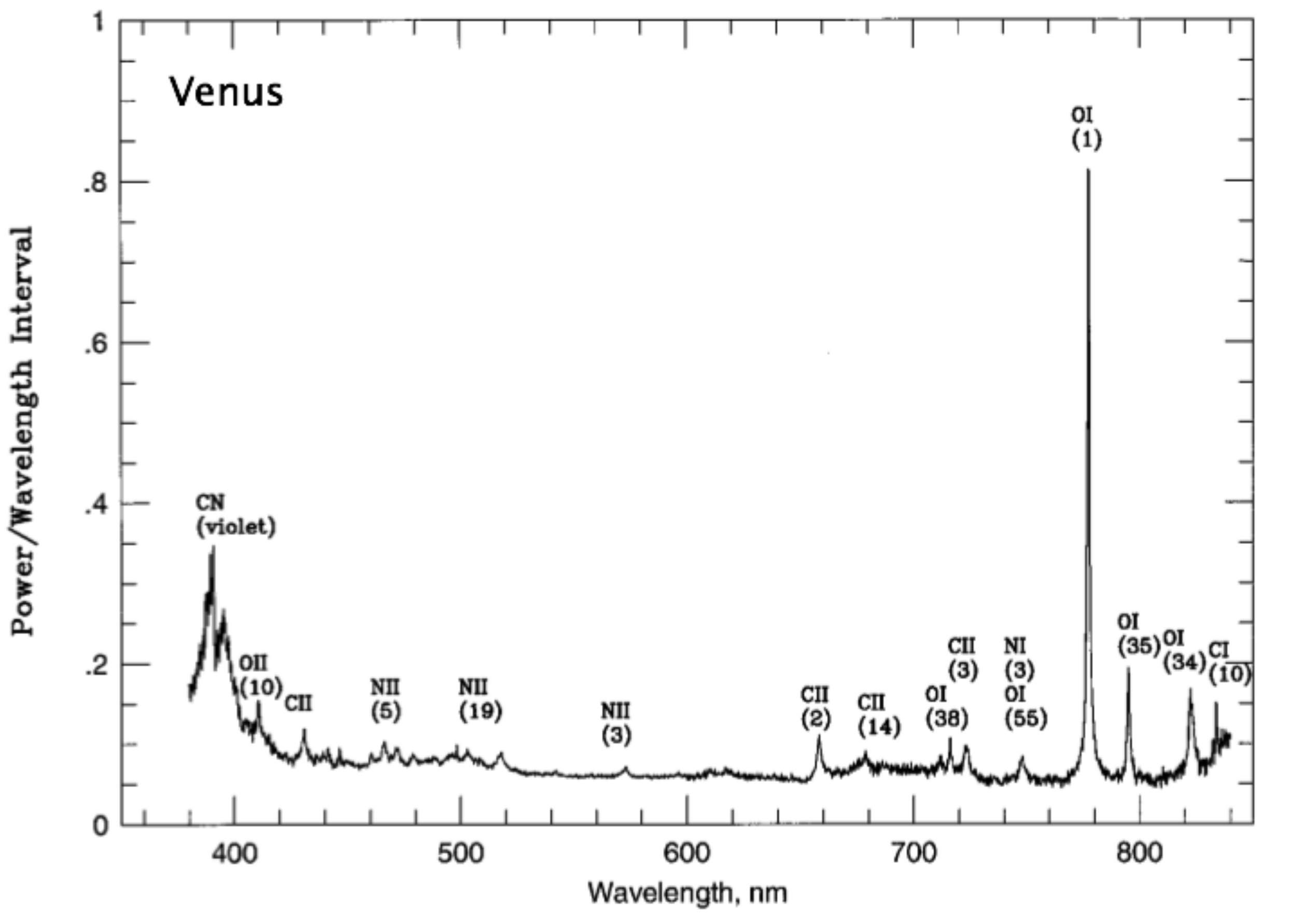} \\
\includegraphics[scale=0.38]{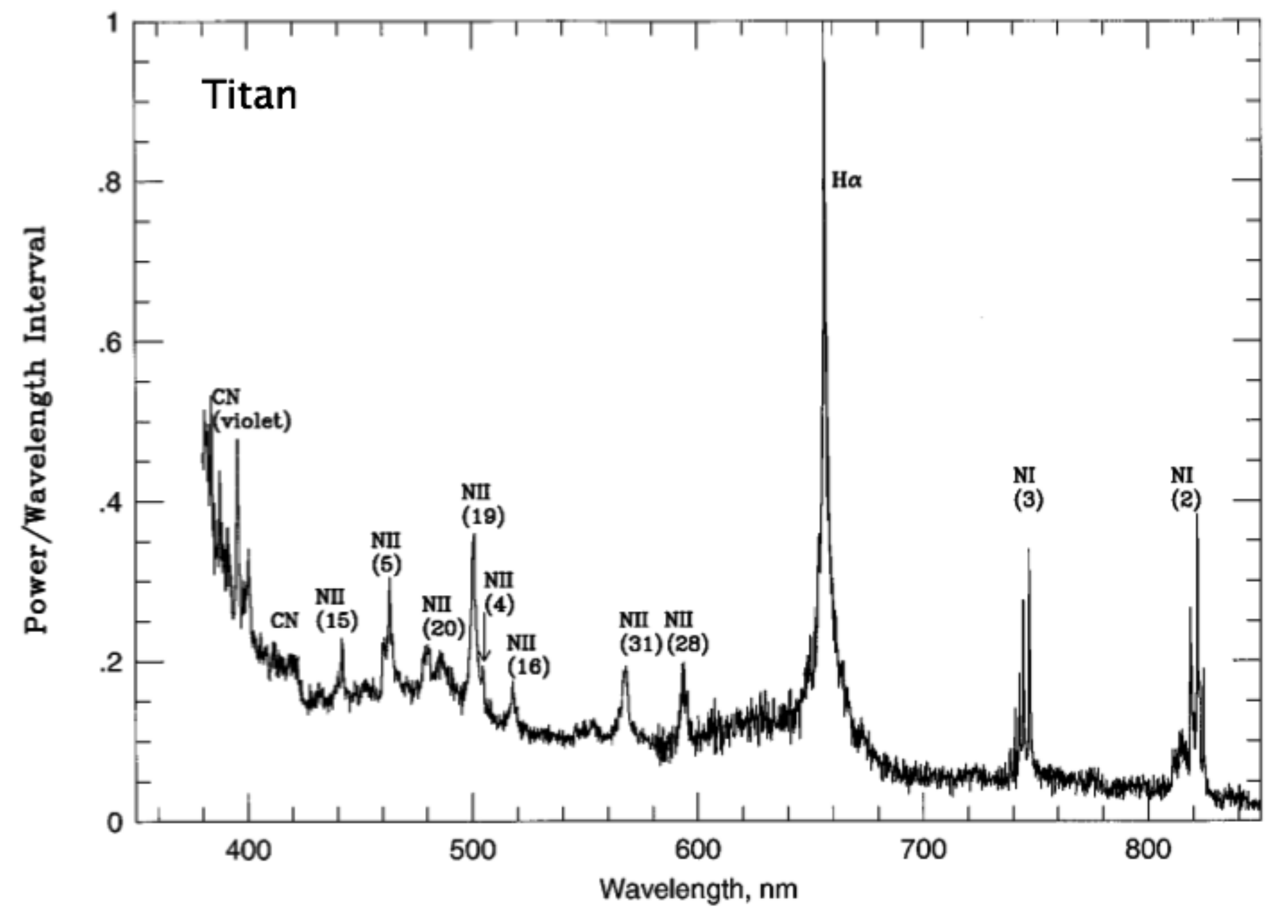} 
\end{center}
\caption{Spectra of simulated lightning in the atmospheres of Jupiter, Venus and Titan (from top to bottom, respectively) at 1 bar. The most prominent features are: H$_\alpha$ at 656 nm (Jupiter, Titan), O(I) at 777 nm and CN at 389 nm (Venus), several nitrogen lines (Titan).  Figures are reproduced with permission from \citet{borucki1996}.}
\label{fig:emission_1}
\vspace{0.8cm}
\end{figure}

On Earth, the energy released in the optical part of the electromagnetic spectrum is about 1-10\% \citep[][p. 334]{borucki1987, hill1979, lewis1984} of the total lightning energy \citep[$\sim 10^9$ J; e.g.][]{yair2008}. %, as we know from Earth. 
\citet{xue2015} measured the spectrum of natural CG lightning discharges (Fig. \ref{fig:2_opt_e}), and noticed that the structure of the spectrum is mostly affected by the current magnitude and duration, but less by the polarity of the discharge and the direction of the current. The emission spectrum of lightning is the result of the high temperatures involved in a discharge process, and hence the creation of a local environment that allows the production of non-equilibrium species. Therefore, the ionizing effect of lightning will produce molecules that are dependent on the local chemical composition. In a direct lightning spectrum, the emission lines of such molecules appear. It is not surprising that spectra presented in early and more recent studies show emission lines and bands mostly from nitrogen and oxygen (Fig. \ref{fig:emission_e}), since these are the most abundant molecules in the Earth atmosphere. One of the most prominent lines in a lightning spectrum is the 777.4 nm line of atomic oxygen \citep{beirle2014, yair2012}. This emission line is used to detect lightning flashes from space, by the LIS and OTD instruments (Chapter \ref{chap:stat}).

\citet{borucki1996} conducted laboratory experiments, comprised of a scanning spectrometer and laser-induced plasma of different composition, to simulate the emission spectra of lightning on Venus, Jupiter and Titan, at 1 bar and 5 bar pressures. The simulated spectra for the three objects at 1 bar are shown in Fig. \ref{fig:emission_1}. \citet{borucki1996} used an atmospheric composition of 96\% CO$_2$, 4\% N$_2$; 87\% H$_2$, 13\% He; and 97\% N$_2$, 3\% CH$_4$ with tracers of CH$_4$ and NH$_3$, by volume for the three objects, respectively. In their experiment, lightning on Venus has a prominent oxygen emission line at 777.4 nm, just like in an Earth lightning spectrum. Venusian lightning would also have weak features of carbon atoms and singly ionized oxygen atoms between 700 and 850 nm. Furthermore, \citet{borucki1996} found a prominent CN line at 388 nm. The Jovian lightning spectrum showed strong Hydrogen Balmer lines, and a Helium line at 588 nm \citep{borucki1996}. \citet{borucki1996} suggested that ammonia and methane would also be present in a Jovian lightning spectrum, however their experiment did not produce detectable quantities. Lightning on Titan would show features of H$_\alpha$ at 656 nm, several nitrogen lines between 400 and 656 nm and two strong lines of nitrogen at 740 and 820 nm. CN emission could also appear at 389 nm, on Titan \citep{borucki1996}. The experiment also showed that at 5 bar (compared to 1 bar) the line radiation is less prominent compared to the continuum, and molecular band radiation increases \citep{borucki1996}.

Similar to \citet{borucki1996}, \citet{dubrovin2010} found that Venusian lightning will have strong 777.4 nm and 656.3 nm features in its spectrum. Cassini observations detected lightning emission concentrated around the 656 nm H$_\alpha$ line on Jupiter \citep{dyudina2004}. \citet{luque2015} observed Jupiter from Earth with the GTC/OSIRIS instrument in H$_\alpha$ but did not detect lightning activity on the planet. \citet{hansell1995} claimed the detection of 6 lightning flashes on Venus at 777.4 nm and one flash at 656.4 nm. The observations were conducted from the ground in Arizona, with a 153-cm telescope. Based on these detections, they estimated the flash rate on Venus to be 1000 times smaller than on Earth, and the energy of lightning flashes to be around $10^8-10^9$ J (roughly the same as on Earth).

%__________________________________________________________________
\subsection{Radio signatures} \label{ssec:radsig}

When electric current is generated by accelerating electrons in a conducting channel, such as it occurs during lightning, the channel will act as an antenna converting the electric power resulting from the time-dependent current into radio waves \citep[e.g.][]{zarka2004}. Lightning radio spectrum covers the few Hz to few hundreds of MHz range \citep{desch2002}. About 1\% of the total lightning energy is released in the radio frequencies between very low and high frequencies \citep{volland1984, farrell2007}. Lightning induced radio emission has been observed from several Solar System planets (Chapter \ref{chap:liginout}). A large part of this thesis focuses on lightning radio signatures, therefore, in this section, I focus on a more detailed explanation and introduction of radio emission from lightning on Earth and other Solar System planets. I also summarize some of the radio background sources and attenuating factors that will affect the observation of radio signals. Lastly, I will shortly introduce non-lightning related radio emission that is produced at similar frequencies to that of lightning radio emission, therefore making it hard to determine the source of such emission based only on the radio signal.

\subsubsection{An overview on types of radio signatures}

\begin{figure}
\begin{center}
\includegraphics[scale=0.42]{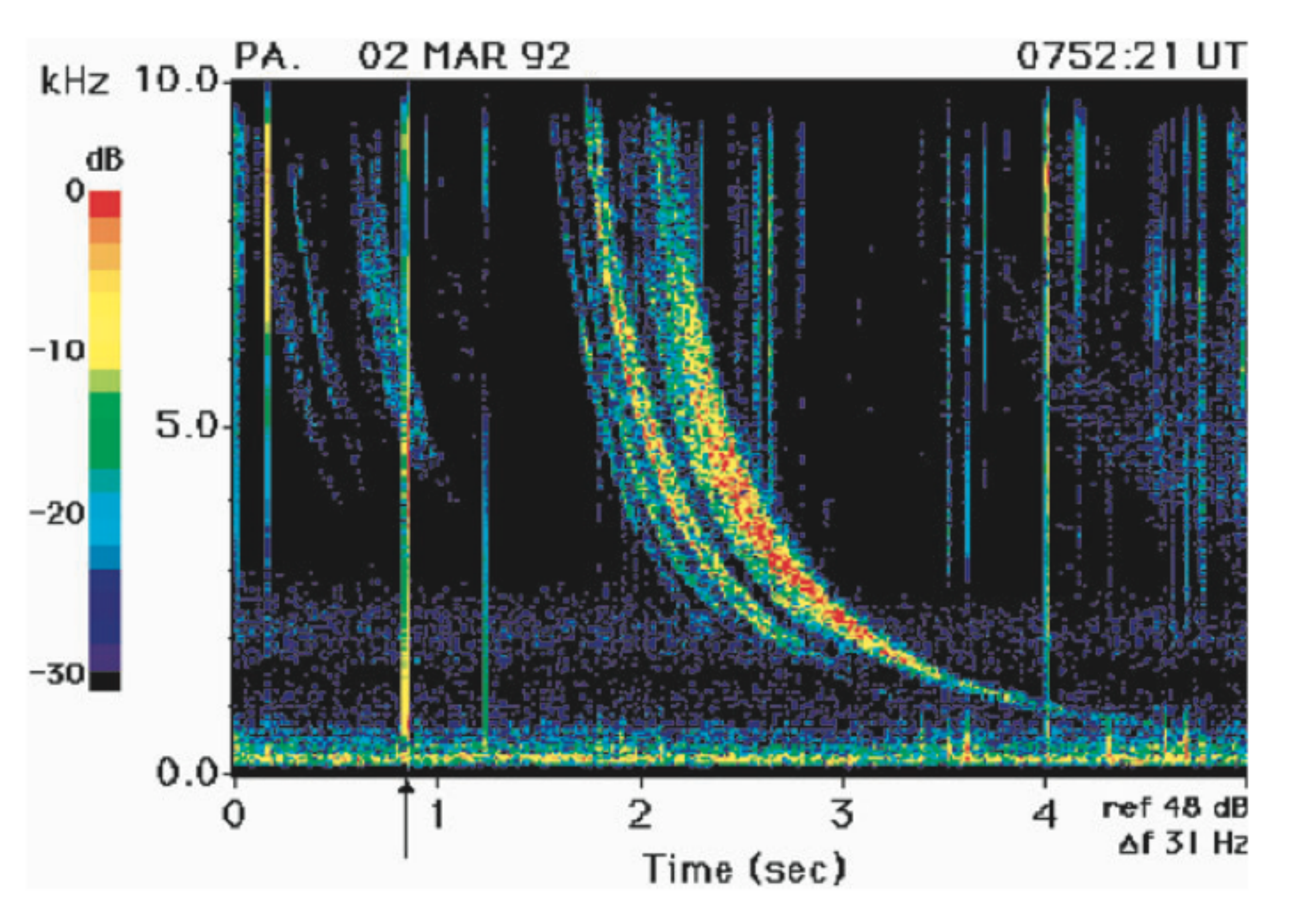} 
\end{center}
\caption{Spectrogram of sferic and whistler originated in North America and observed at Palmer station, Antarctica. Frequency: 0$-$10 kHz. The time axis covers 5 s. The intensity is given in dB, with red indicating the largest values. Arrow labels the sferic of the same lightning flash that produced the whistler (curved, intense feature between $\sim 2$ and 4 s). Reproduced with permission from \citet{desch2002}. Courtesy of Umran Inan (http://nova.stanford.edu/resgroups.html)}
\label{fig:whistler}
\vspace{0.8cm}
\end{figure}

There are four main phenomena related to lightning activity in the radio band. 

a) \textit{Sferics} (or atmospherics; 1 kHz $\le f \le$ hundreds of MHz), in general, are the emission in the low-frequency (LF) range with a power density peak at 10 kHz on Earth \citep{aplin2013} produced by lightning discharges. Since only radio emission in the higher frequency range can penetrate through the ionosphere, high frequency (HF) radio emission caused by lightning on other planets are also called sferics \citep{desch2002}. Sferics are the result of the electromagnetic field radiated by the electric current flowing in the channel of a lightning discharge \citep{smyth1976}. Figure \ref{fig:whistler} shows the dynamic spectrum of a sferic (vertical features). This type of emission is the most probable to be observed coming from other planetary bodies, since it is the only type of lightning radio emission capable of escaping the ionosphere or the magnetosphere of the object. The power spectra of sferics and their dependence on lightning characteristics is discussed on the next pages and in Chapter \ref{chap:model}.

b) \textit{Whistlers} (few Hz $\le f \le$ several kHz) are electromagnetic (EM) waves propagating along magnetic field lines and emitting in the very low-frequency (VLF) range. They were named after the sound they give through a speaker while the waves travel more quickly at higher frequencies and more slowly at lower frequencies \citep{desch2002}. Figure \ref{fig:whistler} shows the dynamic spectrum of a whistler (curved feature). The majority of whistler-type EM signals traverse the ionosphere into the magnetosphere, and after reaching the opposite hemisphere they travers the ionosphere again \citep{rakov2003}. The minority of whistlers propagate through the ionosphere into the magnetosphere and there dissipate \citep{rakov2003}. Radio waves can only propagate along magnetic field lines if their frequency is lower than the electron cyclotron frequency along the same field line \citep{desch2002}. For Earth, the lowest value of cyclotron frequency is $\sim 10$ kHz \citep{desch2002}, therefore whistlers that propagate through the entire field line will have frequencies lower than that. Whistler waves can only be detected, if the receiver is along the path of the same field line as the radio wave travels through.

c) \textit{Schumann resonances} (SRs) are extremely low frequency (ELF) lightning-induced electromagnetic oscillations resonating in the planetary surface-ionosphere cavity, with the wavelength of the planet's circumference \citep{simoes2012}. They have only been observed on Earth. Their existence was predicted by \citet{schumann1952} and they were first observed by \citet{balser1960}. A Schumann-like resonance event at 36 Hz was observed on Titan by the Cassini-Huygens probe in 2005 \citep{grard2006}, however its lightning-related origin is questionable \citep[][see also Sect. \ref{sec:matit}]{beghin2007,simoes2007}. Ideally, the SR modes occur at a discrete frequency determined by the planetary circumference, the thickness of the cavity, and the speed of light in the cavity \citep{simoes2012}. In reality however, the resonance lines are broadened and the observed centre frequencies are shifted from the ideal case. This can be because of the changes in the EM sources, in the properties of the atmosphere, or because of the variability of the upper boundary of the ionosphere, which is the result of the interaction between the solar wind and the magnetosphere/ionosphere \citep{simoes2012, rakov2003}. Due to the nature of the emission, the larger the planet the lower the resonance frequency will be.

\vspace{0.4cm}

%Table 2 - Schumann res.
\begin{table}[ht]
\centering
\footnotesize
\renewcommand{\arraystretch}{1.5}
\caption{The first three modes of Schumann resonances of the major Solar System planets, Titan, and the exoplanets WASP-12b and TRAPPIST-1 e. References: Earth: \citet{simoes2011}; the other Solar System planets and Titan: \citet[][1$^{\rm st}$ mode only]{simoes2012}, \citet{beghin2012} and \citet[][and references therein]{simoes2008}. The values for the exoplanets were calculated applying Eq. \ref{eq:sr}.}
\begin{tabular}{|c|c|c|c|c|c|c|}	
	\hline
	\multirow{2}{*}{Celestial body} & \multicolumn{2}{c|}{1$^{\rm st}$ mode} & \multicolumn{2}{c|}{2$^{\rm nd}$ mode} & \multicolumn{2}{c|}{3$^{\rm rd}$ mode} \\ \cline{2-7} 
	 & f [Hz] & Q & f [Hz] & Q & f [Hz] & Q \\
	\hline \hline
	Venus & $7.9-9.5$ & $4.8-10.5$ & $14.17-16.3$ & $5-11.3$ & $20.37-23.3$ & $5.2-22.7$ \\ 
	Earth & $7.8$ & $5$ & $14.3$ & $5$ & $20.8$ & $5$ \\ 
	Mars & $7.3-14$ & $1.9-4$ & $13-26$ & $1.8-3.8$ & $19.2-38$ & $1.8-40$ \\
	Jupiter & $0.6-0.76$ & $5-10$ & $1.2-1.35$ & $7.2-8.6$ & $1.74-1.93$ & $7.3-8.7$ \\ 
	Saturn & $0.75-0.93$ & $3.5-7.8$ & $1.63$ & $6.8$ & $2.34$ & $6.5$ \\ 
	Titan & $8.2-26$ & $0.92-6$ & $14.3-45$ & $0.8-6$ & $26.7-64$ & $1-4.7$ \\ 
	Uranus & $1-2.5$ & $2-21$ & $1.99-4.27$ & $1.9-19.4$ & $2.96-5.9$ & $0.9-9.5$ \\ 
	Neptune & $1-2.6$ & $1-16$ & $2-4.12$ & $1-9.4$ & $2.96-5.9$ & $0.9-9.5$ \\ 
	WASP-12b & $0.56$ & - & $0.96$ & - & $1.36$ & - \\ 
	TRAPISST-1 d & $11.16$ & - & $19.3$ & - & $27.34$ & - \\
	\hline
\end{tabular}
\label{table:2}
\vspace{0.5cm}
\end{table}
%Table 2

The ideal SR frequencies (eigenfrequency) are given by:

\begin{equation} \label{eq:sr}
f_{N} = \frac{c}{2 \pi R_{\rm p}}\sqrt{N(N+1)},
\end{equation}

\noindent where c is the speed of light, $R_{\rm p}$ is the planetary radius and $N \in \mathbb{N}$. 
In the ideal case the first three frequencies are a few hertz higher than the observed ones \citep{rakov2003}. Table \ref{table:2}. lists the first three eigenfrequencies with their Q-factors for the Solar System planets, Saturn's moon, Titan, and the exoplanets WASP-12b and TRAPPIST-1 e. Mercury is not listed because it does not own an atmosphere, hence surface-ionosphere cavity, therefore cannot form Schumann resonance \citep{simoes2012}. WASP-12b is an inflated hot Jupiter orbiting close to its parent star. Its radius is $1.736$ R$_J$ \citep{chan2011}, which results in slightly lower SRs than the values for Jupiter. The newly discovered TRAPPIST-1 e is about the same size as Earth, R$_p = 0.92$ R$_\oplus$, with a similar equilibrium temperature \citep[T$_{\rm eq} = 251$ K,][]{gillon2017}. Its SR is, therefore, very similar to that of the Earth. When calculating the Schuman resonances for the exoplanets, I do not consider any shifting and broadening effects (I do not give Q-factors). The ideal frequencies of WASP-12b and TRAPPIST-1b were calculated in comparison to the Solar System bodies.  
The Q-factor measures the wave attenuation in the cavity (the decay time) and characterizes the width of the SR signal in the frequency domain \citep{rakov2003}. It represents the invers number of cycles of a propagating VLF wave at frequency $f_N$ before it is fully attenuated. The Q-factor is given by the following equation:

\begin{equation} \label{eq:sr2}
Q_{N} = \frac{f_{N}}{\Delta f_{N}},
\end{equation}
\noindent where $f_N$ is the centre of the frequency and $\Delta f_{\rm N}$ is the width at half-maximum power \citep{rakov2003}. Schumann-resonances only exist if the planet has an ionosphere and can be only observed from within the ionosphere. Therefore, no SR will be observable from extrasolar objects with current technology.

d) Lightning also generates \textit{electromagnetic pulses} (EMPs), most commonly in the form of trans-ionospheric pulse pairs (TIPPs), first detected in 1993 by the Blackbeard payload of the ALEXIS (Array of Low-Energy X-Ray Imaging Sensors) satellite \citep{holden1995}. TIPPs are emissions in the $30-300$ MHz range from lightning EMP and its echo off of the ground \citep{desch2002}. \citet{holden1995} found that TIPPs last for about $10 \mu$s and occur in pairs separated by $\sim$ 50 $\mu$s. Although they did not know what the exact origin of TIPPs was, they observed their appearance to be consistent with thunderstorm activity. \citet{russell1998} found that the TIPPs analysed in their study originated $~8$ km above the ground, possibly from IC discharges. Some of the IC discharges produce very powerful pulses referred to as narrow bipolar pulses \citep{levine1980, jacobson2011}. \citet{zuelsdorf2000} compared the origin of these pulses and TIPPs, and found that they are generated by the same IC discharge process. EMPs are thought to be the sources of ELVE-type transient luminous events \citep[][and Sect. \ref{sec:tle}]{fukunishi1996}.

\subsubsection{Earth-lightning radio spectrum}

\begin{figure}
\begin{center}
\includegraphics[scale=0.6]{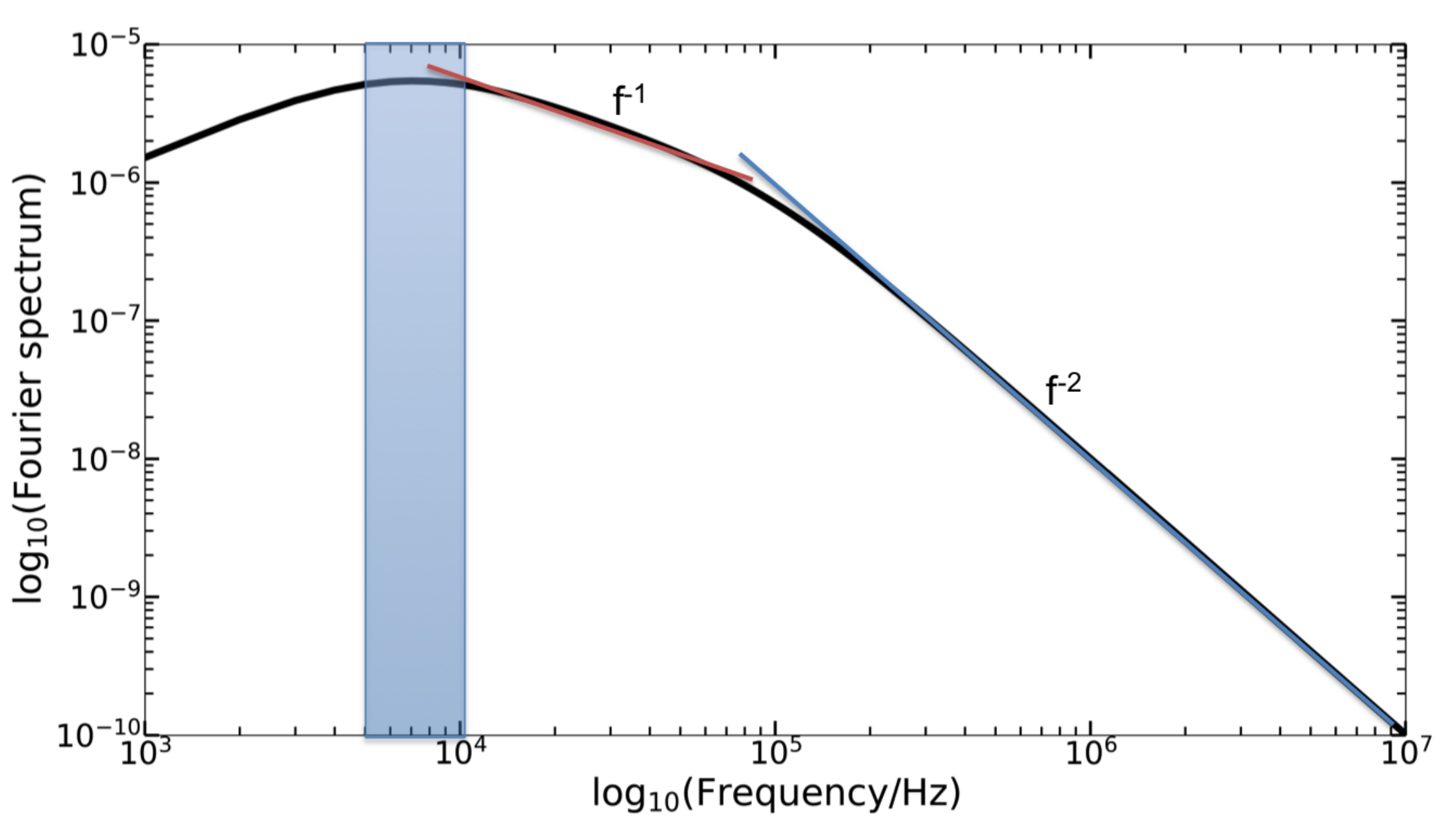} 
\end{center}
\caption{Earth lightning radio spectrum (black). The blue area covers the frequency range where the peak frequency occurs. After that the spectrum decreases with $f^{-1}$ (red line) up to several tens of kHz, while above that up to a few MHz the spectral roll-off becomes $f^{-2}$ (blue line). The lines are for demonstration only and not fitted lines. For further details of lightning frequency spectrum see Chapter \ref{chap:model}.}
\label{fig:2_frspec}
\vspace{0.8cm}
\end{figure}

On Earth, a typical sferic has a broad band spectrum with a peak frequency at $\sim10$ kHz. The radio power spectrum of a simple, non-tortuous return stroke peaks around 7-10 kHz, then drops with $f^{-1}$ up to 40 kHz, between 40 kHz and $\sim 5$ MHz it becomes steeper and drops with $f^{-2}$ \citep{desch2002}. At $\sim 5$ MHz the power spectrum has a second turn-point and it decreases with $f^{-4}$ \citep[e.g.][]{farrell2007}. An example shape of lightning radio frequency spectrum is shown in Fig. \ref{fig:2_frspec}. The observed shape of the frequency spectrum depends on the characteristics of the lightning channel, e.g. its length and tortuosity, the lightning stroke duration and the time profile, and the dispersion and attenuating characteristics of the medium the radio wave propagates through to the receiver \citep{volland1984}. Therefore, the expected radio power spectrum  of lightning originating from different objects, will be different. In Chapter \ref{chap:model}, I will further analyse and discuss possible lightning radio spectra of several Solar System planets, exoplanets and brown dwarfs.

\subsubsection{Escape of radio emission - Role of the ionosphere and other attenuating plasmas}

The ionosphere is the upper part of the atmosphere at the inner boundary of the magnetosphere. It plays a crucial role in radio wave propagation and reflection. On Earth the atmospheric gas is ionized mainly by solar radiation, while on other planets, exoplanets or brown dwarfs, other mechanisms like cosmic ray ionization or collisional ionization may play a bigger role \citep[e.g.][]{rimmer2013,rodriguezbarrera2015}.
The state of ionization of the ionosphere is described in terms of electron number density, $N_e$. $N_e$ has a diurnal variation, also varies with altitude, latitude, and with the variations of the solar UV radiation \citep{macgorman1999}. Electromagnetic waves, including radio waves originating from lightning discharges, cannot propagate through the ionosphere with frequencies below the maximum (or peak) plasma frequency, $f_{\rm pl}$, which is the maximum frequency at which the free electrons in the ionosphere can oscillate \citep{desch2002}. $f_{\rm pl}$ is a function of the electron number density of the ionosphere,

\begin{equation} \label{eq:pl}
f_{\rm pl} = \frac{1}{2 \pi}\sqrt{\frac{N_e e^2}{m_e \epsilon_0}},
\end{equation}

\noindent where $e$ is the elementary charge, $m_e$ is the electron mass and $\epsilon_0$ is the permittivity of the vacuum \citep{lammer2001}. The plasma frequency for Earth is $5-10$ MHz. \citet{zarka2012} suggested that radio observations should be conducted from the night side of the Moon that is shielded from the solar radio waves and occasionally from radio emission from Earth likewise, therefore radio waves with frequencies of a few kHz could reach its surface. The ionosphere does not only block and reflect waves but attenuates the ones traversing it. Waves with frequencies up to a few $f_{\rm pl}$ are affected by the chromatic attenuation effect as a result of electron-neutral collisions \citep{zarka2008,leblanc2008}.
Observing lightning-related radio emission from any object can be a useful tool of characterizing the object's ionosphere. As $f_{\rm pl}$ depends on $N_e$, the frequency at which the lightning emission is observed will given an upper limit to the electron number density in the ionosphere.

The ionosphere is not the only plasma that could block emission with frequencies below its plasma frequency. Extraterrestrial radio waves have to propagate through the space between the source and the observer, which may contain other attenuating or blocking environments. The most prominent one, after the ionosphere, is the interplanetary space composed of the material ejected from the star and carried by the stellar wind. \citet{griessmeier2007} used distance-dependent stellar wind models to estimate the plasma frequency, which they showed to be negligibly small in most analysed cases. It is important to note that radio emission of an exoplanet will propagate through different environments towards the observer, throughout the planet's orbit. For example, when the planet is entering occultation (secondary eclipse), the star and the interplanetary space between the star and the planet will be along the line-of-sight. The plasma frequency in these environments is much higher than at distances further away from the star, therefore the radio emission that might be observable during transit, may be unseen during a secondary eclipse \citep{griessmeier2007}.

\subsubsection{Non-lightning related radio emission and background radio noise} \label{sec:ecme}

Super high frequency radio emission (several GHz) has been observed from late M dwarfs and L-type brown dwarfs on several occasions \citep[e.g.][]{berger2001, berger2006, berger2009, burgasser2005, hallinan2007, hallinan2008, antonova2008, mclean2011}. Recently, T dwarfs have also entered the family of known radio emitting objects \citep{route2012, route2016, kao2016, williams2017}. The detected radio emission is either periodic, 100\% polarized emission attributed to cyclotron maser instability (CMI), or quiescent, moderately polarized, which is probably the result of synchrotron or gyrosynchrotron emission \citep[][and references therein]{helling2014}. Electron cyclotron maser emission (CME) is the result of energized electrons travelling along magnetic field lines and interacting with electromagnetic waves \citep{sprangle1977}. The frequency of the emission is determined by the strength of the magnetic field. According to observations, brown dwarfs are capable of maintaining magnetic fields of kG-strength \citep{berger2006,helling2014}. Similarly, CMI related radio emission has been observed from Earth, Jupiter, Saturn, Uranus, and Neptune in the form of auroral radio emission \citep{zarka1998}. Simultaneous optical and radio spectroscopy suggested the presence of aurorae on a late M dwarf \citep{hallinan2015}, as well. 

Electron cyclotron maser emission has been suggested to be one possible mechanism for observable radio emission from exoplanets too \citep[e.g.][]{lazio2004,griessmeier2007},  where the source of the high energy electrons is the coupling between the stellar wind and the object's magnetic field \citep{lazio2004}. \citet{farrell1999b,zarka2001} and \citet{lazio2004} extended the radiometric Bode's Law of Solar System planets to extrasolar planets. The radiometric Bode's Law is an empirical relation between the solar wind power, the planet's magnetic field strength, and the emitted radio power \citep[e.g.][]{lazio2004}. \citet{lazio2004} argued that most of the exoplanets they considered should have a radio emission at frequencies between 10 and 1000 MHz, and the flux of the emission should reach 1 mJy depending on the distance of the source. \citet{griessmeier2007} estimated the frequency and flux of radio emission from exoplanets, caused by interaction between the planet and the stellar wind. They considered different scenarios as the source of the energy necessary for CMI in the planetary magnetosphere, and showed that the magnetic energy from the interplanetary magnetic field \citep[e.g.][]{farrell2004} would result in the largest radio flux, followed by the energy deposited by stellar coronal mass ejections \citep[e.g.][]{griessmeier2006} on close-in exoplanets, and the kinetic energy of the stellar wind interacting with the planetary magnetic field \citep[e.g.][]{lazio2004}. \citet{griessmeier2007} found that the maximum frequency of such emission would be between 0 and 200 MHz, however above 70 MHz the radio flux would be too low to detect. They suggested that for one of 197 examined exoplanets (GJ436 b) the radio flux can be ideally as high as 5 Jy, however most of the planets show less than 10 mJy flux. A campaign to observe electron cyclotron maser emission has been started \citep[e.g.][and ref. therein]{lecavelier2009, lecavelier2011, lecav2013}; however, it has not been conclusively detected from exoplanets. Electron cyclotron maser emission from exoplanets can cause false positives for lightning detection, as the emission frequencies are in the same range for both mechanism. On the other hand, similarly, a positive detection of radio emission from an exoplanet would not necessarily mean CMI emission, but could be caused by lightning as well (see Chapter \ref{chap:hatp11b}). 

Low frequency radio observations conducted from the surface of the Earth or orbits in the inner Solar System will be affected by radio noise. Here, I only summarize the natural sources originated outside the Earth ionosphere. The major source of radio noise is the Sun. Solar radio bursts, especially type III bursts have the largest contribution to the radio noise between $0.1-10$ MHz \citep{desch1990}. The solar wind has a contribution at frequencies $20-30$ kHz \citep{zarka2012}. The galactic radio background (from diffuse synchrotron emission) is strong above $100$ kHz while quasi-thermal and electrostatic noises have an effect at lower frequencies \citep{zarka2012}.

%__________________________________________________________________
%__________________________________________________________________
\section{Effects of lightning discharges on the local chemistry} \label{sec:chemeff} 

The pressure and temperature changes during a lightning discharge trigger chemistry changes in the local atmosphere. New, non-equilibrium species are produced, which will recombine and form the original molecules once the process causing the changes, i.e. the lightning discharge, terminates and no new triggers are initiated. Two main mechanisms create non-equilibrium molecules during a lightning flash: (i) high energy electrons create new species through electrochemical reactions via colliding with atmospheric particles and transmitting their energy, and (ii) as the temperature and the pressure increases the plasma suddenly expands and creates a shock wave (thunder) which also affects the chemistry \citep{rakov2003}.

The importance of lightning discharges in creating prebiotic molecules, and therefore maybe initiating life-formation, has been recognized early. In their experiment, \citet{miller1953} used electrical discharges instead of UV light that had been previously proposed as the catalizator of the formation of organic molecules \citep{haldane1928,oparin1938}. They observed the formation of amino acids after electrifying the experimental mixture composed of CH$_4$, NH$_3$, H$_2$O, and H$_2$. \citet{johnson2008} used an apparatus built by Miller, but never tested before. The experiment simulated a volatile-rich explosive volcanic eruption and the effect of lightning on the composition of the volcanic plume \citep{johnson2008}. They observed the production of amino acids in the experiment, with amounts comparable to or exceeding those found by \citet[e.g.][]{miller1953}. Their findings are also relevant for exoplanetary studies, since the composition of volcanic plumes may resemble dust clouds of certain exoplanets and brown dwarfs \citep{helling2008b,helling2008}.

The initial chemical composition of the atmosphere will determine the type of molecules that are created by lightning. In an Earth-like atmosphere, mainly composed of N$_2$ and O$_2$, nitrogen oxides \citep[NO$_{\rm x}$; e.g.][]{noxon1976,ott2010} and ozone \citep[O$_3$;][]{zhang2003,lorenz2008} are the most common tracers of lightning activity. O$_3$ and NO$_{\rm x}$ are in disequilibrium: O$_3$ will convert into O$_2$; NO and NO$_2$ (together NO$_{\rm x}$) further contribute to the formation of O$_3$, and may precipitate out of the atmosphere in the form of nitric acid and is absorbed by the soil forming nitrates, important energy sources for plants \citep{lorenz2008}.
 
\citet{krasnopolsky2006} claimed that the appearance of NO in the atmosphere of Venus might be the result of lightning activity. They observed Venus with the TEXES (Texas Echelon Cross Echelle Spectrograph) at NASA IRTF (InfraRed Telescope Facility) on Mauna Kea at the 5.3 $\mu$m band, where three NO lines reside, and found that two lines suggest the presence of NO at the lower atmosphere of the planet. Laboratory experiments conducted by \citet{barnun1975} showed that C$_2$H$_2$ and HCN are potentially produced by lightning in the atmosphere of Jupiter. \citet{podolak1988} obtained lightning-produced CO, HCN, and C$_2$H$_2$ abundances from numerical models, and noted that while the CO and HCN abundances in Jupiters atmosphere are consistent with observations, lightning is not efficient enough to produce the observed amount of C$_2$H$_2$. They suggested that the rest of the C$_2$H$_2$ is produced by a photolytic source. Cassini/VIMS  (Visual-Infrared Mapping Spectrometer) observations in the NIR showed a dark layer around the 1 bar pressure level in Saturn's atmosphere. \citet{baines2009} suggested that these dark cloud layers are vertically conducted, lightning-produced chemicals. Cassini/CIRS observations during and after the 2011 large storm on Saturn \citep{dyudina2013} revealed enhancement of molecules like C$_2$H$_4$ in the atmosphere \citep{hesman2012}, which can be the result of vertical transport of compounds created by lightning in the lower atmosphere \citep{yair2012}. \citet{horvath2009} conducted laboratory experiments to study the effects of corona discharges in the atmosphere of Titan, and noticed that the main product of discharges is C$_2$H$_2$, with traces of C$_2$H$_6$ and HCN. \citet{kovacs2010} carried out a detailed kinetic reaction analysis in a Titan-like atmosphere as a result of lightning activity, and found that HCN is the most important tracer of lightning on Titan.

Chemical changes in the atmosphere and the production of non-equilibrium species is a good indicator of lightning activity. The changes in the concentration of such species can indicate the intensity of lightning storms. Their effect on the spectrum of a planet, even an exoplanet, can be significant (Chapter \ref{chap:hatp11b}). It is important to note, however, that some of the species listed above are produced not only by lightning discharges, but other means as well. On Earth, ozone is mainly produced by the UV dissociation of O$_2$ and recombination to O$_3$ at the top of the atmosphere \citep[in the stratosphere between 15 and 40 km;][and references therein]{staehelin2001}. In the Jovian atmosphere C$_2$H$_2$ and C$_2$H$_4$ can produced by photochemistry induced by solar UV radiation \citep[e.g.][]{moses2005}. Similarly, these species could be produced by various phenomena on extrasolar objects as well. Therefore, it is important to address other formation mechanism of non-equilibrium species, when exploiting non-equilibrium chemistry as a method to detect lightning activity. 

Observations have revealed several spectral features in exoplanetary and brown dwarf atmospheres \citep[e.g.][]{deming2013,pont2013,kreidberg2015}, which features have been shown by models \citep[e.g.][]{benneke2012,barstow2017,kempton2017} to indicate the primary composition of the atmospheres of the objects. Apart from the bulk composition, spectral signatures indicate non-equilibrium chemistry in these atmospheres as well \citep[e.g.][]{knutson2012,macdonald2017}. The two, most well-studied processes causing the enhancement of non-equilibrium species are biological activity and related phenomena, and vertical mixing of the atmosphere \citep{seager2010b,seager2010}. Such processes may cause false indication of lightning activity, e.g. in the case of O$_3$ detection.

Biosignatures are fingerprints in the spectrum which indicate the presence of life on the objects. They are species that were suggested to be biosignatures based on the spectrum of Earth and what features it contains because of life \citep{kaltenegger2002}. For example, O$_2$ and CH$_4$ were suggested to be a good combination to indicate life activity on an object \citep[][and references therein]{seager2010b}: O$_2$ is a highly reactive gas oxidizing its environment, therefore it needs continuous replenishment to keep the large quantities present in Earth's atmosphere. CH$_4$, though can be produced by abiotic processes, is mostly produced by methanogenic bacteria. In a highly oxidizing atmosphere as the Earth's is, CH$_4$ only lives for $\sim$ 9 years, therefore it needs to be constantly replenished as well. Similarly, N$_2$O is considered as a good biosignature, even on its own, though detected in the company of large concentrations of O$_2$, it is a better life indicator \citep{seager2010b}. However, it is produced in very small quantities, therefore it is hard to detect in a spectrum.

In dynamic atmospheres, molecular species are mixed between atmospheric layers casing the appearance of larger abundances of a certain species than what is expected from equilibrium models. Vertical mixing has been observed mostly in brown dwarf atmospheres \citep[e.g.][]{noll1997,saumon2000,sauman2006}, however, lately observations suggest that it occurs on exoplanets as well \citep[e.g.][]{knutson2012}. In hydrogen rich atmospheres, nitrogen chemistry suggests vertical mixing. In thermodynamic equilibrium, the dominant form of nitrogen is NH$_3$, however, quite frequently an enhanced abundance of N$_2$ is observed in the upper part of, mostly, hotter (L, T) brown dwarf atmospheres \citep{seager2010b}, where N$_2$ is in disequilibrium. Another indicator of vertical dynamics in a T and cooler type brown dwarf or exoplanet is the enhanced presence of CO in the upper, cooler part of the atmosphere, where CH$_4$ is the dominating species in thermodynamics equilibrium \citep{seager2010b}. The CO is in equilibrium in the lower, hotter parts of the atmosphere. After it is transported to lower pressures, it slowly converts into CH$_4$, however this timescale is longer than the timescale of vertical mixing.

%__________________________________________________________________
%__________________________________________________________________
\section{Emission caused by secondary events} \label{sec:tle} 

\begin{figure}
  \centering
  \includegraphics[scale=0.5]{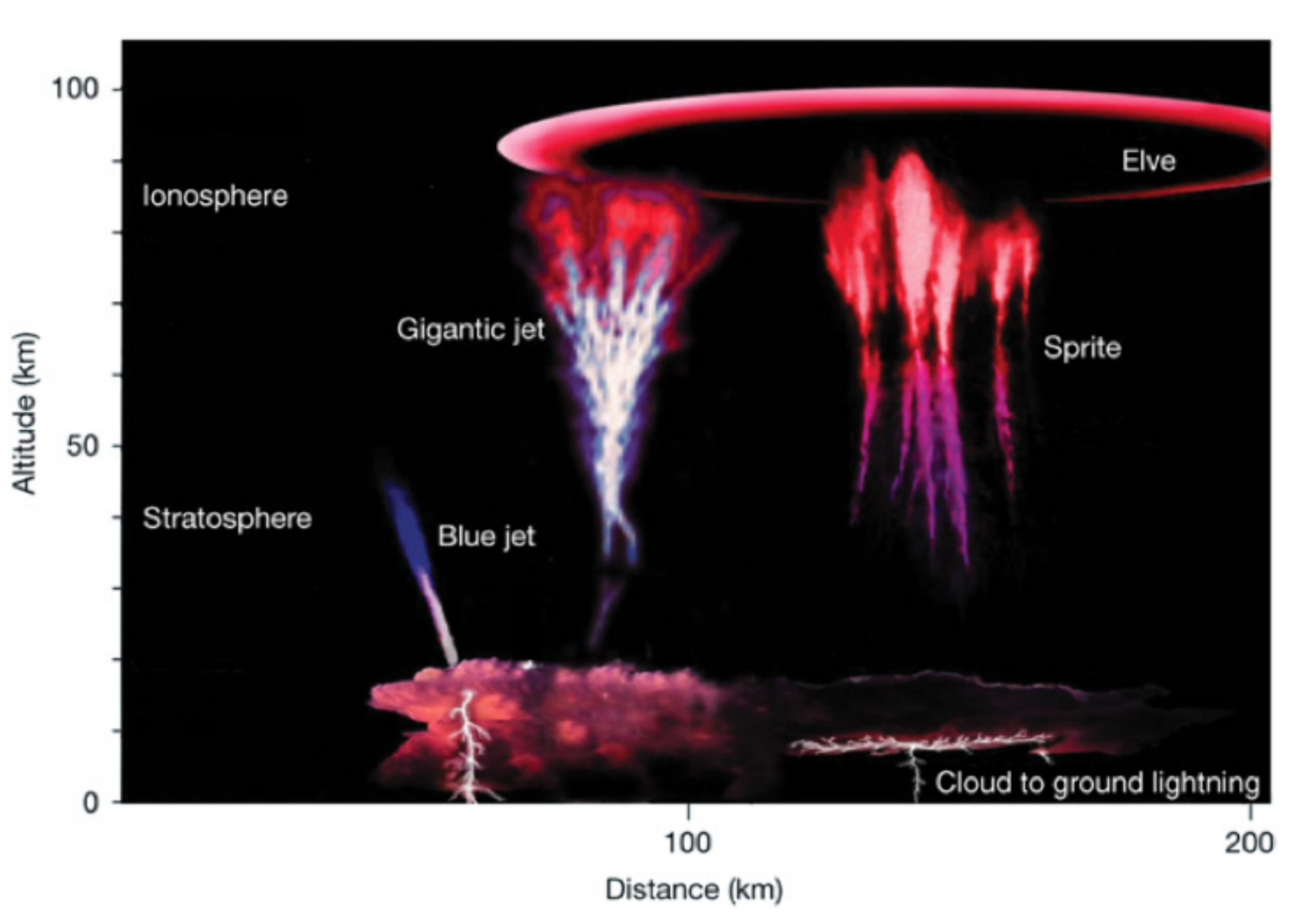}
  \caption{Most common types of transient luminous events (TLEs) appearing after lightning discharges. Figure reproduced with permission from Nature from \citet{pasko2003,pasko2007}.} 
  \label{fig:tle}
\vspace{0.8cm}
\end{figure}

Transient Luminous Events (TLEs) are lightning-related phenomena in the stratosphere and mesosphere of Earth, occurring above thunderclouds between altitudes of 15 to 110 km \citep{pasko2012}. They are caused by the charge imbalance in thunderstorms developing after a lightning flash, or by the EM field generated by the return stroke current \citep{siingh2015}. During the almost thirty years since their discovery in 1989 \citep{franz1990} several different types of TLEs have been observed and studied. The literature on the modelling and observations of TLE phenomena is extensive. Detailed reviews are found in \citet[e.g.][]{pasko2007,pasko2010,pasko2012,surkov2012,siingh2012,siingh2015}. A summary table of the different types of TLE can be found in \citet[][their table 1.]{siingh2015}. Based on the optical appearance and physical properties of TLEs, we can distinguish between the following main types (Fig. \ref{fig:tle}):

\begin{itemize}
\item 
\textit{Sprites} (Fig. \ref{fig:2_sprite}) are predominantly red phenomena \citep{sentman1995}. They appear after an intense positive cloud-to-ground (CG) discharge above the thunderstorm in the altitude range $< 40-90$ km with the brightest part in the $65-85$ km altitude range \citep{rousseldupre2008}. Negative CG discharges may also produce sprites, but very rarely \citep{pasko2012}. Observations indicate that sprites usually occur in clusters of several events \citep[e.g.][]{sentman1995}. Based on the shape, different types of sprites are discerned: "carrot" or "jellyfish" sprites with a heart-shaped body and several tendrils, columniform (C-sprites) sprites appearing as distinct vertical columns, and angle sprites that are bifurcated columns with channels extending diagonally \citep{pasko2012}. They are the most well-studied types of TLE emission.

\begin{figure}
  \centering
  \includegraphics[scale=0.58]{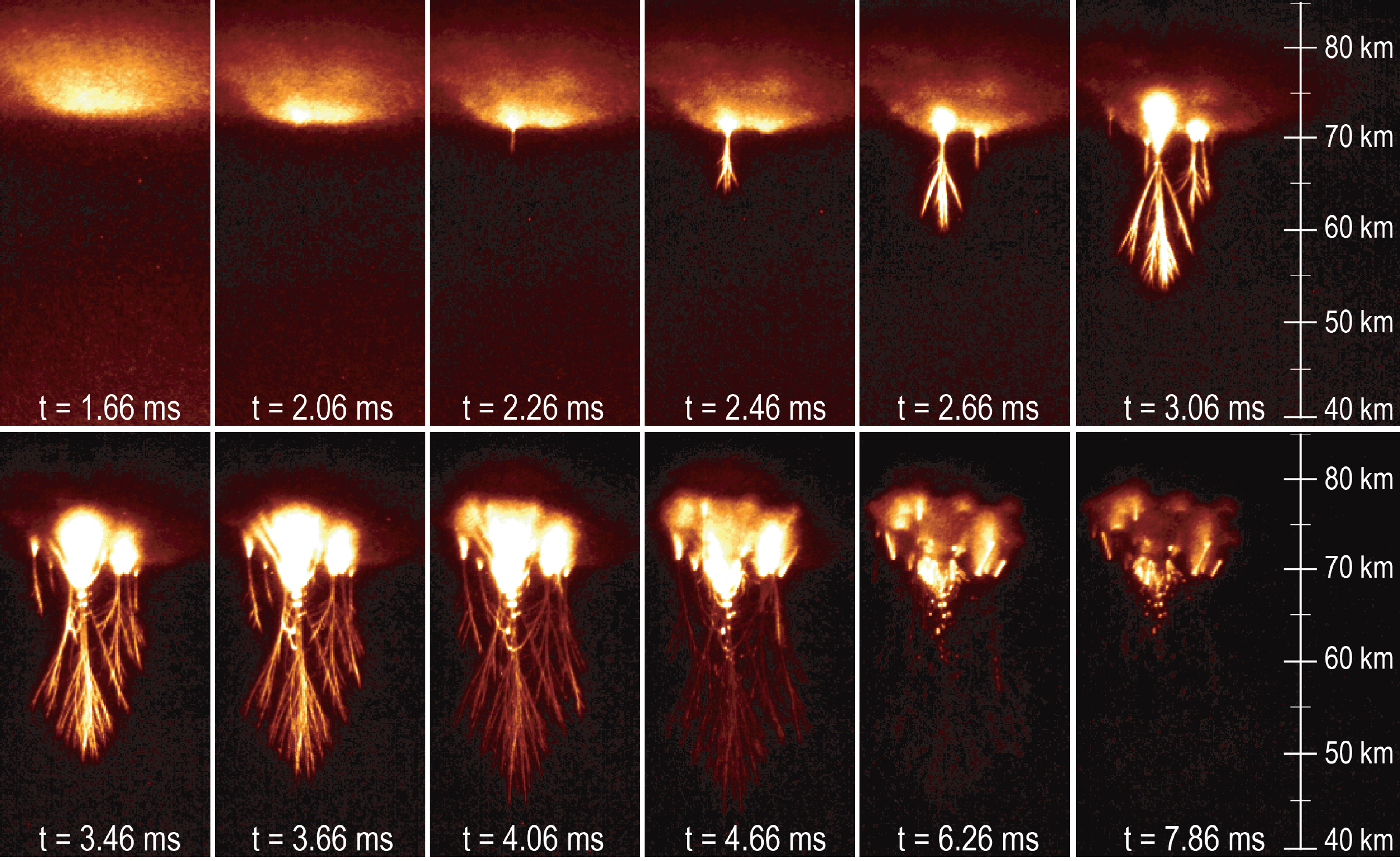}
  \caption{Development of a sprite. Images are labelled with the time after the return stroke initiation. Reproduced with permission from \citet{cummer2006}. Courtesy for the original image to Steven Cummer.}
  \label{fig:2_sprite}
\vspace{0.8cm}
\end{figure}

\item 
\textit{ELVEs} (Emission of Light and Very Low Frequency perturbations due to Electromagnetic Pulse Sources) are diffuse optical flashes with a duration of $< 1$ ms, which occur at $75-105$ km altitude \citep{fukunishi1996}. Their horizontal extension is $\sim100-300$ km. They appear right after the onset of a CG discharge but before sprite initiation \citep{fukunishi1996}. They have specific signatures in the far ultraviolet (FUV) spectral range. \citet{fukunishi1996} suggested that the most likely source for these luminous events is the heating of electrons in the ionosphere by the EMP generated by a lightning discharge. This model explains well the quick horizontal expansion of the phenomenon \citep{frey2005}. 

\item 
\textit{Halos} are diffuse, pancake-like, optical emission occurring at a lower altitude than ELVEs ($\sim70-80$ km). Halos appear before sprites although they are not necessarily followed by sprites \citep{wescott2001}. This unstructured emission lasts for $2-10$ ms \citep{moudry2003}. \citet{barrington-leigh2001} argued that halos are due to a quasi-electrostatic field generated by positive CG lightning discharges.

\item 
\textit{Blue jets} appear mainly with blue colour in the optical. They typically occur in the $40-50$ km altitude range, with a conical shape. Blue jets are thought to occur when a positively charged region is formed on the top of a thundercloud \citep{surkov2012}. \citet{petrov1999} proposed that a blue jet is an upward-propagating positive leader with a streamer on the top, producing the branching structure of the jet. 

\item 
\textit{Gigantic jets} are more energetic than blue jets, they extend to higher altitudes creating a direct link between the troposphere and the lower ionosphere \citep{siingh2012}. Their appearance is more structured than blue jets \citep{pasko2010}. They are believed to appear due to the large amount of negative charges accumulating at the top of thunderclouds \citep{surkov2012}. While blue jets have the same characteristics as a positive leader propagation, gigantic jets show upward-propagating negative leader characteristics \citep{pasko2010,surkov2012}.
\end{itemize}

TLE observations are conducted from the ground \citep[e.g.][]{fukunishi1996, yair2009b} and from space \citep[e.g.][]{yair2006} as well. There were several images take from space shuttles \citep[e.g.][]{yair2003,yair2004} and the International Space Station \citep[e.g.][]{blanc2004, blanc2007,yair2013} too. Important space observations were conducted with the ISUAL (Imager of Sprites and Upper Atmospheric Lightning) instrument on board the FORMOSAT-2 satellite \citep{chern2003}, which was the first instrument dedicated for the investigation of upper atmospheric electrical phenomena \citep{pasko2012}. A major science objective of ISUAL is to investigate the spatial, temporal and spectral properties of TLEs.\footnote{http://www.athena-spu.gr/$\sim$upperatmosphere/index.php?title=FORMOSAT-II\_/\_ISUAL} 
Several microsatellites have been dedicated to the observations of TLEs, such as the Russian Tatjana-1 and Tatiana-2, and the Japanese SPRITE-SAT \citep{pasko2012}.
The previously mentioned (Sect. \ref{sec:highen}) TARANIS satellite and ASIM instrument on board of the ISS will further help the understanding of TLE emission and its relation to thunderstorms in the Earth atmosphere. 

TLEs produce similar EM radiation to lightning discharges occurring inside a thundercloud. Sprites have been observed to produce ELF (extremely low frequency,
$<$ 3 kHz) and VLF (very  low  frequency,  3$-$30 kHz) radio emission, as well as optical emission \citep{surkov2012}.
TLEs in the optical range appear as distinctive red or blue emission depending on the type. Sprites produce strong 1PN$_2$ (first positive band system of N$_2$), 2PN$_2$ (second positive band system of  N$_2$), and 1NN$_2^+$ (first negative band system of N$_2^+$) features, which give the red colour of the event \citep[Table \ref{tab:gab};][]{pasko2007,siingh2015}. Sprites also produce far UV emission in the form of LBH N$_2$ (Lyman-Birge-Hopfield N$_2$ band system) signature. TLEs have a similar effect on the local atmosphere as we have seen for lightning discharges in Sect. \ref{sec:chemeff}. They produce non-equilibrium species, like NO$_x$ in the Earth atmosphere \citep[e.g.][and references therein]{pasko2012}.

TLE emission has only been observed on Earth, though models have predicted their existence on other planets as well \citep[e.g.][]{yair2009,dubrovin2010}. \citet{yair2009} 
modelled sprite occurrence in different planetary atmospheres. They found that on Venus, for an electric charge moment of 500 C km$^{-1}$ in a discharge channel, sprites could occur at 90 km altitude. However, they also mention that due to the supposed low mass content of Venusian clouds, charge separation might not be efficient enough to produce sprite events. For Mars, they found that the conventional breakdown field is exceeded inside dust storms, even for low (10 C) amount of charges at relatively low (10 km) altitudes and not above the dust cloud, suggesting that sprites will not occur in the Martian atmosphere. The results for Titan suggest that lightning could occur inside clouds, however the breakdown field is not reached even hundreds of km above the ground, preventing the development of sprite discharges \citep{yair2009}. \citet{yair2009} also demonstrated that on Jupiter sprits can be expected above lightning-producing, strong convective systems. Depending on the number of charges, the initiation height would be a few hundred km, but still below the ionosphere \citep{yair2009}. \citet{dubrovin2010} conducted laboratory experiments to measure the spectrum of possible sprite events on Jupiter and Venus. They found that Jovian sprites would have a strong continuum spectrum between UV and visible, with higher intensity in the UV part. They also found strong Balmer H$_\alpha$ and H$_\beta$, and various H$_2$ lines (at 575$-$625 nm and at 700$-$800 nm) in the sprite spectrum. Venusian sprites would have strong spectral features from N$_2$-SPS (second positive band of triple heads of N$_2$) in the 300 to 450 nm range \citep{dubrovin2010}. Various CO systems are also found in sprite spectra on Venus at wavelengths between 450 and 670 nm \citep{dubrovin2010}.

\vspace{1cm}
\begin{figure*}[!h]
  \centering
  \includegraphics[scale=0.35]{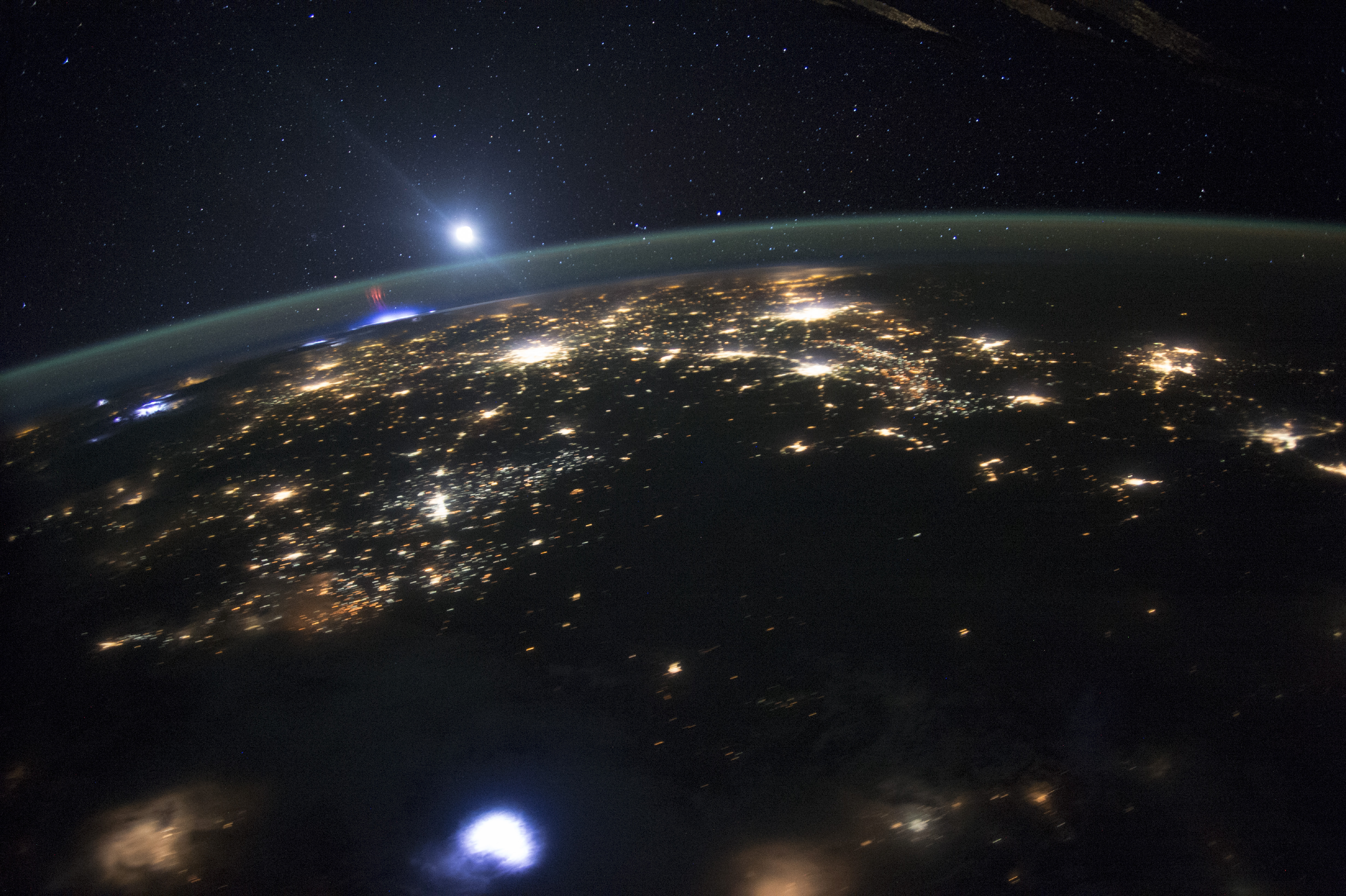}
  \caption{Red sprites above a thunderstorm (left edge of the image) photographed by astronauts on the International Space Station. Another thunderstorm is also visible on the bottom edge of the photo. The big cities are Dallas (to the left) and Houston (to the right). The breath-taking image also shows the atmosphere of the Earth (green airglow) and the bright Moon in the background. Credit: \href{https://earthobservatory.nasa.gov/IOTD/view.php?id=86463}{NASA/NASA Earth Observatory}.}
  \label{fig:space_sprite}
\end{figure*}

%% file: chapters/2c_solsyslightning.tex
\chapter{Lightning inside and outside the Solar System} \label{chap:liginout}

%__________________________________________________________________
%__________________________________________________________________
\section{Extraterrestrial lightning in the Solar System} \label{sec:ligss}

Lightning in the Solar System has been observed on several planets. Optical and radio observations confirmed the presence of lightning on both giant gas planets, Jupiter and Saturn. Radio observations suggest that lightning exists on Uranus, and possibly Neptune and Venus. According to theoretical works, lightning may exist on Mars and Titan as well, however no observations confirmed the theory yet. In this section, I summarize our knowledge on planetary lightning, I do not focus on Earth as such, as the previous chapters (Chapters \ref{chap:ligform}-\ref{chap:ligsig}) describe Earth lightning in much detail. The order I chose to present the information on planetary lightning is intended to reflect the "certainty" of lightning occurring in these atmospheres. On Jupiter (Sect. \ref{sec:int_jup}) and Saturn (Sect. \ref{sec:int_sat}), lightning activity is confirmed by both optical and radio data. It is generally accepted that lightning occurs on Uranus and Neptune (Sect. \ref{sec:int_un}) as well, however apart from Voyager radio data, no other evidence  of lightning activity in these atmospheres has been detected. The research of Venusian lightning (Sect. \ref{sec:int_ven}) is very bi-polar, with scientists arguing both for and against of lightning activity on Earth's twin, with only whistler observations being available currently. Finally, both Mars and Titan (Sect. \ref{sec:matit}) has been suggested to host lightning activity, however no observations have confirmed the theory yet.
For further information on planetary lightning I direct the reader to the following reviews: \citet{desch2002, leblanc2008, yair2008, yair2012, aplin2013}.

%__________________________________________________________________
\subsection{Jupiter} \label{sec:int_jup}

\begin{figure}
\advance\leftskip-1.0cm
\advance\rightskip-1.0cm
\begin{subfigure}[b]{0.6\textwidth}
\includegraphics[width=\textwidth]{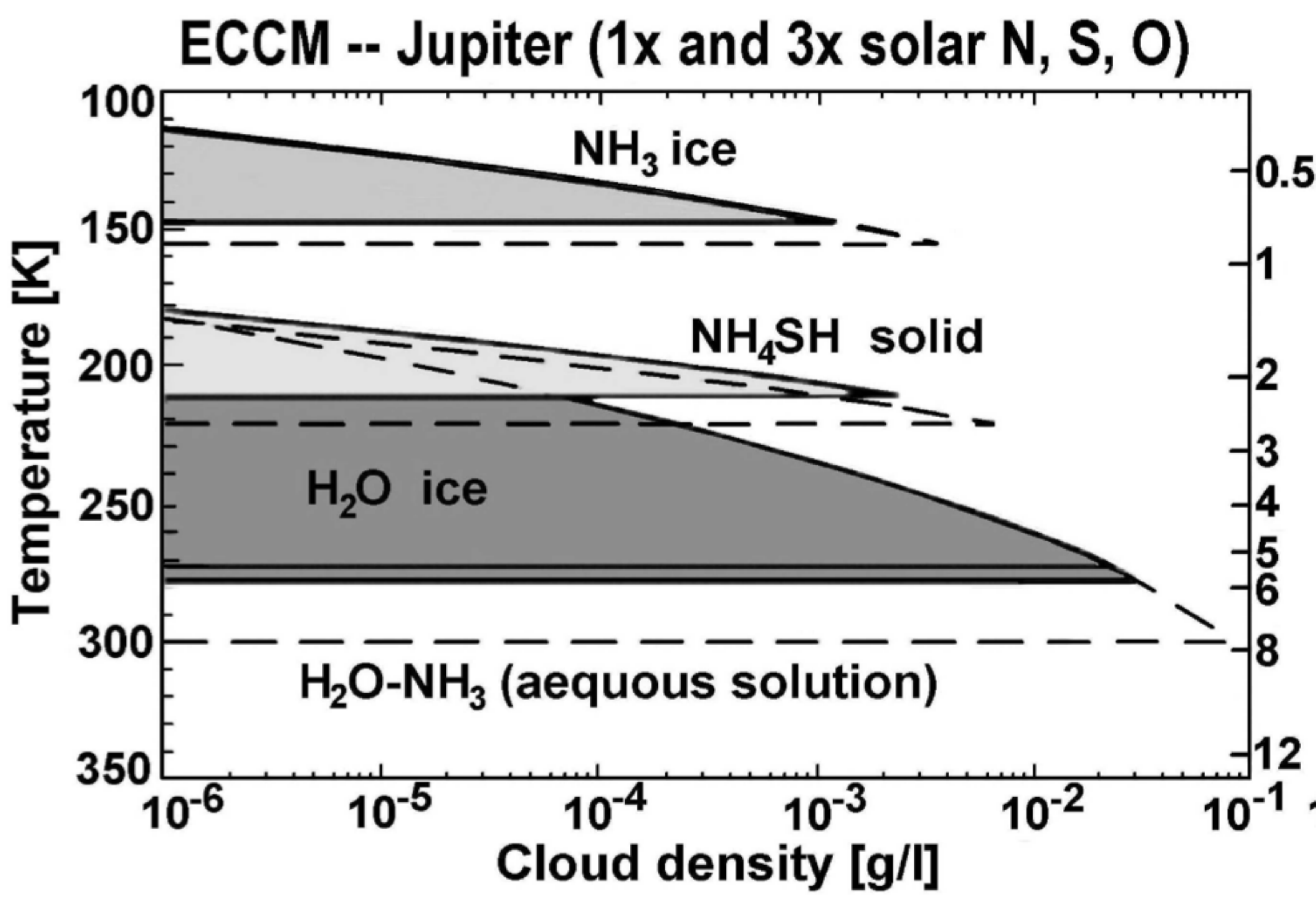}
\end{subfigure} 
\begin{subfigure}[b]{0.6\textwidth}
\includegraphics[width=\textwidth]{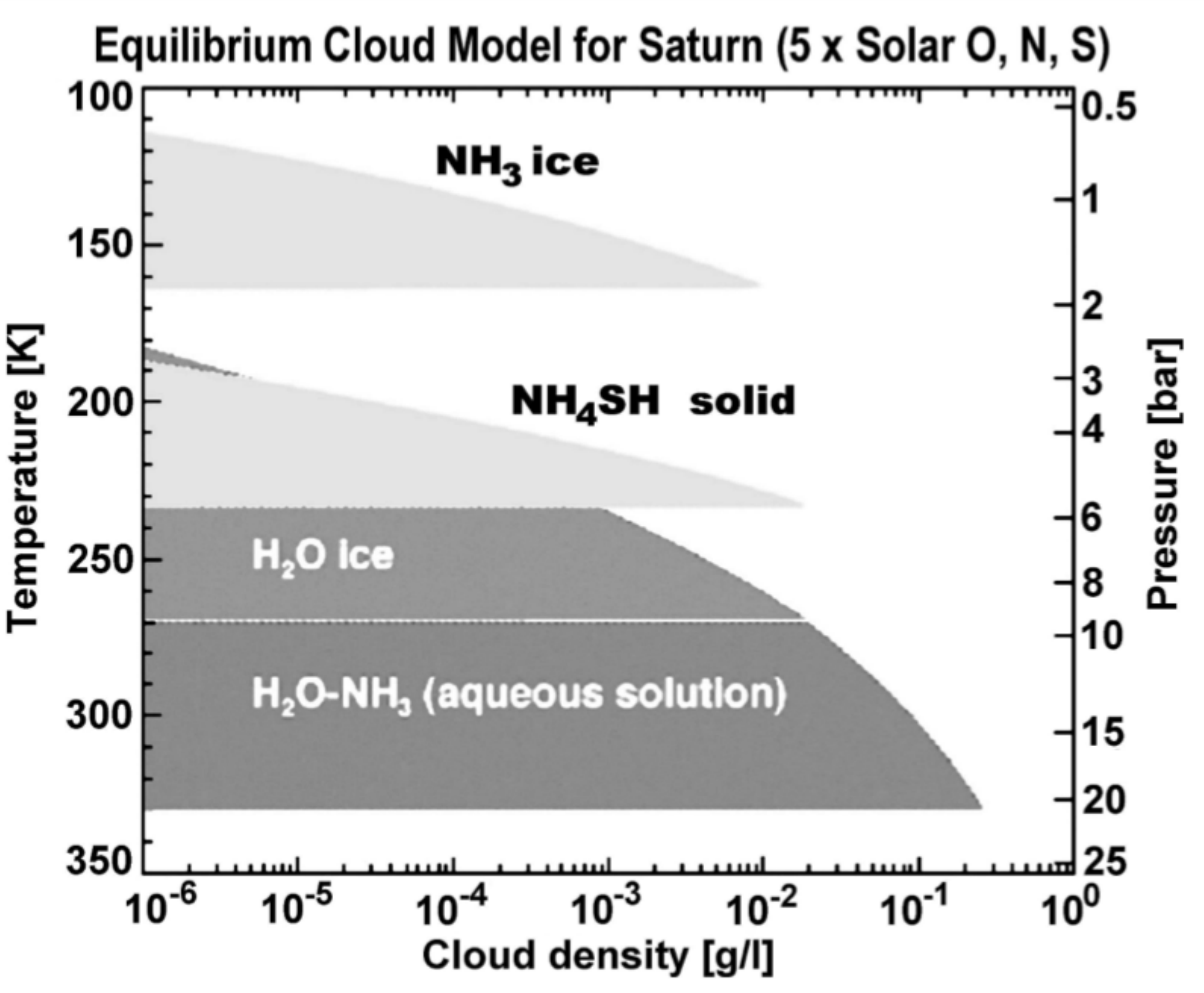} 
\end{subfigure} 
\caption{Equilibrium cloud condensation models (ECCM) for Jupiter (left) and Saturn (right). The models suggest a similar cloud structure for the two giant gas planets. The abundances of condensable volatiles (N, S, O) were taken at 1$\times$ solar (left panel, solid area), 3$\times$ solar (left panel, dashed area), and 5$\times$ solar abundance (right panel). Figures reproduced with permission from \citet{atreya2005}.}
\label{fig:jscloud}
\vspace{0.8cm}
\end{figure}

Lightning was predicted to exist on Jupiter in 1975 based on the observed abundance of molecules such as ethane (C$_2$H$_6$) in the planet's atmosphere \citep{barnun1975}. \citet{barnun1975} suggested that the Great Red Spot hosts an order of magnitude larger electrical activity than the rest of the planet, based on the amount of HCN detected over that area. The first direct evidence of lightning on Jupiter was delivered by the \textit{Voyager 1} and \textit{Voyager 2} spacecraft. Both spacecraft detected optical flashes in the atmosphere \citep{cook1979} and \textit{Voyager 1's} plasma wave instrument observed lightning-induced whistlers in the magnetosphere \citep{gurnett1979,scarf1979}. \citet{magalhaes1991} and \citet{borucki1992}  analysed the Voyager images and found that lightning concentrated at 13.5$^{\circ}$ N and 49$^{\circ}$ N regions, with more brighter spots around 49$^{\circ}$ N. They suggested that the lightning activity was originating from the deep moist convective regions. They did not detect signs of lightning activity on the southern atmosphere of the planet. \citet{yair1995} modelled lightning generation in Jovian water clouds (Fig. \ref{fig:jscloud}), assuming that charge separation works similarly to Earth thunderclouds, with charges accumulating on ice crystals and graupel particles (see Chapter \ref{chap:ligform}, Sect. \ref{sec:separ}). They showed that the electric field in these clouds builds up quickly and exceeds the breakdown field. The lightning flashes in their model produced total energies of the order of 10$^{12}-10^{13}$ J, and optical energies of 10$^{9}-10^{10}$ J. \citet{borucki1996} modelled the optical spectrum of Jovian lightning, which I discuss in Chapter \ref{chap:ligsig} Sect. \ref{sec:optem}. 

\begin{figure}
\begin{center}
\includegraphics[scale=0.25]{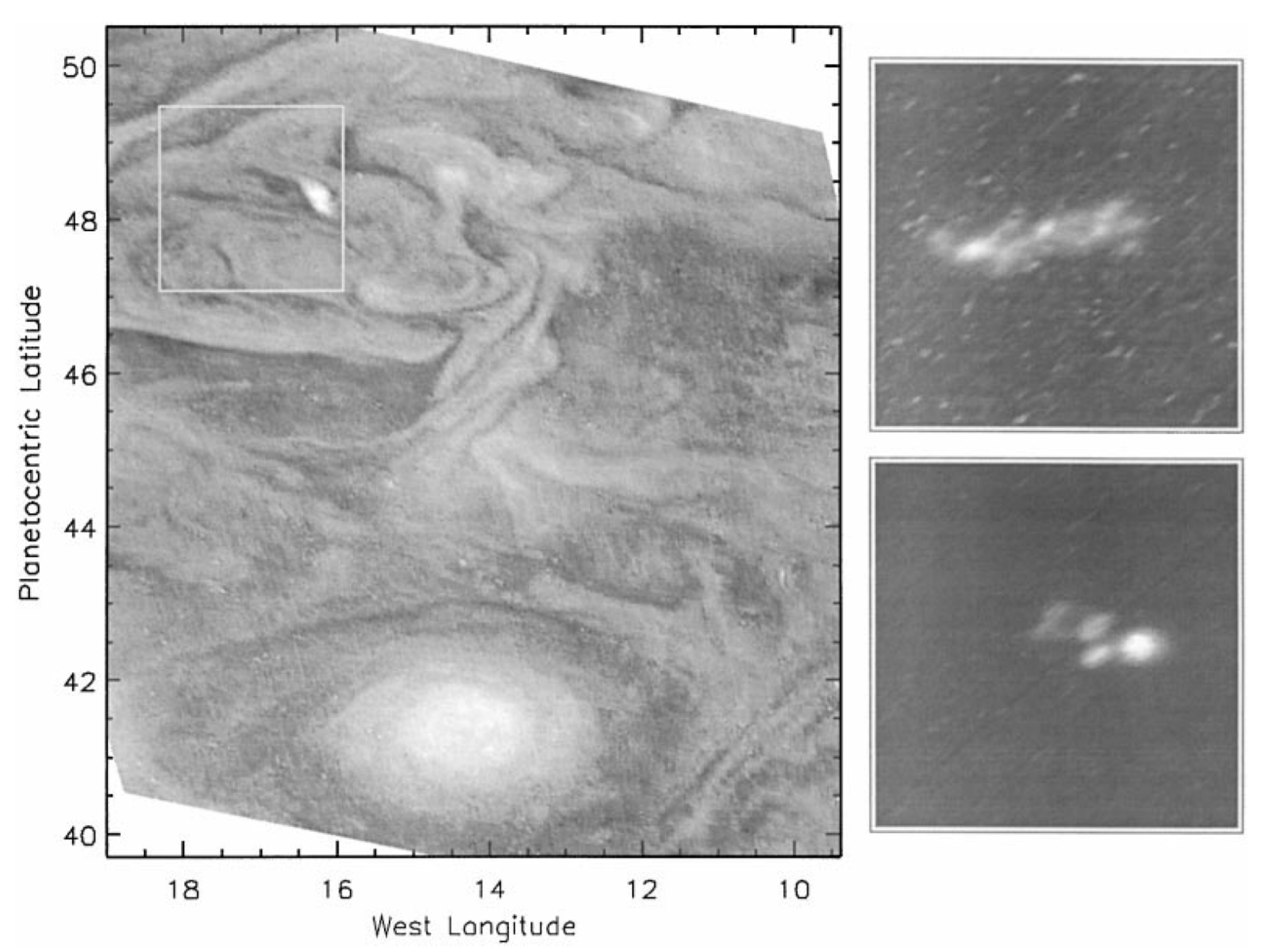} 
\end{center}
\caption{Lightning on Jupiter observed by the \textit{Galileo} spacecraft in 1997. \textbf{Left:} Day-side of the planet. \textbf{Right:} Both panels show the same area marked with a box on the big (left) image, but this time during night time. The bright spots are lightning flashes. Figure reproduced with permission from \citet{little1999}. Courtesy for original figure to Andrew Ingersoll.}
\label{fig:jopt_gal}
\vspace{0.8cm}
\end{figure}

The SSI (Solid State Imager) of the \textit{Galileo} spacecraft observed lightning activity directly on Jupiter (Fig \ref{fig:jopt_gal}) during two orbits in 1997 (C10, E11) and one orbit in 1999 \citep[C20;][]{little1999}. The descending \textit{Galileo} probe also detected lightning induced sferics in 1995 \citep{rinnert1998}. \citet{little1999} estimated the lower limit of flash density on Jupiter, $4.2 \times 10^{-3}$ flashes km$^{-2}$ year$^{-1}$, based on \textit{Galileo} data. This value agrees with the value determined from the \textit{Voyager} measurements  \citep[$4 \times 10^{-3}$ flashes km$^{-2}$ year$^{-1}$,][]{borucki1982}. 
\citet{yair1998} modelled the probability of lightning activity around the entry path of the Galileo probe, and noted that charge separation and lightning generation in these lower-latitude clouds is not intensive. They also found that at mid$-$latitude regions charge separation is more efficient resulting in water-cloud thunderstorms deep in the atmosphere, with positive and negative charged regions at 3 and 4.5 bar levels, respectively. 

Models indicate that Jupiter has three cloud decks made of NH$_3$-ice, NH$_4$SH-ice, and H$_2$O-liquid and -ice extending down from 0.5, 1.3, and 1.6 bar levels, respectively \citep[][Fig. \ref{fig:jscloud}]{atreya2005}. Monte Carlo Radiative Transfer simulations of scattered light in Jovian clouds conducted by \citet{borucki1986} and \citet{dyudina2002} found that the origin of lightning discharges on Jupiter is in the water clouds at 4$-$5 bar pressure levels, or even deeper \citep{dyudina2002}. Lightning was observed on Jupiter by \textit{Cassini} between 2000 and 2001 \citep{dyudina2004}. Correlating lightning flashes with dayside clouds in the \textit{Cassini} data, \citet{dyudina2004} identified the source of lightning on Jupiter to be dense, vertically extended clouds that may contain large particles \citep[$\sim 5 \mu$m,][]{dyudina2001, dyudina2004}, typical for terrestrial thunderstorms. However, they also noted that lightning observed by \textit{Voyager 2} is not always correlated with these bright clouds, meaning that the low number of small bright clouds does not explain the amount of lightning detected by \textit{Voyager 2} that observed fainter flashes at higher latitudes than \textit{Cassini} did \citep{borucki1992, dyudina2004}.

\begin{figure}
\begin{center}
\includegraphics[scale=0.4]{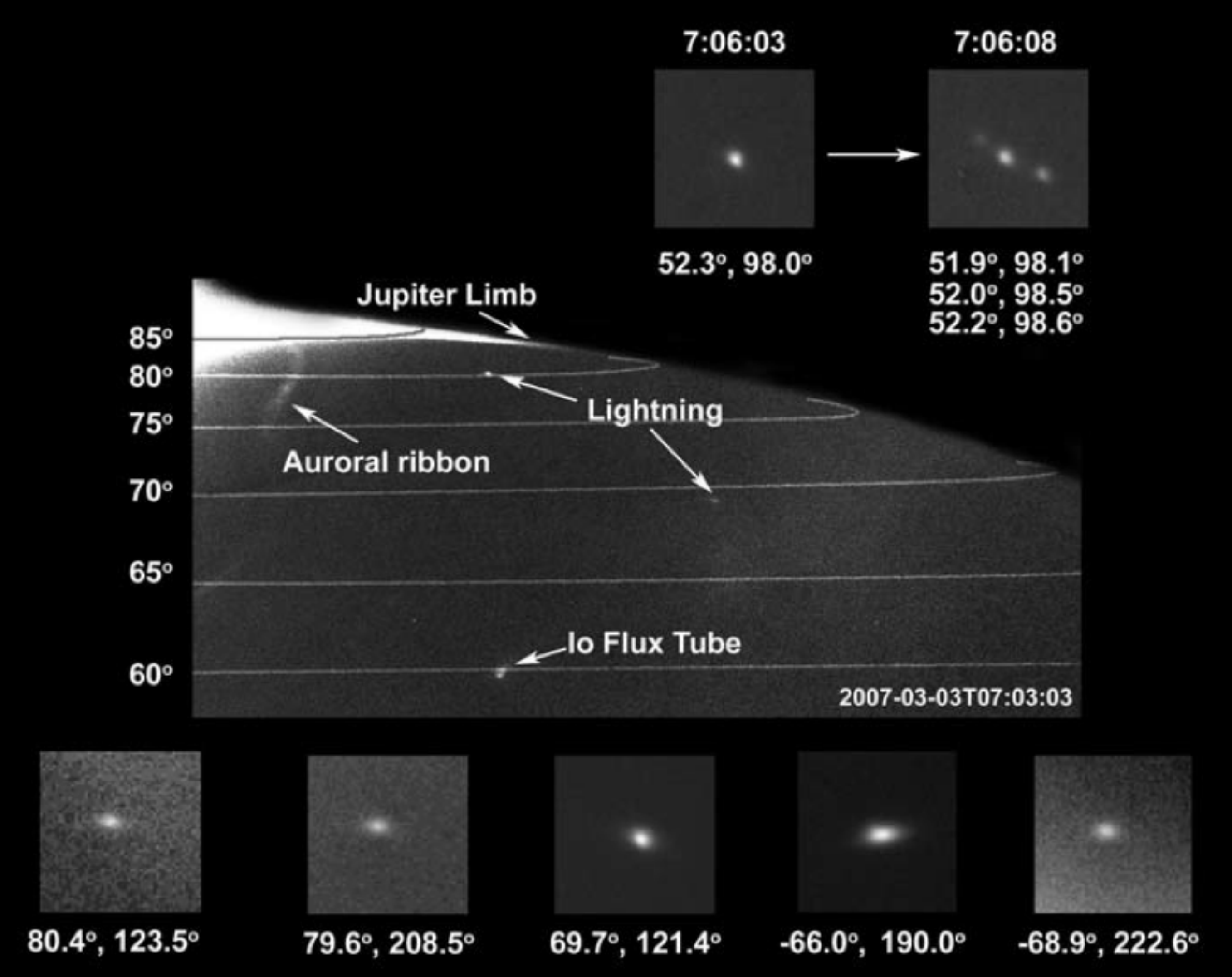} 
\end{center}
\caption{Polar lightning flashes on Jupiter observed by \textit{New Horizons}/LORRI imager in 2007. The spatial extension of the flashes indicates that diffusive aerosols scatter the light originating in the water clouds at 5 bars.
Figure reproduced with permission from \citet{baines2007}.}
\label{fig:jopt_nh}
\vspace{0.8cm}
\end{figure}

\textit{Cassini} detected lightning activity on Jupiter with an $H_\alpha$ filter, around 24$^{\circ}$ N, 34$^{\circ}$ N, and 14$^{\circ}$ S \citep{dyudina2004}. The observations indicated a less intense $H_\alpha$ from lightning compared to what had been expected from laboratory measurements \citep{borucki1996}, which suggests that the source of such flashes is even deeper than 5 bar \citep{dyudina2004,yair2008}. In 2007, \textit{New Horizons} observed sub-polar lightning on Jupiter (Fig. \ref{fig:jopt_nh}), at latitudes $>60^\circ$ north and south, with its broadband camera between 0.35 - 0.85 $\mu$m \citep{baines2007}. Based on these data, \citet{baines2007} calculated almost identical flash rates for the polar regions on both hemispheres (N: 0.15 flashes s$^{-1}$, S: 0.18 flashes s$^{-1}$). 
\citet{luque2015} observed Jupiter from Earth with the GTC/OSIRIS instrument in H$_\alpha$ but did not detect lightning activity on the planet. \citet{luque2014} suggested that faint, transient luminous events analogous to ELVEs on Earth, are produced in the Jovian atmosphere, with energies of $10^8$ J. They addressed TLE emission by modelling and analysing the effects of an upward-propagating electromagnetic pulse produced by a lightning discharge in the planet's atmosphere. 

The JUNO mission reached Jupiter on 5 July, 2016. Its optical camera and plasma wave instrument is capable of detecting optical lightning flashes and lightning$-$induced radio signals in the planet's atmosphere \citep{yair2008}. JUNO will provide further information on cloud dynamics and physical characteristics of the atmosphere\footnote{https://www.nasa.gov/mission\_pages/juno/overview/index.html} down to pressure levels of $\sim 100$ bar \citep{yair2008}.

%__________________________________________________________________
\subsection{Saturn} \label{sec:int_sat}

It has been suggested that Saturn has similar cloud structure to Jupiter's, with a top deck composed of NH$_3$ ice particles, followed by an NH$_4$SH cloud layer, and below that a water ice cloud and liquid water-ammonia \citep[][Fig. \ref{fig:jscloud}]{atreya2005, yair2008}. Saturn has been observed to host large, energetic thunderstorms, which appear infrequently \citep[e.g.][]{fischer2008,dyudina2013,sanchezlavega2016}. \citet{li2015} modelled moist convection in the atmosphere of Saturn, with the aim of explaining the episodic appearance of giant thunderstorms on the planet. They proposed that moist convection is suppressed for decades in Saturn's atmosphere due to the large molecular weight of water in such H$_2$-He rich atmospheres. They suggested that the quasi-periodic thunderstorm activity can be explained by an oscillation produced by the interaction between moist convection and radiative cooling in the troposphere. Such oscillations produce giant storms with a period of $\sim$ 60 years \citep{li2015}.

\begin{figure}
\begin{center}
\includegraphics[scale=0.45]{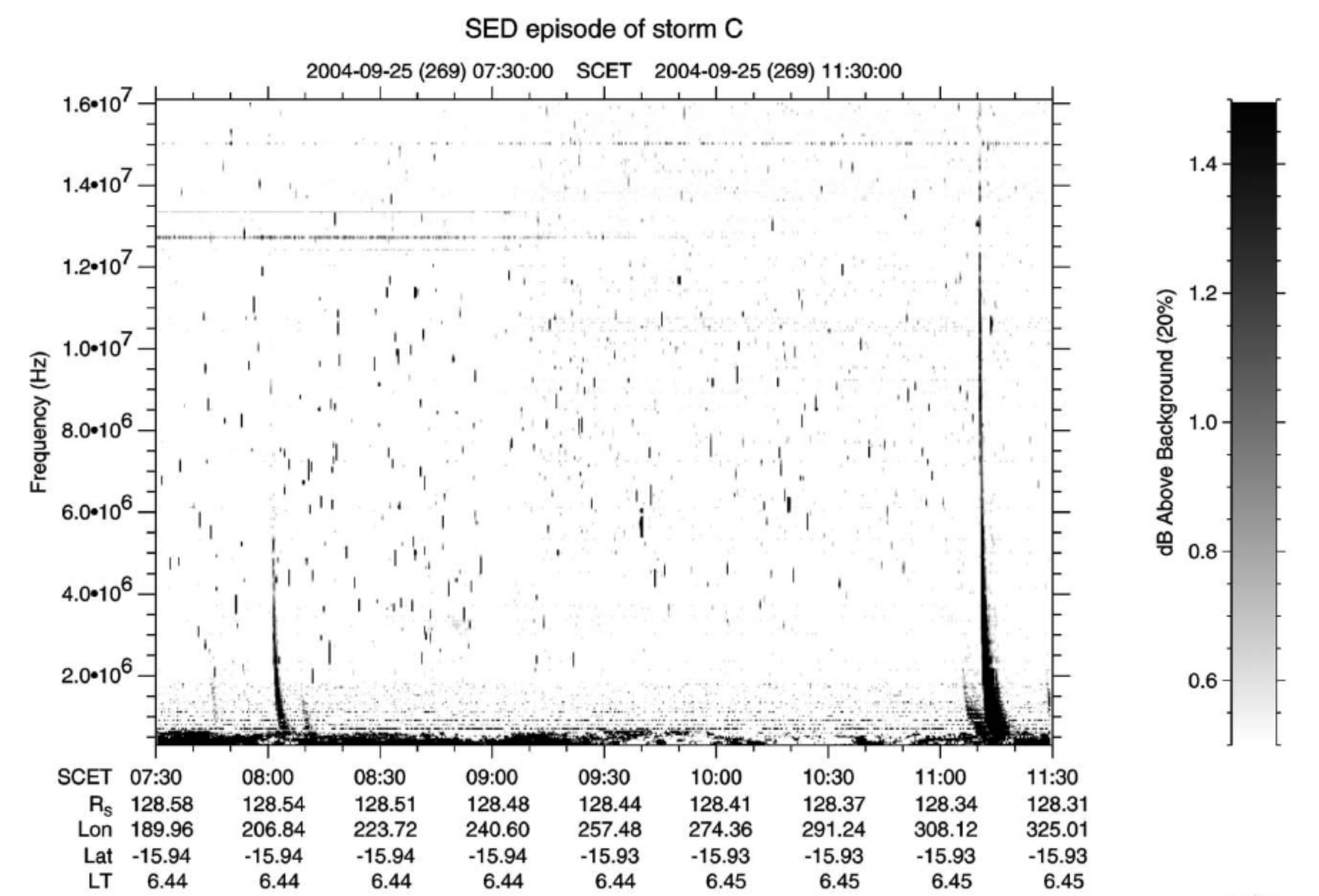} 
\end{center}
\caption{Dynamic spectrum (frequency vs time) of Saturn Electrostatic Discharges (SED; dark black spikes) recorded by the RPWS instrument of the \textit{Cassini} spacecraft in 2004. Spacecraft  event  time  (SCET),  distance  in  Saturn  radii,
western longitude and latitude, and local time of the spacecraft are given along the time axis. 
Figure reproduced with permission from \citet{fischer2006}.}
\label{fig:4_sed}
\vspace{0.8cm}
\end{figure}

\begin{figure}
\begin{center}
\includegraphics[scale=0.5]{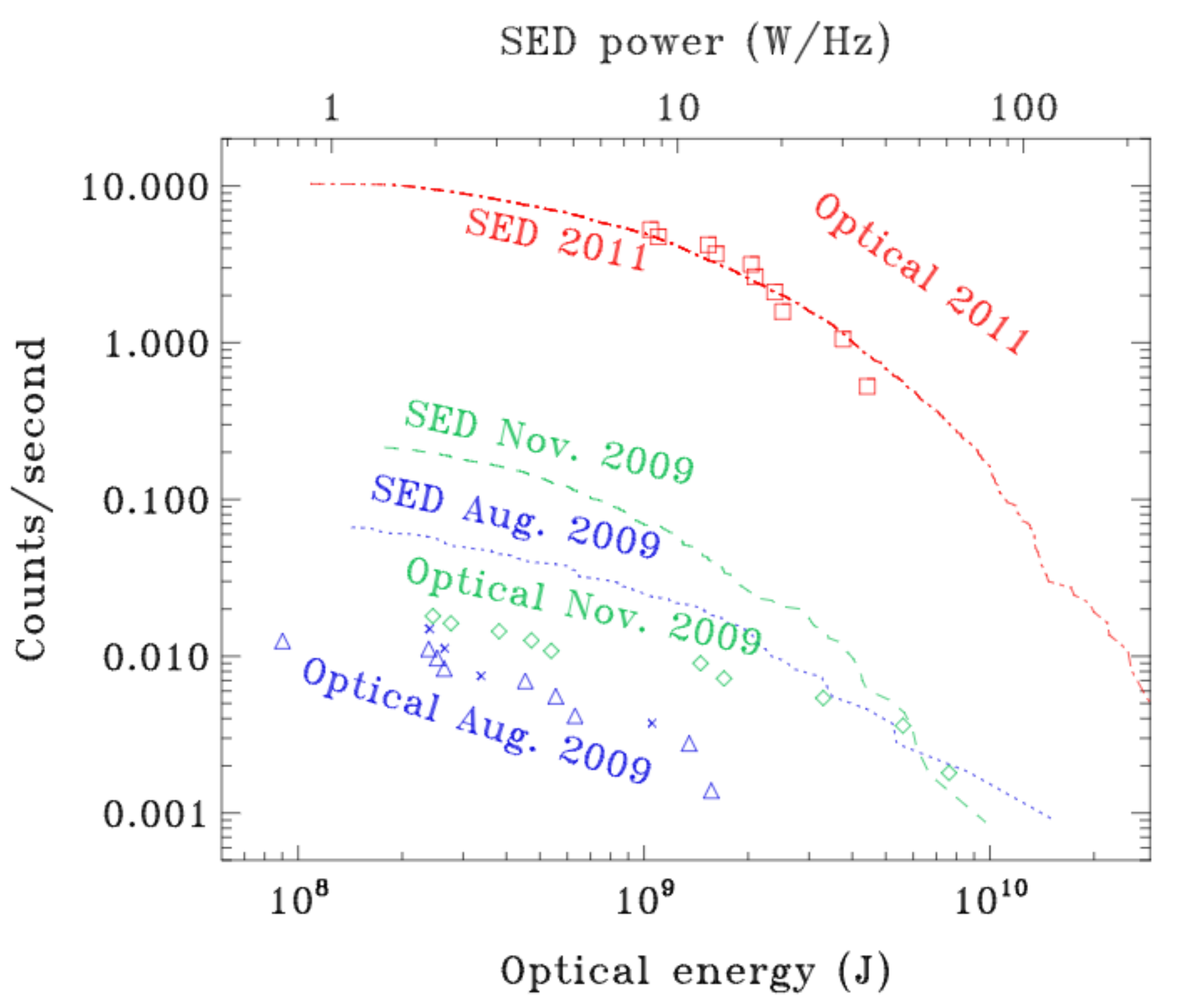} 
\end{center}
\caption{Cumulative distribution of optical lightning energies (bottom x-axis; open plane symbols) and spectral powers of 35-ms long SEDs (top x-axis; dotted, dashed, and dot-dashed lines). The figure indicates that \textit{Cassini} observed the high-energy tale of lightning flashes, assuming the distribution of lightning energies is similar to Earth lightning (see Fig. \ref{fig:endist} in Chapter \ref{chap:stat}). Figure reproduced with permission from \citet{dyudina2013}.}
\label{fig:soptl}
\vspace{0.8cm}
\end{figure}

Lightning-induced sferics named Saturn Electrostatic Discharges (SEDs) were first observed by Voyager 1 during its close approach in 1980 \citep{warwick1981}. The short, strong radio bursts from Saturnian thunderstorms were detected again by the RPWS (Radio and Plasma Wave Science) instrument of the \textit{Cassini} spacecraft in 2004 \citep[][Fig. \ref{fig:4_sed}]{fischer2006}. \citet{fischer2006} and \citet{fischer2007} analysed the occurrence rate of SEDs during the 2004-2006 storms and obtained SED rates that generally vary between 30$-$87 h$^{-1}$, with two storms with SED rates much higher, 367 h$^{-1}$. From Cassini data \citet{fischer2006} estimated the total energy of $10^{12}-10^{13}$ J of Saturnian lightning flashes (Fig. \ref{fig:soptl}), based on the assumption that SED energy output is proportional to the total energy the same way as is for Earth lightning. \citet{farrell2007} suggested that SEDs are much shorter compared to terrestrial discharges and hence the flash should be less energetic, of the order of $10^7$ J. \citet{fischer2011} reported the detection of a giant storm that erupted in December 2010, and examined its SED occurrence. They identified the largest SED rates ever detected on Saturn, to be 36000 SED h$^{-1}$, $\sim 98$ times larger than the SED rate of the largest episode in 2006. 

\begin{figure}
\begin{center}
\includegraphics[scale=0.37]{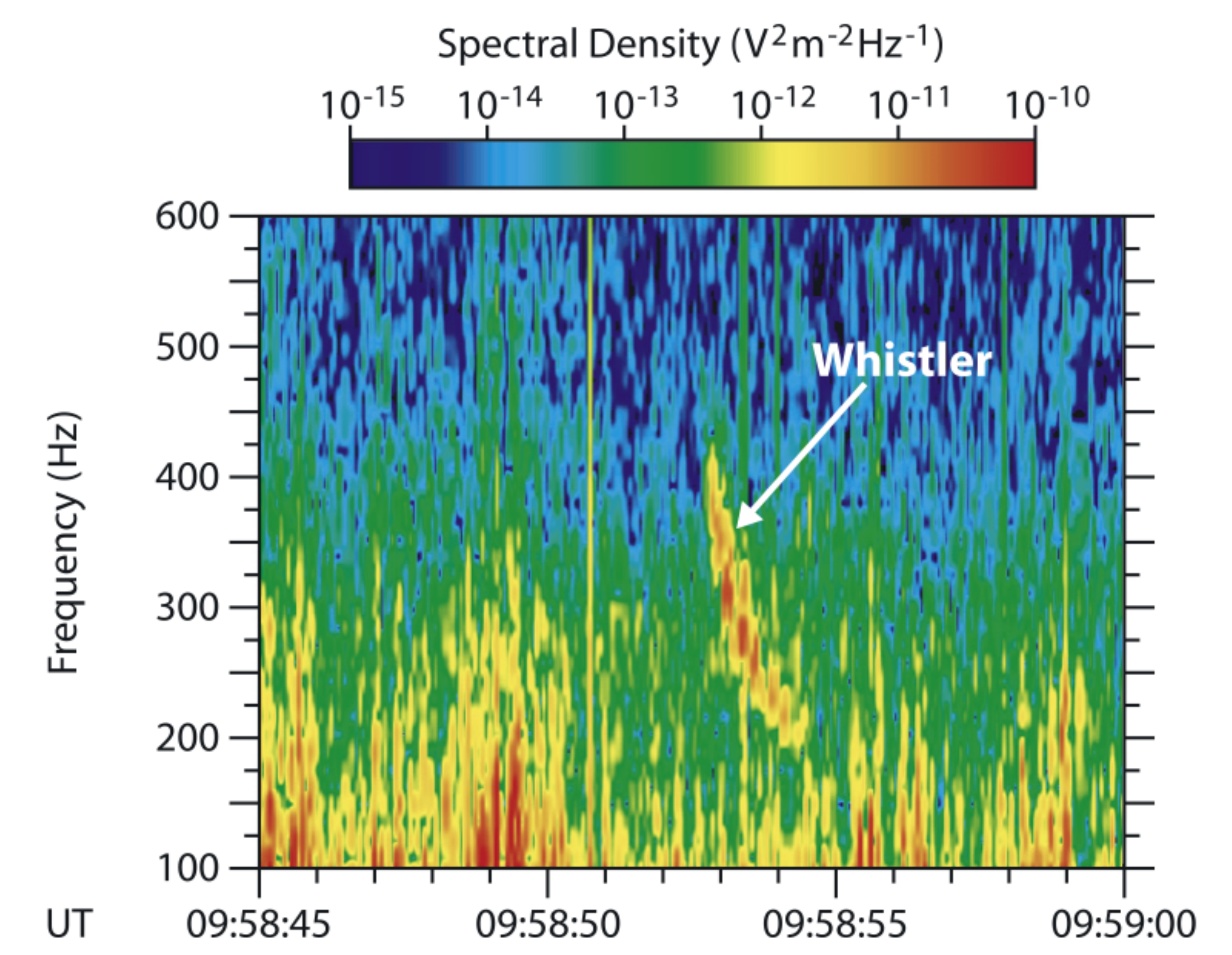} 
\end{center}
\caption{Spectrogram of a whistler event detected by \textit{Cassini}/RPWS. Frequency: 100$-$600 kHz. The time axis covers 15 s. The intensity is given in V$^2$m$^{-2}$Hz$^{-1}$, with red indicating the largest values. Arrow labels the curved feature of the whistler. For comparison to Earth whistlers see Fig. \ref{fig:whistler} in Chapter \ref{chap:ligsig}. Figure reproduced with permission from \citet{akalin2006}.}
\label{fig:swhi}
\vspace{0.8cm}
\end{figure}

SEDs were confirmed to be a signature of lightning activity by the \textit{Cassini} spacecraft, when, based on its data, \citet{dyudina2007} associated the radio emission with clouds visible on the images. \citet{baines2009} detected these clouds with \textit{Cassini}/VIMS, and realized that they appear dark in the NIR. They suggested that lightning at 10 bar level, the base of water clouds \citep[][Fig. \ref{fig:jscloud}, right panel]{atreya2005} produces material, mostly carbon, that is transported to the 1 bar level regime where they obscure the clouds in the spectral range 0.8 to 4.1 $\mu$m. \citet{bjoraker2011} and \citet{hesman2011, hesman2012} found enhanced amounts of C$_2$H$_2$, C$_2$H$_4$ and C$_2$H$_6$ in the atmosphere of Saturn during the 2010/2011 thunderstorm, which species could have been produced by lightning in the deeper cloud regions and transported to lower pressures by updraft \citep{yair2012}.

\citet{akalin2006} reported the first detection of a whistler event in Saturn's magnetosphere. Fig. \ref{fig:swhi} illustrates the radio signal, which shows the same pattern as Earth whistlers do (Fig. \ref{fig:whistler}). Based on \textit{Cassini}/RPWS data, \citet{akalin2006} proposed that the radio signal originated from lightning on the northern hemisphere. \citet{moghimi2012} suggested the detection of a whistler event at 430 to 200 Hz (higher frequency detected first) in \textit{Cassini} data from 2004. They concluded that the source of the emission was lightning that occurred at 66.85$^{\circ}$ N, and suggested an internal energy source for lightning activity.

\begin{figure}
\begin{center}
\includegraphics[scale=0.45]{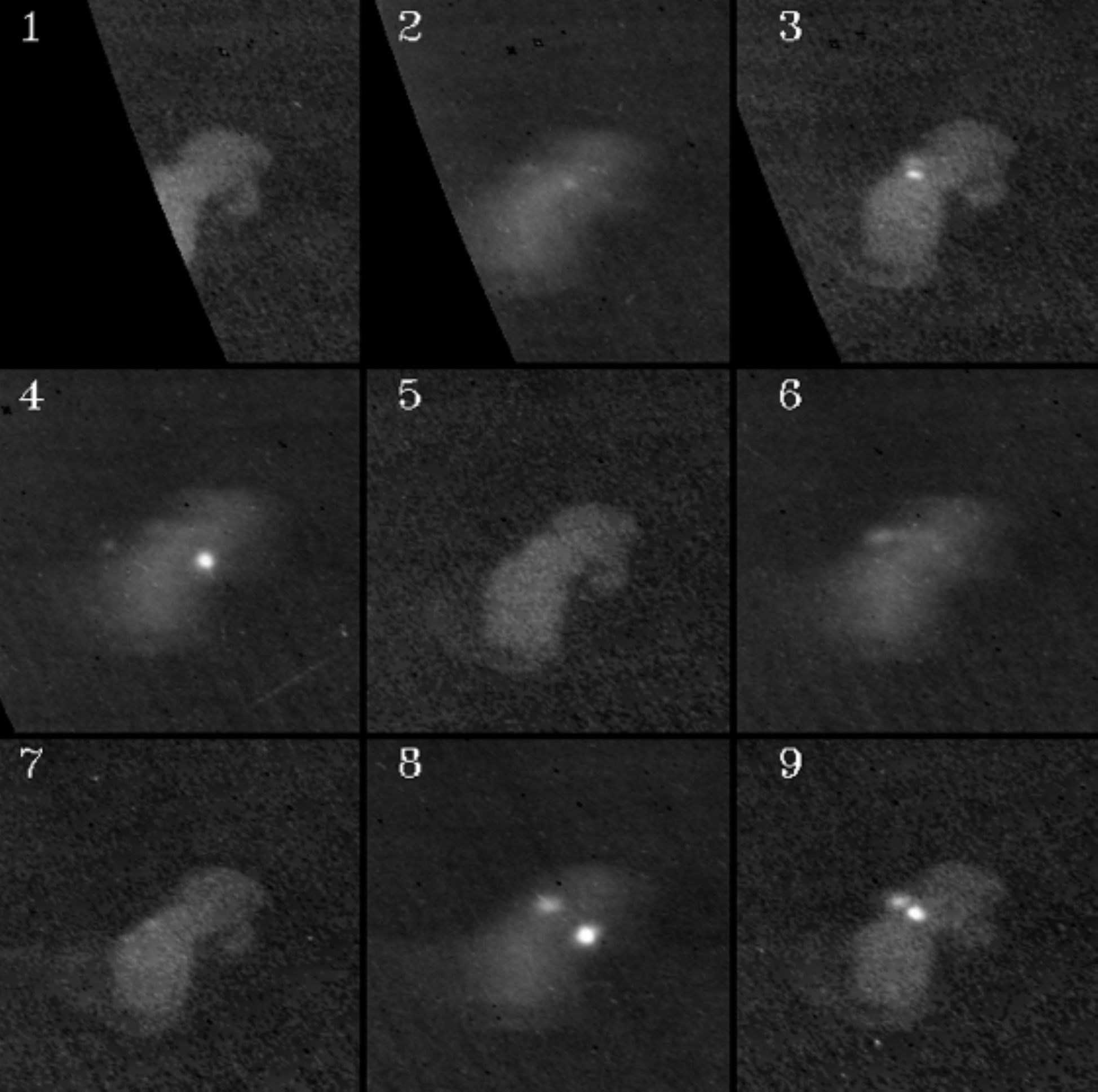} 
\end{center}
\caption{Optical lightning flashes (bright spots) on the night side of Saturn taken by \textit{Cassini} on 30 November 2009. The images are shown in chronological order (following the numbers) as they were taken during a 16-min observation time. The light-grey cloud with diameter of $\sim3000$ km, was illuminated by Saturn's rings and did not change during the observations. 
Figure reproduced with permission from \citet{dyudina2013}.}
\label{fig:soptl2}
\vspace{0.8cm}
\end{figure}

The first Saturnian lightning detection in the visible range was reported by \citet{dyudina2010}. They detected optical lightning flashes on the night side of the planet in the \textit{Cassini} wide- and narrow-angle camera data taken in 2009. \citet{dyudina2010} treated the lightning spots as light sources on top of the cloud emitting isotropically up and down, and estimated the optical energy of a single lightning flash to be 10$^9$ J (Fig. \ref{fig:soptl}). This way they confirmed the high total energy output of lightning on Saturn as was originally inferred from SED data \citep[Fig. \ref{fig:soptl};][]{fischer2006}. \citet{dyudina2010} also mentioned that these flashes maybe the most energetic ones occurring on Saturn, as the detection limit of the cameras is $10^8$ J in terms of lightning optical energy. Based on the observed diameter of the lightning spots (200 km), \citet{dyudina2010} inferred a source altitude of 125-250 km below cloud tops, which is above the liquid water-ammonia cloud base, probably in the NH$_4$SH cloud or in the H$_2$O ice cloud (Fig. \ref{fig:jscloud}, right panel). \citet{dyudina2013} reported further lightning detections (Fig. \ref{fig:soptl2}) on the dayside by \textit{Cassini} at latitude $35^{\circ}$ N, from a new, much stronger storm than previous storms, observed in February, 2011 \citep[also reported in][]{fischer2011}. 

\citet{zarka2004} estimated the detectability of planetary lightning with the state-of-the-art radio array, LOFAR (Low Frequency ARray), and suggested that SED activity could be monitored by the instrument, and due to the sporadic occurrence of Saturnian thunderstorms, preferably on a regular basis. \citet{griessmeier2010} and \citet{griessmeier2011} presented the results of ground-based search for lightning on Saturn using three arrays, LOFAR, UTR-2 (Ukrainian T-shaped Radio telescope), and the Dutch WSRT (Westerbork Synthesis Radio Telescope). \citet{zakharenko2012} described the simultaneous observations of Saturnian lightning activity with UTR-2 at frequencies 12 to 33 MHz, and the \textit{Cassini} spacecraft at 1.8 to 16 MHz. They noted a good coincidence between the data of the two instruments and reported the first ground-based detection of SEDs. \citet{mylostna2013} conducted further observations of SEDs with UTR-2. They obtained high time-resolution data ($\mu$s resolution), and found that SEDs are composed of 100 $\mu$s bursts with varying intensity, and no finer structure was observed, which is consistent with a high energy release of $10^{12}-10^{13}$ J, since a discharge with such duration and observed radio power release \citep[$\sim 50$ W Hz$^{-1}$,][]{fischer2006} has to be very energetic \citep{farrell2007}. They also reported the low frequency ($<$200 kHz) power spectrum of SEDs, which showed an intensity peak around 17 kHz, and a spectral variation of $f^{-2}$ between 20 and 200 kHz. \citet{konovalenko2013} summarized the earliest ground-based detections of Saturnian lightning with the UTR-2.

\citet{dubrovin2014} modelled the effects of lightning activity on the bottom of the ionosphere (1000 km) of Saturn, and showed that a conservative estimate of charge moment ($10^4-10^5$ C km) produced during a lightning flash could result in the production of transient luminous events in the form of halos and sprites. However, if the ionosphere is lower (600 km), a large (10$^6$ C km) moment is needed to produce such events. They also suggested that the blue/UV emission from such TLEs would be very faint and not detectable by \textit{Cassini}. \citet{luque2014} found that lightning induced electromagnetic pulses could carry energies of the order of $10^7-10^{10}$ J to the ionosphere and produce ELVE-like TLEs with energies of 10$^8$ J.

%__________________________________________________________________
\subsection{Uranus and Neptune} \label{sec:int_un}

\begin{figure}
\begin{center}
\includegraphics[scale=0.3]{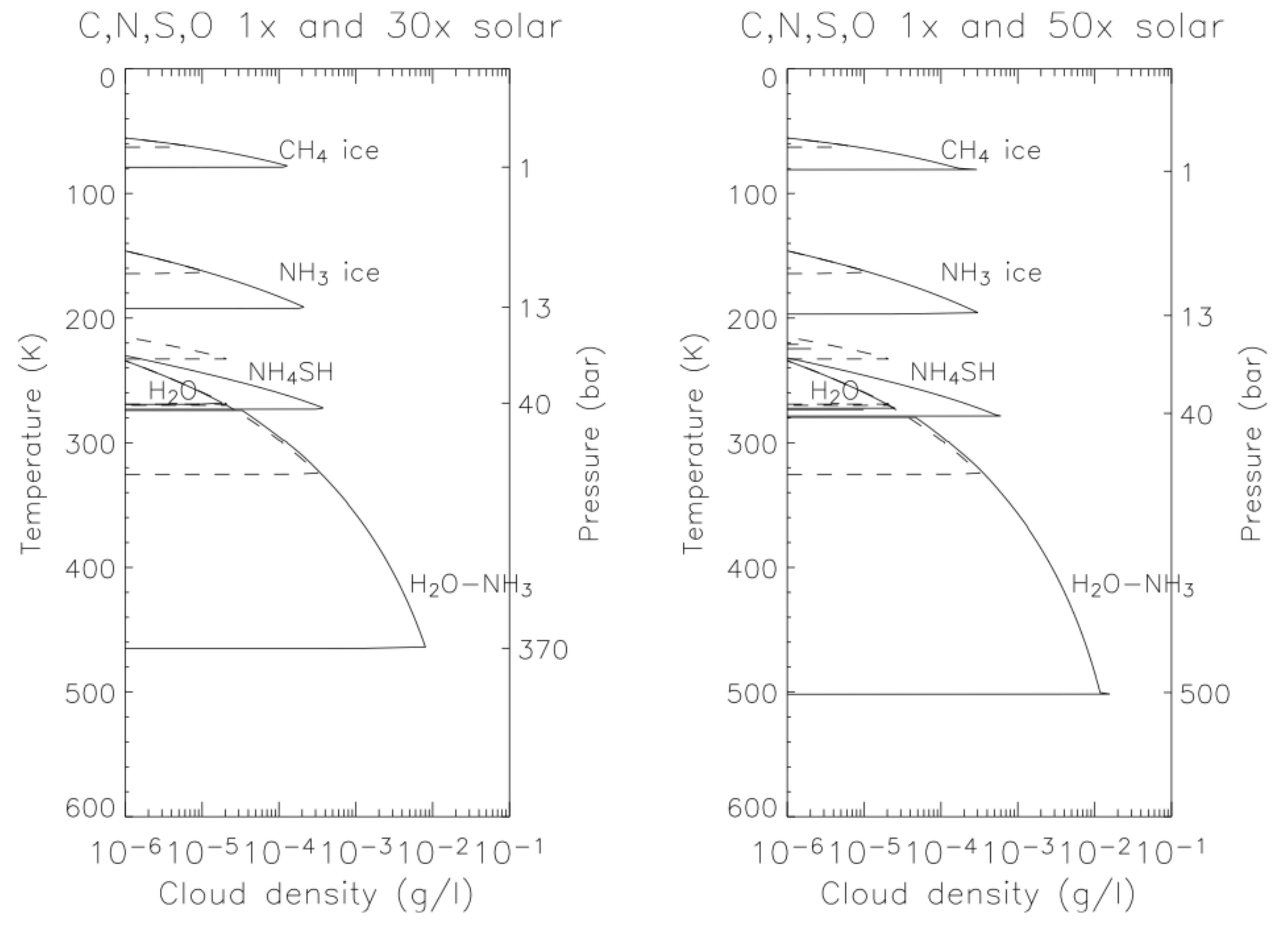} 
\end{center}
\caption{Equilibrium cloud condensation models (ECCM) for Neptune. The cloud structure for Uranus is similar because of the similar thermal structure and atmospheric density. The abundances of condensable volatiles (C, N, S, O) were take at 1$\times$ solar (dashed line, both panels), 30$\times$ solar (left panel, solid area), and 50$\times$ solar abundance (right panel, solid area). The cloud density is represented on the x-axis, and shows an upper limit. Figure reproduced with permission from \citet{atreya2005}.}
\label{fig:nucloud}
\vspace{0.8cm}
\end{figure}

Uranus and Neptune host several cloud layers including methane (CH$_4$) at high altitudes, around 1 bar pressure, followed by H$_2$S$-$NH$_3$, NH$_4$SH, and water layers \citep[][Fig. \ref{fig:nucloud}]{depater1991,gibbard1999}. Due to the lower gravitation and smaller temperature gradient, water clouds condense at higher pressures on Uranus and Neptune than on Jupiter and Saturn \citep{yair2008}. \citet[][Fig. \ref{fig:nucloud}]{atreya2005} demonstrated with cloud models that the base of water clouds on the ice giants is around 40 bar pressure level.

\citet{zarka1986} reported the detection of SED-like radio burst from the atmosphere of Uranus. In 1986, \textit{Voyager 2}/PRA (Planetary Radio Astronomy) experiment observed the bursts with duration from 100 to 300 ms, in the frequency range $\sim 900$ kHz $-$ 40 MHz. \citet{zarka1986} pointed out that the emission is different from what was previously observed from the planet's magnetosphere by \textit{Voyager 1}. They attributed the bursts to lightning activity and adopted the term UED (Uranian Electrostatic Discharges) as an analogue to SEDs (Saturnian Electrostatic Discharges). They found the UED spectrum to decrease smoothly from low to high frequencies, with $f^{-2}$. They obtained an average power density in the low and high frequency bands of 2 and 60 W Hz$^{-1}$, and calculated a total radiated power of 10$^8$ W for UEDs. Assuming an average flash duration of 120 ms, \citet{zarka1986} found the average radiated energy of UEDs to be $10^7$ J.

On Neptune, four possible sferics \citep{kaiser1991} and sixteen whistler events \citep{gurnett1990} were recorded by \textit{Voyager 2}. \citet{gurnett1990} reported the detection of the whistler events at frequencies between 6.1 and 12 kHz, at the closest approach of \textit{Voyager 2} at low magnetic latitudes. They suggested that the origin of the whistlers is lightning on the day-side of the planet, and that the signal was bouncing back-and-forth from the ionosphere until it reached a frequency higher than the plasma frequency and escaped along a magnetic field line. \citet{kaiser1991} analysed the four sferic-like events detected by the PRA on \textit{Voyager 2}. The events were very close to the noise level, one at 15 MHz and three around 20 MHz. Several other signals were also present in the data, however they were very contaminated by spacecraft noise, therefore were eliminated from the analysis process \citep{kaiser1991}. There can be several reasons why so few sferics were detected from Neptune: lightning can be less energetic than on Jupiter or Earth; it could be much slower resulting in less power released at higher frequencies; or absorption due to the ionosphere \citep{kaiser1991}. \citet{borucki1992b} conducted an optical search for Neptunian lightning in the \textit{Voyager 2} wide- and narrow-angle imager data. The analysed images covered 94\% of the surface of the planet, but no lightning spots were detected. The sensitivity of the imagers was known to be good enough to observe lightning spots on Jupiter at similar distances, which suggest that lightning on Neptune is at least 1/4 less frequent than on Jupiter, assuming the same cloud-top brightness \citep{borucki1992b}. \citet{gibbard1999} applied a charge separation model to investigate possible lightning activity on Neptune, and found that lightning most likely occurs in the top H$_2$S$-$NH$_3$ cloud layer (Fig. \ref{fig:nucloud}) if collisional charge transfer is sufficiently large. They also suggested that it is less possible to produce lightning discharges in the lower NH$_4$SH and water cloud decks because of the high pressure. 

Though the reports on lightning induced radio emission from the two ice giants are convincing, lightning activity on Uranus and Neptune has not been confirmed by other instruments yet \citep{aplin2013}. \citet{zarka2004} estimated the detectability of planetary lightning by LOFAR, and found that lightning on Uranus and Saturn are the best and most promising candidates for follow-up observations from the ground, because they have a high enough flux density, they last for long enough, and their spectrum reaches high enough frequencies for LOFAR detections. Lightning on Neptune, however, will remain undetectable even with LOFAR from the ground \citep{zarka2004}.

%__________________________________________________________________
\subsection{Venus} \label{sec:int_ven}

\begin{figure}
\begin{center}
\includegraphics[scale=0.35]{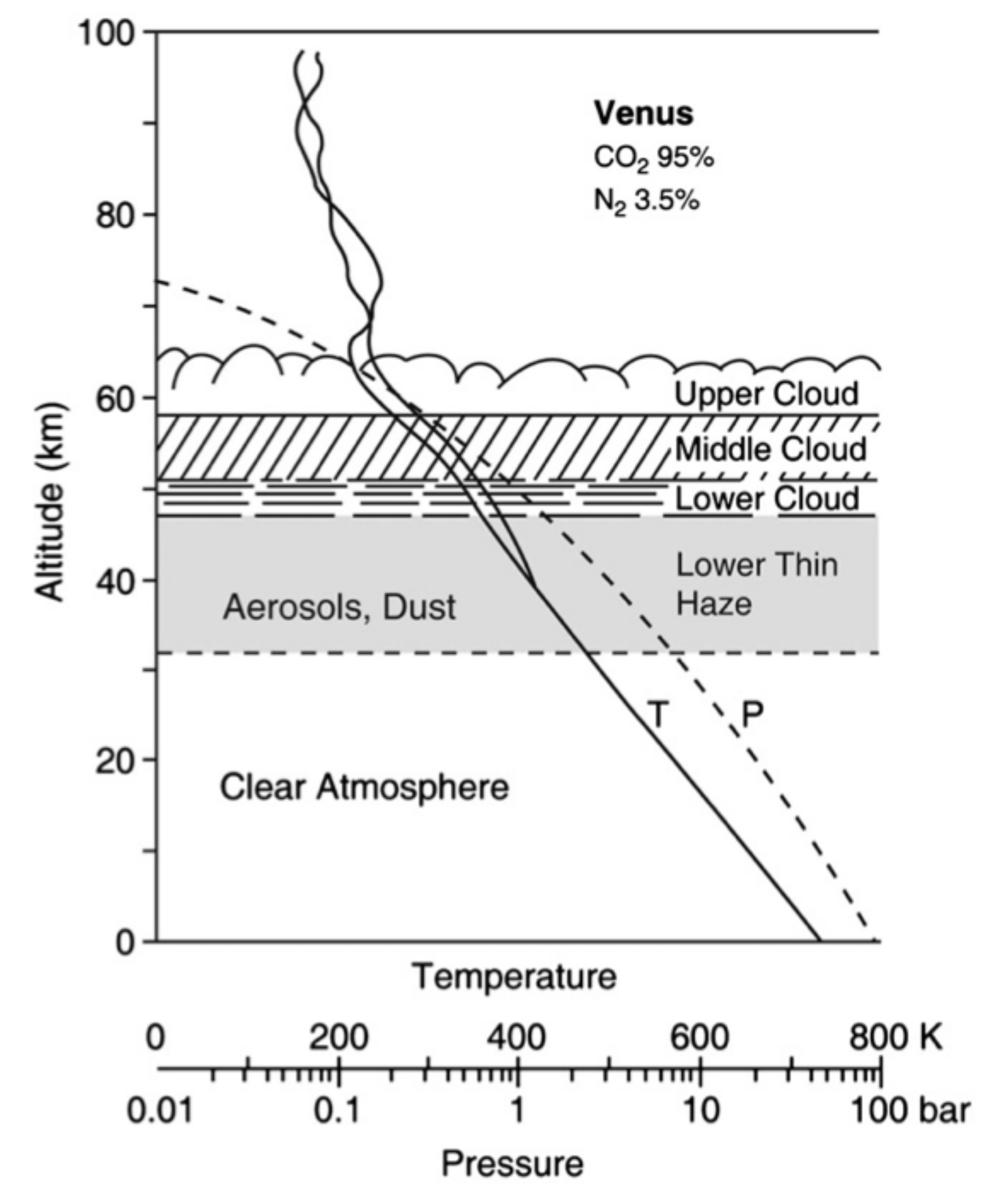} 
\end{center}
\caption{Cloud structure, and temperature and pressure profile of the Venusian atmosphere. Figure reproduced with permission from \citet{russell2011}.}
\label{fig:vcloud}
\vspace{0.8cm}
\end{figure}

Venus is fully covered by a super-rotating (rotating quicker than the planet itself) cloud system, composed of small sulphuric acid ice crystals \citep{yair2012}. The system is built up of three cloud decks between 45 and 75 km altitudes and H$_2$SO$_4$ haze above the cloud tops \citep[Fig. \ref{fig:vcloud}][]{yair2012}.  The cloud particles are thought to be mostly charged by cosmic rays \citep{michael2009}. \citet{michael2009} investigated the accumulation of charges in cloud particles in the Venusian atmosphere, and found the ratio of negative and positive charges to be large in the middle and upper cloud layers. If lightning on Venus exists, it may appear as intra-cloud or cloud-to-cloud discharge because of the high atmospheric pressure, which would not allow cloud-to-ground discharges to occur unless the electric field becomes extremely high \citep{yair2008}.

Though the work of \citet{michael2009} indicated that the low aerosol abundance and high conductivity found in the 40 to 70 km altitude region rules out lightning activity in this part of the Venusian atmosphere, the presence of lightning on Venus has been suggested by multiple observations since the late 1970s. \citet{krasnopolski1980, krasnopolski1983} announced the detection of optical signatures of lightning with the spectrometer of the \textit{Venera 9} spacecraft. \citet{ksanfomaliti1980} reported lightning detection based on the radio data gathered by the \textit{Venera 11} and \textit{Venera 12} landers. \citet{scarf1980} presented whistler detections by the \textit{Pioneer Venus Orbiter} (PVO). However, these early observations were not widely accepted. \citet{taylor1987}, for example, interpreted the VLF radio signals as interplanetary magnetic field/solar wind related perturbations appearing around the PVO spacecraft. Since then several attempts have been made to detect lightning on Venus, and the controversy of the existence of lightning on the planet has not yet been resolved. 

The \textit{Galileo} spacecraft flew by Venus in 1990 and scanned the planet's night side looking for lightning signatures. \citet{gurnett1991} reported the detection of 9 pulse-like radio events with intensities slightly above the detector's noise level, between frequencies 100 kHz and 5.6 MHz. They suggested that the 9 events were caused by lightning activity, as the observed values tend to show decreasing intensity with increasing frequency, and are comparable with what would have been expected if Earth-lightning was observed at similar conditions (i.e. spacecraft-planet distance).  \citet{hansell1995} observed the night-side of Venus with the 153-cm telescope at Mt Bigelow, Arizona, and claimed the detection of 7 lightning flashes on Venus, 6 at 777.4 nm and 1 at 656.4 nm. They estimated the flash rate on Venus to be 1000 times smaller than on Earth, and the energy of lightning flashes to be around $10^8-10^9$ J. The \textit{Cassini} spacecraft made two close fly-bys of Venus in 1998-99, but did not detect lightning induced radio emission in the low frequency range above 1 MHz \citep{gurnett2001}. Based on the non-detection, \citet{gurnett2001} calculated a lower limit to the flash rate of 70 s$^{-1}$, which is almost twice the average global flash rate on Earth \citep[44 s$^{-1}$,][]{christian2003}. \citet{borucki1996} simulated the spectra of lightning on Venus, which I discuss in Chapter \ref{chap:ligsig} Sect. \ref{sec:optem}. \citet{krasnopolsky2006} inferred a flash rate of 90 s$^{-1}$ with discharge energy of 10$^9$ J from lightning-related NO detections in the infrared spectra of Venus. Other attempts of optical observations were conducted by \citet{garcia2011} to observe the 777 nm O emission line with several advanced instruments such as the Caral Alto 3.5-m telescope and the 10.4-m Gran Telescopio Canarias, however no detection was reported, which suggest a rare Venusian lightning occurrence or at least that it is less energetic than Earth lightning.

\begin{figure}
\begin{center}
\includegraphics[scale=0.45]{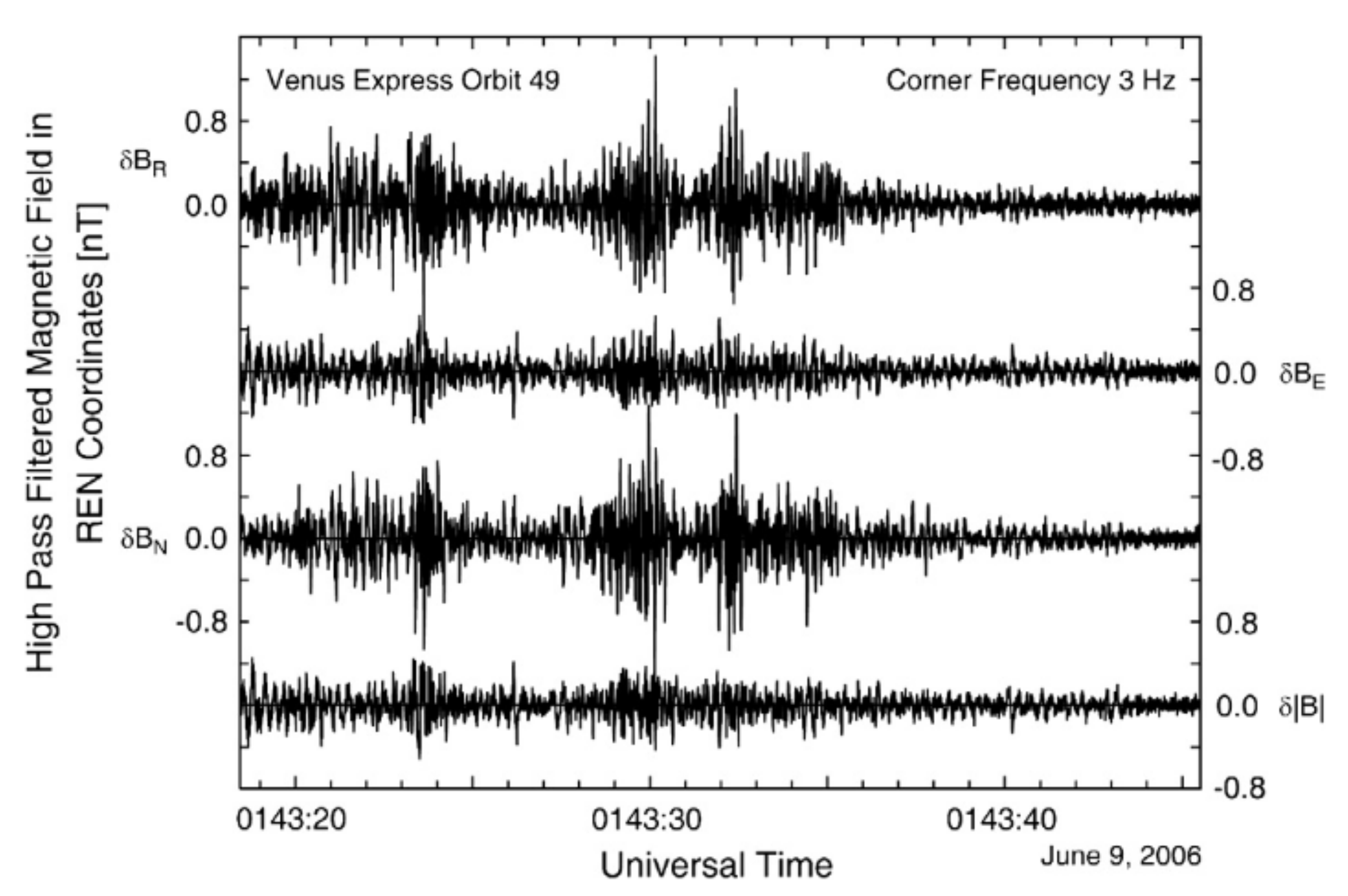}
\includegraphics[scale=0.45]{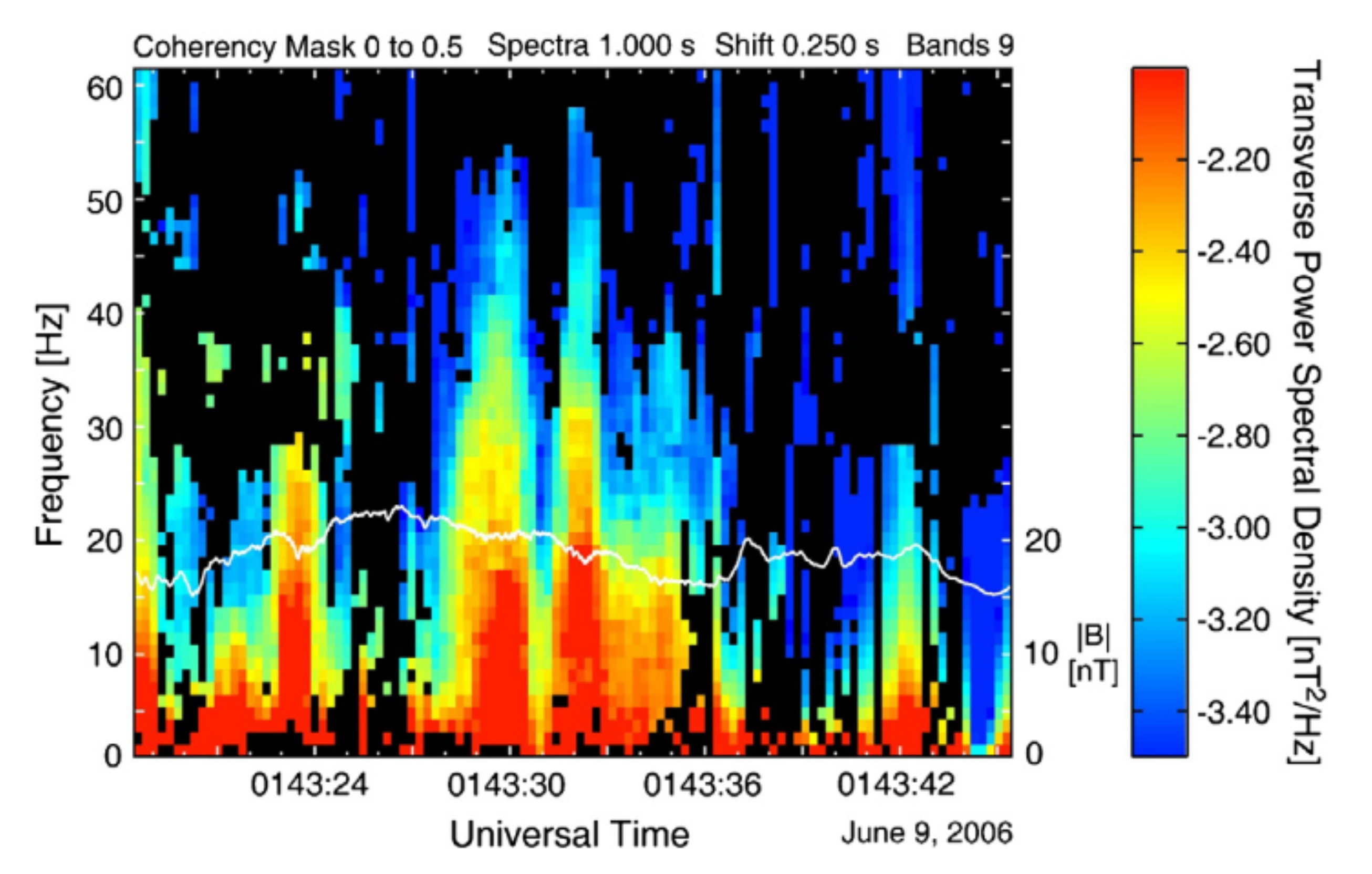} 
\end{center}
\caption{\textit{Venus Express} magnetic data of whistler mode emission on June 9, 2006, from 0143:18.5 to 0143:45.5 UT. \citet{russell2013} suggested that the whistler bursts (e.g. at 0143:30 on both panels) were emitted by lightning in Venus' atmosphere. \textbf{Top:} High-pass filtered magnetic field measurements. \textbf{Bottom:} Dynamic spectrum of the transverse power spectral density. The white line shows the magnetic field strength. Figure reproduced with permission from \citet{russell2013}.}
\label{fig:vexpr_1}
\vspace{0.8cm}
\end{figure}

In 2006, when the \textit{Venus Express} reached Venus, a new gate to lightning explorations opened \citep[e.g.][]{russell2008, russell2011, daniels2012, hart2014b, hart2014}. \citet{russell2008} reported whistler detections by \textit{Venus Express} near the Venus polar vortex from 2006 and 2007, which they associated with lightning activity and inferred a stroke rate of 18 s$^{-1}$. \citet{russell2013} found two ELF emission sources in the Venusian atmosphere. One appeared to be generated at the ionopause-solar wind boundary and showed emission below 20 Hz \citep{russell2013}. They attributed the other emission that went up to 64 Hz, to lightning activity (Fig. \ref{fig:vexpr_1}). This latter emission was right-handed circularly polarized whistler emission, with slower group velocities and higher amplitudes than of whistler emission on Earth. These characteristics are expected for Venusian lightning whistlers as the planet's magnetosphere is weaker than Earth's \citep{russell2013}. The MAG (Magnetometer) on board of \textit{Venus Express} detected lightning induced whistlers in 2012 and 2013 too \citep[e.g.][]{russell2014,hart2015b}. \citet{russell2014} found that the number of detected whistler signals increased throughout the \textit{Venus Express} mission, as the spacecraft orbited closer and closer to the planet. The data were analysed by \citet{hart2015}, who confirmed the whistler events with dynamic spectra. 

\citet{bagheri2016} investigated the possibility of detecting gamma-ray flashes from lightning on Venus, in analogy to Earth TGFs. They conducted a Monte Carlo simulation using the Runaway Electron Avalanche Model and found that if gamma-rays are produced between 58 and 70 km altitudes with similar properties to that of TGFs, then they should be detectable by low-orbit spacecraft. As the electrons are propagated through the medium, their model considered energy losses of electrons due to ionization and atomic excitation; scattering of particles, positron annihilation, and bremsstrahlung radiation. \citet{bagheri2016} suggested the name VGFs, Venusian Gamma-ray Flashes for these events. 

In 2010, the Japanese \textit{Akatsuki} or {Venus Climate Orbiter, (VCO)} was launched towards Venus. Its science mission focuses on the understanding of Venusian atmospheric circulation, including cloud distribution and lightning activity \citep{nakamura2007,takahashi2008}. Though in 2010 the first attempt to insert the spacecraft on orbit around Venus failed, in 2015 it successfully entered an elliptical orbit \citep{nakamura2016}, which orbit was modified in 2016, so that the spacecraft settled on a 9-day orbit around the planet. The Lightning and Airglow Camera on board the spacecraft was reported to be fully functional and started its data gathering in 2016 \citep{takahashi2016}. Due to the modified orbit, it maps the planet's night side for 30 minutes every 9 days \citep{takahashi2016}. The Japanese spacecraft may be the first to settle the long-lasting argument between scientists on the existence of Venusian lightning.

%__________________________________________________________________
\subsection{Mars and Titan} \label{sec:matit}

Both Mars and Saturn's moon Titan have been suggested to host lightning activity, however the observational data is scarcer and even more controversial than that of Venus. Review on Martian dust electrification: \citet{harrison2016}. Review on Titan's storms and climate: \citet{griffith2009,mitchell2016,horst2017}

Mars has a very low density, thin but dynamic atmosphere. The importance of the effects of dust on the climate of the planet is large, as the small dust grains can be easily picked up by winds, which will form dust storms and dust devils on the surface \citep{aplin2006}. \citet{farrell2001} modelled and studied the possibility of the presence of a global electric circuit in the Martian atmosphere, driven by dust storms connecting the surface to the ionosphere. They showed that the global electric field will largely depend on the Martian season and will vary between hundreds of Vm$^{-1}$ and 1 Vm$^{-1}$. Based on electric field measurements, dust storms and dust devils are suggested to produce lightning discharges on Earth (Chapter \ref{chap:ligform}, Sect. \ref{sec:envir}). Similarly, on Mars these environments thought to be electrified due to dust-dust collisions and triboelectrification, while charge separation is thought to occur due to gravitational settling \citep{yair2012}. \citet{farrell2004b} conducted measurements of desert dust devils in order to address the electrification of Martian dust devils, as part of the MATADOR (Martian ATmosphere And Dust in the Optical and Radio) project. They found that dust devils act as large electric dipoles, with oppositely charged regions concentrating at the foot and the top of the devil. They concluded that dust electrification in Martian devils could be similar to what was found on Earth, with the difference that the Martian surface possesses a larger conductivity due to lower ambient neutral density. The large conductivity creates a large dissipation current into the Martian atmosphere, which would be a separate channel drawing charges away from the triboelectric currents of dust storms and dust devils \citep{farrell2003}. This would result in a slower charge build-up in dust devils on Mars compared to their Earth counterparts \citep{farrell2004b}.

There has been no detection of Martian lightning activity, though attempts have been made to detect discharge signatures. \citet{ruf2009} observed Mars with ground-based Deep Space Network's DSS-13 antenna, and reported the detection of non-thermal radio emission, which they observed when a dust storm occurred on the surface of the planet. They suggested that the emission was produced by large-scale discharges, since their frequencies coincide with the first three modes of estimated Martian Schumann-resonances. \citet{andreson2012} monitored Mars with the Allen Telescope Array in 2010 with the purpose of repeating the observations of \citet{ruf2009}. Though several small-scale dust storms occurred during the observations, they did not detect the same SR-like, non-thermal radio emission that was reported by \citet{ruf2009}. \citet{gurnett2010} reported the non-detection of Martian lightning throughout a 5-year observational campaign conducted with the \textit{Mars Express} spacecraft at frequencies 4.4$-$5 MHz. During the observations 2 major and several smaller dust storms appeared in the view of the spacecraft, and the detection threshold was several orders of magnitude lower than a signal expected from Earth lightning for the same observing conditions \citep{gurnett2010}. The non-detection suggests that lightning, if exists on Mars, is much less energetic than Earth discharges.

\citet{krauss2003} simulated Martian dust environments and observed signatures that they interpreted as discharges due to triboelectrification. \citet{aplin2011} conducted laboratory experiments to produce discharges in Martian-analogue environment. They suggested the detection of radio emission from discharges produced in 9 mbar CO$_2$ composition, due to the triboelectric charging of the medium. \citet{aplin2012} reproduced the experiment of \citet{krauss2003} and found that the effects found by that group can be explained by the wall of the tank, and concluded that such experiments cannot resemble Martian environments. 

\citet{fischer2012} reported fluctuations in the Martian ionosphere detected by the Mars Global Surveyor magnetometer, which they interpreted as either discharge-induced SRs, or resonances created by the solar wind's interaction with the crustal magnetic field, or spacecraft noise. \citet{espley2014} presented the non-detection of lightning-induced radio emission by Mars Global Surveyor. The joint ESA/Roscosmos (Russian Space Agency) mission, ExoMars2016 had a great promise for new electric field measurements on the surface of Mars. It was launched in March 2016 and the orbiter and lander units arrived in October, 2016. The lander carried the DREAMS (Dust characterization, Risk assessment and Environment Analyser on the Martian Surface) unit, which had the MicroARES subsystem, the first sensor dedicated to measure electric fields on the surface of Mars \citep{esposito2012,deprez2014,deprez2015}. However, on 16 October 2016 the lander crashed into the surface of Mars\footnote{https://www.nytimes.com/2016/10/22/science/mars-crash-landing-site-explosion.html?\_r=0}.

Titan shows similar atmospheric dynamics to Earth. It hosts methane (CH$_4$) clouds, extensive tropical clouds \citep[][and references therein]{yair2008}, and convective clouds \citep{rodriguez2007, schaller2009, turtle2011}. The latter types are the most likely to host lightning activity \citep{yair2008}. \textit{Cassini's} VIMS (Visual and Infrared Mapping Spectrometer) instrument detected convective methane clouds during the 2004-2006 fly-bys \citep{rodriguez2007}. \citet{rodriguez2007} found that the clouds appeared around the south pole, the southern 40$^{\circ}$ latitudes, and the northern 50$-$60$^{\circ}$ latitudes, and suggested that the cloud systems are latitudinally and seasonally controlled by convective cells due to global atmospheric circulation. \citet{schaller2009} further detected convective tropospheric clouds in the tropical regions of Titan. In these regions it is more difficult to form convection, therefore, \citet{schaller2009} suggested that the air is locally heated around the surface, which helps convection and the formation of small clouds. 

Lightning has been suggested to occur on Titan by models and experiments, however no observations have confirmed its presence yet. In their experiment, \citet{gupta1981} studied the effects of different energy sources, such as UV radiation, lightning discharges, $\gamma$-ray radiation, on the organic chemistry in the atmosphere of Titan, simulated by a gas mixture of N$_2$ and CH$_4$. Their results regarding electrical discharges show that lightning can be accounted for all the compounds in Titan's atmosphere (e.g. HCN, C$_2$H$_2$), observed by Voyager 1. \citet{borucki1988} conducted laser-induced plasma experiments to address lightning activity and its chemical effects in Titan. They suggested that lightning is an adequate process to explain observed ethylene in the moon's atmosphere. \citet{desch1990b} inferred an upper limit of total energy of lightning on Titan from the non-detection of radio signals by Voyager 1. They found that lightning should have energies three orders of magnitude lower, $10^6$ J, than typical Earth lightning flashes of $10^9$ J. \citet{borucki1996} modelled the spectrum of lightning on Titan, which I summarize in Chapter \ref{chap:ligsig} Sect. \ref{sec:optem}. \citet{tokano2001} performed a theoretical study of thundercloud formation on Titan, taking into account methane condensation and ion chemistry, and considering different cloud charging mechanisms. Though cloud formation by convection is difficult on Titan, they found that when it occurs, cloud particles quickly attach free floating charges, and in rare cases the cloud is capable of maintaining a temporary large electric field to produce lightning discharges. They suggest that such discharges are likely to consist of 20 km long cloud-to-ground flashes. \citet{navarrogonzalez2001} introduced simulated lightning discharges into a gas mixture of  N$_2$ and CH$_4$ with various mixing ratios. They observed the production of several hydrocarbons and determined the freeze out-temperature of these species, the temperature at which the equilibrium mixing ratios of species freeze out in the lightning channel. More recently \citet{horvath2009} analysed the impact of corona discharges on Titan-like atmosphere and observed the production of C$_2$H$_2$, HCN and tholins. They also found that in the N$_2$-CH$_4$ mixture the onset voltage of the discharge increased and the discharge current decreased. \citet{kovacs2010} used a detailed chemical kinetic network to explore the chemical evolution of the atmosphere of Titan after the onset of lightning, by evaluating the temporal temperature profile of a gas parcel affected by the discharge. They confirmed the findings of the previous works, that lightning produces traceable amounts of several different hydrocarbons. 

Between 2004 and 2016, \textit{Cassini} approached Titan several times to take images and conduct scientific observations of the moon. Previously, \citet{lammer2001} estimated \textit{Cassini's} capability of detecting Titanian lightning, and found the RPWS instrument should be able to detect lightning radio signals at frequencies between 500 kHz and 1 MHz, up to 200 Titan radii away from the planet, if lightning on Titan is similar to typical Earth cloud-to-ground discharges. However, despite of \textit{Cassini}-RPWS being able to detect lightning radio signatures from Earth, Jupiter and Saturn, it did not detect such signatures from Titan's atmosphere \citep{fischer2007b,fischer2011c}. 

In 2005, Cassini deployed the Huygens Probe above Titan, which took measurements throughout its descent and from the surface. \citet{fulchignoni2005} reported the results of, amongst others, the electricity measurements, and found that the largest conductivity in the atmosphere is at $\sim 60$ km altitude. They also reported the detection of a high-amplitude pulse at 36 Hz, which they suggested to be a possible Schumann-resonance wave generated by lightning activity. \citet{simoes2007} modelled Schumann-resonances on Titan, and found that the 36 Hz signal can match the second eigen frequency of SR, however they also list several other sources that could have produced such emission, like instrumental, and temperature and haze effects. These sources are discussed and addressed in more depth in \citet{beghin2007}, who concluded that the source of the 36 Hz signal was most probably not lightning. In 2008-2009 two groups reanalysed the data of the Huygens probe and argued about whether it actually reflects lightning activity on Titan or not. Works by \citet{morente2008,morente2009,morente2009b} claimed the detection of week resonances in the data, which they suggested to be a clear evidence of atmospheric electricity on Titan, however \citet{hamelin2009,hamelin2011} argued that those results were obtained through errors in the analysis methods.

%__________________________________________________________________
%__________________________________________________________________
\section{Lightning beyond the Solar System - exoplanets and brown dwarfs} \label{sec:ligos}

In principle, lightning, or lightning-like discharges, can occur in any environment which is capable of hosting charged seed particles and producing large-scale charge separation. Though this is most prominent for planetary-like atmospheres, such as that of exoplanets and brown dwarfs, lightning-like discharge events have been suggested to occur in other, less obvious environments, like protoplanetary disks \citep[e.g.][]{desch2000,nuth2012}, pulsars \citep{katz2017}, and black holes \citep[e.g.][]{aleksic2014,eisenacher2015}. In this section I will summarize studies focusing on exoplanetary and brown dwarf lightning. Though, lightning has not been detected from outside the Solar System, the theory of "exo-lightning" formation is a developing field, which suggest that in the future observation of lightning on exoplanets and brown dwarfs might be possible. For further information on atmospheric electrification (inside and) outside the Solar System see \citet{helling2016b}.

Exoplanets analysed through transit spectroscopy are observed to have clouds in their atmospheres, most likely made of silicate particles \citep[e.g.][]{kreidberg2014, sing2009, sing2013, sing2015}. These findings are supported by kinetic cloud models as in \citet{helling2008b, helling2011, helling2011b}. Various authors demonstrated that atmospheric circulation leads to the formation of zonal jets and local vortices as known from Jupiter and Saturn \citep[e.g.][]{dobbsdixon2012, dobbs-dixon2013, mayne2014}. E.g. \citet{zhang2014} showed that strong internal heating and weak radiative dissipation results in the formation of large-scale jets. \citet{lee2015} modelled local and global cloud patterns on the planet HD 189733b, a tidally locked hot Jupiter orbiting a K star. Their dust opacity, grain size distribution and albedo maps indicate that cloud properties change significantly from dayside to night side forming a spot-like cloud pattern driven by a latitudinal wind jet around the equator. As we have seen in the previous sections, such dynamic environments host lightning activity in the Solar System. The charging of extrasolar atmospheres has been suggested to be the result of particle collisional ionization \citep{helling2011b, helling2013b}, cosmic ray ionization \citep{rimmer2013}, and ionization due to the internal \citep{rodriguezbarrera2015} and external \citep{batygin2011} heating of the object. Other processes, such as Alfven ionization, may also contribute to the production of charged particles in magnetic environments \citep{stark2013}. 

\citet{helling2011} studied electron avalanches initiated by dust collision-induced, local charge-inequilibrium in dust clouds of brown dwarfs. They argued that thermal$-$ and dust collisional processes alone will result in a globally neutral atmosphere. However, stochastic ionization in such atmospheres may occur on a short time scale, resulting in enough free electrons to form electron avalanches and eventually intra-cloud discharges. \citet{helling2013} estimated electric breakdown characteristics in dusty atmospheres and found that the breakdown field depends on the local gas-phase chemistry, the effective temperature, and the primordial gas-phase metallicity. They found that charged particles will gravitationally settle resulting in large-scale charge separation. They suggest that different discharge processes will dominate at different atmospheric pressures, such as small-scale sparks at higher pressures and gas densities, and large-scale discharges near and above cloud tops. The critical electric field in such dusty atmospheres varies between $10^{-7}$ and $10^7$ V cm$^{-1}$, and the critical number of charges per dust surface per cm$^3$ varies between $<1.6 \times 10^{14}$ and $1.6 \times 10^4$ C \citep{helling2013}. \citet{bailey2014} modelled large-scale discharges and derived discharge properties using scaling-laws for giant gas planets and brown dwarfs. The properties they obtained include the breakdown field, the minimum number of charges needed to overcome the electric breakdown, the initiation height of the discharge, the total discharge length, the total energy dissipated (based on a simple scaling law), and the total discharge volume. I will further discuss the findings of \citet{bailey2014} in Chapter \ref{chap:model}. \citet{rimmer2016} introduced a chemical kinetics network for extrasolar atmospheres, which can be used to study the chemical affects of lightning discharges in various atmospheric compositions. \citet{rimmer2016b} used the same chemical network to study lightning-induced chemistry in super-Earth atmospheres, and found that amino acids are produced if the atmosphere is sufficiently reductive, but with very small redox ratios, lightning does not support such chemical changes. \citet{hodosan2017} estimated radiated energies and radio powers of lightning in certain giant gas planetary and brown dwarf atmospheres, with input parameters taken from \citet{bailey2014}, which is also presented in Chapter \ref{chap:model}.

To date, only a handful of studies have presented estimates of detectability of extrasolar lightning. \citet{zarka2012} analysed the possibility of detecting lightning emission from extrasolar planets, by up-scaling the radio emission observed from Jupiter and Saturn. They concluded that flashes $10^5$ times stronger than Jovian or Saturnian lightning from a distance of $10$ pc, with a bandwidth of $1-10$ MHz, integration time of $10-60$ min would be possible to detect. Although, as they point it out, propagation effects will strongly affect the radio emission below a few MHz. \citet{vorgul2016} modelled the conductivity in extrasolar atmospheres caused by flash ionization processes, such as lightning, and found that such events have an effect on the signature of electron cyclotron maser emission (CME) causing a pulse-like amplification of the signal. They suggest that, if such signatures are observed in CME, one could infer the properties of underlying flash events in the atmosphere. The newest research conducted regarding observability of "exo-lightning" is part of this thesis. \citet{hodosan2016} estimated lightning occurrence on the exoplanet HAT-P-11b based on previous radio observations carried out by \citet{lecav2013}. The extended version of this work is discussed in Chapter \ref{chap:hatp11b}. \citet{hodosan2016b} carried out a first statistical study of lightning occurrence on extrasolar planets based on observed occurrence of lightning in the Solar System. This lightning climatology study is discussed in Chapter \ref{chap:stat}. In an on-going study, I also estimate observability of lightning radio signatures based on the results of \citet[][also in Chapter \ref{chap:model}, Chapter \ref{chap:concl}]{hodosan2017}.

%__________________________________________________________________
%__________________________________________________________________
\section{Summary} 

Lightning is a common phenomenon in the Solar System. On Jupiter and Saturn it has been suggested to occur inside the deep water clouds through similar processes known from Earth \citep{yair2008,yair2012}. The optical and radio emission from the two planets seem to be consistent with higher lightning energy release on the gas giants, than on Earth, with a total energy of $\sim 10^{12}-10^{13}$ J \citep{yair1995,gurnett2005}. Lightning activity on Uranus and Neptune is less studied. Because water clouds form on very high pressures (around 40 bar), it was suggested that on the ice giants lightning occurs in the H$_2$S$-$NH$_3$ clouds closer to the top of the atmosphere \citep{gibbard1999}. Lightning on Venus most probably occurs in the form of intra cloud lightning due to the high pressures close to the surface \citep{yair2008}. In extrasolar planets so far dust cloud charging has been modelled (but not yet observed) in more details, and it has been suggested that such mineral clouds will allow large enough charge separation through gravitational settling to overcome the breakdown threshold and produce large-scale discharges \citep[e.g.][]{helling2013}.

%% file: chapters/3_lightningstat.tex
\chapter{Lightning statistics and climatology} \label{chap:stat}

\section{Introduction}

In the following chapters, I discuss my own research in the context of exoplanetary sciences.
This chapter presents an analysis of lightning surveys on Earth, Venus, Jupiter and Saturn. Our planetary system provides opportunities to compare different environments where lightning occurs, and therefore, provides guidance for the large diversity of exoplanets and their atmospheres. I compare lightning climatology from the above mentioned Solar System planets and use these statistics as a guide for a first consideration of lightning activity on extrasolar objects. I use lightning climatology maps to find patterns in the spatial distribution of lightning strikes, such as increased lightning activity over continents than over oceans, and calculate flash densities (flashes km$^{-2}$ year$^{-1}$ and flashes km$^{-2}$ hour$^{-1}$) and flash rates (flashes unit-time$^{-1}$) in order to estimate the total number of events at a certain time over a certain surface area. Estimating the number of lightning flashes and their potential energy distribution is essential for follow-up studies such as lightning chemistry \citep[e.g.][]{rimmer2016} in combination with 3D radiative hydrodynamic models \citep{lee2016}. 

The chapter is organized in three main parts. Section \ref{sec:earth} analysis Earth lightning data in the optical (direct lightning detection) and radio (low frequency (LF) emission). The data were obtained by several Earth-based stations (STARNET, WWLLN) and Earth-orbiting satellites (OTD/LIS). I compare the data by exploring the detection limits, general trends and differences between the data sets. In Sect. \ref{sec:solsys}, I explore lightning observations on Venus, Jupiter and Saturn by summarizing and analysing data from various spacecraft and by creating lightning distribution maps. In Section \ref{sec:exopl}, I use the lightning climatology data to address potential lightning occurrence on the diverse population of exoplanets. Specific exoplanets are discussed and brown dwarfs are also included in this section.
Section \ref{sec:con} summarizes this chapter.

%__________________________________________________________________
%__________________________________________________________________
\section{Lightning data from Earth} \label{sec:earth}

Earth is the most well-known planet we can learn from and apply as an analogue for exoplanetary sciences. Both observational and theoretical works that used Earth as a guide have been conducted to analyse different features of exoplanets. \citet{palle2009}, for example, compared the transmission spectrum of Earth taken during a lunar eclipse and the spectrum of the Earthshine, which is the reflection spectrum of Earth. They used the transmission spectrum as an analogue for a primary transit of Earth as seen from outside the Solar System, while the reflection spectrum is an indicator of a directly imaged exo-Earth after removal of the Sun's features. Similar studies of Earth as an exoplanet, such as looking for vegetation or other signatures caused by biological activity, were conducted by e.g. \citet{montanes2010, arnold2002, sterzik2009, kaltenegger2007}.

Lightning detection and statistics on Earth are very important because of the hazards (e.g. forest fires, large scale power outage, fatalities) it causes. Lightning detecting networks are set up on the surface of the planet while satellites monitor the atmosphere for lightning events. Earth measurements provide the largest data set due to the continuous observations and the high spatial coverage of the instruments. Data used here were provided by the \textit{Lightning Imaging Sensor} (LIS)/\textit{Optical Transient Detector} (OTD) instruments on board of satellites in the optical, and two ground based radio networks, the \textit{Sferics Timing and Ranging Network} (STARNET) and \textit{World Wide Lightning Location Network} (WWLLN). WWLLN and STARNET detect strokes\footnote{events with discrete time and space} while LIS/OTD observe flashes\footnote{events with duration and spatial extent; one flash contains multiple strokes} \citep{rudlosky2013}. Table \ref{table:instr} lists relevant properties of the lightning detecting instruments and networks.

OTD was in operation between 1995 and 2000 on board the Microlab-1 (OV-1) satellite orbiting 735 km above the terrestrial surface on an orbit with inclination of $70^{\circ}$ with respect to the Equator, allowing the monitoring of the whole globe, but excluding the polar regions \citep{boccippio2000}. LIS was in operation between 1997 and 2015 on board the Tropical Rainfall Measuring Mission (TRMM)\footnote{http://trmm.gsfc.nasa.gov/}. The satellite's orbit was restricted to the tropical region, between $\pm 38^{\circ}$ latitude, 350 km above the Earth \citep{beirle2014}. Both OTD and LIS detected lightning flashes by monitoring the 777.4 nm oxygen line in the lightning spectrum \citep{beirle2014}. The optical observations allow the detection of cloud-to-ground (CG), intra-cloud (IC), and cloud-to-cloud discharges from space. The composite, gridded data set of OTD/LIS gives information on the location and time of occurrence of individual flashes, including the number of events (pixels exceeding the intensity background threshold) and groups (events occurring in adjacent pixels within the same integration time) that the flashes (groups occurring within 330 ms and within 15.5(OTD)/6.5(LIS) km) are composed of \citep{beirle2014}. The OTD/LIS data used here were obtained on 18 July 2014\footnote{http://thunder.nsstc.nasa.gov/data/data\_lis-otd-climatology.html} (Daniel Cecil, private com.) for the period of $1995-2013$. The downloaded data include different types of flash rates such as mean annual flash rates, annual cycles of flash rates and daily time series of flash rate, raw flash counts and flash counts scaled by detection efficiency.

STARNET is an Earth-based radio network currently composed of 11 antennas operating in the very low frequency (VLF) range ($7 - 15$ kHz). STARNET has been in operation since 2003 in Africa and since 2006 in Brazil and the Caribbean (previously operating as a test network in the United States between 1993-1998). STARNET has integrated the European ZEUS lightning network, which was operating until 2005\footnote{http://www.zeus.iag.usp.br/index.php?lan=en}. The publicly available STARNET data is composed of monthly and daily processed sferic information including the time of observation (date and time to milliseconds), location of the origin (latitude, longitude), arrival time difference (ATD) error in ms, and quality control. The point of origin of sferics is determined by using the ATD technique that involves the measurement of the time difference between the detection of the individual sferics with different antennas. For this technique to work, at least four antennas have to observe the radio signal \citep{morales2014}.The data were obtained on 17 July 2014\footnote{http:/www.zeus.iag.usp.br/index.php? lan=en} (Carlos Augusto Morales Rodrigues, private com.) for years 2009 and 2013. It is important to note that STARTNET is not a global network, its best coverage is over Central and South America (Fig. \ref{fig:4a}. Therefore, when comparing the network with other lightning detection instruments, I only consider the area that STARNET covers.) 

WWLLN is a developing lightning location network that observes VLF ($3 - 30$ kHz) sferics. WWLLN includes $\sim 70$ stations all around the world \citep{hutchins2013}. It detects both IC and CG discharges (individual strokes in flashes) but is more sensitive to the CG flashes since they are stronger than the IC ones \citep{rudlosky2013}. WWLLN data were obtained\footnote{http://www.wwlln.com/} (Robert H. Holzworth, private com.) in Aug-Sep 2014 for the years 2009 and  2013, however, there are 15 days missing from the 2009 series (first part of April). The data files include, amongst others, locations of strokes (latitude, longitude), times of observations and energy estimates. A separate file contains the relative detection efficiencies in the form of maps for each hour of each day.

%__________________________________________________________________

\subsection{Detection efficiency} \label{subs:de}

%Table - Instrumental properties
\begin{table*} 
\resizebox{\columnwidth}{!}{
\begin{threeparttable}
 %\small 
 \begin{center}
% \begin{minipage}{110mm}
 \caption{Properties of instruments used for lightning detection on Earth. (FoV = Field of View.) OTD: \citet{boccippio2000, boccippio2002, beirle2014}. LIS: \citet{christian2003, cecil2014, beirle2014, christian2000}. STARNET: \citet{morales2014}. WWLLN: \citet{abarca2010, hutchins2012, hutchins2013}.}
  \begin{tabular}{@{}lccllc@{}}	
	\hline
	Instrument/Network & Spatial resolution & Temporal resolution & Detection threshold & FoV/Coverage & Detection Efficiency \\
	\hline
	OTD & $10 - 11$ km & $\sim 2$ ms & 9-21 $\mu$J m$^{-2}$ sr$^{-1}$ & 1300 km $\times$ 1300 km\tnote{(1)} & \vtop{\hbox{\strut day: 40\%}\hbox{\strut night: 60\%}} \\
	LIS & $4 - 6$ km & 2 ms & 4-11 $\mu$J m$^{-2}$ sr$^{-1}$ & 600 km $\times$ 600 km & \vtop{\hbox{\strut day: 70\%}\hbox{\strut night: 90\%}} \\
	STARNET & $5-20 $ km & 1 ms\tnote{(2)} & - (no information) & \vtop{\hbox{\strut South America}\hbox{\strut Caribbean}\hbox{\strut SW-Africa}} & \vtop{\hbox{\strut day: 45\%}\hbox{\strut night: 85\%}}\\
	WWLLN & $\sim 5$ km & $\sim 15 \mu$s & \vtop{\hbox{\strut Space, time and}\hbox{\strut station dependent}} & Full Earth & $\sim 2-13$\% \\
	\hline
  \label{table:instr}
  \end{tabular}
  \begin{tablenotes}
	\item[1] http://thunder.msfc.nasa.gov/otd/  
	\item[2] http://www.starnet.iag.usp.br/index.php?lan=en     
  \end{tablenotes}
% \end{minipage}
 \end{center}
\end{threeparttable}
}
\vspace{0.8cm}
\end{table*}
%Table 1

The detection efficiency (DE) is the detected percentage of the true number of flashes \citep{chen2013}. It depends on the sensitivity threshold of the instrument, geographic location, and time of the observation \citep{cecil2014}. Seen from an astronomical perspective, the DE is extremely well determined for Earth, however, less so for the Solar System planets. Therefore, I use the knowledge from Earth to discuss the impact of the DE on the lightning data, in order to understand the limits, but also the potentials of the available data for exoplanetary research.

For LIS/OTD the DE is determined by two different approaches. \citet{boccippio2000} cross-referenced individual flash detections with the U.S. National Lightning Detection Network data, which provides an empirical estimate on the DE. \citet{boccippio2002} used independent measurements of pulse radiance distributions to model the DE. The estimated DEs for OTD and LIS are listed in Table \ref{table:instr}.

The DE for STARNET is determined by comparing detections with other networks (e.g. with WWLLN) in the regions where STARNET operates (i.e. Central and South America). According to the comparison studies conducted by \citet{morales2014} STARNET detects $\sim 70\%$ of lightning strokes, however this value depends on the antennas in use and it has a diurnal pattern (85\% day, 45\% night DE). Two different WWLLN DEs are quoted in the literature: relative DE (RDE) and absolute DE (ADE). The RDE is determined by the model given in \citet{hutchins2012} that is based on the detected energy per stroke: once the energy distribution of observed samples is known, the missing energies (and amount of lightning) can be estimated. The RDE compensates for the uneven distribution of sensors on Earth and variations in VLF radio propagation and allows representing the global distribution of strokes as if it was observed by a globally uniform network \citep{hutchins2012}. The ADE was determined by comparing WWLLN data with other networks. \citet{abarca2010} cross-correlated stroke locations with detections of the National Lightning Detection Network (NLDN) data and found that WWLLN DE is highly dependent on the current peak and polarity of the lightning discharge and varies between $\sim 2 - 11 \%$. \citet{rudlosky2013} showed the improvement of WWLLN DE between 2009 and 2013 compared to LIS observations (up to $\sim 10 \%$), while \citet{hutchins2012} found the ADE to be $\sim 13$\%. In the calculations, following \citet{rudlosky2013}, the WWLLN DE was taken to be 9.2\% for 2012 under the assumption that LIS was 100\% efficient. 

The DE is an important parameter of the lightning detecting instruments, however, it cannot be determined perfectly and unambiguously. It introduces an uncertainty in the measurements, it is estimated based on models and/or comparison studies. Models include estimates \citep[e.g. see the models of][]{boccippio2002}, and comparison studies assume a lightning detecting network/satellite to be, ideally, 100\% efficient. Since the true value of the DE of an instrument or network is unknown, the obtained flash densities are only a lower limit of the total number of flashes occurring on Earth at a certain time. No DEs are yet available for the lightning observations on Venus, Jupiter and Saturn. Therefore, it seems justified to conclude that the Solar System data, including Earth, are a lower limit for lightning occurrence on these planets.

%__________________________________________________________________

\subsection{Lightning climatology on Earth} \label{sec:e_data}

In this section, I derive and compare flash densities for the different networks and satellites based on already published, extensive data from Earth.

\begin{figure*}
  \begin{center}
  \includegraphics[trim=0cm 0cm 0cm 0cm, scale=0.55]{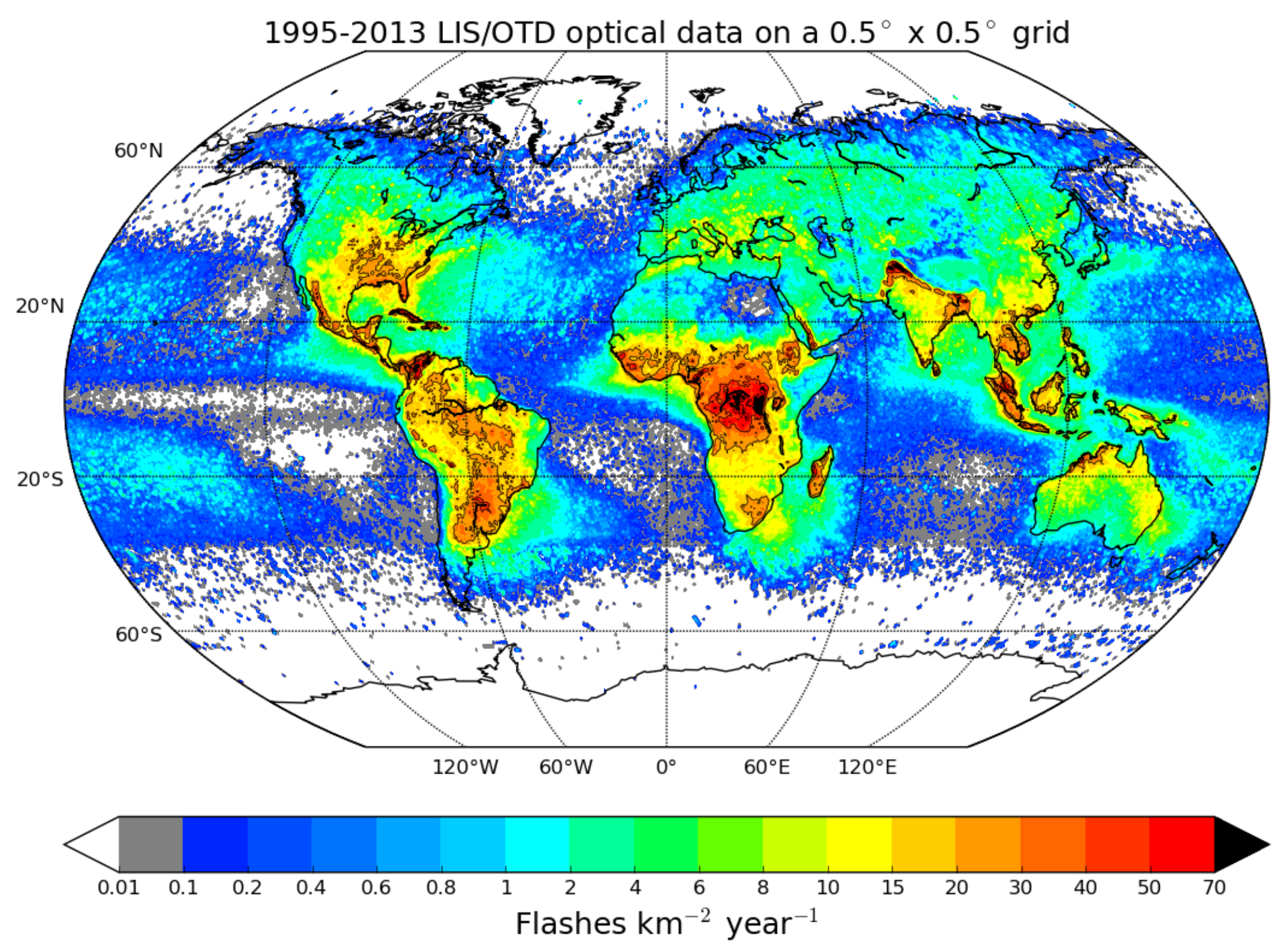}

  \vspace{0.5cm}
  \includegraphics[trim=0cm 0cm 0cm 0cm, scale=0.55]{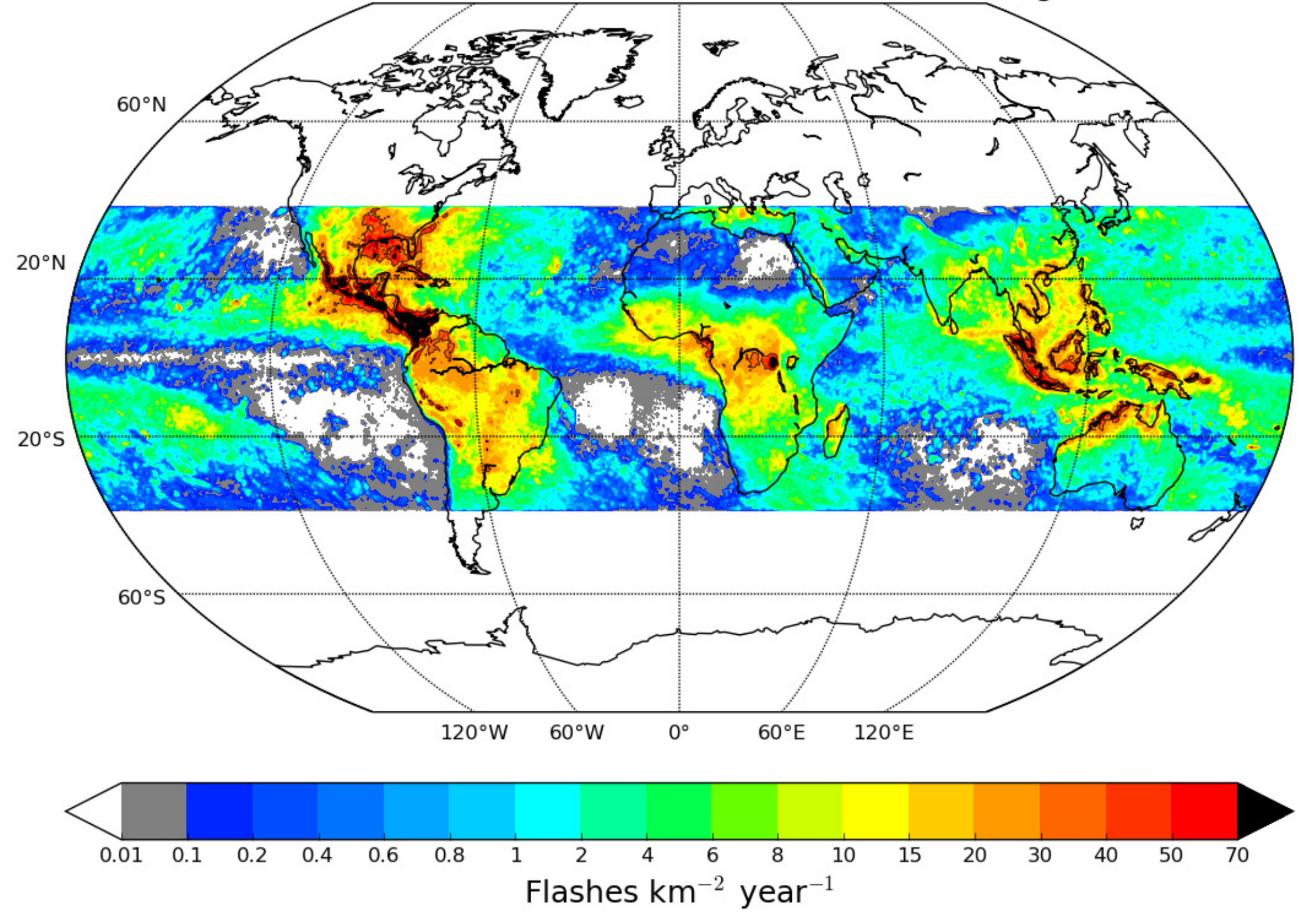}
  \end{center}
  \caption{\textbf{Top:} Mean annual flash density from optical LIS/OTD data averaged on a $0.5^{\circ} \times 0.5^{\circ}$ geographical grid across Earth's surface \citep[for description of the data see section on "High resolution flash climatology (HRFC)" in][]{cecil2014}. LIS covers the area between $\pm 38^{\circ}$ in latitude and the years 1998-2013, while OTD monitored the whole globe (excluding polar regions) in the period of 1995-2000 \citep{cecil2014}. The map shows the differences between continents and oceans. Most of the lightning activity was recorded over continents, especially on low-latitudes.
\textbf{Bottom:} WWLLN mean annual flash density on a $0.5^{\circ} \times 0.5^{\circ}$ grid across the LIS field of view. WWLLN data were scaled by DE and strokes were converted into flashes to match the LIS observations. Comparing it to the figure on the top, I find that WWLLN detects fewer flashes than LIS/OTD.}
  \label{fig:1}
  \vspace{0.8cm}
\end{figure*}

\begin{figure*}
  \begin{center}
  \includegraphics[trim=0cm 0cm 0cm 0cm, scale=0.43]{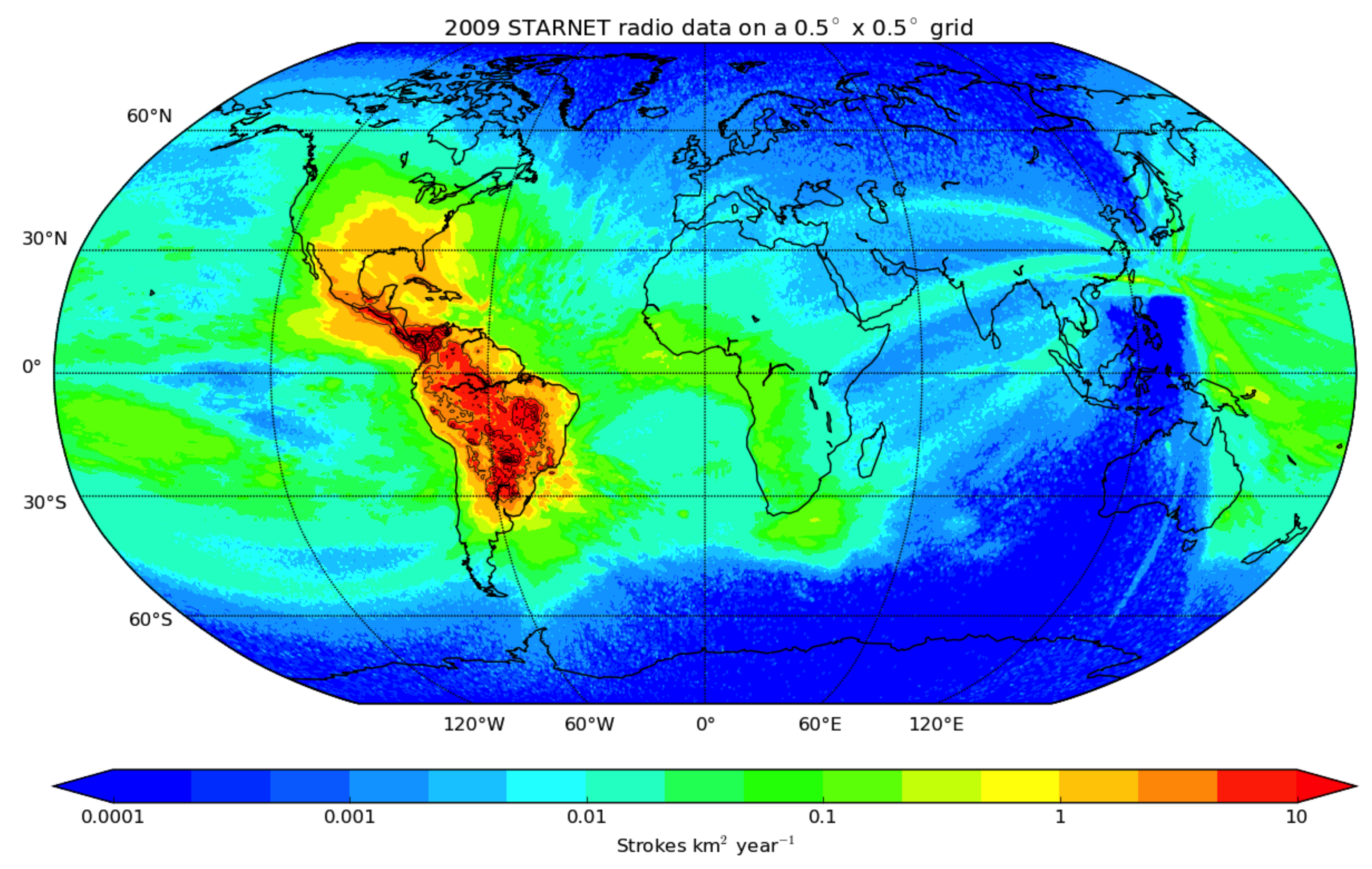}

  \vspace{0.5cm}
  \includegraphics[trim=0cm 0cm 0cm 0cm, scale=0.43]{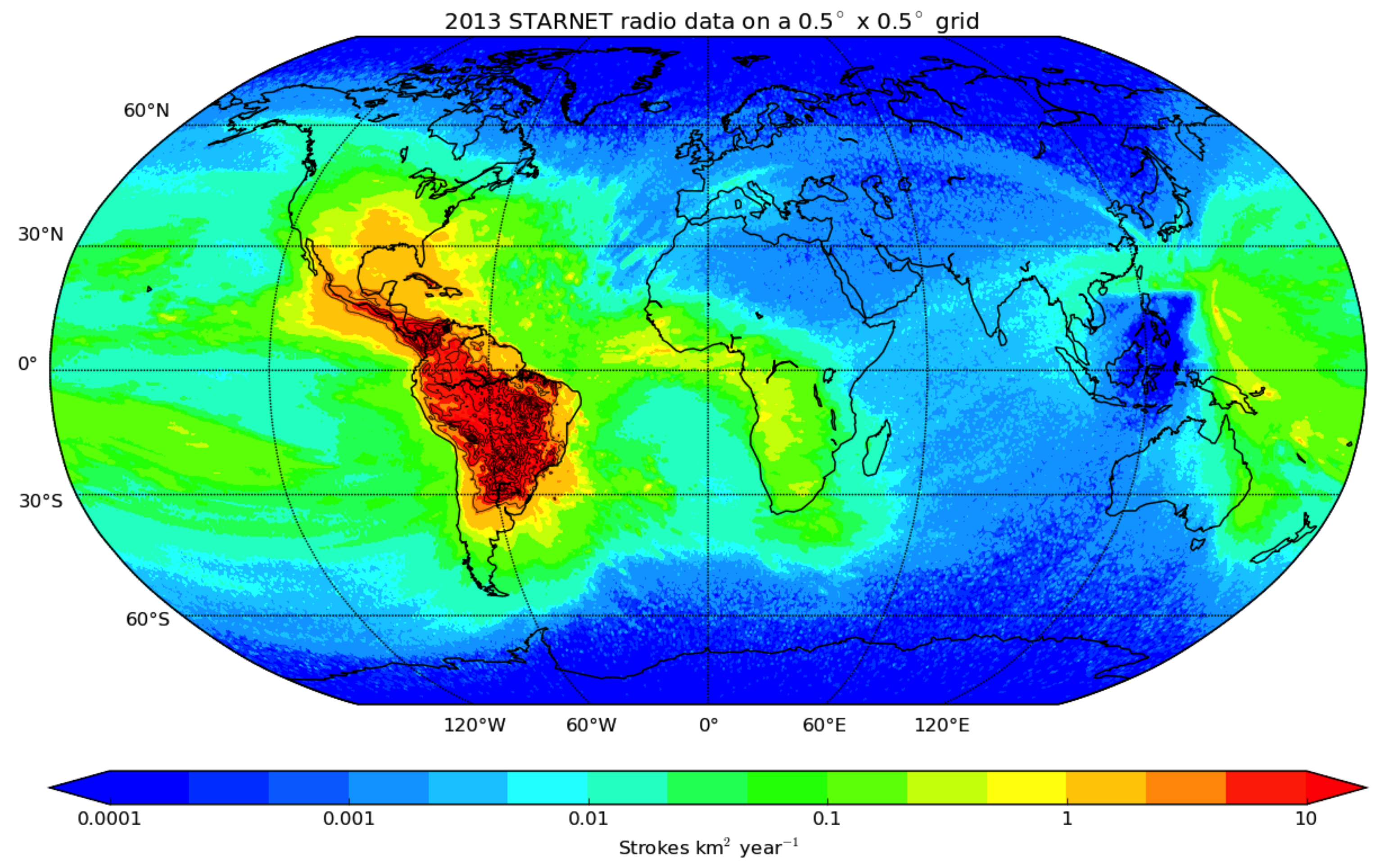}
  \end{center}
  \caption{Mean annual stroke density on a $0.5^{\circ} \times 0.5^{\circ}$ grid, created from very-low frequency STARNET ($7 - 15$ kHz) radio data. \textbf{Top:} 2009, \textbf{Bottom:} 2013. In 2009 the Sun was close to the minimum of its 11-year cycle, while 2013 was close to solar maximum. Comparing the two maps, they show more lightning in 2013 than in 2009, just like the WWLLN maps (Figure \ref{fig:4b}). The arch-like trend above the Indian Ocean and Asia is most probably a numerical or observational artefact.}
  \label{fig:4a}
  \vspace{0.8cm}
\end{figure*}

\begin{figure*}
  \begin{center}
  \includegraphics[trim=0cm 0cm 0cm 0cm, scale=0.43]{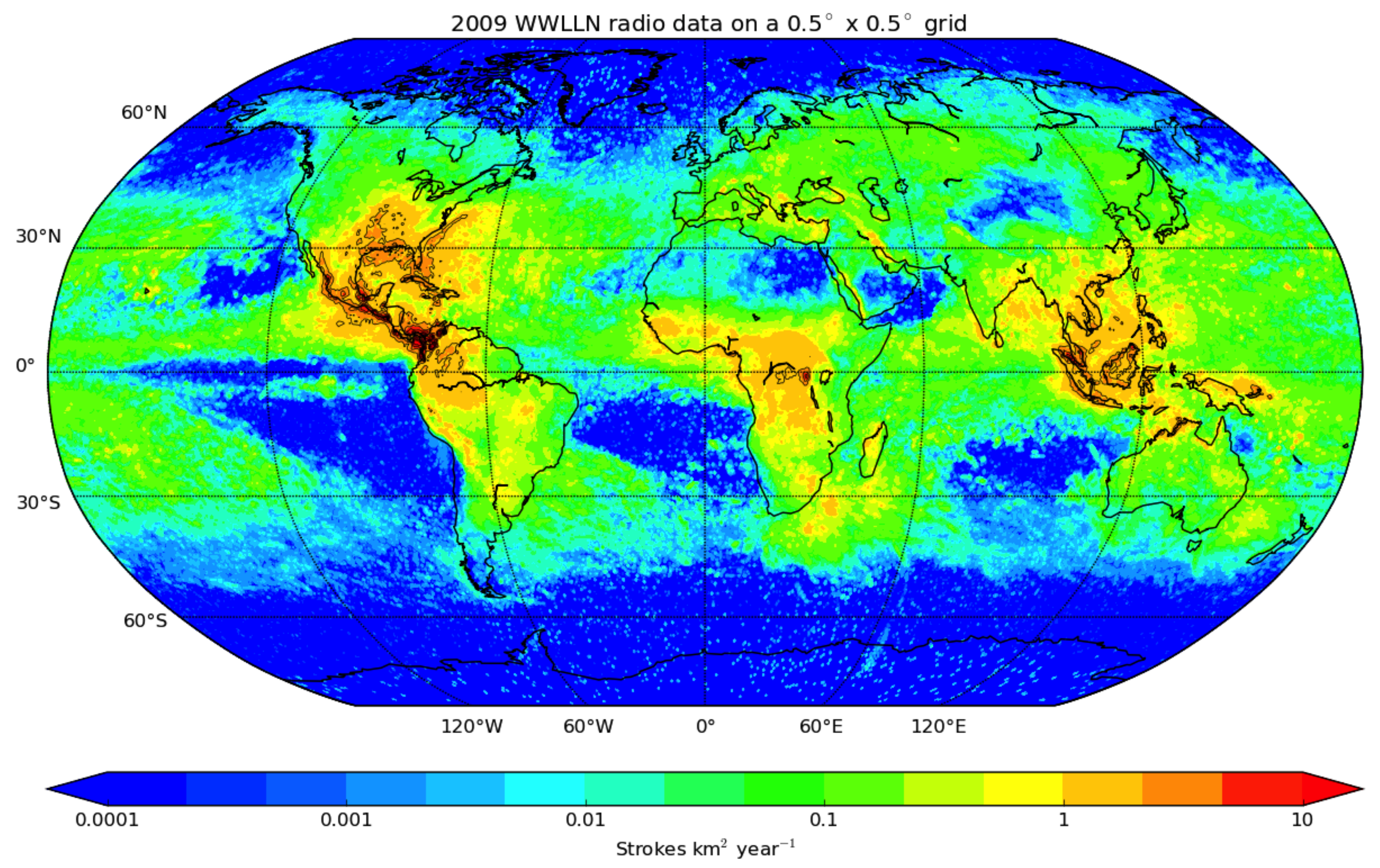}

  \vspace{0.5cm}
  \includegraphics[trim=0cm 0cm 0cm 0cm, scale=0.43]{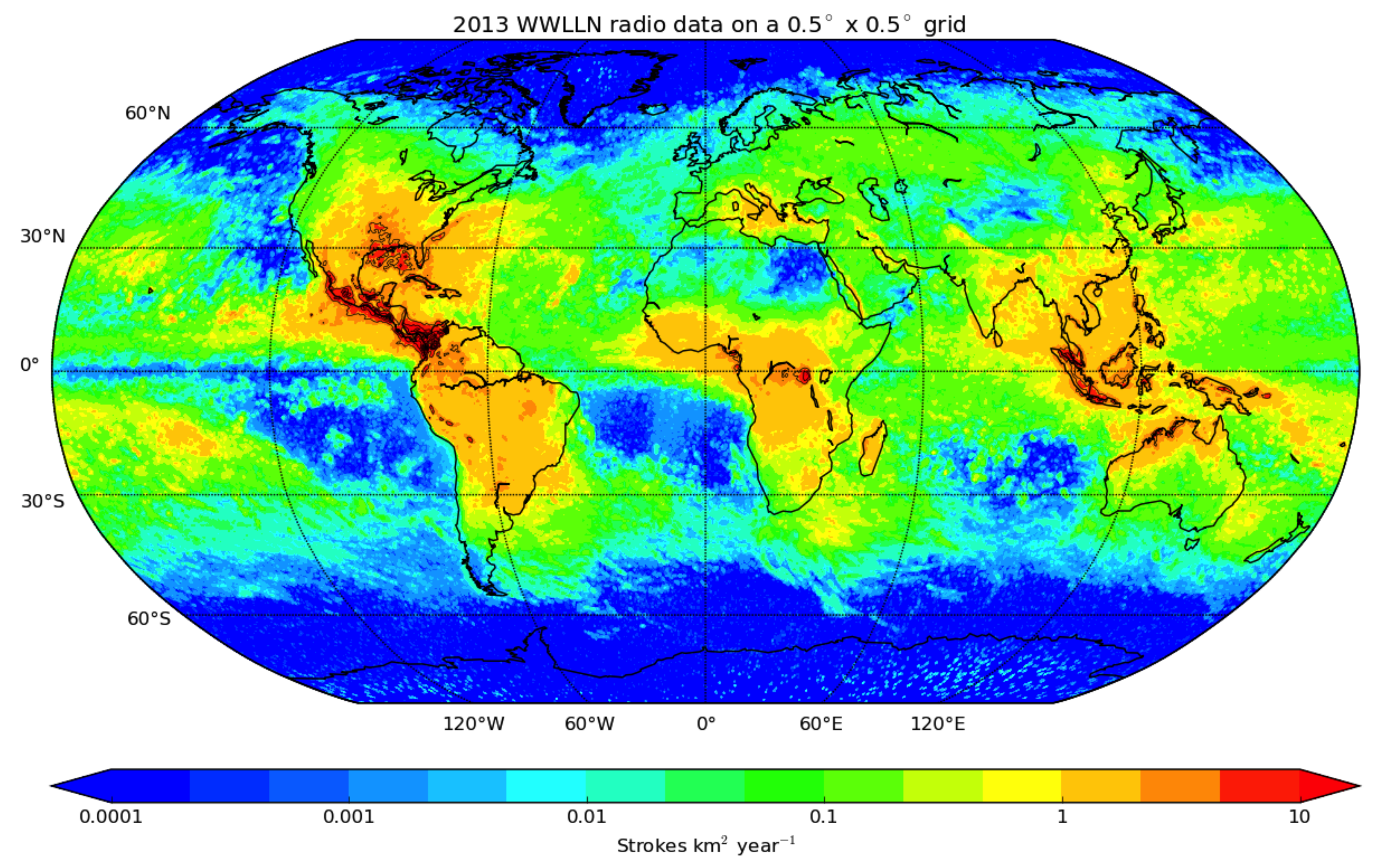}
  \end{center}
  \caption{Mean annual stroke density on a $0.5^{\circ} \times 0.5^{\circ}$ grid, created from very-low frequency WWLLN ($3 - 30$ kHz) radio data. \textbf{Top:} 2009, \textbf{Bottom:} 2013. In 2009 the Sun was close to the minimum of its 11-year cycle, while 2013 was close to solar maximum. Comparing the two maps, they show more lightning in 2013 than in 2009, just like the STARNET maps (Figure \ref{fig:4a}).}
  \label{fig:4b}
  \vspace{0.8cm}
\end{figure*}

Figures \ref{fig:1}$-$\ref{fig:4b} show flash densities averaged and plotted on a $0.5^{\circ} \times 0.5^{\circ}$ geographical grid.
The top panel of Fig. \ref{fig:1} shows the mean annual flash densities \citep[flashes km$^{-2}$ year$^{-1}$,][]{cecil2014} based on LIS/OTD data in the period of 1995-2013. The LIS/OTD data show lower flash densities over oceans and dry regions than continents. Fewer flashes are detected at high latitudes (e.g. Canada, Siberia, etc.), than  at lower latitudes. \citet{cecil2014} derived the global average flash density from the $0.5^{\circ} \times 0.5^{\circ}$ high resolution data set to be 2.9 flashes km$^{-2}$ year$^{-1}$ and the peak value to be 160 flashes km$^{-2}$ year$^{-1}$. Their results are reproduced here from the original data to be $\sim 2$ flashes km$^{-2}$ year$^{-1}$ for the annual average (Table \ref{table:plan}) and $\sim 163$ flashes km$^{-2}$ year$^{-1}$ for maximum values.

Fig. \ref{fig:4a} shows maps with annual stroke densities (strokes km$^{-2}$ year$^{-1}$) from STARNET data for the years 2009 (top) and 2013 (bottom). For these years STARNET had a coverage over the Caribbean, South America and western Africa. Fig. \ref{fig:4b} shows the mean annual stroke density maps for 2009 (top) and 2013 (bottom) from WWLLN data (missing 15 days from Apr 2009). WWLLN shows similar stroke distribution pattern to LIS/OTD, more lightning over continents than oceans, although WWLLN finds the maximum of lightning strokes (km$^{-2}$ year$^{-1}$) over Central-America, while LIS/OTD shows the most lightning over Africa (Fig. \ref{fig:1}, top).  

The effects of different DEs are seen in Figs \ref{fig:4a} and \ref{fig:4b} for STARNET and WWLLN data. If we choose one of the years, e.g. 2009, and focus on the South-American region, it is clearly seen that STARNET detects more strokes than WWLLN. STARNET operates more radio antennas in this region, than WWLLN, which increases the DE of the network. Data from two years (2009, 2013) are plotted in Figs \ref{fig:4a} and \ref{fig:4b}. The two years were chosen in order to represent different phases of solar activity: there was a solar minimum in 2009, while in 2013 the Sun was very active\footnote{http://www.climate4you.com/Sun.htm - Climate4you developed by Ole Humlum}. Comparing the data for the two years in Figs \ref{fig:4a} and \ref{fig:4b} leads to the conclusion that more lightning strokes were observed in 2013 ($\sim$ solar maximum) than in 2009 ($\sim$ solar minimum). However, in case of WWLLN the increase of detected lightning strokes may be the reason of increased DE between 2009 and 2013 \citep{rudlosky2013}, hence the correlation with solar activity remains uncertain. (A more detailed comparison between solar activity and lightning activity is discussed in Sect. \ref{sec:stelact}.) The maps from the two years can be correlated with El Ni\~no events. El Ni\~no was observed in 2009, however not in 2013\footnote{http://ggweather.com/enso/oni.htm - by Jan Null}. Interestingly, both Fig. \ref{fig:4a} and Fig. \ref{fig:4b} show more lightning activity in 2013, on the contrary to what is expected from previous studies showing slightly larger lightning activity during El Ni\~no periods over tropical and sub-tropical continental regions \citep[e.g.][]{satori2009, siingh2011}.

WWLLN strokes were scaled by the DE and converted into flashes to match the LIS data by assuming 1.5 strokes/flash \citep{rudlosky2013}. The bottom panel of Fig. \ref{fig:1} demonstrates that WWLLN detects fewer flashes in Africa than LIS (Fig. \ref{fig:1}, left). This suggests that the difference between the detections is caused by the lower WWLLN DE in Africa. Flashes may contain more than 1.5 strokes \citep{rakov2003}, in which case the WWLLN would detect even fewer flashes than the LIS satellite.

The obtained flash densities are summarized in Table \ref{table:plan} and their potential application to exoplanets is discussed in Sect. \ref{sec:exopl}.

%__________________________________________________________________
\subsection{Lightning in volcano plumes} \label{sec:volc}

%Table - Volcano lightning
\begin{table*}
\resizebox{\columnwidth}{!}{
\begin{threeparttable}
 %\small 
 \caption{Volcano eruptions investigated in this study, their characteristics and calculated lightning flash densities. I determined the flash densities based on reported observations as described in Sect. \ref{sec:volc}. The values are used to estimate lightning occurrence on the exoplanets Kepler-10b and 55 Cnc e, and the brown dwarf Luhman-16 B (Sect. \ref{sec:casest}).}
  \begin{tabular}{@{}lllllc@{}}	
	\hline 
	N$^{\rm o}$ & Volcano & Eruption date & Information & Reference & \vtop{\hbox{\strut Average flash densities}\hbox{\strut [flashes km$^{-2}$ hour$^{-1}$]}} \\
	\hline \hline
	$[1]$ & \multirow{6}{*}{Eyjafjallaj\"okull} & 14-19 Apr 2010 & \vtop{\hbox{\strut Electrically active for $\sim90$ h}\hbox{\strut 171 strokes observed}\hbox{\strut Standard deviation of location: 4.8 km}} & \multirow{6}{*}{\citet{bennett2010}} & 0.1 \\
	 $[2]$ & & 11-20 May 2010 & \vtop{\hbox{\strut Electrically active for $\sim235$ h}\hbox{\strut 615 strokes observed}\hbox{\strut Standard deviation of location: 3.2 km}} & & 0.32 \\ \hline
	$[3]$ & \multirow{5}{*}{Mt Redoubt} & 23 Mar 2009\tnote{(1)} & \vtop{\hbox{\strut Electrically active for 20.6 min}\hbox{\strut 573 flashes observed}\hbox{\strut Farthest sources from the vent: 28 km}} & \multirow{5}{*}{\citet{behnke2013}} & 12.04 \\ 
	 \vtop{\hbox{\strut $[4]$}\hbox{\strut $[5]$}} & & 29 Mar 2009\tnote{(1)} & \vtop{\hbox{\strut Phase 1: 100 flashes min$^{-1}$ per 3 km$^{2}$}\hbox{\strut Phase 2: 20 flashes min$^{-1}$ per 11 km$^{2}$}} & & \vtop{\hbox{\strut 2000.0}\hbox{\strut 109.0}}\\ 
	\hline
  \label{table:vol}
  \end{tabular}
  \begin{tablenotes}
	\item[1] One of the twenty-three episodes occurring in March-April 2009 \citep{behnke2013}    
  \end{tablenotes}
\end{threeparttable}
}
\vspace{0.8cm}
\end{table*}
%Table 

Electrical activity has long been associated with large-scale, explosive volcanic eruptions \citep{james2008, mather2006}. There are records on lightning events from 1650, occurring at a volcanic eruption near Santorini, Greece \citep{fouque1879}. Eye-witnesses reported electrical phenomena, which coincided with the eruption of the Krakatoa in Indonesia in 1883 \citep{symons1888}. The modern era has produced a high number of volcanic lightning observations, after volcanic eruptions like, e.g., Etna in 1979, 1980; Mt St Helens in 1980, 1983; Gr\'imsv\"otn in 1996, 1998, 2004; or Hekla in 2000; etc. \citep[for references and an extended list see][]{mather2006}. 

In this section, I analyse statistics from two volcanic eruptions: the Icelandic Eyjafjallaj\"okull's eruption from 2010 and the Mt Redoubt eruption in Alaska, 2009 (Table \ref{table:vol}). I derive flash densities (Table \ref{table:vol}), which I use to estimate lightning activity in rocky exoplanet and brown dwarf atmospheres (Sect. \ref{sec:exopl}). The composition of volcanic plumes may reflect the composition of dust clouds on these extrasolar objects.

The Eyjafjallaj\"okull eruption had two main phases: 14-19 April 2010, with 171 strokes occurring in $\sim$90 hours, and 11-20 May 2010, a more intensive one with 615 strokes in about 235 hours \citep{bennett2010}. The standard deviation of the location of the lightning events was 4.8 and 3.2 km, respectively \citep{bennett2010}. I use this information to estimate the influenced area, assuming that the area is a circle with the diameter of the standard deviation. I calculate the stroke density for the two phases to be 0.1 strokes km$^{-2}$ h$^{-1}$ and 0.32 strokes km$^{-2}$ h$^{-1}$, respectively. \citet{bennett2010} measured the multiplicity of the flashes, the number of strokes occurring in one flash, and found that only 14 flashes had 2 strokes, while all other flashes were composed of single strokes. Based on this information, I assume that the flash densities during the 2010 Eyjafjallaj\"okull eruption are equal to the calculated stroke densities (14-19 April 2010: 0.1 km$^{-2}$ h$^{-1}$; 11-20 May 2010: 0.32 km$^{-2}$ h$^{-1}$).

\citet{behnke2013} analysed various episodes of the 2009 Mt Redoubt eruption. I used the information on two episodes: the 23 March 2009 episode, which resulted in the occurrence of 573 lightning flashes in 20.6 minutes (0.34 hours); and the 29 March 2009 episode with two main phases, the first with a flash rate of 100 flashes min$^{-1}$ over a 3 km$^2$ area and the second with 20 min$^{-1}$ over 11 km$^2$. During 23 March 2009 the farthest sources were located 28 km from the vent \citep{behnke2013}, which suggest that vent dynamics may not be the primary driver for this lightning. Assuming that the affected area can be approximated by a rectangle of sizes 28 km $\times$ 5 km \citep[][fig. 6]{behnke2013}, the total affected area would be 140 km$^2$. The obtained average flash density for the 23 March 2009 episode is 12.04 km$^{-2}$ h$^{-1}$. The episode 29 March 2009 show much larger flash densities, with 2000 km$^{-2}$ h$^{-1}$ for the intensive first phase and 109 km$^{-2}$ h$^{-1}$ for the longer second phase.

\citet[][table 3]{mather2006} list flash densities based on \citet{anderson1965} for volcano plumes to be between 0.3 and 2.2 km$^{-2}$ min$^{-1}$, which is 18 and 132 km$^{-2}$ h$^{-1}$, respectively. The large lightning storm on 29 March 2009 around Mt Redoubt shows comparable flash densities during its second phase. The obtained flash densities (Table \ref{table:vol}) are used to estimate lightning occurrence on rocky exoplanets without water surfaces, and on brown dwarfs, since clouds on these types of objects may resemble volcano plumes. I note that lightning statistics are not well studied in case of volcano eruptions. The values listed in Table \ref{table:vol} (last column) are guides and may only be used under certain assumptions as I outline in Sections \ref{sec:casest} and \ref{sec:flashdens}.

%__________________________________________________________________
%__________________________________________________________________
\section{Lightning on other Solar System planets} \label{sec:solsys}

\subsection{Lightning on Venus?} \label{sec:venus}

Chapter \ref{chap:liginout}, Section \ref{sec:int_ven} summarizes the observations of Venusian lightning and gives an overview of modelling and experimental work. Early studies derived flash occurrence for Venus based on in-direct measurements: \citet{gurnett2001} calculated a lower limit of flash rate from the non-detection of lightning by the \textit{Cassini} spacecraft to be 70 s$^{-1}$. \citet{krasnopolsky2006} detected NO in the spectra of Venus, which they related to lightning activity in the lower atmosphere of the planet and inferred a flash rate of 90 s$^{-1}$, which seems very high considering that it is almost twice the flash rate on Earth and \textit{Cassini} did not detect any signals from Venus but did detect lightning on Earth. \citet{russell2008} reported whistler detections by \textit{Venus Express} in 2006 and 2007. They suggested that the whistlers originated from lightning discharges and inferred a stroke rate of 18 s$^{-1}$. 

\textit{Venus Express} detected lightning-induced whistlers between 2012 and 2013 as well. The data were analysed by \citet{hart2015}, who confirmed the whistler events with dynamic spectra. However, since the magnetic field around Venus is not yet fully understood, the field lines cannot be traced back to their origin, therefore, the coordinates of the source of the lightning events are unknown. Although exact locations are not available, one can estimate preliminary statistics from the number of bursts\footnote{\citet[][priv. com.]{hart2014b} defined a burst as an event of at least one second in duration and separated from other events by at least one second.} observed by \textit{Venus Express}. \citet[][priv. com.]{hart2014b} counted 293 bursts in total with varying duration during three Venus-years (between 2012 and 2013). Obtained flash densities and their possible applicability are shown in Table \ref{table:plan} and discussed in Sect. \ref{sec:exopl}.

%__________________________________________________________________
\subsection{Giant gas planets} \label{sec:ggp}

Optical and radio observations confirmed the presence of lightning on both giant gas planets, Jupiter and Saturn (Chapter \ref{chap:liginout}, Section \ref{sec:int_jup}$-$\ref{sec:int_sat}). Due to the position of the spacecraft, the existing data are limited to specific latitudes and observational times for both planets. Bearing in mind these limitations, i.e. we do not have data from the whole surface of the planet or from continuous observations for a longer period of time (e.g. a year), I use the available data to estimate flash densities for the whole globes of these planets (Table \ref{table:plan}), assuming that at least a similar lightning activity can be expected inside their atmospheres.

\citet{little1999} estimated a lower limit for flash densities on Jupiter to be $4.2 \times 10^{-3}$ flashes km$^{-2}$ year$^{-1}$ based on \textit{Galileo} observations. This value agrees well with the values estimated from the \textit{Voyager} measurements  \citep[$4 \times 10^{-3}$ flashes km$^{-2}$ year$^{-1}$,][]{borucki1982}. \citet{dyudina2004} analysed the same data set and complemented it with \textit{Cassini} observations. In 2007 \textit{New Horizons} observed polar (above $60^\circ$ latitude south and north) lightning on Jupiter with its broadband camera (0.35 - 0.85 $\mu$m bandpass). From these data, \citet{baines2007} found almost identical flash rates for the polar regions on both hemispheres (N: 0.15 flashes s$^{-1}$, S: 0.18 flashes s$^{-1}$).

On 17 August 2009 images of Saturn's night side were taken by \textit{Cassini}. Lightning flashes were located on a single spot of the surface at $\sim -36^{\circ}$ latitude \citep{dyudina2010}. On 30 November 2009 flashes were observed at about the same latitude as before. The flash rate from these observations is 1$-$2 min$^{-1}$ \citep{dyudina2013}. \citet{dyudina2013} reported further lightning observations on the dayside by \textit{Cassini} at latitude $35^{\circ}$ north. A new, much stronger storm was observed on 26 February 2011 between latitudes $30^{\circ}-35^{\circ}$ north. A flash rate of 5 s$^{-1}$ was estimated for this storm \citep{dyudina2013}. In the meantime, simultaneous SED observations were conducted with the \textit{Cassini}-RPWS instrument between $\sim 2-16$ MHz (the first value is the low cut-off frequency of Saturn's ionosphere, while the second one is the instrumental limit). SED rates and flash rates vary for the three storms. Radio (SED) observations were previously carried out in 2004-2006 by the RPWS instrument as well. The different storms were observed in different antenna mode, which have different sensitivity. When calculating the SED rates, \citet{fischer2006} took into account the instrument mode. The storms and SED episodes are listed in \citet{fischer2006}, their table 1. They found SED rates varying between 30$-$87 h$^{-1}$. Two more SED storms (D and E) were observed in 2005 and 2006 with SED rates much higher than before \citep[367 h$^{-1}$,][]{fischer2007}. \citet{fischer2011} analysed the SED occurrence of the 2011-storm that started in early December 2010, and found the largest SED rates ever detected on Saturn, to be 10 SED s$^{-1}$. This results in, on average, 36000 SED h$^{-1}$, $\sim 98$ times larger than the SED rate of the largest episode of storm E from 2006.

%__________________________________________________________________
\subsubsection{Lightning climatology on Jupiter and Saturn} \label{subs:climatjs}

%Table - Exoplanet types -- flash densities
\begin{table*}
\resizebox{\columnwidth}{!}{
\begin{threeparttable}
 \centering
 \caption{Lightning flash densities ($\rho_{\rm flash}$) from four Solar System planets. Exoplanetary examples are also listed under six categories where the flash densities were considered. All values are based on observations. Flash densities are calculated over a year defined in Earth-days, and an hour. Hourly densities are used for estimating lightning activity on exoplanets and brown dwarfs. Yearly $\rho_{\rm flash}$ are calculated in Earth-, Venusian-, Jovian-, or Saturnian-years explained in the text (Sect. \ref{subs:climatjs}).}
  \begin{tabular}{@{}lllllll@{}}	
	\hline 
	Planet & Region & Instrument\tnote{(1)} & \vtop{\hbox{\strut Average yearly $\rho_{\rm flash}$}\hbox{\strut [flashes km$^{-2}$ year$^{-1}$]}} & \vtop{\hbox{\strut Average hourly $\rho_{\rm flash}$}\hbox{\strut [flashes km$^{-2}$ hour$^{-1}$]}} & Exoplanet type & Example \\
	\hline \hline
	\multirow{5}{*}{Earth} & global & LIS/OTD & 2.01 & $2.29\times 10^{-4}$ & Earth-like planet & Kepler-186f \\
	 & continents & LIS-scaled WWLLN & 17.0 & $1.94 \times 10^{-3}$ & \multirow{2}{*}{\vtop{\hbox{\strut Rocky planet with}\hbox{\strut no liquid surface}}} & \multirow{2}{*}{\vtop{\hbox{\strut Kepler-10b}\hbox{\strut 55 Cnc e}}} \\
	 & & LIS/OTD & 28.9 & $3.30 \times 10^{-3}$ & & \\
	 & oceans & LIS/OTD & 0.3 & $3.42 \times 10^{-5}$ & Ocean planet & Kepler-62f \\
	 & & LIS-scaled WWLLN & 0.6 & $6.85 \times 10^{-5}$ & & \\
	\hline
	Venus & global\tnote{(2)} & Venus Express & $2.12\times 10^{-7}$ & $3.64 \times 10^{-11}$ & Venus-like planet & Kepler-69c \\
	\hline
	\multirow{3}{*}{Jupiter} & \multirow{3}{*}{global} & \multirow{3}{*}{\vtop{\hbox{\strut Galileo\tnote{(3)}}\hbox{\strut New Horizons}}} & \multirow{3}{*}{\vtop{\hbox{\strut $2.46 \times 10^{-2}$}\hbox{\strut 0.15}}} & \multirow{3}{*}{\vtop{\hbox{\strut $2.37 \times 10^{-7}$}\hbox{\strut $1.43 \times 10^{-6}$}}} & giant gas planets & \vtop{\hbox{\strut HD 189733b}\hbox{\strut GJ 504b}}\\ \cdashline{6-7}
	 & & & & & brown dwarfs & Luhman-16B \\
	\hline
	\multirow{3}{*}{Saturn} & \multirow{3}{*}{global} & \multirow{3}{*}{\vtop{\hbox{\strut Cassini (2009)}\hbox{\strut Cassini (2010/11)}}} & \multirow{3}{*}{\vtop{\hbox{\strut $1.53 \times 10^{-2}$}\hbox{\strut 1.31}}} & \multirow{3}{*}{\vtop{\hbox{\strut $8.20 \times 10^{-8}$}\hbox{\strut $5.09 \times 10^{-6}$}}} & giant gas planets & \vtop{\hbox{\strut HD 189733b}\hbox{\strut GJ 504b}}\\ \cdashline{6-7}
	 & & & & & brown dwarfs & Luhman-16B \\
	\hline 
  \label{table:plan}
  \end{tabular}
  \begin{tablenotes}
	\item[1] Flash densities are calculated from the data gathered by these instruments
	\item[2] Based on whistler observations, assuming 1 whistler/flash  
	\item[3] Excluding detections during the C20 orbit \citep{gierasch2000, dyudina2004}. 
  \end{tablenotes}
\end{threeparttable}
}
\vspace{0.8cm}
\end{table*}
%Table 2

\begin{figure}
  \begin{center}
  \includegraphics[trim=0cm 0cm 0cm 0cm, scale=0.4]{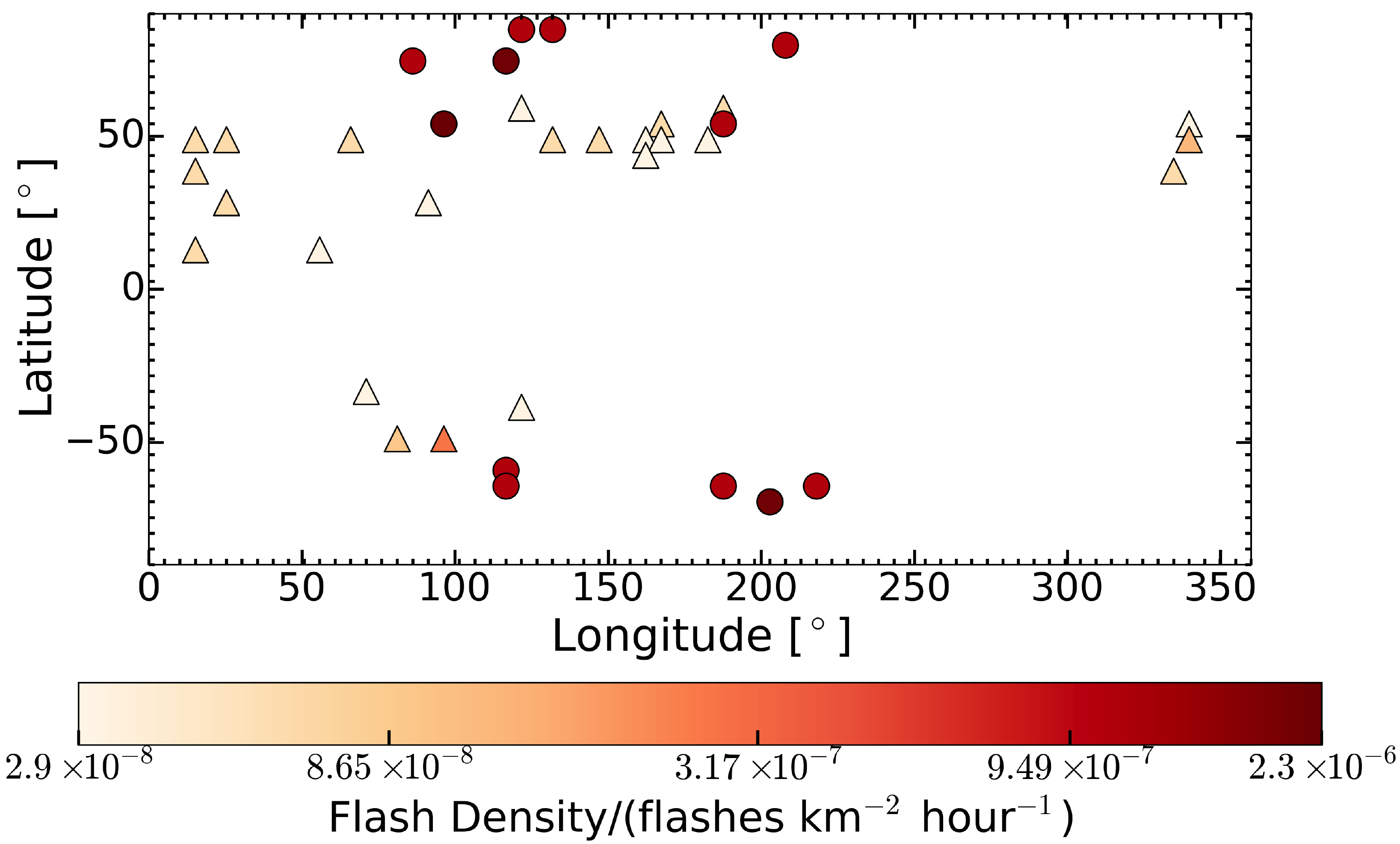}
  \end{center}
  \caption{Jovian lightning occurrence. The colours show the number of flashes averaged in a $5^{\circ} \times 5^{\circ}$ area box on the surface of the planet in an hour on a logarithmic scale. Triangles: \textit{Galileo} data \citep[year: 1997,][]{little1999}, Circles: \textit{New Horizons} data \citep[year: 2007,][]{baines2007}. The 10-year gap between the two data sets implies that the plotted lightning flashes are from two different storms.}
  \vspace{0.8cm}
  \label{fig:jup}
\end{figure}

\begin{figure}
  \begin{center}
  \includegraphics[trim=0.5cm 0cm 0cm 0cm, scale=0.4]{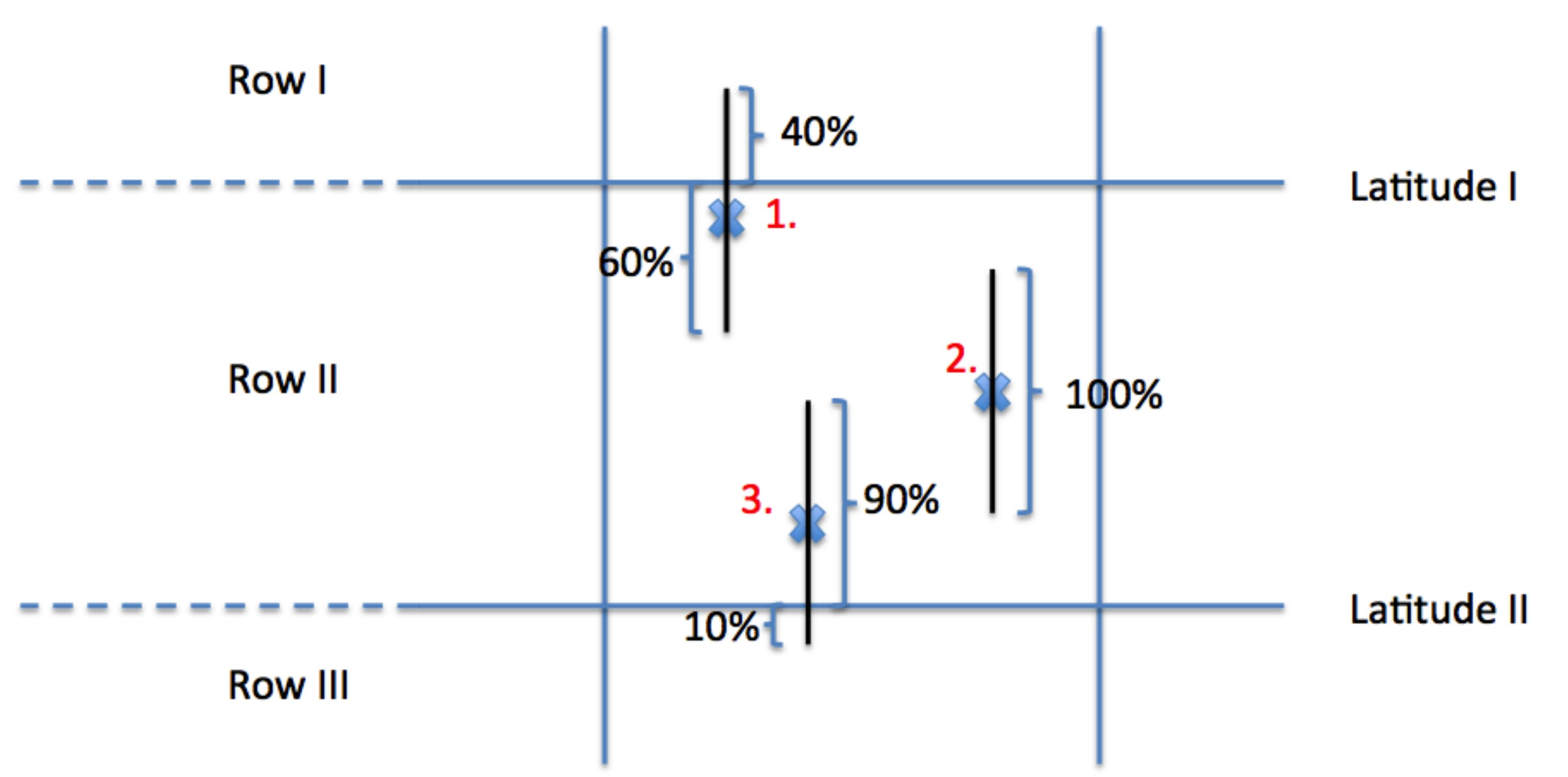}
  \end{center}
  \caption{Sketch of grid cells and correction of lightning flash locations. Latitudes I and II define the top and bottom boundaries of a cell. Blue $\times$ signs show the position of the lightning flash with latitude and longitude coordinates. The black error bars are calculated from the spatial resolution of the instrument. The correction is based on the length of the error bar. When counting the flashes in one grid cell the individual flashes are summed up based on what portion of the full error bar is in the particular cell. E. g. flash 1 is counted 0.6 times in the Row II cell and 0.4 times in the Row I cell; flash 2 is counted as 1 in Row II; flash 3 adds 0.9 times to Row II and 0.1 times to Row III. Adding up, there are 0.4 flashes in Row I, $1 + 0.6 + 0.9 = 2.5$ flashes in Row II and 0.1 flashes in Row III, in this example. (The error bars on the figure are for illustration only and do not represent real proportionality to the grid cells.)}
  \vspace{0.8cm}
  \label{fig:grid}
\end{figure}

%% ---- JUPITER
Data for Jupiter were taken from \citet[][table 1]{little1999}, \citet[][table 1]{dyudina2004} and \citet[][table 1]{baines2007}. The \textit{Galileo} spacecraft observed lightning activity on Jupiter during two orbits in 1997 (C10, E11) and one orbit in 1999 (C20). The surveyed area covers more than half of the surface of the planet \citep{little1999}. \citet{dyudina2004}, their table 1, also lists lightning detections from \textit{Galileo}'s C20 orbit partly based on \citet{gierasch2000}. However, there is no information on the occurrence rate of lightning from this orbit, or the coordinates of the observed flashes. Therefore, I did not include these detections in my study. Similarly, no observed coordinates, or flash number estimates are given for the lightning storms observed by \textit{Cassini}, listed in \citet{dyudina2004}, which are also omitted from this study. I summarize these observational data in Figure \ref{fig:jup}, which shows the total number of flashes in an hour (logarithmic scale), averaged in $5^{\circ} \times 5^{\circ}$ area boxes over the surface of Jupiter. As explained in Fig. \ref{fig:grid}, and below, I corrected the spatial (latitudinal) coordinates of the flashes from the \textit{Galileo} data with the pointing error of the instrument calculated from the spatial resolutions given in \citet{little1999}.\footnote{We note that the spatial resolution of the \textit{Galileo} satellite is much finer than the grid set up by us. However, flashes close to the grid edges may overlap two grid cells if the error bars are considered, as described in Fig. \ref{fig:grid}, in which case it is worth applying these error calculations.} The same correction was done for the \textit{New Horizons} data based on spatial resolutions from \citet{baines2007}.

%% ---- SATURN

\begin{figure}
  \begin{center}
  \includegraphics[trim=0cm 0cm 0cm 0cm, scale=0.37]{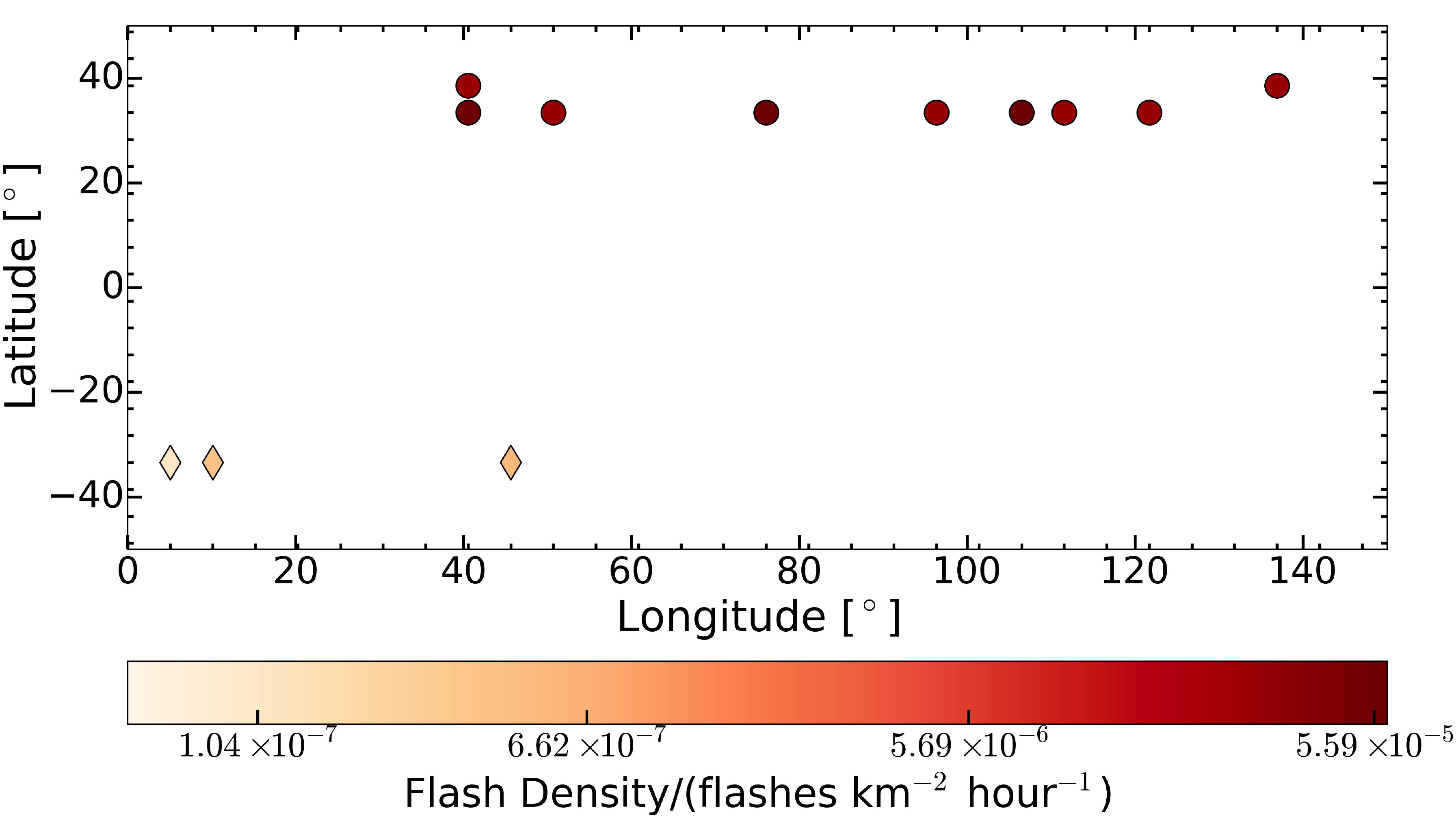}
  \includegraphics[trim=0cm 0cm 0cm 0cm, scale=0.37]{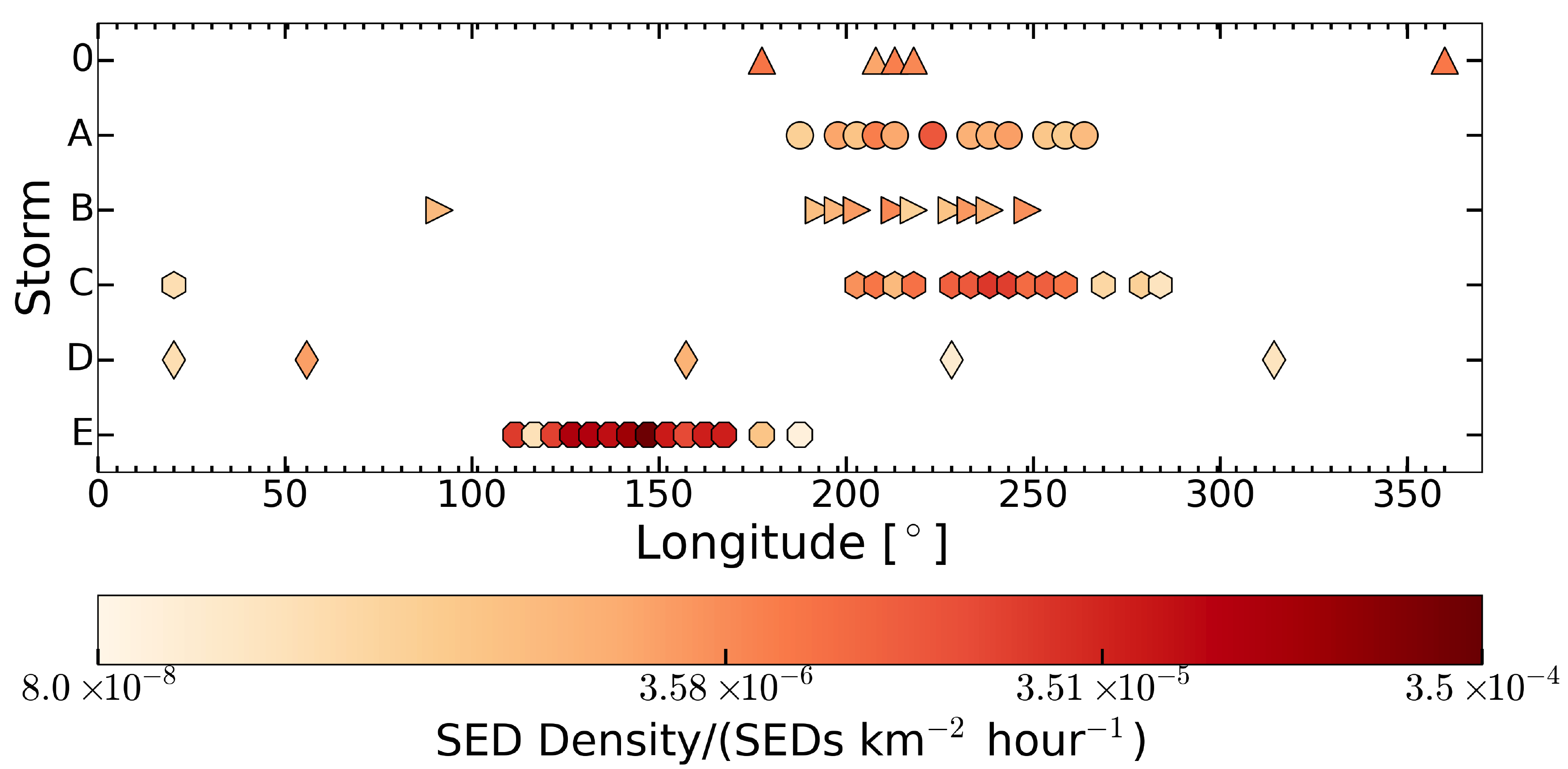}
  \end{center}
  \caption{\textbf{Top}: Saturnian optical lightning occurrence from \textit{Cassini} observations. Data \citep{dyudina2013} are from the years 2009 (diamonds) and 2011 (circles). The data shown are from two different storms. Shown surface region: $\pm45^{\circ}$ latitude, $0^{\circ}-150^{\circ}$ longitude. \textbf{Bottom}: Radio lightning emission, SED, occurrence on Saturn from \textit{Cassini}-RPWS observations in $2004-2007$ \citep{fischer2006, fischer2007}. All SED storms shown on the plot appeared at the $-35^{\circ}$ latitude region. The figure shows the SED density of each storm plotted against the planetary longitude. The colours on both images show the flash/SED densities averaged in a $5^{\circ} \times 5^{\circ}$ area box on a logarithmic scale.
}
  \vspace{0.8cm}
  \label{fig:7}
\end{figure}

Saturnian optical data were taken from \citet[][table A1]{dyudina2013}. They list, amongst others, latitudes, longitudes, times of observations, exposure times and spatial resolution. The top panel of Fig. \ref{fig:7} shows the spatial distribution of lightning flashes observed on Saturn in 2009 (diamonds) and 2011 (circles), between latitudes $\pm45^{\circ}$  and longitudes $0^{\circ}-150^{\circ}$. The concentration around $\pm 35^{\circ}$ latitudes is clearly seen. The spatial coordinates of the data were corrected with the spatial resolution of the instrument taken from \citet[][Supplement]{dyudina2013} as explained below. To illustrate SED occurrence, I used data taken from \citet{fischer2006} and \citet{fischer2007}. The bottom panel of Fig. \ref{fig:7} shows the SED density on Saturn for 6 different storms, which all appeared on $-35^{\circ}$ latitude. SED observations were reported from the 2011 storm \citep{dyudina2013}, but because of the lack of the spatial coordinate information, I do not plot them in Fig. \ref{fig:7}.

The above-mentioned corrections were applied because the observing instruments have pointing errors, which result in an uncertainty of the location measurement for lightning flashes. To correct for the uncertainty in latitudes in the \textit{Galileo} data I used \citep{little1999}:

\begin{equation} \label{aeq:1}
{\rm error} = \frac{{\rm point}_{\rm err}}{\cos({\rm lat})} \times \frac{360}{462000 \times \cos({\rm lat})},
\end{equation}

\noindent where point$_{\rm err} = {m} \times {res}$ is the pointing error in km, $m$ is 20 pixels or 40 pixels depending on the observing mode \citep{little1999}, $res$ is the image resolution in km, lat is the latitude at which the lightning flash was observed, and $360/462000$ converts km to degrees for Jupiter. The first term of Eq. \ref{aeq:1} (${\rm point}_{\rm err} / \cos({\rm lat})$) is the spatial resolution of the instrument. Similar calculations can be applied to the New Horizons and the Cassini data. The spatial resolution of these data sets, given in km, is taken from \citet{baines2007} and \citet[][Supplement]{dyudina2013} and it is converted into degrees to get the latitude correction. In case of Saturn and the Cassini data, the conversion factor is $360/378680$.

The spacecraft observing Jupiter (e.g. \textit{Voyager}, \textit{Galileo}) have found that Jovian lightning activity has a local maximum near $50^{\circ}$ N \citep[Fig. \ref{fig:jup}; see also][]{little1999}. This might be a consequence of the increasing effect of internal heating compared to solar heating at this latitude. Here, convection is more effective producing thunderclouds with lightning \citep{baines2007}. Solar heating would suppress this effect. \citet{zuchowski2009} modelled the meridional circulation in stratospheric and tropospheric heights of Jupiter's atmosphere, and found an upwelling in the zones and downwelling in the belts in stratospheric levels. However, at lower atmospheric heights upwelling was found in the belts, which allows the formation of water clouds and lightning discharges, just like observations indicate \citep{little1999, ingersoll2000, zuchowski2009}. \citet{dyudina2013} found that on Saturn lightning occurs in the diagonal gaps between large anticyclones. These gaps are similar to Jovian belts, composed of upwelling, convective thunderstorms \citep[Fig. \ref{fig:7};][]{dyudina2013, read2011}. I do not attempt to compare lightning occurrence via longitudes, since due to the drift of the storms that would not be a valid approach without correcting for this drift.

The results in Table \ref{table:plan} include hourly and yearly average flash densities obtained for the Solar System planets.\footnote{The results in Table \ref{table:plan} are based on positive detections of lightning. This is important especially on Saturn, where most of the time no storm was observed resulting in 0 flash densities \citep{fischer2011b}.} Yearly flash densities were calculated for a year defined in Earth-days (24-hour days), and they represent the length of a year on the appropriate planet. For example: when calculating flash rates (flashes year$^{-1}$) for Jupiter, I used a Jovian-year of 4330 days and not 365-Earth days (apart from Earth lightning flash rates). Similarly I define Venusian- and Saturnian-years too. Global flash densities were estimated for all of the planets (Table \ref{table:plan}). For Earth, I distinguish between continental and oceanic rates. The values in Table \ref{table:plan} for the two latter regions are calculated from LIS/OTD (larger value) and LIS-scaled WWLLN (lower value) data. Similarly, the larger values for Jupiter are estimated from  \textit{New Horizons} data, while lower ones are based on \textit{Galileo} data. For Saturn, the larger values are based on data from the giant storm in 2011, while the lower ones are from the 2009-storm.

I calculated flash rates (flashes year$^{-1}$ or flashes hour$^{-1}$; $R_{\rm flash}$) for Jupiter and Saturn for each of the images taking into account the exposure times as given by:

\begin{equation} \label{eq:stat1}
R_{{\rm flash},i} = \frac{n_i}{t_{{\rm exp},i}} C,
\end{equation}
\noindent where $n$ is the number of flashes detected in image $i$, $t_{\rm exp}$ is the exposure time of the image in seconds, and $C$ is a unitless scaling factor, which converts the time units from seconds to hours or years.\footnote{I do not analyse flash rates. For more details about flash rates see \citet{dyudina2013}, their table 2.} The flash density (flashes unit-time$^{-1}$ km$^{-2}$, $\rho_{\rm flash}$), is calculated from Eq. (\ref{eq:stat2}), with $R_{\rm flash}$, given by Eq. (\ref{eq:stat1}).

\begin{equation} \label{eq:stat2}
	\rho_{\rm flash} = \frac{\sum_{i=1}^{i=N}{R_{{\rm flash},i}}}{A_{\rm surv}},
\end{equation}
\noindent where $N$ is the total number of images and $A_{\rm surv}$ is the total surveyed area: $A_{\rm surv}^{\rm Galileo} = 39.5 \times 10^9$ km$^2$ \citep{little1999}, $A_{\rm surv}^{\rm New Horizons} = 8.0 \times 10^9$ km$^2$ \citep{baines2007}\footnote{Calculated based on \citet{baines2007}, information on image resolution in footnote 15 and surveyed latitude range in figure 1.}. $A_{\rm surv}$ for the 2009 storm on Saturn is the 30\% of Saturn's surface area \citep[][Supplement]{dyudina2010}, and $A_{\rm surv}$ for the 2011 storm is the total area of Saturn based on the fact that the \textit{RPWS} instrument detected only one SED storm on the whole planet at a time \citep{fischer2006, fischer2007}. The flash density given by Eq. \ref{eq:stat2} can be considered as a global flash density, even though it is calculated using observations restricted to part of the planet. The underlying assumption is that the flash density calculated from a storm, on average, can be extrapolated to the whole globe. This is applied for all investigated planets, Earth, Jupiter, Saturn, and Venus.

The flash densities for Jupiter derived here are different from previously published values \citep[$\sim 4 \times 10^{-3}$ flashes km$^{-2}$ year$^{-1}$,][]{little1999, borucki1982}, which is the result of converting exposure times, which are given in seconds, to years. For example, from the \textit{Galileo} data I obtain a flash density of $2 \times 10^{-2}$ flashes km$^{-2}$ year$^{-1}$ when I take the length of a Jovian year to be the number of days Jupiter orbits the Sun, 4330 days. This way I get a flash density an order of magnitude higher than previously estimated \citep[e.g.][]{little1999}. However, when I determine the flash rate (flashes year$^{-1}$) considering a year to be 365 days long, the way it is done in \citet{little1999}, and divide it by the \textit{Galileo} survey area, the result becomes the same order of magnitude but twice lower than the one in \citet{little1999}, or $2 \times 10^{-3}$ flashes km$^{-2}$ year$^{-1}$ compared to $4 \times 10^{-3}$ flashes km$^{-2}$ year$^{-1}$. This factor of two is a reasonable difference, since I do not consider over-lapping flashes in my work (U. Dyudina, private communication). \citet{little1999} calculated flash densities saying that on average there were 12 flashes detected in one storm. They multiplied this by the number of storms observed (26, their table I) and divided by an exposure time of 59.8 s and the total survey area of $39.5 \times 10^9$ km$^2$. In my approach, I took the data from table I in \citet{little1999} and table 1 of \citet{dyudina2004}, counted the flashes on each frame, assuming that one "lightning spot" in table 1 of \citet{dyudina2004} corresponds to one lightning flash, then divided that number with the exposure time (in years or hours, with 1 year on Jupiter being $3.73 \times 10^8$ s) of the frame. After summing up these flash rates, I divided the result by the total surveyed area of $39.5 \times 10^9$ km$^2$. In summary, the differences between previously calculated flash densities and flash densities listed in Table \ref{table:plan} are the result of converting exposure times to years. However, for my purposes I only use hourly flash densities, which do not depend on the length of a year.

The above derived formulas and the resulting values listed in Table \ref{table:plan} involve various uncertainties, which also affect the comparability. The flash rate, $R_{\rm flash}$, depends on the number of detected flashes ($n_i$) at a certain time determined by the exposure time ($t_{{\rm exp},i}$). $n_i$ is affected by instrumental sensitivity, the time of the survey (seasonal effects on lightning occurrence) and the place of the survey (different lightning occurrence over different latitudes and surface types, Figs. \ref{fig:1}-\ref{fig:4b}). The flash density, $\rho_{\rm flash}$, is derived from $R_{\rm flash}$ (Eq. \ref{eq:stat2}). Uncertainties also rise from the not-precise determination of total surveyed area. Bearing in mind these limitations of the data and uncertainties in the values in Table \ref{table:plan}, I apply the results of flash densities on exoplanets and brown dwarfs in Sect. \ref{sec:exopl}.

%__________________________________________________________________
\subsection{Energy distribution} \label{subs:endis}

\begin{figure*}
  \begin{center}
  \includegraphics[trim=0cm 0cm 0cm 0cm, scale=0.09]{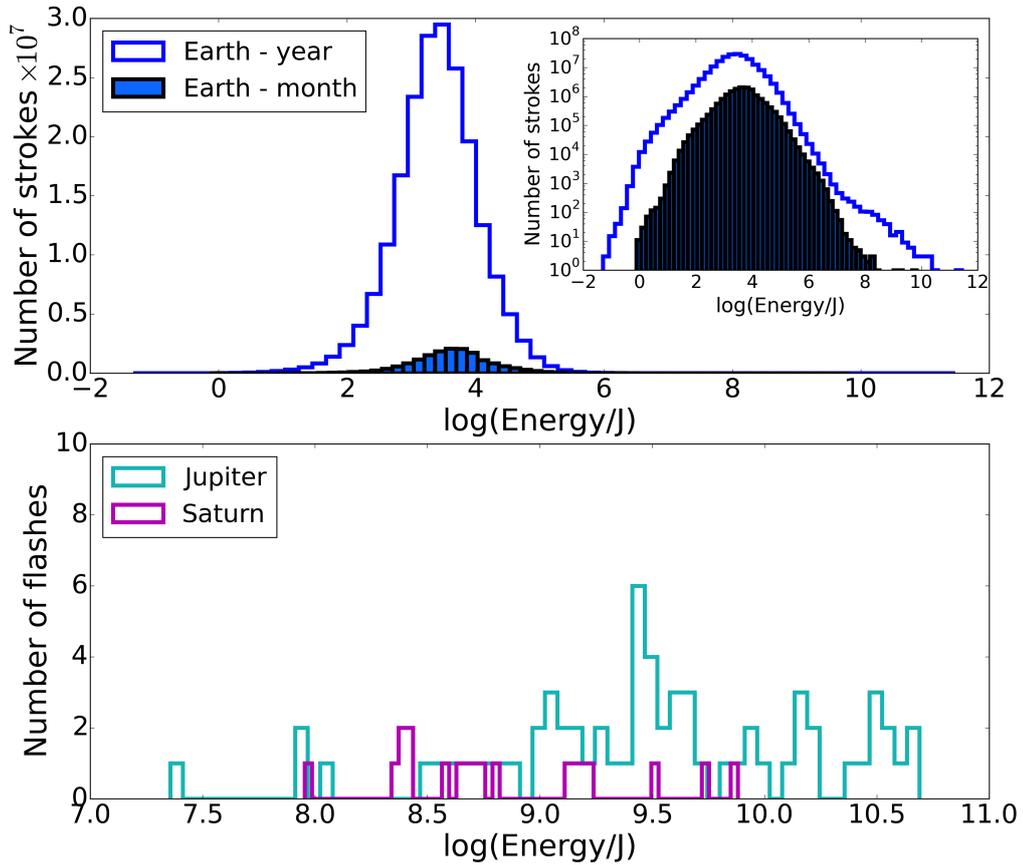}
  \end{center}
  \caption{Radio energy distribution of lightning strokes from Earth (WWLLN, 2013; top) based on data from the whole year compared to a month (2013, December), which do not distinguish between IC and CG lightning; and optical energy distribution (calculated from measured powers, see Sect. \ref{subs:endis}) of Jovian and Saturnian flashes (bottom). The data from Jupiter (\textit{Galileo}, 1997; \textit{New Horizons}, 2007) and Saturn (\textit{Cassini}, 2009, 2011) are both from less than an hour of observations (about 50 minutes for Jupiter and 10 minutes for Saturn \citep[exposure times in][]{little1999, baines2007, dyudina2013}). The number-energy distribution of the Earth-data seems to be self-similar in time as it has the same shape if plotted for a month or for a year. The inset plot (top panel) shows a log-log scale of the Earth data.
}
  \label{fig:endist}
  \vspace{0.8cm}
\end{figure*}

Figure \ref{fig:endist} summarizes the number distribution of stroke energies for Earth (top), and number distribution of flash energies for Jupiter and Saturn (bottom). For Earth I used WWLLN data from 2013, while for the outer planets I included all data from \textit{Galileo}, \textit{New Horizons} and \textit{Cassini}. \citet{dyudina2004} lists the power [$W = {\rm J} {\rm s}^{-1}$] of lightning as observed by the \textit{Galileo} probe (their table 1, column 11). Following the procedure in \citet[][eq. 1]{dyudina2013} where they treated storms as continuously flashing steady light sources and each flash as a patch of light on a Lambertian surface, I converted the measured power values to energies by multiplying them with the exposure time. On Earth most of the strokes have radio energies of the order of $10^3-10^{3.5}$ J. This indicates that less energetic lightning flashes, due to their large number, are likely to be more significant for chemically changing the local gas in large atmospheric volumes. However, a detailed modelling of the structure and size of discharge channels are required for drawing more definite conclusions.

We need to be careful with over-interpretation of the directly accessible data; however, the knowledge gained about their limitations is useful when discussing lightning observability. Due to instrumental limitations (detection threshold), only the most energetic lightning events are detectable. This is particularly prominent in the Saturnian and Jovian data (Fig. \ref{fig:endist}, bottom panel). It seems impossible to find the peak of the energy distribution, being lower than the detection limit, on Saturn and Jupiter just by extrapolating the limited number of data points. However, one may assume that most of the lightning flashes will cluster around one energy also for Jupiter and Saturn, and that this peak in flash numbers will move to higher energies compared to Earth. This expectation is based on the fact that the underlying physics (i.e. electron avalanches develop into streamers in an electric potential gradient)  is only marginally affected by the chemical composition of the atmospheric gas \citep[e.g.][]{helling2013}, and the fact that Jupiter's and Saturn's clouds have a larger geometrical extension and, hence, a larger potential difference than on Earth. \citet{bailey2014} showed that a larger surface gravity, like on Jupiter compared to Earth, leads to larger geometrical extension of a discharge event with higher total dissipation energies. \citet{dyudina2004} suggested that their lightning power values derived from observations are underestimates, as $25\%$ of the lightning spots are saturated in the \textit{Galileo} images. They do not consider the scattered light on clouds, which may dim the flashes by a couple of orders of magnitude \citep{dyudina2002}. This suggests that the observed energies on Jupiter are most likely exceeding the largest lightning energies observed on Earth. From this, one may assume that the peak of the energy distribution of lightning flashes on the gas giant planets also shifts to higher energies. \citet{dyudina2004} analysed the power distribution of optical lightning flashes on Jupiter, considering only flashes recorded by \textit{Galileo}'s clear filter\footnote{385 - 935 nm \citep{little1999}.} (their fig. 7). They showed that the number of flashes with high power is small, which is similar to observations for Earth (similarly: Fig. \ref{fig:endist}, top panel). However, observations result in low detected flash numbers. Moreover, lightning observations in the Solar System have biases towards higher energy lightning. Therefore, \citet{dyudina2004} concluded that lightning frequencies at different power levels cannot be predicted unequivocally.

We also note that \citet{farrell2007} suggested that Saturnian discharges might not be as energetic as they were thought to be ($\sim 10^{12}$ J). They assumed a shorter discharge duration, which would result in lower discharge energies. Their study shows the importance of exploring the parameter space that affects lightning discharge energies and radiated power densities, in order to interpret possible observations of not yet fully explored planets. I present such a parameter study in Chapter \ref{chap:model}.

%__________________________________________________________________
%__________________________________________________________________
\section{Discussing lightning on exoplanets and brown dwarfs} \label{sec:exopl}

\begin{figure}
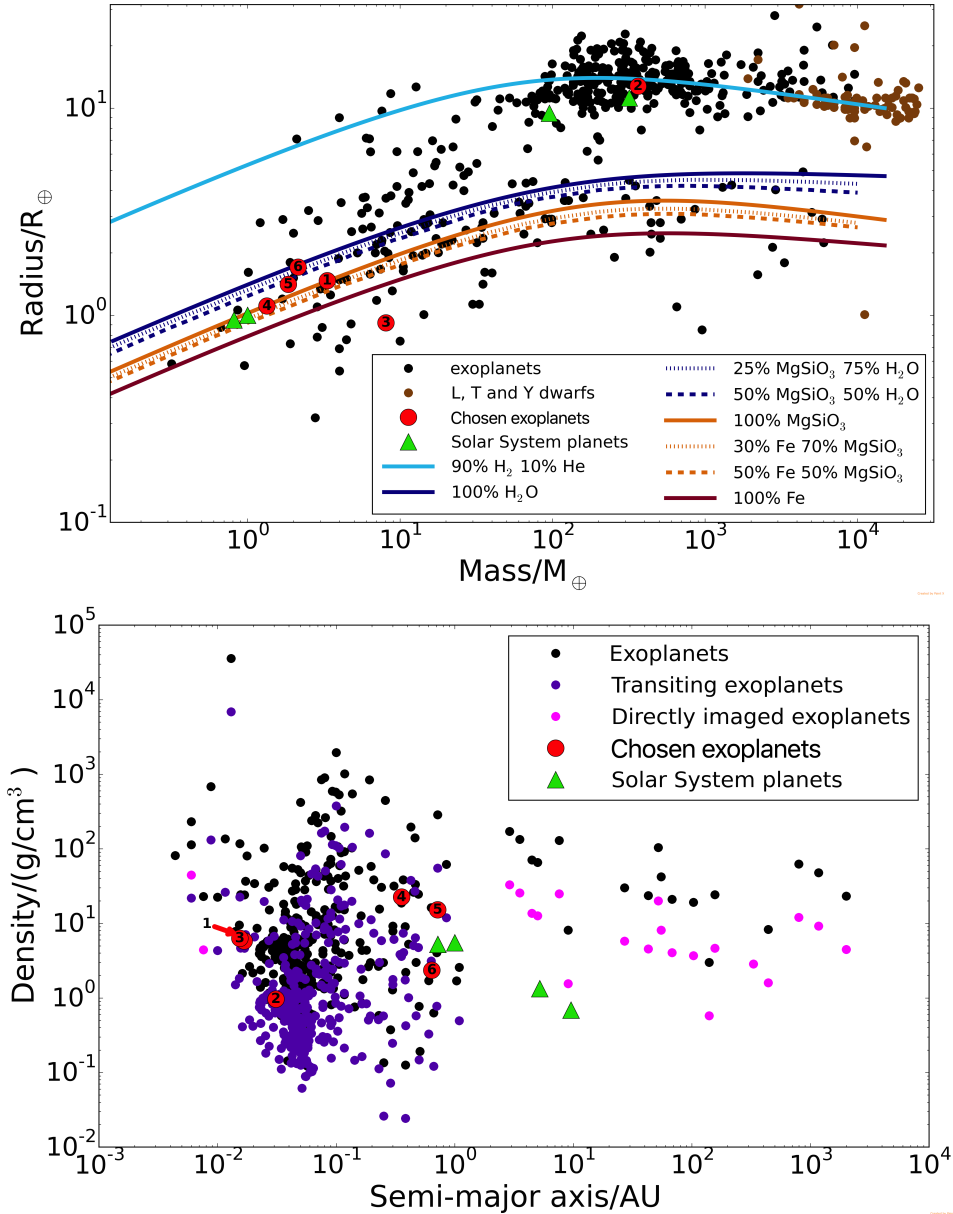

  \begin{center}
  \includegraphics[trim=0cm 0cm 0cm 0cm, scale=0.066]{./chapters/3_figs/09a_exopl_massradius_v4.pdf}
  \includegraphics[trim=0cm 0cm 0cm 0cm, scale=0.0665]{./chapters/3_figs/09b_exopl+transdir_au-density_loglog_eu_6.pdf}
  \end{center}
  \caption{Diversity of known exoplanets and brown dwarfs. Red circles with numbers represent the exoplanet examples used in the case studies (Sect. \ref{sec:casest}): 1 - Kepler-10b, 2 - HD 189733b, 3 - 55 Cnc e, 4 - Kepler-186f, 5 - Kepler-62f, 6 - Kepler-69c. GJ 504b is not on the plots, since no radius is available for this planet. Green triangles indicate Venus, Earth, Jupiter and Saturn.
\textbf{Top}: Relation between mass and radius (in Earth-values: M$_{\oplus}$, R$_{\oplus}$). Black and brown dots represent exoplanets and brown dwarfs, respectively. The lines show mass-radius relationships for various bulk compositions. I note that for some cases the uncertainties in mass and radius are large enough to move the planet from one compositional region to the other. The uncertainties are especially large for Kepler-62f, for which only the upper mass limit is known (Table \ref{table:planet}). However, \citet{kaltenegger2013} estimated the mass of Kepler-62f to be, on average, $\sim 1.85$ M$_{\oplus}$, and I use this value on the figure.
\textbf{Bottom}: Average density ($\rho_{\rm bulk}$ [g/cm$^3$]) vs semi-major axis ($a$ [AU]) of exoplanets. Blue dots indicate transiting planets, magenta dots show directly imaged planets. The density of Kepler-186f, Kepler-62f (see Sect. \ref{sec:exopl}) are mean densities calculated from the radius and mass, while the density of Kepler-10b \citep{dumusque2014}, 55 Cnc e \citep{demory2015} and HD 189733 b \citep{torres2008} are from Markov Chain Monte Carlo analysis of photometric data.}
  \label{fig:mr}
  \vspace{0.8cm}
\end{figure}

The Solar System planets, especially Earth, have been guiding exoplanetary research for a long time. Models have been inspired, for example, for cloud formation \citep[e.g.][]{lunine1986, ackerman2001, helling2008, kitzmann2010} and global atmospheric circulation \citep[e.g.][]{dobbs-dixon2013, mayne2014, zhang2014}, and have been used for predictions that reach far beyond the Solar System. Habitability studies \citep[e.g.][]{kaltenegger2007, betremieux2013} have been conducted based on signatures, called biomarkers \citep{kaltenegger2002}, appearing in Earth's spectra.

In this chapter I use lightning climatology studies from Solar System planets for a first discussion on the implications of potential lightning occurrence on exoplanets and brown dwarfs. Though the data discussed in Sections \ref{sec:earth} and \ref{sec:ggp} are limited to radio and optical observations, these studies are also useful for better understanding the evolution of extrasolar atmospheres through, for example, changes in the chemistry as a result of lightning discharges \citep{rimmer2016}.

Figure \ref{fig:mr} shows the diversity of extrasolar planetary objects with respect to their mean composition (top panel) and their distance from the host star (bottom panel). Figure \ref{fig:mr} also includes the Solar System planets discussed in this chapter and the exoplanets considered in the next section (green triangles and red circles, respectively). I include L, T and Y brown dwarfs, for which the masses and radii were taken from a brown dwarf list.\footnote{johnstonsarchive.net/astro/browndwarflist.html - by Wm. Robert Johnston. Several brown dwarf lists can be found on the internet, though most of them do not include size and mass parameters. A well-composed, continuously updated list of brown dwarfs can be found on https://jgagneastro.wordpress.com/list-of-ultracool-dwarfs/ by J. Gagne, where coordinates, identifiers, proper motions, etc. are listed, however no radius and mass information are added.}

The top panel of Fig. \ref{fig:mr} includes density curves for different bulk compositions, including pure water, iron and enstatit (MgSiO$_3$) and the mix of these. The line for a 90\% H$_2$ 10\% He composition is also included.\footnote{These lines were calculated by solving the equations for hydrostatic equilibrium and the mass of a spherical shell. For all compositions except H$_2$/He, a modified polytrope was assumed for the equation of state, $\rho = \rho_0 + cP^n$ with the parameters ($\rho_0,c,n$) taken from \citet{seager2007}. For H$_2$/He, the equation of state from \citet{militzer2013} was used.} The density lines visualize the diversity of the global chemical composition of extrasolar bodies. The gas giants and brown dwarfs line up around the H$_2$/He line (light blue line), possible water words and Neptune-like planets follow the lines with H$_2$O content (dark blue lines), while rocky planets, super-Earths are found around the MgSiO$_3$ composition lines (orange lines). A populated region above the pure H line includes the inflated hot Jupiters, whose radii are larger due to the close vicinity to the host star (Fig. \ref{fig:mr}, top panel). Figure \ref{fig:mr} (bottom) further illustrates that many of the presently confirmed exoplanets reside considerably closer to their host star than any of the Solar System planets. Therefore, the characteristics of the host star will also be of interest for my purpose of discussing potential candidates for further theoretical and observational lightning studies.

The diversity of observed extrasolar planets implies a large variety of atmospheric chemistry and dynamics. Some planets will have atmospheric chemical compositions similar to brown dwarfs, others will be more water or methane dominated and therefore, may be more comparable to the Solar System planets. The basic physical processes that lead to the formation of clouds (nucleation, bulk growth/evaporation, gravitational settling, element depletion) will be the same, independent of the local chemistry, though their efficiency might differ \citep[e.g.][]{helling2014b}. According to transit spectrum observation, extrasolar planets form clouds in their atmosphere \citep[e.g.][]{sing2009, sing2013, sing2015}, and \textit{Hubble Space Telescope} and \textit{Spitzer} observations have suggested that these atmospheres are very dynamic \citep[e.g.][]{knutson2008, knutson2012, buenzli2014, buenzli2015}. The study of possible cloud particle ionization has only begun in the context of extrasolar planets and brown dwarfs (see Chapter \ref{chap:liginout}, Sect. \ref{sec:ligos}). \citet{helling2013} have demonstrated, based on data by \citet{sentman2004}, that the electric field breakdown, which initializes a lightning discharge does not very strongly depend on the chemical composition of the gas (e.g. their fig. 5). I, therefore, suggest that the Solar System lightning statistics presented here can be used as a first representation of lightning occurrence on extrasolar planets and brown dwarfs. I also note that the Solar System flash rates and densities carry uncertainties as presented in Sect. \ref{subs:climatjs}.

In order to apply the results of the previous sections on lightning climatology, I group the extrasolar planetary objects into several categories (Sect. \ref{sec:casest}). Bearing in mind the diversity of exoplanets, I choose specific examples for each category, which are discussed in more details to demonstrate why they might be suitable candidates for lightning activity. Figure \ref{fig:mr} shows where these planets (red circles) lie in the (M$_p$, R$_{\rm p}$)-plane and in the (a, $\rho_{\rm bulk}$)-plane compared to the whole ensemble of known exoplanets and brown dwarfs. Section \ref{sec:flashdens} presents the flash densities estimated for the extrasolar category examples. Section \ref{sec:stelact} discusses the challenges arising from the stellar activity of the host stars of planets, and also how this activity may favour the production of lightning on planets. However, I note that more fundamental modelling of the 3D cloud forming, radiative  atmosphere structure like in \citet{lee2015} and \citet{helling2016}, possibly in combination with kinetic gas-phase modelling like in \citet{rimmer2016} is required to provide quantitative results. In the following, I make a first qualitative attempt of selecting possible candidates for future studies.

%__________________________________________________________________
\subsection{Case-study categories} \label{sec:casest}

%%Table - Planetary parameters
\begin{table*} 
\resizebox{\columnwidth}{!}{
\begin{threeparttable}
 \caption{Properties of exoplanets and the brown dwarf, Luhman 16B, listed in this chapter as examples for further study of lightning activity.}
  \begin{tabular}{@{}lllllll@{}}	
	\hline
	\vtop{\hbox{\strut super-Earth}\hbox{\strut size planet}} & \vtop{\hbox{\strut Mass}\hbox{\strut (M$_p$/M$_\oplus$)}} & \vtop{\hbox{\strut Radius}\hbox{\strut (R$_{\rm p}$/R$_\oplus$)}} & \vtop{\hbox{\strut Density}\hbox{\strut ($\rho_p$/(g cm$^{-3}$))}} & \vtop{\hbox{\strut Semi-major axis}\hbox{\strut (a/AU)}} & \vtop{\hbox{\strut Calculated temperature\tnote{(1)}}\hbox{\strut (T$_{\rm cal}$/K)}} & Reference \\
	\hline
	Kepler-186 f & $0.31-3.77$ & $1.11\substack{+0.14 \\ -0.13}$ & & $0.356 \pm 0.048$ & & \citet{quintana2014} \\
	Kepler-62 f & $ < 35$ & $1.41 \pm 0.07$ & & $0.718 \pm 0.007 $ & $208 \pm 11$ (T$_{\rm eq}$) & \citet{borucki2013} \\
	Kepler-10 b & $3.33 \pm 0.49$ & $1.47\substack{+0.03 \\ -0.02}$ & $5.8 \pm 0.8$ & $0.01685 \pm 0.00013$ & $2169\substack{+96 \\ -44}$ (T$_{\rm eq}$) & \citet{dumusque2014} \\
	55 Cnc e & $8.08 \pm 0.31$ & $1.92 \pm 0.08$ & $6.3\substack{+0.8 \\ -0.7}$ & $0.01544 \pm 0.00009$ & $\sim 2400$ (T$_{\rm eq}$) & \citet{demory2015} \\
	Kepler-69 c & 2.14 & $1.71\substack{+0.34 \\ -0.23}$ & 2.36 & $0.64\substack{+0.15 \\ -0.11}$ & $299\substack{+19 \\ -20}$ (T$_{\rm eq}$) & \vtop{\hbox{\strut \citet{barclay2013}}\hbox{\strut \citet{kane2013}}} \\
	\hline
	\vtop{\hbox{\strut Jupiter}\hbox{\strut size planet /}\hbox{\strut brown dwarf}} & \vtop{\hbox{\strut Mass}\hbox{\strut (M$_p$/M$_{\rm Jup}$)}} & \vtop{\hbox{\strut Radius}\hbox{\strut (R$_{\rm p}$/R$_{\rm Jup}$)}} & \vtop{\hbox{\strut Density}\hbox{\strut ($\rho_p$/(g cm$^{-3}$))}} & \vtop{\hbox{\strut Semi-major axis}\hbox{\strut (a/AU)}} & \vtop{\hbox{\strut Calculated temperature\tnote{(1)}}\hbox{\strut (T$_{\rm cal}$/K)}} & Reference \\
	\hline
	HD 189733 b & $1.14 \pm 0.06$ & $1.14 \pm 0.03$ &  $0.75 \pm 0.08$ & $0.0309 \pm 0.0006$ & $1201 \pm 13$ (T$_{\rm eq}$) & \citet{torres2008} \\
	GJ 504 b & $4.0\substack{+4.5 \\ -1.0}$ & - & - & 43.5 & $510\substack{+30 \\ -20}$ (T$_{\rm eff}$) & \citet{kuzuhara2013} \\
	Luhman 16B & $20-65$ & - & - & - & $1280 \pm 75$K (T$_{\rm{eff}}$) & \citet{faherty2014} \\
	\hline
  \label{table:planet}
  \end{tabular}
  \begin{tablenotes}
	\item[1] T$_{\rm eff}$: effective temperature; T$_{\rm eq}$: equilibrium temperature     
  \end{tablenotes}
\end{threeparttable}
}
\vspace{0.8cm}
\end{table*}
%%Table 

%%Table - Stella parameters
\begin{table*} 
\resizebox{\columnwidth}{!}{
\begin{threeparttable}
 \caption{Properties of the stars hosting the example exoplanets listed in Chapter \ref{chap:stat}.}
  \begin{tabular}{@{}llllll@{}}	
	\hline 
	Star & Spectral type & \vtop{\hbox{\strut Effective Temperature}\hbox{\strut estimate (T$_{\rm eff}$/K)}} & \vtop{\hbox{\strut Mass}\hbox{\strut (M$_*$/M$_\odot$)}} & \vtop{\hbox{\strut Radius}\hbox{\strut (R$_*$/R$_\odot$)}} & Reference \\
	\hline	
	Kepler-186 & M1V\tnote{(1)} & $3790$ & $0.478 \pm 0.055$ & $0.472 \pm 0.052$ & \citet[][Supplement]{quintana2014} \\
	Kepler-62 & K2V & $4930$ & $0.69 \pm 0.02$ & $0.64 \pm 0.02$ & \citet[][Supplement]{borucki2013} \\
	Kepler-10 & G & $5710$ & $0.910 \pm 0.021$ & $1.065 \pm 0.009$ & \citet{dumusque2014} \\
	55 Cnc & K0IV-V & $5200$ & $0.905 \pm 0.015$ & $0.943 \pm 0.01$ & \citet{vonbraun2011} \\
	Kepler-69 & G4V & $5640$ & $0.81^{+0.09}_{-0.081}$ & $0.93^{+0.18}_ {-0.12}$ & \citet{barclay2013} \\
	HD 189733 & K2V\tnote{(2)} & $5050$ & $0.82 \pm 0.03$ & $0.76 \pm 0.01$ & \citet{bouchy2005} \\
	GJ 504 & G0V & $6230$ & $1.22 \pm 0.08$ & & \citet{kuzuhara2013} \\
	\hline
  \label{table:star}
  \end{tabular}
  \begin{tablenotes}
	\item[1] http://simbad.u-strasbg.fr/simbad/sim-basic?Ident=kepler-186
	\item[2] \citet{fares2013}        
  \end{tablenotes}
\end{threeparttable}
}
\vspace{0.8cm}
\end{table*}
%%Table 

Transiting planets like Kepler-186f, Kepler-62f, Kepler-10b, 55 Cancri e, Kepler-69c and HD 189733b, directly imaged planets such as GJ 504b, and brown dwarfs like Luhman 16B, are some of the best candidates for the detection of lightning or its effects on the atmosphere.

The spectrum of a transiting exoplanet may contain a considerable amount of information on the atmosphere of the planet, possibly including signatures of lightning. These signatures may be emission or absorption lines either caused by lightning or by non-equilibrium species as a result of lightning activity \citep[Chapter \ref{chap:ligsig}; e.g.][]{barnun1985,krasnopolsky2006,kovacs2010,bailey2014}\footnote{Lightning may occur anytime throughout a planet's orbit, and its signatures could appear in any observational technique good enough to pick them up. However, currently transiting exoplanets offer the largest numbers of detected exoplanets with techniques related to transit- or occultation-observations being one of the most successful ones in characterizing these objects.}. Directly imaged planets are another category of good candidates for lightning-hunting. They are far enough from their host star, so that the stellar light can be blocked by coronagraphs and the planet's disc can be observed directly. These planets, being far from stellar effects, are comparable to non-irradiated brown dwarfs \citep[e.g.][]{kuzuhara2013, janson2013}. Brown dwarfs are much closer to us than most of the exoplanets and, in most of the cases, no host star will outshine their signal. Therefore, brown dwarfs are among the best candidates from the sample of objects that we have available (see Fig. \ref{fig:mr}) to detect lightning in their spectrum (e.g. radio, or other suitable means).

Lightning may be an indicator of potentially habitable environments, since it may be essential for the formation of prebiotic molecules and because it carries information about cloud dynamics. Some of the planets that I examine below are suggested to reside in the Habitable Zone (HZ) of their host star. The HZ is usually defined as the region where the incident flux of the star is enough for liquid water to be maintained on the surface of a planet with adequate atmospheric pressure \citep[e.g.][]{kasting1993, kasting2014, kopparapu2013, kopparapu2014}. Habitability is a very hot topic of exoplanetary research, resulting in various studies and concepts of the HZ. Some researchers apply the "water loss" and "maximum greenhouse" limits \citep[e.g.][]{kasting1993,kopparapu2013}, others define the boundaries between "recent Venus" and "early Mars" limits \citep[e.g.][]{kasting2014}, and some use an even more extended HZ concept \citep[e.g.][]{seager2013}. These various HZ definitions show the uncertainty in the precise definition of a habitable planet, which allows us to develop a wider concept of planets with lightning. 

Below, I define six categories (not exclusive) guided by the availability of lightning observations from the Solar System planets. I use lightning climatology results from Sections \ref{sec:earth} and \ref{sec:solsys} in order to provide a first estimate of potential lightning occurrence on extrasolar planetary bodies. The objects listed under each category are examples of a larger number of planets/brown dwarfs as demonstrated in Fig. \ref{fig:mr}. The chosen examples have been observed with different techniques before. The properties of the planets considered below and the properties of their host stars are summarized in Tables \ref{table:planet} and \ref{table:star}. 

\begin{itemize}
\item \textbf{Earth-like planets}: Planets with similar continent-ocean fraction as Earth. Studies have shown that, in principle, it is possible to estimate the ocean-land ratio of the surface of the planet by detecting diurnal variability in the photometric light curve of the planet \citep[e.g.][]{ford2001, kawahara2010}. \citet{ford2001} built a model, which considers Earth as an exoplanet and analysed its light curve with and without clouds. They found significant, potentially detectable, changes in the light curve as the different surfaces (ocean, land, desert) rotated into view. \citet{kawahara2010} developed a method to reconstruct the surface of a planet using variations in its scattered-light curve. This model was shown to work for an Earth-like surface, however, several assumptions were made, such as cloudlessness or lack of atmospheric absorption. \citet{kawahara2011} used simulated exoplanet light curves from Earth observations by the \textit{EPOXI} mission and demonstrated that the inversion of the light curves recovers the cloud coverage of the planet. By subtracting the cloud features they also showed that the residual maps created from the data trace the continental distribution of Earth. Knowing the ratio of continent-ocean coverage of an exoplanet would help to estimate the lightning occurrence on such planets, however, based on above mentioned studies, it seems that retrieving land-ocean fractions on planets needs improvement in observational instrumentation. Regardless, once the tools are available, either continent-ocean surface mapping of a planet may help lightning detections or, vice versa, lightning signal distribution may help the surface mapping of an extrasolar object, assuming that the same dynamics act on the planet as on Earth. I choose the candidate planet for this category based on previous studies. I used the global average flash density from Earth for these planets. 

\begin{itemize}
\item[$\ast$] Example: \textit{Kepler-186f} \citep[][and number 4 in Fig. \ref{fig:mr}]{quintana2014}.
\end{itemize}

The Kepler-186 planetary system is composed of five planets, all with sizes smaller than $1.5$ R$_\oplus$ (Earth radius) \citep{quintana2014}. \citet{quintana2014} reported the discovery of Kepler-186f, the only planet of the five in the system lying in the HZ of the host star. According to their modelling the mass of Kepler-186f can range from $0.31$ M$_\oplus$ (M$_\oplus$: Earth mass) to $3.77$ M$_\oplus$ depending on the bulk composition (from pure water/ice to pure iron composition). In case of an Earth-like composition its mass would be $1.44$ M$_\oplus$. \citet{torres2015} found that Kepler-186f has a 98.4\% chance of being in the HZ of the host star. \citet{bolmont2014} showed that with modest amount of CO$_2$ and N$_2$ in its atmosphere, the surface temperature can rise above 273 K and the surface of the planet could maintain liquid water permanently. If Kepler-186f indeed has an Earth-like composition as Fig. \ref{fig:mr} suggests, it may host an atmospheric circulation and convectively active clouds just as Earth, which makes it an interesting candidate of hosting lightning activity.

\item \textbf{Water worlds (Ocean planets)}: Planets with surfaces fully covered by water or very small continent-to-water ocean fractions. The irradiation from the host star can drive strong winds, which may cause the formation of intermittent clouds. Lightning flash density over the Pacific Ocean is used in this analysis. 

\begin{itemize}
\item[$\ast$] Example: \textit{Kepler-62f} \citep[][and number 5 on Fig. \ref{fig:mr}]{borucki2013}.
\end{itemize}

Using Ca H\&K emission index, \citet{borucki2013} concluded that Kepler-62, a K-type main-sequence star, is inactive. Kepler-62f is the outermost planet in the 5-planet system. By calculating the incident flux, \citet{borucki2013} found that the super-Earth is within the HZ of the host star. \citet{kane2014b} arrived to the same conclusion by analysing the HZ boundaries based on stellar parameter uncertainties, showing that planet "f" is 99.4\% likely to be in the HZ. \citet{kaltenegger2013} assumed, based on the packed system of Kepler-62 with solid planets, that Kepler-62f was formed outside the ice line, indicating water or ice covered surface of the planet depending on the atmospheric pressure of CO$_2$. Based on the assumption that Kepler-62f is indeed a water planet, and using the observed radius, \citet{kaltenegger2013} found that the planet's mass would be $1.1-2.6$ M$_\oplus$. \citet{bolmont2015} used their \textit{Mercury-T} code to study the evolution of the Kepler-62 system. They found that Kepler-62f potentially have a high obliquity and a fast rotation period, which would result in seasonal effects and both latitudinal and longitudinal winds on the planet. The possible seasonal and latitudinal changes may result in a diverse weather system on the planet, therefore, Kepler-62f may host a quite variable lightning activity.

\item \textbf{Rocky planets with no liquid surface}: These planets supposedly do not have permanent liquid oceans on their surface. However, they still may host a chemically active atmosphere that forms clouds and produces lightning. Lightning production on these planets may also be caused by volcanic activity or electrostatic discharges caused by dust collision (e.g. in dust devils). \citet{schaefer2009} and \citet{miguel2011} modelled different types of potential atmospheres, created by the outgassing of the lava-oceans on the surface of the planet, of hot, volatile-free, rocky super-Earths, and found them to be composed mostly of Na, O, O$_2$, SiO \citep{schaefer2009} and at temperatures $\le 2000$K Fe and Mg \citep{miguel2011}. \citet{ito2015} considered these "mineral atmospheres", evaluated their temperature profiles and investigated their observability via occultation spectroscopy. They considered four rocky planets, CoRoT-7b, Kepler-10b, Kepler-78b, and 55 Cnc e and showed that IR absorption features of K, Na and SiO could be detected in case of Kepler-10b and 55 Cnc e with future missions like the \textit{James Webb Space Telescope}. Such atmospheres would be close to the composition of volcano plumes on Earth and may host lightning activity. I use volcanic lightning flash densities evaluated in Sect. \ref{sec:volc}. The various values in Table \ref{table:vol} (last column) represent various activity stages of eruptions. For example, if I assume that the surface of these planets is covered by almost constantly erupting volcanoes, the flash densities could be very high, like during Phase 1 of the Mt Redoubt eruption. However, the surface is still covered by volcanoes, but they do not erupt as frequently, or the frequency of explosive eruptions is less, then a smaller flash density can be used, like during the eruptions of Eyjafjallaj\"okull. I also used continental flash density from Earth, though, I note that this value likely underestimates the actual electric activity compared to pure visual inspection of lightning in volcanoes \citep[e.g. Eyjafjallaj\"okull, Sakurojima, Puyehue; see also][]{mcnutt2000}.

\begin{itemize}
\item[$\ast$] Example: \textit{Kepler-10b} \citep[][and number 1 on Fig. \ref{fig:mr}]{batalha2011}.
\end{itemize}

Stellar chromospheric activity measurements (using the Ca II H\&K index) conducted by \citet{dumusque2014} indicate that Kepler-10 is less active than the Sun, which is in accordance with the star's old age (10.6 Gyr). According to \citet{ito2015}, Kepler-10b, a hot, tidally locked rocky super-Earth \citep{dumusque2014}, may host an atmosphere mostly composed of Na, O, O$_2$, SiO and K outgassed from the lava-surface of the planet. The bulk density (Table \ref{table:planet}) of the planet indicates a composition similar to Earth \citep{dumusque2014}.

\begin{itemize}
\item[$\ast$] Example: \textit{55 Cancri e} \citep[][and number 3 on Fig \ref{fig:mr}]{mcarthur2004, vonbraun2011}.
\end{itemize}

The second candidate for a rocky planet is 55 Cancri e (55 Cnc e), which recently has been reported to be a planet with possible high volcanic activity \citep{demory2015}. The super-Earth orbits the K-type star 55 Cnc on a very close orbit, resulting in a high equilibrium temperature (Table \ref{table:planet}), which may result in the loss of volatiles of the planet. Multiple scenarios have been proposed for its composition including a silicate-rich interior with a water envelope, and a carbon-rich interior with no envelope \citep[see][and ref. therein]{demory2015}. A recent study suggests that 55 Cnc e is rather a volcanically very active planet \citep{demory2015}. A large number of volcanic eruptions, especially explosive eruptions, may result in increased lightning activity on the planet due to the large number of volcano plumes. This would allow the production of lightning discharges without the necessity of cloud condensation. \citet{kaltenegger2010} studied the observability of such volcanic activity on Earth-sized and super-Earth-sized exoplanets. They found that large explosive eruptions may produce observable sulphur dioxide in the spectrum of the planet. Similarly to Kepler-10b, 55 Cnc e may host an atmosphere composed of minerals, as a result of the outgassing of the lava on its surface.
 
Combining the findings of studies such as \citet{kaltenegger2010} and observational signatures of lightning, one may confirm a high volcanic activity on terrestrial, close-in exoplanets like Kepler-10b and 55 Cnc e, making these planets interesting candidates for future lightning observations.

\item \textbf{Venus-like planets}: Venus and Earth, though they are similar in size and mass, are very different from each other. Due to Venus' thick atmosphere, the runaway greenhouse effect increases the surface temperature of the planet to uninhabitable ranges. Such exoplanets, Earth- or Super-Earth-size rocky planets with very thick atmospheres, may be quite common \citep{kane2014}. For these planets I use flash density based on radio observations from Venus. 

\begin{itemize}
\item[$\ast$] Example: \textit{Kepler-69c} \citep[][and number 6 on Fig \ref{fig:mr}]{barclay2013}.
\end{itemize} 

%Kepler-69c is a super-Earth sized planet orbiting close to the HZ of a star very similar to the Sun \citep{barclay2013}. 
\citet{barclay2013} analysed the place of Kepler-69c, a super-Earth sized planet, in its system, and the stellar irradiation, and found that Kepler-69c is very close to the HZ of the star or, depending on model parameters, it may lie inside the HZ. They investigated the equilibrium temperature boundaries that Kepler-69c may have, using different albedo assumptions. They found that the temperature of the planet may be low enough to host liquid water on the surface, if not considering an atmosphere. However, a thick atmosphere may increase the temperature high enough (with a low albedo) to prevent water to stay in liquid form \citep{barclay2013}. \citet{kane2013} estimated that Kepler-69c most probably does not lie in the conservative HZ, but rather at a distance equivalent to Venus's distance from the Sun. Also taking into account the stellar flux the planet receives (which is very similar to the incident flux Venus receives, $\sim2600$ W m$^{-2}$), they defined Kepler-69c as a "super-Venus" rather than a super-Earth \citep{kane2013, kane2014}. The low bulk density calculated by \citet{kane2013} may suggest a silicate and carbonate dominated composition of the planet. In case the planet acquired water during or after its formation, and the evolution of the planet's atmosphere was similar to Venus', then the planet may host a thick CO$_2$ atmosphere \citep{kane2013}. On a Venus-like planet, such as Kepler-69c, lightning activity may be the result of on-going volcanic activity, or, in the presence of strong atmospheric winds, the electrostatic activity of dust-dust collision.

\item \textbf{Giant gas planets}: In this category I consider planets with sizes (mass and/or radius) in the range of Saturn's to several Jupiter-sizes. Large variety of exoplanets have been discovered, which fall into this category, from close-in hot Jupiters mostly detected by the transit or the radial velocity technique, to young, cool planets hundreds of AU far from their stars detected by direct imaging. I calculate flash densities for the candidate planets based on Saturnian and Jovian flash densities.
	\begin{itemize}
	\item[$\bullet$] \underline{Transiting planets}: Most of the gas giant planets discovered by the transit technique lie within $\sim1.6$ AU from the host star\footnote{Based on data from exoplanet.eu, 29/Jul/2015}. A large number of these planets are found within 0.5 AU, creating a new (not-known from the Solar System) type of exoplanet category called "warm-" or "hot-Jupiters", latter ones lying within 0.1 AU \citep{raymond2005}. 

\begin{itemize}
\item[$\ast$] Example: \textit{HD 189733b} \citep[][and number 2 on Fig \ref{fig:mr}]{bouchy2005}.
\end{itemize}
	\end{itemize}
HD 189733 is a K-type star with a hot-Jupiter ("b") in its planetary system. \citet{wright2004} measured Ca H\&K line strength and found the star to be relatively active. Stellar activity due to star-planet interaction has been observed in X-ray \citep[e.g.][]{pillitteri2014} and FUV \citep{pillitteri2015} spectra at certain times of the planetary transit. Namely, after the secondary eclipse, X-ray flares appeared in \textit{XMM-Newton} \citep{pillitteri2014} and \textit{Swift} \citep{lecavelier2012} data, while a brightening in the FUV spectrum was also seen \citep{pillitteri2015}. \citet{pillitteri2015} explained the FUV features by material accreting onto the stellar surface from the planet. \citet{see2015} investigated exoplanetary radio emission variability due to changes in the local stellar magnetic field. They found potential variations up to 3 mJy. The frequency of magnetospheric radio emission \citep[$<40$ MHz,][]{zarka2007} coincides with the radio emission range that lightning may produce \citep[$<\sim100$ MHz,][see also Chapter \ref{chap:ligsig}, Sect. \ref{ssec:radsig}]{desch2002}. The magnetic radio emission may potentially be a background radio noise source in lightning radio observations. A slope in the IR transmission spectrum of HD 189733b has been measured by several groups \citep[e.g.][]{pont2008, sing2011}, which was interpreted as a feature caused by cloud-induced Rayleigh scattering in the atmosphere. \citet{mccullough2014} found prominent water features in the NIR transmission spectrum of HD 189733b and simultaneously reinterpreted the slope in the spectrum. They suggested that the slope can be produced by a clear planetary atmosphere and unocculted star spots. \citet{lee2015}, however, supported the finding that HD 189733b is covered by a thick layer of clouds. The atmosphere of HD 189733b may host lightning activity due to cloud convection and charge separation due to gravitational settling. This well-studied (see references above) exoplanet is a good candidate for lightning observations, because other effects, like stellar activity, can be modelled easier than for less known systems.

	\begin{itemize}
	\item[$\bullet$] \underline{Directly imaged planets}: Planetary objects detected by direct imaging are way fewer in numbers than e.g. transiting exoplanets. These objects, due to the selection effect of the technique, lie far from the host star, from $\sim10$ to thousands of AU. Though I list these objects under the category of gas giant planets, I note the ambiguity in the classification due to the uncertainty in the definition of the mass limit between brown dwarfs and planets \citep{perryman2011}. This category may include brown dwarfs, planets, and objects with masses on the borderline \citep[][table 7.6]{perryman2011}. 

\begin{itemize}
\item[$\ast$] Example: \textit{GJ 504b} \citep{kuzuhara2013}.
\end{itemize}
	\end{itemize}

GJ 504, a young, $160^{+350}_{-60}$ Myr old, solar-type star shows X-ray activity, typical property of such stars \citep{kuzuhara2013}. \citet{kuzuhara2013} investigated the colour of GJ 504b and found it to be colder (Table \ref{table:planet}) than previously imaged planets. The place of the object on the colour-magnitude diagram suggests that it is rather a late T-type dwarf with a mostly clear atmosphere \citep{kuzuhara2013}. \citet{janson2013} detected strong methane absorption in the atmosphere of GJ 504b, which also indicates that the object is a T-type brown dwarf. Though the atmospheres of these objects are considered to be clear, studies showed that it is possible to form clouds, e.g. made of sulphides \citep{morley2012} in T-type dwarf atmospheres. The potential sulphide clouds make GJ 504b a good candidate of hosting electric discharges in its atmosphere. The object is far enough from the host star so that its internal heating suppresses the external one, which could result in extensive convective patterns, just as lightning hosting clouds may form on Jupiter \citep{baines2007}. Convection and gravitational settling can be viewed as preconditions for lightning to occur.

\item \textbf{Brown dwarfs}: Brown dwarfs have masses from several M$_{\rm Jup}$ (Jupiter mass) to several tens of M$_{\rm Jup}$ and temperatures low enough for cloud formation \citep[e.g.][]{helling2014}. L type brown dwarfs are fully covered by clouds. The variability of L/T transition and most probably T type brown dwarfs is explained by patchy cloud coverage \citep{showman2013, helling2014}. \citet{helling2008} modelled cloud formation on brown dwarfs and derived grain size distributions and chemical composition through the entire atmosphere. They found that the particle size in these atmospheres is of the order of 0.01 $\mu$m (upper layers) to 1000 $\mu$m (deep layers). For a brown dwarf with log $g = 3$ or $5$ and $T_{\rm{eff}} = 1800$ or $1300$ K, the $5 \mu$m particle size range, the assumed thundercloud particle size for Jupiter (see Sect. \ref{sec:int_jup}), is in the atmospheric layers with $\sim 1300$ K local temperatures, which correspond to the mid layers of the atmosphere \citep[][fig. 4, fourth panel]{helling2008}. They also found the clouds to be made of mixed mineral cloud particles that change their size according to atmospheric height. Volcano plumes, producing lightning flashes, are mostly made of dust. These plumes may resemble brown dwarf clouds. I use flash densities obtained in Sect. \ref{sec:volc}, to estimate lightning occurrence on brown dwarfs. However, these statistics are based on a few eruptions, which does not provide a general idea about volcanic lightning densities, but provide several scenarios with more electrically active and less electrically active dust clouds. I also use flash densities from Jovian and Saturnian thunderclouds, because the particle sizes may resemble brown dwarf dust particles, and basic physical processes of cloud formation are fundamentally same in these environments, though their efficiency may vary, as I discuss it in the following sub-section.

\begin{itemize}
\item[$\ast$] Example: \textit{Luhman 16B} \citep{luhman2013}.
\end{itemize}

Luhman 16B (or WISE J104915.57-531906.1B) is the secondary component of the closest brown dwarf binary system discovered so far, with a distance of 2 pc from the Sun. It is a late L, early T type object representing the L/T transition part of the brown dwarf family \citep{luhman2013}. \citet{crossfield2014} monitored the brown dwarf during its one rotational period \citep[4.9-hour,][]{gillon2013} and mapped its surface using Doppler imaging. They interpreted the revealed features as dust clouds in the atmosphere of the object. \citet{buenzli2015} found the variability of Luhman 16B to be relatively high, up to more than 10\%. Their cloud structure model showed that the variability is caused by varying cloud layers with different thickness, rather than varying cloudy and clear parts of the atmosphere. This means, we see into various levels of cloud regions, making the possibility of detecting lightning inside the atmosphere higher. Similarly to GJ 504b, Luhman 16B may host intensive lightning activity, because cloud formation, convection and gravitational settling determine its atmosphere. Different cloud layers have been detected on Luhman 16B, which may cause similar dynamic structures to occur like on Jupiter and Saturn and may allow the observer to detect lightning inside the atmosphere of the brown dwarf. 

\end{itemize}

%__________________________________________________________________
\subsection{Flash densities for extrasolar objects} \label{sec:flashdens}

%Table - Flash numbers through the length of the transit 
\begin{table}
\begin{center}
\setlength{\tabcolsep}{5.5pt} 
 \footnotesize
 \caption{Estimated total flash/SED numbers during a transit over the disc of the planet calculated from flash densities in Table \ref{table:plan}. As the values in Table \ref{table:plan} are lower limits, the flash numbers given here represent lower limits too. The bottom four lines of the table present Venus, Earth, Jupiter and Saturn as transiting planets (with inclinations of $90^{\circ}$). Transit duration was calculated based on \citet[][equations 6.2 and 6.3]{perryman2011}. Since the determined eccentricity ($e$) for these planetary orbits is low (largest is $\sim 0.1$), for the calculation of the transit time I assumed $e$ to be 0 for all objects. Here I use the full length of the transit (from 1$^{\rm{st}}$ contact to 4$^{\rm{th}}$ contact).}
  \begin{tabular}{@{}lcc@{}}	
	\hline 
	Planet & Transit duration [h] & \vtop{\hbox{\strut Total number of flashes}\hbox{\strut during transit}} \\
	\hline \hline
	Kepler-186f & 6.25 & $4.51 \times 10^5$ \\
	Kepler-62f & 7.72 & $1.34 \times 10^5$ \\
	Kepler-10b & 1.85 & $2.67 \times 10^6$ \\
	55 Cancri e & 1.57 & $4.16 \times 10^6$\\
	Kepler-69c & 11.78 & $3.2 \times 10^{-1}$ \\
	HD 189733b & 1.89 & \vtop{\hbox{\strut $6.57 \times 10^4$}\hbox{\strut (Jupiter)}} \\
	HD 189733b & 1.89 & \vtop{\hbox{\strut $2.04 \times 10^5$}\hbox{\strut (Saturn)}} \\
	\hline
	Venus & 11.15 & $9.0 \times 10^{-2}$ \\
	Earth & 13.11 & $7.67 \times 10^5$\\
	Jupiter & 32.59 & $8.34 \times 10^5$ \\
	Saturn & 43.46 & $2.39 \times 10^6$ \\
	\hline
  \label{table:exo}
  \end{tabular}
\end{center}
\vspace{0.8cm}
\end{table}
%Table 

%Table - Flash numbers through the length of the transit, volcano 
\begin{table}
	\setlength{\tabcolsep}{5.5pt}
\begin{center}
 \footnotesize
 \caption{Estimated total flash numbers during a transit over the disc of Kepler-10b and 55 Cancri e calculated from volcanic flash densities in Table \ref{table:vol} (used flash densities are marked with the row number). See also the caption of Table \ref{table:exo}.}
  \begin{tabular}{@{}lccc@{}}	
	\hline 
	Planet & \vtop{\hbox{\strut Transit duration}\hbox{\strut [h]}} & \vtop{\hbox{\strut Total number of}\hbox{\strut flashes during transit}} & \vtop{\hbox{\strut Row,}\hbox{\strut Table \ref{table:vol}}} \\
	\hline \hline
	\multirow{5}{*}{Kepler-10b} & \multirow{5}{*}{1.85} & $1.02 \times 10^8$ & $[1]$ \\
	 & & $3.26 \times 10^8$ & $[2]$ \\
	 & & $1.23 \times 10^{10}$ & $[3]$ \\
	 & & $2.04 \times 10^{12}$ & $[4]$ \\
	 & & $1.11 \times 10^{11}$ & $[5]$ \\
	\hline
	\multirow{5}{*}{55 Cancri e} & \multirow{5}{*}{1.57} & $1.59 \times 10^8$ & $[1]$  \\
	 & & $5.07 \times 10^8$ & $[2]$ \\
	 & & $1.91 \times 10^{10}$ & $[3]$ \\
	 & & $3.17 \times 10^{12}$ & $[4]$ \\
	 & & $1.73 \times 10^{11}$ & $[5]$ \\
	\hline
  \label{table:exovol}
  \end{tabular}
\end{center}
\vspace{0.8cm}
\end{table}
%Table 

Table \ref{table:plan} lists extrasolar objects with their Solar System counterparts. Based on the data available, I arrange these objects into six groups (Sect. \ref{sec:exopl}). From Earth, I obtained three flash densities using LIS/OTD flash observations and WWLLN and STARNET sferics detections. Strokes detected by WWLLN and STARNET were converted to flashes \citep[assuming 1.5 sferics/flash;][]{rudlosky2013}. I assume that a planet with a similar surface to Earth's, in the HZ of the star has the same flash density, as the global value on Earth. Kepler-186f is the candidate for these conditions. However, depending on the continent-ocean fraction and the amount of insolation of the planet, the flash density may vary. I consider Kepler-10b and 55 Cnc e to be rocky planets with no liquid surface. Although it is arguable whether these planets host an atmosphere, in case they do, lightning activity may be similar to the activity over Earth-continents. Both planets may also be good candidates for volcanically active planets, resulting in lightning discharges in volcano plumes. Similarly, I used flash densities from oceanic regions in order to simulate lightning statistics on Kepler-62f, a presumed ocean planet. Earth is the most well studied planet, resulting in the most accurate flash densities obtained. Uncertainties raise, however, from the accuracy with which one can determine the similarities between the exoplanet and Earth or a Terran environment. The arguments for my approach, such as similarities between Earth and the exoplanets in size, composition or cloud occurrence, are discussed in Sections \ref{sec:exopl} and \ref{sec:casest}.

From an astrophysical perspective, Jupiter and Saturn have the same flash densities within an order of magnitude (Table \ref{table:plan}). Two types of gas giant planets are studied, HD 189733b a hot Jupiter, and GJ 504b a fairly cold giant planet in the outer regions of the stellar system, which has been suggested to be comparable to a brown dwarf of spectral type T. These planets represent the two edges of giant planetary bodies, the former being a highly insolated one, while for the latter, internal heating has a higher contribution to global temperatures and cloud formation. For Luhman 16B, representing the L/T transition brown dwarfs with most probably patchy cloud coverage, Jupiter was considered as a good analogue. The flash densities obtained for Jupiter and Saturn carry relatively large uncertainties, due to the fact that these planets are much less studied than Earth. The observations have been carried out for shorter time, less frequently, with less sensitive instruments. However, the flash densities listed in Table \ref{table:plan} serve well as lower statistical limits for these planets and their extrasolar counterparts. I support my approach of using Solar System, in this case Jovian and Saturnian, lightning statistics as guidance for extrasolar studies, with the fact that the basic physical processes of cloud formation are fundamentally the same in every environment, though their efficiency may vary. It also has been shown that the electric field breakdown initializing a lightning discharge does not depend strongly on the chemical composition of the gas \citep[][see also Chapter \ref{chap:liginout}, Sect. \ref{sec:ligos}]{helling2013}. Therefore, I suggest that Solar System lightning flash densities provide good first estimates for extrasolar lightning occurrence.

Table \ref{table:vol} lists flash densities of various eruptions of two volcanoes, Eyjafjallaj\"okull and Mt Redoubt. I suggest that these occurrence rates may resemble several scenarios on rocky exoplanets with no water surfaces (Kepler-10b, 55 Cnc e) and on brown dwarfs (Luhman 16B). Such scenarios may include a surface fully covered by volcanoes erupting very frequently. In this case, flash densities may be as high as it was during Phase 1 \citep["explosive phase",][]{behnke2013} of the 29 March 2009 eruption of Mt Redoubt. Volcanically very active surfaces, but with not that frequent explosive eruptions, may have flash densities of the order of the Eyjafjallaj\"okull values. Dust charging in brown dwarf atmospheres may be similar to charging in volcano plumes, and could produce flash densities similar to Eyjafjallaj\"okull densities and the values of Phase 2 \citep["plume phase",][]{behnke2013} of the 29 March 2009 Mt Redoubt eruption.
 
The majority of extrasolar planets was discovered by the transit method\footnote{http://exoplanet.eu/ (on 26/01/2016)}. Transit observations and measurements taken during the transit or the occultation of the planet are the most successful techniques in characterizing exoplanets and their atmospheres. Therefore, it is interesting and informative to see how much lightning could occur during a planet's transit. This information will further allow us to determine observable signatures of lightning coming from these planets. Also, it is a good example to show how the obtained hourly flash densities of this chapter can be used for scientific predictions. Table \ref{table:exo} lists the transiting exoplanets introduced in the previous sections and Venus, Earth, Jupiter and Saturn as a transiting planet. Table \ref{table:exovol} lists flash densities calculated for two transiting planets, Kepler-10b and 55 Cnc e, based on volcanic lightning densities listed in Table \ref{table:vol}. These tables summarize how many flashes could be present during a transit on the disc of the planet observed from 1$^{\rm{st}}$ to 4$^{\rm{th}}$ contact \citep{perryman2011}. The projected surface area (disc) of a planet is given by $2r^2\pi$, where $r$ is the mean radius of the planet. The total number of flashes during the transit is calculated from the hourly flash densities (flashes km$^{-2}$ h$^{-1}$) given in Tables \ref{table:plan} and \ref{table:vol} by multiplying these values by the area (km$^2$) of the planetary disc and the length of the transit (h). The gas giant, HD 189733b is listed twice in Table \ref{table:exo} indicating that flash densities from both Jupiter and Saturn (both averaged from the values in Table \ref{table:plan}) were used to estimate lightning occurrence on this planet. Since the two Solar System planets show similar densities (Table \ref{table:plan}), I obtain similar results for HD 189733b for the two cases. If HD 189733b should develop a storm feature similar to Saturn's gigantic 2010/11 storm with its extremely high flash density, then  this might also produce potentially observable signatures on HD 189733b even during its short transit time. Comparing the values for Kepler-10b and 55 Cnc e in Tables \ref{table:exo} and \ref{table:exovol}, it is clearly seen that volcano eruptions produce much higher lightning activity, than thunderclouds. However, it is important to note that, while I took average values for continental thundercloud lightning activity, there are extreme values from certain eruptions that might not resemble average flash densities of volcano plumes. The numbers listed in the third column of Tables \ref{table:exo} and \ref{table:exovol} are guides to a lightning flash count one can expect for the listed transiting planets. These flash densities suggest a relatively high lightning activity on these planets when they are observed during a full transit. This increases the probability of measuring signals resulting from lightning discharges.  

For example, let us consider HD 189733b. During its 2-hour long transit, about $10^{5}$ lightning flashes occur on the projected surface according to my calculations. Assuming the average total energy content of these flashes is $\sim 10^{12}$ J \citep[p. 43]{leblanc2008}, based on Jovian lightning optical efficiency calculations \citep{borucki1987}, then the total energy dissipated from lightning discharges during the transit of HD 189733b is of the order of $10^{17}$ J, or $10^{5}$ TJ. For comparison, on Earth a typical lightning flash releases energy of the order of $10^{9}$ J \citep{maggio2009}. Energy measurements of Earth lightning suggest that about $1-10$\% of the total energy is released in optical \citep[][p. 334]{borucki1987, hill1979, lewis1984} and $\sim 1$\% in the radio \citep{volland1984, farrell2007}. This leaves us with a bit less than 90\% of energy going into mechanical and thermal release, affecting the local chemistry of the atmosphere, which will produce yet unexplored observable spectral signatures. Going back to the example, during the transit of HD 189733b, $9 \times 10^{4}$ TJ energy would affect the atmosphere of the planet. This example benefits from previous lightning energy estimates; however, these estimates are based on Earth lightning properties. Once the energy release from lightning in various extrasolar planetary atmospheres is studied, one can estimate, based on my lightning climatology statistics, how much energy is released not just into observables (optical and radio emission) but to energy affecting the local chemistry. This energy and the caused chemical changes can be further explored and determined whether it is enough to produce observable emission lines in the spectrum of the planet, or the stellar light and planetary thermal emission would suppress these transient signatures.

%__________________________________________________________________
\subsection{Observational challenges: Effects of stellar activity} \label{sec:stelact}

Apart from technical issues (such as instrumental limits, detection thresholds, etc.), there are natural effects limiting observations, mostly coming from stellar activity. Cool dwarf stars, G K and M spectral types, are in more favour of exoplanet surveys, than hotter ones. G and K stars are the targets of scientists looking for an Earth twin orbiting a Sun-like star. M dwarfs, apart from being the most widespread stars in the Galaxy, are small stars making it easier to detect variation caused by planets in their light curves (the planet-star size ratio can be large enough to detect small planets around the M dwarf) or spectra (variations caused by smaller planets around a small star can be detected easier). However, \citet{vidotto2013} and \citet{see2014} showed that M dwarfs might not be good candidates for the search of Earth-like habitable planets because of their high stellar activity. This activity significantly reduces the size of a planetary magnetosphere exposing the planetary atmosphere to erosive effects of the stellar wind. G and K stars have similar activity cycles to the Sun's (11-year cycle), younger stars being rapidly rotating and more active, than older ones \citep{baliunas1995}. Early M dwarfs (M3 and earlier) have radiative cores and outer convective zones indicating similar dynamo processes to the Sun's \citep{west2008}. Later type M dwarfs are fully convective, therefore no solar-like dynamo can operate in them, which result in the change of magnetic field structure \citep{donati2009}. Later M dwarfs in general are more active than earlier type ones, keeping their activity for longer, probably due to this change in magnetic field structure and rapid rotation. As stars age, their rotation slows down and they become close to inactive \citep{west2008}.

However, the activity of the star may support lightning activity in close-in planets. Studies suggest a correlation between solar activity and the number of thunder days. \citet{pintoneto2013} analysed data of a $\sim 60$-year period in Brazil looking for 11-year cycle variations in thunderstorm activity correlated to solar activity. They suggested that the anti-correlation they found is the result of solar magnetic shielding of galactic cosmic rays, which have a large effect on lightning production. \citet{romps2014} suggested a link between global warming over the United States and flash rate variability. Their results showed an increase of flash numbers due to an increase of global precipitation rate and of the convective available potential energy (CAPE), a proxy of lightning activity. \citet{scott2014} found a correlation between the arrival of high-speed solar wind streams at Earth, following an increase in sunspot number and decrease in solar irradiance, and lightning activity. They measured the correlation based on lightning occurrence over the United Kingdom using UK Met Office radio observations. This correlation may be the result of increasing number of solar energetic particles reaching the upper atmosphere (coming from the solar wind), which triggers discharges and may increase the number of lightning events. \citet{siingh2011} compared different studies (Brazil, USA, India) and concluded that the relation between lightning activity and sunspot numbers is complex, since data showed correlation in the USA and Brazil and anti-correlation in the Indian Peninsular \citep[see][fig. 6]{siingh2011}. The STARNET and WWLLN data analysed in my study also show more lightning from 2013, close to solar maximum, than from 2009, when the Sun was at its minimum of activity (Figs \ref{fig:4a}-\ref{fig:4b}). However, \citet{rudlosky2013} showed an improvement of 10\% of the WWLLN DE between 2009 and 2013, which may also be the cause of more stroke detections in 2013. These studies suggest that lightning in the astrophysical context  will depend on internal heating and stellar irradiation that will affect the local atmospheric temperature, which determines where clouds form. Consequently, lightning activity on a planet will be affected by the age, and hence the magnetic activity of the host star, and by the distance of the planet from the star. In the case of brown dwarfs, it is the age of the object that counts most as this determines its total energy household including magnetic activity driven by rotation. If the brown dwarf resides in a binary system \citep[e.g.][]{casewell2012, casewell2013, casewell2015}, the characteristics of the companion may also play a role in the production of lightning discharges.

%__________________________________________________________________
%__________________________________________________________________
\section{Summary} \label{sec:con}

This chapter uses Solar System lightning statistics for a first exploratory study of potential lightning activity on exoplanets and brown dwarfs. I presented lightning flash densities for Venus, Earth, Jupiter and Saturn, based on optical and/or radio measurements. I also included lightning- and lightning energy- distribution maps for the gas giant planets and Earth, based on observational data. The obtained information in Sections \ref{sec:earth} and \ref{sec:solsys} was used to estimate lightning occurrence on extrasolar planetary objects. My sample of extrasolar objects contains transiting planets (Kepler-186f, Kepler-62f, Kepler-10b, 55 Cancri e, Kepler-69c and HD 189733b), directly imaged planets (GJ 504b) and brown dwarfs (Luhman 16B). Transmission spectra are relatively easy to take and may contain signatures of lightning activity. Directly imaged planets are far enough from their parent stars to be observed directly, and the effects of stellar activity are less prominent, such as in the case of non-irradiated brown dwarfs. Brown dwarfs, because they are close to us, are one of the most promising candidates for lightning-hunting. I defined six categories of extrasolar bodies, with one or two examples, in analogy to Solar System planets. All of these candidates potentially host an atmosphere with clouds, based on either observations or atmospheric models (Sect. \ref{sec:casest}). These examples were chosen because they have common features with Solar System planets or lightning hosting environments (e.g. Kepler-62f being an ocean planet, 55 Cnc e hosting extreme volcanic activity, etc.), or because they represent a specific object type, such as hot Jupiters (HD189733b), Jupiter-sized planets at large distances from the star (GJ 504b), or brown dwarfs (Luhman 16B), which also have a great potential for lightning activity \citep{helling2013, bailey2014}. I suggest that these objects could be potential candidates for lightning activity in their atmospheres based on their characteristics and on our knowledge on lightning forming environments. However, I note that the flash densities estimated in this study are affected by several uncertainties, mostly due to instrumental limits and, in case of Jupiter and Saturn, the lack of temporally and spatially extensive data sets. Regardless, the obtained flash densities give a first guidance for the study of extrasolar lightning.

The best data coverage was from Earth (using data from the LIS/OTD optical satellites, and the STARNET and WWLLN radio networks), which resulted in more accurate flash densities than from the other planets. Earth provides us with three different options: for ocean planets flash densities from over the Pacific ocean were used; for rocky planets with no water surface, where mineral clouds may form as was shown both by models \citep[e.g.][]{miguel2011} and observations \citep[e.g.][]{kreidberg2014, sing2009, sing2015}, values from over continents were used; while I considered Earth-twins with similar continent/ocean coverage and with a global flash density from Earth. Data for Jupiter and Saturn were taken from published papers, these include \textit{Galileo} \citep{little1999, dyudina2004}, \textit{New Horizons} (Jupiter) \citep{baines2007} and \textit{Cassini} (Saturn) \citep{dyudina2013} observations. The derived flash densities were used to represent giant gas planets and brown dwarfs. The special case of Venus (only whistler observation with no coordinates for flashes) allowed me to estimate flash densities but not to create a lightning climatology map as it was done for the three other Solar System planets (Figures \ref{fig:1}-\ref{fig:4b}, \ref{fig:jup}, and \ref{fig:7}). I also considered volcanic lightning flash densities in case of Kepler-10b, 55 Cnc e and Luhman 16B. These densities are guides for special scenarios discussed in the previous sections.

Table \ref{table:plan} summarizes my findings of planetary flash densities, while \ref{table:vol} shows flash densities of example volcanic eruptions. All numbers are expected to be higher because the guiding data provide lower limits as only the most powerful events in the optical and radio wavelengths are detected. No other spectral energies were taken into account here. Using these flash densities, I estimated the global and regional distribution of lightning in space and time. Most of the planets listed under the defined categories are transiting objects, with the potential of taking their transmission spectra, hence possibly observing lightning spectral features. In Tables \ref{table:exo} and \ref{table:exovol} I list the total number of flashes that might occur on these planets during their full transit. I find that volcanically very active planets would show the largest lightning flash densities if lightning occurred at the same rate on these planets as it does in volcano plumes on Earth. It is also prominent that the exoplanet HD 189733b would produce high lightning occurrence even during its short transit, if it had a large storm occurring in its atmosphere, like the one on Saturn in 2010/11. The findings of this chapter can be applied to estimate lightning occurrence for future observing campaigns, such as I demonstrate it in Chapter \ref{chap:danish}.

%% file: chapters/4_hatp11b.tex
\chapter{Lightning on HAT-P-11b? - A case study} \label{chap:hatp11b}

\section{Introduction}
\label{sec:intro}

\begin{figure}
\begin{center}
\includegraphics[scale=0.45]{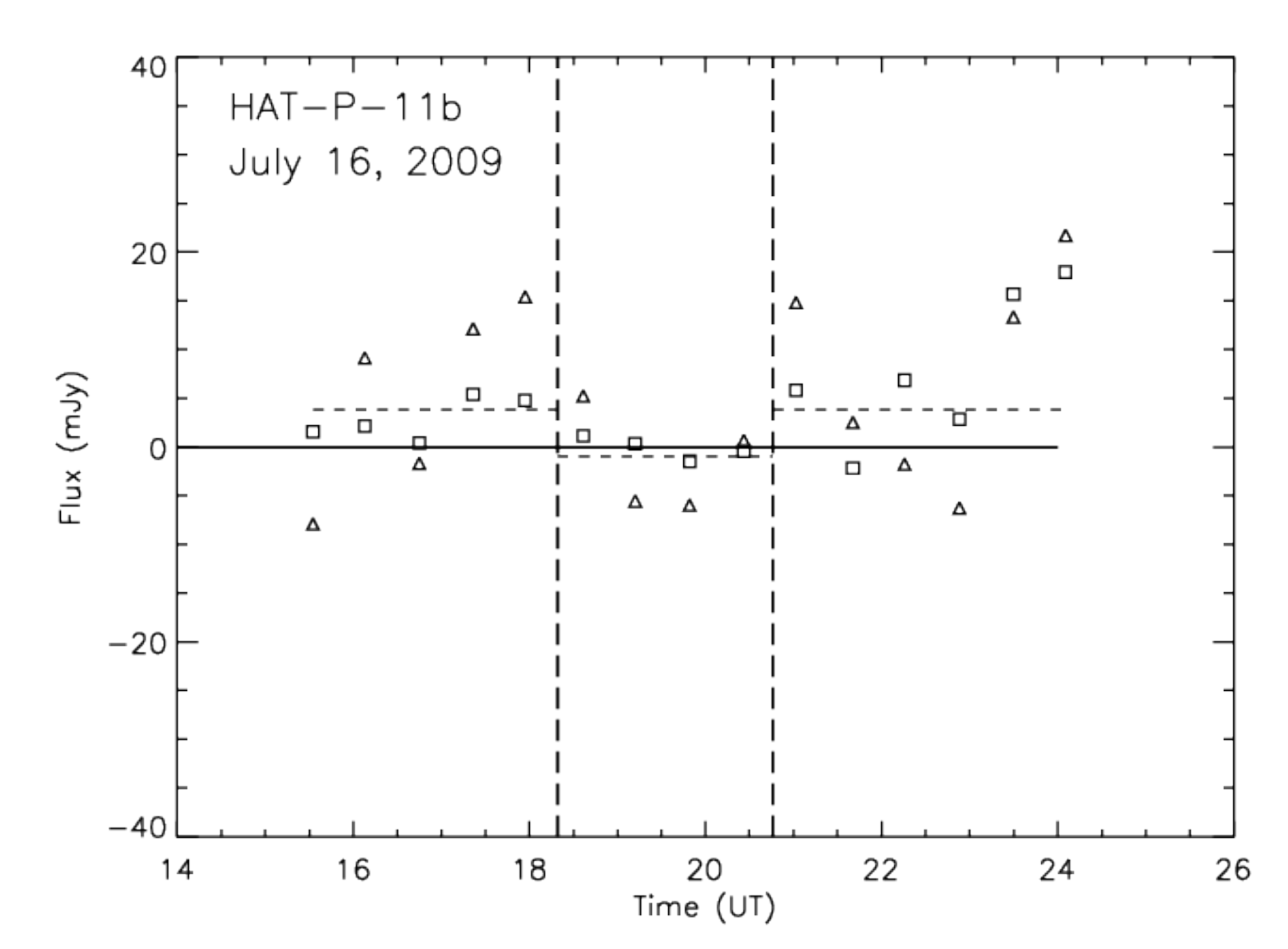} 
\end{center}
\caption{Radio light curve measured from the direction of HAT-P-11b in 2009. The data were binned to 36 minutes. The observations were conducted with the Giant Meterwave Radio Telescope in two polarization mode, RR and LL (R is right-handed, L is left-handed polarization; expressed by the stokes parameters: RR = I + V, LL = I - V; Farnes 2012, p. 31.). Vertical dashed lines: beginning and end of the secondary eclipse. Horizontal dashed line: fitted box-shaped light curve model. Figure reproduced with permission from \citet{lecav2013}. Courtesy of Alain Lecavelier des Etangs for providing the original figure.}
\label{fig:lecav}
\vspace{0.8cm}
\end{figure}

Lightning radio emission is one of the most prominent lightning signatures (Chapters \ref{chap:ligform} and \ref{chap:ligsig}). It has been observed on several Solar System planets, including Earth, Jupiter, and Saturn (Chapter \ref{chap:liginout}).
Recently, radio observations have opened new paths to study properties of extrasolar objects, such as brown dwarfs \citep[e.g.][]{williams2015}, which are only a step away from giant gas planet detections in the radio wavelengths. \citet[][]{lecav2013} \citepalias[hereafter][]{lecav2013} presented a tentative detection of a radio signal from the exoplanet HAT-P-11b. This planet is estimated to have a radius of 4.7 R$_{\oplus}$ (R$_{\oplus}$: Earth radius), a mass of 26 M$_{\oplus}$ (M$_{\oplus}$: Earth mass), and is at a distance of $\sim 0.053$ AU from its host star \citep{bakos2010, lopez2014}. In 2009, \citetalias{lecav2013} observed a radio signal at 150 MHz with an average flux of 3.87 mJy that vanished when the planet passed behind its host star. They re-observed the planet with the same instruments in 2010, but no signal was detected this time. Assuming that the 150 MHz signal from 2009 is real and comes from the exoplanet, the non-detection in 2010 suggests that it was produced by a transient phenomenon. \citetalias{lecav2013} suggested that the obtained radio signal is the result of interactions between the planetary magnetic field and stellar coronal mass ejections or stellar magnetic field. If the radio signal is real, it is unlikely to be due to cyclotron maser emission, because this type of emission is generally polarized \citep{weibel1959, vorgul2016}, and Fig. \ref{fig:lecav} shows a non-detection of polarization in the data. If the mJy radio emission were caused by cyclotron maser emission, a large planetary magnetic field of 50 G would be required \citepalias{lecav2013}. For comparison, the strength of the surface magnetic field of the Solar System planets are between $\sim10^{-4}$ G (Mars) and $\sim 4$ G \citep[Jupiter;][]{russell1993}. Therefore, based on the non-detection of polarization in combination with the possible transient nature of the observed radio emission, I tentatively hypothesize that the emission on HAT-P-11b is caused by lightning discharges. 

HAT-P-11b is much closer to its host star than the Solar system planets with lightning, resulting in a stronger irradiation from the star. 3D simulations of irradiated giant gas planets have demonstrated that a very strong circulation of the atmosphere results from the high irradiation \citep[e.g.][]{heng2015}, and it seems reasonable to expect similar effects for Neptune-like planets. Works like \citet{lee2015} and \citet{helling2016} further suggest that highly irradiated atmospheres will form clouds in very dynamic environments, which may host high lightning activity, because triboelectric charging in combination with gravitational settling allows lightning discharge processes to occur in extrasolar clouds \citep{helling2013}. (Furthermore, studies have investigated the relation between lightning activity on Earth and Solar activity, which I discuss in Chapter \ref{chap:stat} Sect. \ref{sec:stelact}.) \citet{fraine2014} took the transmission spectra of HAT-P-11b and interpreted the data with a clear atmosphere model. However, \citet{line2016} found that in the case of HAT-P-11b, patchy clouds could explain these transmission spectra. HAT-P-11b, orbiting its host star closely, likely has a dynamic atmosphere that will impact the observable cloud distribution. The resulting patchy clouds could focus potential lightning activity to a certain region, maybe at certain times covering a large fraction of the planet. These potentially large, dynamical cloud systems could support the occurrence of high lightning rates in particular regions.

In this chapter, I estimate lightning flash densities [flashes km$^{-2}$ h${-1}$] of a hypothetical thunderstorm on HAT-P-11b, which could produce the observed 3.87 mJy radio signal. I include a parameter study, to address several potential scenarios of lightning activity. The method and results are presented in Sect. \ref{sec:radio}. In Sect. \ref{sec:other}, I approximate the optical emission of the previously determined thunderstorm. Section \ref{sec:chem} evaluates the chemical effects of the storm in the form of HCN production in the atmosphere of HAT-P-11b. I summarize the Chapter in Sect. \ref{sec:summary}.

%__________________________________________________________________
%__________________________________________________________________
\section{Radio signal strength and lightning frequency}  \label{sec:radio}

In this section, I calculate the radiated power spectral density, $P/\Delta f$ [W Hz$^{-1}$], of lightning at  $f = 150$ MHz, the frequency at which HAT-P-11b was observed. This allows me to estimate the radio flux of one lightning flash at this frequency. I aim to derive a lower limit for the lightning flash density, $\rho_{\rm fl}$ [flashes km$^{-2}$ h$^{-1}$], that would be needed to reproduce the intermittent radio emission from HAT-P-11b. In Sect. \ref{sec:hatprad}, I summarize the assumptions I made and collect the necessary formulas for the calculations. Then, in Sects \ref{sec:hatpfld} and \ref{sec:hatppar}, I estimate $\rho_{\rm fl}$ for a single parameter combination in order to demonstrate the process, and conduct a parameter study, respectively. In Sect. \ref{sec:hatpdis}, I discuss the results of this section.

%__________________________________________________________________
\subsection{Lightning radio emission and flash density model} \label{sec:hatprad}

I assume that the radio signal measured by \citetalias{lecav2013} was real and was originating from HAT-P-11b. Because of the transient nature of the signal, and because it does not show clear polarization (Fig. \ref{fig:lecav}), I assume that it was produced by a thunderstorm that was present over the observed disc of the planet, continuously producing lightning discharges throughout the observations. I also assume that lightning on HAT-P-11b has the same physical and radiating properties that we know from Saturn. 
The goal is to determine how much lightning could produce the observed radio flux of $3.87$ mJy.

First, I determine the radio power spectral density, $P/\Delta f$ [W Hz$^{-1}$], radiated by one lightning discharge at frequency $f$ [Hz]: 

\begin{equation}
\label{eq:hat1}
\dfrac{P}{\Delta f} = \dfrac{P_0}{\Delta f} \Bigg(\dfrac{f_0}{f}\Bigg)^{\!\! n},
\end{equation}

\noindent where $P_0/\Delta f$ [W Hz$^{-1}$] is the peak power spectral density at a peak frequency $f_0$ [Hz], and $n$ is the spectral roll-off at high frequencies \citep{farrell2007}. The spectral irradiance of a single lightning flash, $I_{\rm \nu, fl}$, from distance $d$ is obtained from Eq. \ref{eq:2b} through $P/\Delta f$: 

\begin{equation}
\label{eq:2b}
I_{\rm \nu, fl} = \frac{(P/\Delta f)}{4 \uppi d^2} \times 10^{26},
\end{equation}

\noindent where 1 W Hz$^{-1}$ m$^{-2} = 10^{26}$ Jy. The observed spectral irradiance, $I_{\rm \nu, obs}$ will be the contribution of all the lightning flashes occurring during the observation:

\begin{equation}
\label{eq:2}
I_{\rm \nu, obs} = I_{\rm \nu, fl} \frac{\tau_{\rm fl}}{\tau_{\rm obs}} n_{\rm tot,fl},
\end{equation}

\noindent where $\tau_{\rm fl}$ [h] is the characteristic duration of the lightning event, $\tau_{\rm obs}$ [h] is the time over which the observations were taken, and $n_{\rm tot,fl}$ is the total number of flashes contributing to $I_{\rm \nu, obs}$. Equation (\ref{eq:2}) gives the total spectral irradiance resulting from lightning flashes from over the projected disc of the planet ($2\uppi R_p^2$). A lightning flash has a much shorter duration than the observation time, therefore it cannot be considered as a continuous source. As a result, the contribution of one lightning flash ($I_{\rm \nu, fl}$) to the observed spectral irradiance ($I_{\rm \nu, obs}$) has to be weighted by its duration time ($\tau_{\rm fl}$) over the observation time ($\tau_{\rm obs}$).

I am interested in the number of lightning flashes producing an average 3.87 mJy radio flux, which was observed from the direction of HAT-P-11b \citepalias{lecav2013}. This is given by $n_{\rm tot,fl}$, which I obtain from Eq. \ref{eq:2}. Then, I convert the result into flash density, $\rho_{\rm fl}$ [flashes km$^{-2}$ h$^{-1}$], which can be compared to lightning occurrence rate observed in the Solar System:

\begin{equation}
\label{eq:2a}
\rho_{\rm fl} = \frac{n_{\rm tot, fl}}{2\uppi R_p^2 \tau_{\rm obs}},
\end{equation}

\noindent where $R_p$ [km] is the radius of the planet. $\rho_{\rm fl}$ carries a statistical information on the occurrence of lightning in space [km] and time [h] (Chapter \ref{chap:stat}).

%__________________________________________________________________
\subsection{Flash density for representative parameters $-$ $\rho_{\rm fl,1}$} \label{sec:hatpfld}

To demonstrate the method, I use example parameters for the calculations. These are $n=3.5$ and $\tau_{\rm fl}=0.3$ s. The rest of the parameters are taken to be the same for all the estimates, in the parameter study as well.
In order to estimate the flash density that would result in a radio signal like the one obtained by \citetalias{lecav2013}, I assume that lightning on HAT-P-11b has the same energetic properties as lightning on Saturn. \textit{Cassini-RPWS} measured the radiated power spectral density of lightning on Saturn to be $P/\Delta f =$ 50 W Hz$^{-1}$ at $f =$ 10 MHz \citep{fischer2006, farrell2007}. I use Eq. \ref{eq:hat1} and the values observed by the \textit{Cassini} probe for $P/\Delta f$ and $f$ to obtain a peak spectral power density of $P_0/\Delta f = 1.6 \times 10^{12}$ W Hz$^{-1}$, for $f_0 = 10$ kHz and $n = 3.5$\footnote{The Earth value $f_0 = 10$ kHz \citep{rakov2003} and a gentler spectral roll-off, $n = 3.5$ ($n = 4$ for Earth) were used because these values are not known for any other Solar system planet. These values are used for modelling lightning on Jupiter or Saturn \citep[e.g.][]{farrell2007}.}. Next, by applying $P_0/\Delta f$ to Eq. \ref{eq:hat1}, I estimate the radiated power spectral density at the source of a single lightning flash at $f = 150$ MHz, frequency at which the HAT-P-11b radio signal was observed \citepalias{lecav2013}, to be $P/\Delta f = 3.9 \times 10^{-3}$ W Hz$^{-1}$.

\begin{figure*}
\begin{center}
\includegraphics[scale=0.35]{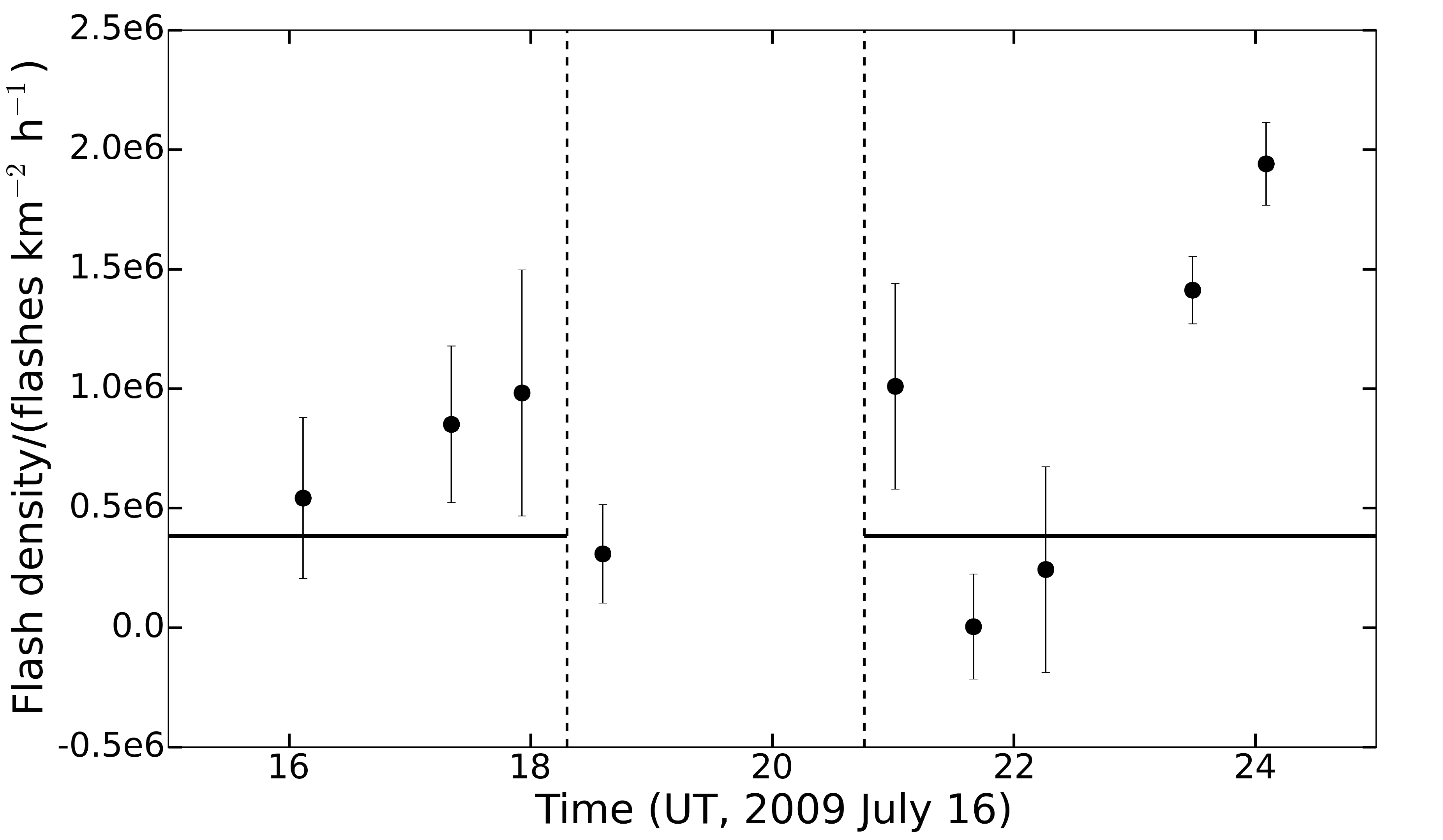}
\includegraphics[scale=0.35]{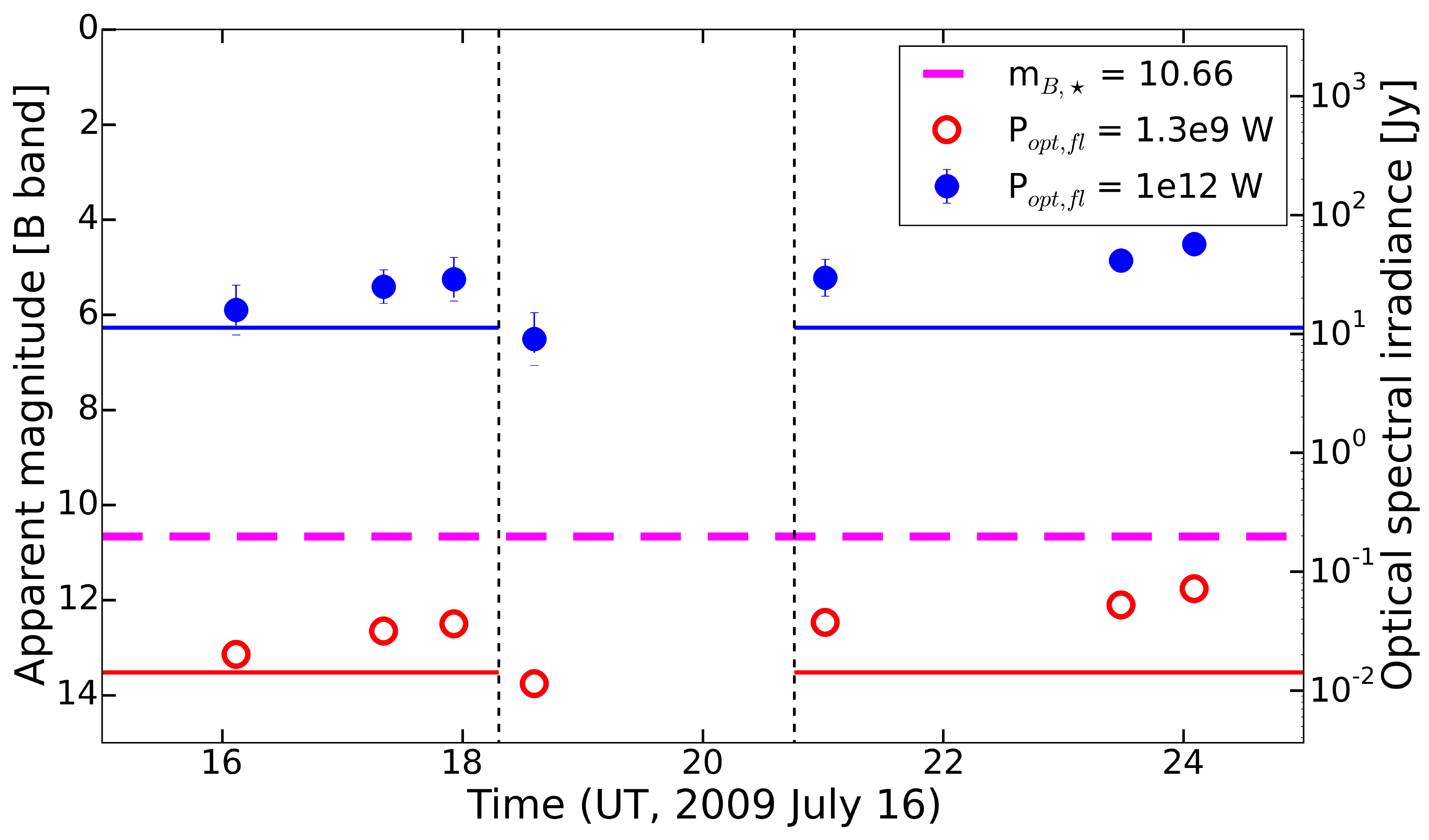}
\end{center}
\caption{Lightning flash densities (top) and apparent magnitude of the lightning flashes (bottom) that would produce the radio fluxes observed by \citetalias{lecav2013} for HAT-P-11b (Fig. \ref{fig:lecav}) for parameters $n=3.5$ and $\tau_{\rm fl}=0.3$ s. Horizontal solid lines: average values for the average observed radio flux of 3.87 mJy outside eclipse. Vertical dashed lines: beginning and end of the secondary eclipse of the planet. I show the mean results for the range of observed values per time from \citetalias[][]{lecav2013} (see Fig. \ref{fig:lecav}). \textbf{Bottom:} Results for two different optical powers, Saturnian ($1.3 \times 10^{9}$ W; red) and terrestrial super-bolt ($10^{12}$ W; blue). Magenta dashed line: apparent B magnitude of the host star HAT-P-11.}
\label{fig:rflux}
\vspace{0.8cm}
\end{figure*}

Using equation (\ref{eq:2b}) and the distance of HAT-P-11, $d = 38$ pc, I obtain the spectral irradiance for a single lightning flash to be $I_{\rm \nu, fl} = 2.2 \times 10^{-14}$ Jy. \citetalias{lecav2013} found the average observed spectral irradiance, $I_{\rm \nu, obs}$, to be  3.87 mJy. Solving equation (\ref{eq:2}) for $n_{\rm tot,fl}$, the total number of lightning flashes needed to explain the observed spectral irradiance, with an average event duration, $\tau_{\rm fl} = 0.3$ s \citep[the largest event duration on Saturn according to][]{zarka2004}, I obtain a value of $n_{\rm tot,fl} \approx 1.3 \times 10^{15}$ flashes. The integration time for a single data point in \citetalias{lecav2013} (their fig. 2) is $\tau_{\rm obs} = 36$ min, and the radius of HAT-P-11b is $R_p \approx 0.4$ R$_{\rm J}$ \citep[R$_{\rm J}$: Jupiter radius;][]{bakos2010}. Substituting these values and the derived $n_{\rm tot,fl}$ into equation (\ref{eq:2a}) I obtain a flash density of $\rho_{\rm fl,1} \approx 3.8 \times 10^5$ flashes km$^{-2}$ h$^{-1}$. Figure \ref{fig:rflux} (top) shows the flash densities that would be needed to produce a radio flux comparable to observations in \citetalias{lecav2013} (Fig. \ref{fig:lecav}), using the example $n$ and $\tau_{\rm fl}$ values, and assuming that the flux is from the planet and is entirely produced by lightning.

%__________________________________________________________________
\subsection{Parameter study} \label{sec:hatppar}

I conduct a parameter study to investigate the effects of the spectral roll-off, $n$, and the discharge duration, $\tau_{\rm fl}$, on the resulting lightning flash densities, $\rho_{\rm fl}$. I do not consider the dependence on the peak frequency, $f_0$, but assume, that it is the same in all planetary atmospheres considered here, as it does not have an effect on the final result. 

\begin{figure}
\centering
\includegraphics[width=0.75\textwidth]{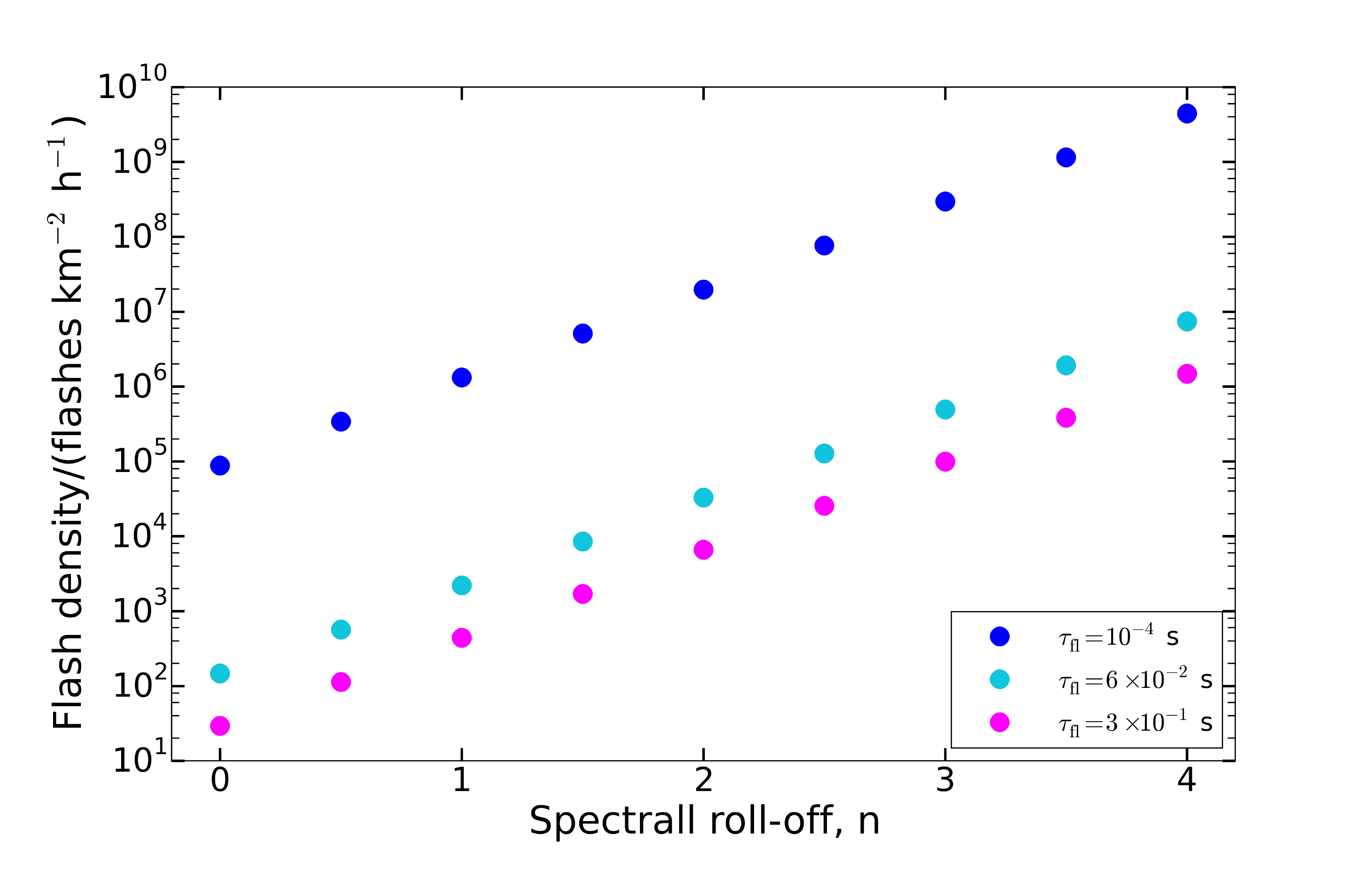}
\caption{Lightning flash density $\rho_{\rm fl}$ [flashes km$^{-2}$ h$^{-1}$] needed to emit the 3.87 mJy radio flux measured by \citetalias{lecav2013}, depending on the spectral roll-off, $n$, used in Eq. \ref{eq:hat1}. The figure also shows the dependence of $\rho_{\rm fl}$ on the discharge duration, $\tau_{\rm fl}$ (Eq. \ref{eq:2}). For all the calculations, I assumed that the radiated power spectral density of lightning on HAT-P-11b is $P/\Delta f = 50$ W Hz$^{-1}$ at $f = 10$ MHz, as was measured on Saturn \citep{fischer2006b}. The results show that the flatter the radiated power spectrum (the smaller $n$), the fewer flashes are needed to produce the observed radio flux ($\rho_{\rm fl}$ is smaller). For the same spectral roll-off, $n$, the slower discharges produce larger amount of power than quicker discharges.}
\label{fig:rpar}
\vspace{0.8cm}
\end{figure}
 
\citet{zarka1983} found the SED spectrum to be relatively flat ($n\approx 0.0$) above a few MHz. \citet{fischer2006b} found the roll-off of the SED spectrum to be $n=0.5$. The largest roll-off for lightning on Earth was found to be $n=4$ \citep{rakov2003}. Therefore, I conduct the study for $n = 0\,\ldots\,4$ with increments of 0.5. For each $n$ I calculate the flash density applying a discharge duration of $\tau_{\rm fl} = 10^{-4}, 6 \times 10^{-2}$, and $3 \times 10^{-1}$ s. 100 $\mu$s is the average stroke duration on Earth \citep{volland1984}. 60 ms is the average flash duration on Saturn \citep{zarka2006}, while 300 ms is the largest flash duration on Saturn \citep{zarka2004}. The results of the parameter study are shown in Fig. \ref{fig:rpar}. The results show that the flatter the radiated power spectrum (the smaller $n$), the fewer flashes are needed to produce the observed radio flux ($\rho_{\rm fl}$ is smaller). Smaller $n$ means that the radiated radio power decreases much slower, therefore, one lightning flash will produce larger radio powers at higher frequencies, than what it would produce if the spectrum was steeper ($n$ was larger). For the same spectral roll-off, slower discharges release more power than quicker ones. In a best case scenario with $n = 0$ and $\tau_{\rm fl} = 0.3$ s, a flash density $\rho_{\rm fl,2} = 29$ flashes km$^{-2}$ h$^{-1}$ would be enough to produce the observed radio flux of 3.87 mJy .

%__________________________________________________________________
\subsection{Comparison of flash density results} \label{sec:hatpdis}

\begin{table}
\caption{Obtained lightning flash densities for HAT-P-11b and examples from across the Solar System. Values are from Chapter \ref{chap:stat}, Tables \ref{table:vol} and \ref{table:plan}, apart from (*), which was obtained from \citet{huffines1999}, and the Saturn values, which are estimated in this chapter. The top table lists values according to planets, the bottom one shows volcanic eruptions on Earth.}
\begin{center}
%\small
\footnotesize
\begin{tabular}{lll}
	\hline 
	Planet & Flash density [km$^{-2}$ h$^{-1}$] & Comment \\
	\hline
	\textbf{HAT-P-11b} ($\rho_{\rm fl,1}$) & $\mathbf{3.8 \times 10^5}$ & $n=3.5$, $\tau_{\rm fl} = 0.3$; This work \\
	\textbf{HAT-P-11b} ($\rho_{\rm fl,2}$) & $\mathbf{29}$ & $n=0.0$, $\tau_{\rm fl} = 0.3$; This work \\	
	Earth & 0.1 & largest average in the USA$^{(*)}$ \\
	 & $2.29 \times 10^{-4}$ & global average from LIS/OTD data \\
	Saturn & $8.4 \times 10^{-7}$ & from SED rates of \citet{fischer2011} for the globe \\
	 & $9 \times 10^{-3}$ & same as above, but for one storm (see Sect. \ref{sec:hatpdis}) \\
	Jupiter & $1.43 \times 10^{-6}$ & from New Horizons (2007) data \\
	\hline
\end{tabular}

\vspace{0.5cm}

\begin{tabular}{lll}
	\hline 
	Volcano & Flash density [km$^{-2}$ h$^{-1}$] & Eruption \\
	\hline
	Eyjafjallaj\"okull & 0.1 & 2010 Apr 14$-$19 \\
	 & 0.32 & 2010 May 11$-$20 \\
	Mt Redoubt & 12.04 & 2009 Mar 23 \\
	 & $2 \times 10^3$ & 2009 Mar 29 (Phase 1) \\
	\hline
\end{tabular}
\label{table:hatp1}
\end{center}
\vspace{0.8cm}
\end{table}

The results suggest that the necessary flash density to produce the observed 3.87 mJy radio flux spans a large range of values depending on the parameters I choose for the calculations. The example values used in Sect. \ref{sec:hatpfld}, represent a roughly average flash density in the parameter study (Sect. \ref{sec:hatppar}), $\rho_{\rm fl,1} = 3.8 \times 10^5$ flashes km$^{-2}$ h$^{-1}$. For a best-case scenario an $\rho_{\rm fl,2}$ as low as $29$ flashes km$^{-2}$ h$^{-1}$ is enough to be maintained throughout the time of the observations on the disc of the planet. Table \ref{table:hatp1} lists the result for HAT-P-11b in comparison to a few examples of flash densities that were observed in the Solar System (Chapter \ref{chap:stat}). I only list and compare the best case and the example case for HAT-P-11b.

\citet{fischer2011} analysed the SED (Saturnian Electrostatic Discharges) occurrence of a giant storm that occurred on Saturn in 2010/2011. They found an SED rate of 10 s$^{-1}$, which is 36000 SED h$^{-1}$. This is the largest rate observed on Saturn. Since no other storms were observed during this period \citep{dyudina2013}, I apply this flash rate (assuming that one SED originates from one flash) for the whole surface area of the planet. This results in a flash density of $8.4 \times 10^{-7}$ flashes km$^{-2}$ h$^{-1}$ for Saturn.

The observed signal on HAT-P-11b would require a storm with the example flash density, $\rho_{\rm fl,2} = 3.8 \times 10^5$ flashes km$^{-2}$ h$^{-1}$, $\sim 4.5 \times 10^{11}$ times greater, and a storm with the best-case value $\rho_{\rm fl,1} = 29$ flashes km$^{-2}$ h$^{-1}$, $\sim 3.5 \times 10^{7}$ times grater than observed on Saturn. However, one may argue that a planet can host multiple thunderstorms at the same time, so the SED rates are only true for the specific storm and not for the whole planet. Considering an average storm size of 2000 km on Saturn\footnote{Also, similar storm size was observed in December 2010 \citep{fischer2011}} \citep{hurley2012}, the flash density based on the average SED rate of the 2010/2011-storm is $9 \times 10^{-3}$ flashes km$^{-2}$ h$^{-1}$. This flash density is $\sim 4.2 \times 10^{7}$ times smaller than the calculated example $\rho_{\rm fl,2}$, and $3.2\times10^{3}$ times smaller than the best-case $\rho_{\rm fl,1}$ on HAT-P-11b.

On Earth, one of the highest flash densities observed, $\sim 0.1$ flashes km$^{-2}$ h$^{-1}$ \citep{huffines1999}, is produced in thunderstorms within the United States (USA). The 3.87 mJy signal from $\rho_{\rm fl,2} \approx 3.8 \times 10^5$ flashes km$^{-2}$ h$^{-1}$ on HAT-P-11b, would require a sustained global storm with flash densities $\sim 3.8 \times 10^6$ times greater than observed within the USA. However, the best-case flash density $\rho_{\rm fl,1} \approx 29$ flashes km$^{-2}$ h$^{-1}$ is only two orders of magnitude larger than the above listed Earth value. The most intense lightning activity, however, is not found in thunderstorms on Earth, but rather in volcano plumes after explosive volcanic eruptions. The largest value in Table \ref{table:hatp1} was observed during the 29 Mar 2009 eruption of Mt Redoubt (Chapter \ref{chap:stat}, Sect. \ref{sec:volc}). Its $\rho_{\rm fl} = 2000$ flashes km$^{-2}$ h$^{-1}$ is only two orders of magnitude smaller than the average example $\rho_{\rm fl,2}$, and two orders of magnitude larger than the best-case $\rho_{\rm fl,1}$.

This comparison shows that, a thunderstorm with a lightning occurrence rate obtained for the hypothetical storm on HAT-P-11b ranges between thunderstorms with flash densities of the same order of magnitude as the Mt Redoubt eruption showed in 2009 Mar 23, and thunderstorms never seen in the Solar System before. However, we have to remind ourselves, that the Jovian and Saturnian values are based on data from spacecraft, which can only observe the most energetic flashes from the planet (Chapter \ref{chap:stat}). Secondly, I assumed that lightning on HAT-P-11b produces the same amount of energy and radio power that is known from the Solar System. In Chapter \ref{chap:model}, I show that lightning can be more energetic and produce 4$-$10 orders of magnitude more radio power on hot exoplanets ($T_{\rm eff} = 1500 \dots 2000$ K) and brown dwarfs, than lightning on Earth. HAT-P-11b orbits the host star on a very close orbit, resulting in high atmospheric temperatures \citep[T$_{\rm eq}$ between 630 K and 950 K,][]{huber2017}, and a planetary object that the Solar System does not contain. Therefore, it is reasonable to think that if lightning exists on HAT-P-11b, it is more energetic and more frequent than lightning in the Solar System.

%__________________________________________________________________
%__________________________________________________________________
\section{Lightning detection in the optical range} \label{sec:other}

The emitted power of lightning has been measured in other wavelengths on Jupiter and Saturn. In this section, I apply the flash densities derived in Sect. \ref{sec:hatpfld} and shown in Fig. \ref{fig:rflux} (top), and estimate the emitted optical flux of the lightning storm that could produce the observed radio flux on HAT-P-11b. I do not consider the effects of varying parameters, but apply the same parameters as were used to calculate the example average value of flash density, $\rho_{\rm fl,2} \approx 3.8 \times 10^5$ flashes km$^{-2}$ h$^{-1}$, and the results in Fig. \ref{fig:rflux} (top). 

\citet[][table 2]{dyudina2013} lists the survey time (1.9 s), the total optical power ($1.2 \times 10^{10}$ W) and the optical flash rate (5 s$^{-1}$) of the large thunderstorm on Saturn in 2011. Based on this information the average optical power released by a single flash of this Saturnian thunderstorm is $P_{\rm opt,fl} \approx 1.3 \times 10^9$ W. Assuming that flashes on HAT-P-11b produce the same amount of power as Saturnian flashes and using equation (\ref{eq:opt}) I obtain an optical irradiance from a single flash to be $I_{\rm opt,fl} = 1.13 \times 10^{-14}$ Jy.

\begin{equation}
\label{eq:opt}
I_{\rm opt,fl} = \frac{P_{\rm opt,fl}/f_{\rm eff}}{4 \uppi d^2} \times 10^{26},
\end{equation}

\noindent where $P_{\rm opt,fl}$ is the optical power of a single flash and $f_{\rm eff}\approx 6.47 \times 10^{14}$ Hz is the effective frequency of \textit{Cassini}'s blue filter. The total optical irradiance of flashes is obtained from $I_{\rm opt,fl}$ and $n_{\rm tot,fl}$, the total number of flashes hypothetically producing the same radio flux as was observed by \citetalias{lecav2013}. This optical irradiance is of the order of $10^{-2}$ Jy or brightness of $\sim 13$ mag (B band) as shown in Fig \ref{fig:rflux}, bottom panel (red). The star HAT-P-11 has an apparent B magnitude = 10.66 \citep{hog2000}, which is $\sim 0.2$ Jy in the B band. The optical emission resulting from lightning, therefore, would be slightly lower than that of the star. 

I carried out the same calculations to determine the planetary and stellar flux ratio, in case lightning on HAT-P-11b emitted a power of the order of super-bolt power on Earth, $P_{\rm opt,fl} \approx 10^{12}$ W \citep[][p. 164]{rakov2003}. The produced optical flux densities and the corresponding magnitude scale are shown in Fig. \ref{fig:rflux}, bottom panel (blue). The ratio of the planetary lightning flux and the flux of the star (with lightning flux density of $\sim 10$ Jy, Fig. \ref{fig:rflux}, bottom panel) is $\sim 10^2$. If every single lightning flash on HAT-P-11b would emit $\sim 10^{12}$ W, the optical emission of lightning that would produce the radio emission, would outshine the host star by two orders of magnitude.

%__________________________________________________________________
%__________________________________________________________________
\section{Lightning Chemistry} \label{sec:chem}

Lightning produces non-equilibrium species affecting the composition of the local atmosphere (Chapter \ref{chap:ligsig}, Sect. \ref{sec:chemeff}). It is important, therefore, to consider what the chemical effects of lightning activity on HAT-P-11b would be on the local atmosphere, and on the spectrum of the planet. Here, the effect of large, Saturn-like lightning storms on HCN chemistry is considered. \citet{lewis1980} estimated that lightning produces HCN at a rate of $2 \times 10^{-10}$ kg/J. Their model was set up for Jupiter, and is applicable for any hydrogen-dominated atmosphere with roughly solar composition. \citet{farrell2007} estimated the dissipative energy of a single lightning flash on Saturn, with peak frequency, $f_0 = 10$ kHz, and spectral roll-off, $n = 3.5$, to be:

\begin{equation}
E_d \approx 260 \, {\rm J} \; \Bigg(\dfrac{f_{\rm Sat}}{f_0}\Bigg)^{\!\!n},
\end{equation}

\noindent where $f_{\rm Sat} = 10$ MHz is the frequency at which the lightning on Saturn was observed. One can multiply the dissipative energy by the flash density of 1 flash km$^{-2}$ h$^{-1}$. Multiplying the lightning energy density by the production rate of HCN, it is estimated that $5 \times 10^{-7}$ kg m$^{-2}$ s$^{-1}$ of HCN is produced, of the order of $10^9$ greater than the estimate of \citet{lewis1980} for Jupiter. Accepting the energetics arguments from \citet{lewis1980}, the resulting HCN will achieve a volume mixing ratio of $\sim 10^{-6}$ within the mbar regime of the atmosphere. \citet{moses2013} found that similar mixing ratios (their fig. 11) should have significant observational consequences in the \textit{L} ($3.0-4.0 \mu$m) and \textit{N} ($7.5-14.5 \mu$m) IR bands, which they show in their fig. 16 comparing their model spectra both with and without HCN.

\begin{figure}
\center
\includegraphics[width=0.7\linewidth]{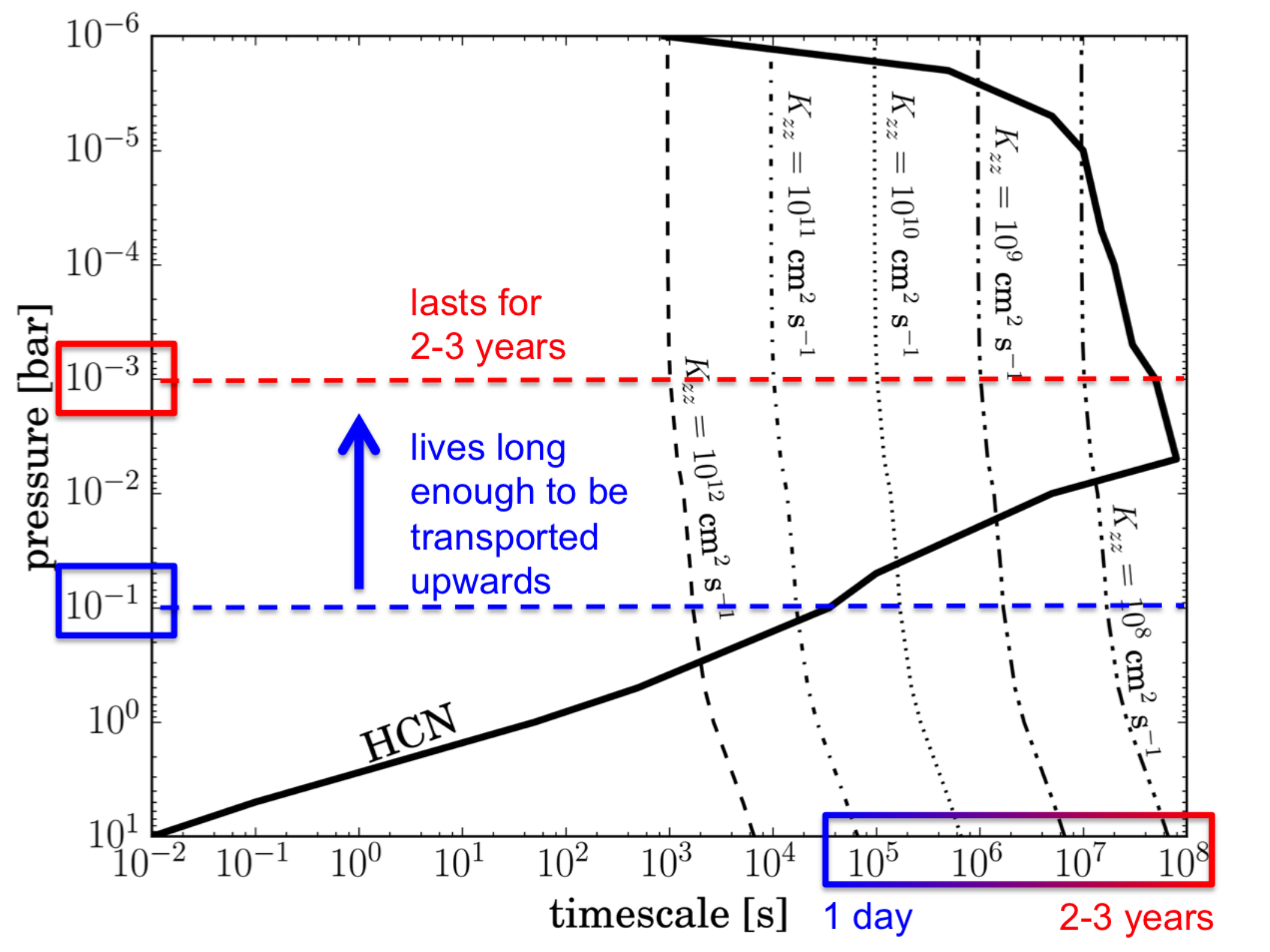}
\caption{The chemical lifetime of HCN [s] plotted vs. pressure [bar], along with various dynamical timescales with eddy diffusion coefficients ranging $K_{zz} = 10^8 \,\ldots\, 10^{12}$ cm$^2$ s$^{-1}$. If HCN is produced in the $10^{-1}$ bar pressure level in the atmosphere, than it will live long enough (days) to be transported to the higher atmosphere, where it will be present for a couple of years after the storm have occurred.}
\label{fig:hcn}
\vspace{0.8cm}
\end{figure}

In order to estimate the chemical timescale\footnote{The amount of time necessary for the atmospheric abundance of HCN to achieve thermochemical equilibrium.} for HCN on HAT-P-11b, a semi-analytical pressure$-$ temperature profile is developed, appropriate for the object using the method of \citet{hansen2008}. The parameters for HAT-P-11b (mass, radius, distance from host star) given by \citet{bakos2010} and \citet{lopez2014}, and the stellar temperature from \citet{bakos2010} are used for this profile. To determine the XUV flux impinging on the atmosphere, a spectrum appropriate for a K4-type star was taken, the X-Exospheres synthetic spectrum for HD 111232 \citep{sanz-Forcada2011}. It is assumed that the atmosphere is hydrogen rich, and the atmospheric gas of HAT-P-11b is at roughly solar metallicity with respect to C, N and O, i.e. that the primordial concentrations of these elements in HAT-P-11b is solar and that there is no elemental depletion into clouds. The atmospheric chemistry is calculated using the semi-analytic temperature profile and synthetic XUV flux, with the STAND2015 chemical network and the ARGO diffusion-photochemistry model from \citet{rimmer2016}. Then a variety of locations in the atmosphere is examined, injecting HCN at a mixing ratio of $10^{-6}$, and evolving the atmospheric chemistry in time to determine the chemical timescale for HCN as a function of pressure, shown in Fig. \ref{fig:hcn}. The dynamical timescale for vertical mixing is overlaid on the top of this plot, assuming a range of constant eddy diffusion coefficients.

The chemical timescale for HCN ranges from 100 milliseconds at the bottom of the model atmosphere (10 bar), to 2.5 years at 5 mbar (Fig. \ref{fig:hcn}). At pressures less than 5 mbar, the timescale for HCN slowly drops down to about 4 months at 10 $\mu$bar, and then drops precipitously at lower pressures, until at 1 $\mu$bar it achieves a timescale of about 30 minutes (Fig. \ref{fig:hcn}). These results can be compared to the dynamical timescale of the atmosphere, represented approximately by the eddy diffusion coefficient \citep[see][for details]{lee2015}. If the chemical timescale for HCN is smaller than the dynamical timescale, then the HCN would be destroyed before it is transported higher into the atmosphere. If the chemical timescale for HCN is larger than the dynamical timescale, then the HCN will survive long enough to reach other parts of the atmosphere, where it will survive longer. Assuming that lightning takes places on HAT-P-11b at pressures of $\lesssim 0.1$ bar, the produced HCN will survive long enough to be transported into the mbar regime, where it will survive for $2-3$ years before being chemically destroyed. If, on the other hand, HCN is formed much below the 0.1 bar level, at pressures of $\gtrsim 1$ bar, the chemical timescale is too short for the HCN to escape, and it will be rapidly destroyed before it could be observed.
This suggests that to confirm the presence of lightning in an atmosphere, future radio observations can be followed up by IR observations looking for molecules that are the result of lightning chemistry.

%__________________________________________________________________
%__________________________________________________________________
\section{Summary} \label{sec:summary}

In this chapter, I presented an interpretation of the radio observations of HAT-P-11b made by \citetalias{lecav2013} under the assumption that these transient radio signals are real and were caused by lightning on HAT-P-11b. I estimated that the flash density necessary to explain the average radio signal ranges between 29 and $2 \times 10^9$ flashes km$^{-2}$ h$^{-1}$, depending on the parameters I use. These values range between the same order of magnitude as terrestrial volcanic eruptions show, and thunderstorms never seen in the Solar System before. I also examined the optical emission such a storm would generate, as well as the impact of this storm on the atmospheric chemistry, assuming a hydrogen-rich atmosphere.

In summary, I found that

\begin{enumerate}
\item the radio emission of a few mJy at 150 MHz, at the distance of HAT-P-11b, requires unrealistically high flash densities if this lightning is like in the Solar System. However, if we let the parameters of spectral roll-off and lightning duration vary, the flash density can be as low as values seen during volcano eruptions. Nevertheless, a large part of the parameter space requires extremely large lightning activity, therefore lightning produced radio emission most probably cannot be observed by current radio telescopes at frequencies of 150 MHz or higher, from distances of several tens of pc.

\item The optical counterpart of the enormous lightning storm would be as bright as the host star itself.

\item The amount of HCN produced by lightning at pressures $\leq 0.1$ bar, in a hydrogen-rich atmosphere of an irradiated exoplanet with strong winds, may in some cases yield detectable quantities that linger in the atmosphere for $2-3$ years after the advent of the lightning storm. If the lightning occurs much deeper in the atmosphere, the HCN will react away before it can diffuse into the upper atmosphere, and will probably not be observable.
\end{enumerate}

The results show that the radio emission on HAT-P-11b is unlikely to be caused by lightning, if lightning properties similar to the Solar System ones are assumed. However, intermittent, powerful thunderstorms are not unprecedented in the Solar System: In 2010/11 a huge storm was observed in Saturn, producing the largest flash densities ever observed, with a total power comparable to Saturn's total emitted power \citep{fischer2011}. Such large, or even more powerful storms may occur on exoplanets. The current study also shows a new interpretation that could be applied to high frequency (up to $\sim30-50$ MHz) radio observations, where it is more probable to observe lightning, because of its radiating properties (Eq. \ref{eq:hat1}). The calculations explained in this chapter can be also applied to determine the minimum storm size detectable within an exoplanetary atmosphere using current or future radio instruments. The recommendation to observers who detect radio emission in the frequency range of a few tens of MHz, especially if it is unpolarized, from an exoplanet would be to follow up these observations with infrared observations made in the \textit{L} and \textit{N} bands when possible, in order to look for HCN emission, which should be observable for $2-3$ years if lightning occurs around the 0.1 bar level of an atmosphere with reasonably large vertical convective velocities. If HCN is detected at that time, and if both the radio emission and the HCN turn out to be transient, this would be strong evidence for lightning on an exoplanet.

%% file: chapters/4b_danishtel.tex
\chapter{Looking for lightning on the closest brown dwarfs}	\label{chap:danish}

%__________________________________________________________________
%__________________________________________________________________
\section{Introduction} \label{sec:intt}

In the previous chapter, I introduced a method to estimate lightning activity on extrasolar objects from observed radio emission. Based on the results I estimated the optical emission a thunderstorm on the exoplanet HAT-P-11b could produce if its radio emission was the observed one in \citet{lecav2013}. In this chapter, I turn the method the other way around and ask the question: what is the optical flux of lightning from a target object, if we assume that lightning on that body has the same statistical and energetic properties that is known from the Solar System. The purpose of this short project is to estimate whether lightning optical emission can be observed with the 1.54-m Danish Telescope in La Silla, Chile\footnote{Information about the telescope can be found here: http://www.ls.eso.org/lasilla/Telescopes/2p2/D1p5M/}. The estimates were used as part of an observing proposal for the telescope. The telescope is equipped with a CCD and an EMCCD camera, and standard Johnson-Cousins System filters, of which I estimated observability in I, V and U bands. The sensitivity of the CCD camera of the Danish Telescope is $\sim90$\% in I, 80\% in V and 20\% in U bands, while the EMCCD is $\sim90$\% sensitive in all bands (Uffe G. Jorgensen, private communication).

Electromagnetic radiation propagates as $r^{-2}$, therefore, the closer the object to us the larger the chance to receive electromagnetic flux from the same source on the object. The targets of this study are the three closest brown dwarf systems observable from Chile: Luhman 16, $\epsilon$ Indi, and SCR 1845-6357. Luhman 16 is a binary brown dwarf system \citep{luhman2013}, $\epsilon$ Indi is composed of a K5 star and a brown dwarf binary separated by $\sim 1450$ AU from the primary \citep{scholz2003,volk2003}. The third target, SCR 1845-6357, is a brown dwarf orbiting an M8.5 red dwarf \citep{biller2006}. Table \ref{table:bds} lists the intrinsic properties of the brown dwarfs, which are important for lightning flux estimates, and for planning observations.

\begin{table}
\resizebox{\columnwidth}{!}{
\begin{threeparttable}
\center
\footnotesize
\caption{Properties of the three brown dwarf systems used in this study. Data were gathered using \textsc{Simbad} \citep{wenger2000}. Only I, V, and U magnitudes are listed where available.}
  \begin{tabular}{lccccc}
	\hline 
	Brown Dwarf & Distance [pc] & Spectral Type & Apparent magnitude & Right Ascension & Declination \\ %& Comment \\
	\hline
	\vtop{\hbox{\strut Luhman 16AB}\hbox{\strut \citep{luhman2013}}} & 2 & L7.5 and T0.5\tnote{(1)} & I: 14.95\tnote{(1)} & $10^{\rm h}$ $49^{\rm m}$ $18.9^{\rm s}$\tnote{(1)} & $-53^{\circ}$ $19'$ $10.1''$\tnote{(1)} \\
	\vtop{\hbox{\strut $\epsilon$ Indi \citep{scholz2003}}\hbox{\strut \citep{volk2003}}} & 3.6 & T1 and T6\tnote{(2)} & \vtop{\hbox{\strut V: 24.12\tnote{(2)}}\hbox{\strut I: 15.60}} & $22^{\rm h}$ $04^{\rm m}$ $10.5^{\rm s}$\tnote{(2)} & $-56^{\circ}$ $46'$ $57.7''$\tnote{(2)} \\
	\vtop{\hbox{\strut SCR 1845-6357}\hbox{\strut \citep{biller2006}}} & 3.85 & T6\tnote{(3)} & J: 13.29\tnote{(3)} & $18^{\rm h}$ $45^{\rm m}$ $05.5^{\rm s}$\tnote{(3)} & $-63^{\circ}$ $57'$ $46.3''$\tnote{(3)} \\
	\hline 
  \end{tabular}
  \begin{tablenotes}
	\item[1] http://simbad.u-strasbg.fr/simbad/sim-id?Ident=2MASS\%20J10491891-5319100
	\item[2] http://simbad.u-strasbg.fr/simbad/sim-id?Ident=*+eps+Ind+B
	\item[3] http://simbad.u-strasbg.fr/simbad/sim-id?Ident=SCR+J1845-6357B
  \end{tablenotes}
\label{table:bds}
\end{threeparttable}
}
\vspace{0.8cm}
\end{table}

In Sect. \ref{sec:met}, I shortly summarize the method used to calculate optical fluxes. The equations first appear in Chapter \ref{chap:hatp11b}, however for better understanding, I repeat them here as well. In the same section, I introduce the parameters I used for my calculations. In Sect. \ref{sec:res}, I present the results and discuss them. I conclude the chapter in Sect. \ref{sec:concc}.

%__________________________________________________________________
%__________________________________________________________________
\section{Method} \label{sec:met}

In this section, I summarize the necessary equations for estimating the optical flux of lightning originating in the example brown dwarf atmospheres. I also list and discuss the input parameters, which affect the results. I do not consider radiation effects inside the atmosphere of the observed object. I assume that the emitted power of the lightning flash will contribute entirely to the lightning flux.
 
The total observed flux ($I_{\rm obs}$ [Jy]) from a transient event like a lightning storm can be expressed the following way:

\begin{equation} \label{eq:dan1}
I_{\rm obs} = I_{\rm opt,fl} \frac{\tau_{\rm fl}}{\tau_{\rm obs}} n_{\rm tot,fl},
\end{equation}

\noindent where $I_{\rm opt,fl}$ [Jy] is the optical flux from a single lightning discharge (as I am interested in the optical emission this time), $\tau_{\rm fl}$ [s] is the duration of the discharge, $\tau_{\rm obs}$ [s] is the observation or exposure time, and $n_{\rm tot,fl}$ is the total number of flashes occurring during the observation contributing to the obtained total flux, $I_{\rm obs}$. $I_{\rm opt,fl}$ is given by:

\begin{equation} \label{eq:dan2}
I_{\rm opt,fl} = \frac{P_{\rm opt,fl}}{f_{\rm eff}} \frac{10^{26}}{4 \pi d^2},
\end{equation}

\noindent where $P_{\rm opt,fl}$ [W] is the optical power of a single lightning flash, $f_{\rm eff}$ [Hz] is the effective frequency of the filter used for the observation, $d$ [m] is the distance of the brown dwarf system, and 1 W Hz$^{-1}$ m$^{-2} = 10^{26}$ Jy. The total number of flashes, $n_{\rm tot,fl}$, over the disc of the object during the observation time, is derived from the flash density, $\rho_{\rm fl}$ [flashes km$^{-2}$ h$^{-1}$]:

\begin{equation} \label{eq:dan3}
n_{\rm tot,fl} = \rho_{\rm fl} 2 \pi R^2 \tau_{\rm obs},	
\end{equation}

\noindent where $R$ [km] is the radius of the brown dwarf, and $\tau_{\rm obs}$ is given in hours.

%__________________________________________________________________
\subsection{Input parameters} 

Most of the input parameters in Eqs \ref{eq:dan1}-\ref{eq:dan3} are dependent on whether we assume that lightning has similar statistical and radiating properties to lightning on Earth, Jupiter, and Saturn, or these properties are different. Lightning on exoplanets and brown dwarfs might be very different from what we know from the Solar System (Chapter \ref{chap:model}), however, for a first estimate I consider Solar System-like properties in my calculations.

\subsubsection{Intrinsic properties of the brown dwarfs}
Two main properties of the brown dwarf systems are important for the calculations: size (radius), $R$, and distance, $d$. These will be input parameters for Eq. \ref{eq:dan2} and Eq. \ref{eq:dan3}.
Brown dwarf size estimates are based on evolutionary models, which suggest that old objects ($>\sim 2$ Gyr) have a radius roughly $R_{\rm p} = 1 R_{\rm Jup} {\rm(}= 6.99 \times 10^4$ km) \citep[e.g.][]{burgasser2006}. The three brown dwarf systems are the closest ones to Earth, that are visible from Chile, the location of the Danish telescope. Luhman-16 is 2 pc, SCR 1845-6357 is 3.861 pc, and $\epsilon$ Indi is 3.626 pc away from us (Table \ref{table:bds}).

\subsubsection{Lightning flash density} \label{ssec:fld}
To estimate the total number of flashes, $n_{\rm tot,fl}$ (Eq. \ref{eq:dan3}), that could be expected on a brown dwarf one has to estimate lightning occurrence rates on the object. For this, I use five different flash densities, $\rho_{\rm fl}$, from Chapter \ref{chap:stat}, Tables \ref{table:vol} and \ref{table:plan}:  
\begin{itemize}
\setlength\itemsep{0.5em}
\item[i.] \underline{Global Earth flash density:} $2.29 \times 10^{-4}$ flashes km$^{-2}$ h$^{-1}$. Earth is the most well studied planet ever known. Lightning statistics are the best explored here. However, Earth as a global environment is very different from a brown dwarf, which may result in very different lightning occurrence in a brown dwarf atmosphere.

\item[ii.] \underline{Global Jovian flash density (New Horizons data):} $1.43 \times 10^{-6}$ flashes km$^{-2}$ h$^{-1}$. From the Solar System planets, Jupiter is the closest in size and composition to a brown dwarf. Lightning data from Jupiter, however, is very incomplete, surveys are not long-term and do not cover the whole surface of the planet (Chapter \ref{chap:stat}). On the other hand, only the most energetic lightning flashes would be detectable from any object, just like on Jupiter, therefore flash densities obtained for this planet may resemble a lower estimate of brown dwarf flash densities.

\item[iii.] \underline{Global Saturnian flash density (Cassini data, 2010/11 storm):} $5.09 \times 10^{-6}$ flashes km$^{-2}$ h$^{-1}$. Similarly to Jupiter, Saturn provides a lower limit of flash density that can be expected from an extrasolar planetary object with similar atmospheric properties to Saturn's. Therefore, though the data are incomplete due to the nature of surveys, it is reasonable to apply Saturnian flash densities to estimate lightning activity on a brown dwarf.

\item[iv.] \underline{Eyjafjallaj\"okull eruption, (2010 Apr $14-19$):} 0.1 flashes km$^{-2}$ h$^{-1}$. Volcano plumes may best resemble brown dwarf dust clouds \citep{helling2008b}. Explosive eruptions like the 2010 Eyjafjallaj\"okull one show intense volcanic activity with flash densities from less than one up to several thousand flashes km$^{-2}$ h$^{-1}$. 

\item[v.] \underline{Mt Redoubt eruption, (2009 Mar 29):} 2000 flashes km$^{-2}$ h$^{-1}$. During its Phase 1 eruption Mt Redoubt showed an extremely large lightning activity, which will serve as an upper limit in our calculations.
\end{itemize}

\subsubsection{Lightning optical power output}
To obtain the optical flux of one lightning flash, $I_{\rm opt,fl}$, we require multiple components (Eq. \ref{eq:dan2}): the distance, $d$, of the brown dwarf system (see above), the effective frequency, $f_{\rm eff}$, of the filter that is used for the observations, and the optical power of a lightning flash. Here, I discuss the optical power, which depends on the radiating properties of the lightning flash. I use two approaches to estimate the optical power output of lightning: 

\noindent A) \citet{bailey2014} estimated the total energy dissipating from lightning discharges in gas giant and brown dwarf atmospheres. From their figure 11, I estimate the largest total energy released from lightning on a brown dwarf to be on the order of $10^{12}$ J. The lightning energy radiated into optical wavelength is around 1\% on Earth \citep{volland1984}. To convert the energy to power, I use \textsc{Power = Energy/time}, and take into account the duration of the discharge, $\tau_{\rm fl}$. However, this is not a unique property of lightning either. I use two examples: 
\begin{itemize}
\item[a)] $\tau_{\rm fl} = 10^{-4}$ s, the average duration of Earth lightning stroke \citep{volland1984}
\item[b)] $\tau_{\rm fl} = 0.3$ s, the duration of the slowest flashes observed on Saturn \citep{zarka2004}. 
\end{itemize}

\noindent B) In the second approach, I use the optical power observed by the \textit{Cassini} spacecraft on Saturn, which was also used to calculate the optical emission of a hypothetical lightning-storm on HAT-P-11b in Chapter \ref{chap:hatp11b}. \citet{dyudina2013} lists the survey time, 1.9 s, the total measured optical power of a storm during this survey, $1.2 \times 10^{10}$ W, and the optical flash rate of this storm, 5 s$^{-1}$. From this information the optical power of a single lightning flash is estimated to be $1.26 \times 10^9$ W.. 

In summary, I use the following optical powers to calculate the flux of one lightning flash:
\begin{itemize}
\item[A)] \underline{From \citet{bailey2014}}:
	\begin{itemize}
	\item[a)] $\tau_{\rm fl} = 10^{-4}$ s: $P_{\rm opt,fl} = 10^{15}$ W.
	\item[b)] $\tau_{\rm fl} = 0.3$ s: $P_{\rm opt,fl} = 3.33 \times 10^{11}$ W.
	\end{itemize}

\item[B)] \underline{From \citet{dyudina2013}}: $P_{\rm opt,fl} = 1.26 \times 10^9$ W.
\end{itemize}

%Table 1 - Filter information
\begin{table*}  
 \begin{center}
\begin{threeparttable}
 \small
 \caption{Parameters of the \textit{I}-, \textit{V}-, and \textit{U}-band Johnson-Cousins System filters.}
  \begin{tabular}{@{}lllll@{}}	
	 & \textit{I} & \textit{V} & \textit{U} & \\
	\hline
	$\lambda_{\rm eff}$ [nm] & 806 & 551 & 365 & \citet[][table 2.1]{binney1998} \\
	$f_{\rm eff}$ [Hz] & $3.72 \times 10^{14}$ & $5.44 \times 10^{14}$ & $8.21 \times 10^{14}$ & $f_{\rm eff} = \frac{c}{\lambda_{\rm eff}}$\tnote{(1)} \\
	$m_{\rm zp}$ & 0.443 & 0.008 & 0.79 & \citet[][table 3]{bessell1990} \\
	$F_{\rm zp}$ [Jy] & 2416 & 3636 & 1790 & \citet{bessell1998}\tnote{(2)} \\
	\hline
   \label{table:1}
  \end{tabular}
   \begin{tablenotes}
	\item[(1)] c = seed of light
	\item[(2)] http://www.astronomy.ohio-state.edu/$\sim$martini/usefuldata.html
   \end{tablenotes}
  \end{threeparttable}
 \end{center}
\vspace{0.8cm}
\end{table*}
%%Table 

The above estimates of the optical power may be different for different wavelengths. The estimate from \citet{bailey2014} assumes that 1\% of the total energy is radiated into optical wavelengths. However, it does not take into account the different amount of energy radiating into e.g. the \textit{I}, \textit{V}, and \textit{U} bands, which are the used filters on the Danish telescope. Similarly, \textit{Cassini} detected optical flashes with a filter centred on the $H_{\alpha}$ line at 656.28 nm \citep{dyudina2013}. In this work, I assume that the power radiated by lightning into all parts of the optical spectrum is the same.

\subsubsection{Effective frequency of observing filter}
The observed optical flux of a lightning flash depends on the effective frequency, $f_{\rm eff}$, of the filter that is used for observations. The Danish telescope uses standard Johnson-Cousins filters in the \textit{I}, \textit{V}, and \textit{U} bands. I obtain $f_{\rm eff}$ from the effective wavelength, $\lambda_{\rm eff}$ \citep{binney1998}, %wiki: https://en.wikipedia.org/wiki/Photometric_system
by dividing the speed of light with $\lambda_{\rm eff}$. $\lambda_{\rm eff}$ and $f_{\rm eff}$ for the three filters are listed in Table \ref{table:1}

\subsection{Total optical flux of lightning and apparent magnitudes}
Substituting Eq. \ref{eq:dan3} into Eq. \ref{eq:dan1}, one finds that the observational times, $\tau_{\rm obs}$, cancel out, therefore $I_{\rm obs}$ only depends on the duration of the lightning discharge, $\tau_{\rm fl}$. I use the two, previously introduced values in order to calculate the total optical flux according to Eq.\ref{eq:dan1}: a) $\tau_{\rm fl} = 10^{-4}$ s, and b) $\tau_{\rm fl} = 0.3$ s. 

The total optical flux of lightning can be converted to magnitudes or can be compared to the flux of the brown dwarf. The Pogson's Formula (e.g Eq. \ref{eq:dan4}) gives the relation between the apparent magnitude of two objects and their fluxes. To convert the flux of a known object to apparent magnitudes, a zero-point scale has to be defined, which, for the Standard Filters, is based on the apparent magnitude of Vega in different wavelengths \citep{bessell1990}.

\begin{equation} \label{eq:dan4}
m_1 = -2.5 \log_{10}{\frac{F_1}{F_{\rm zp}}} + m_{\rm zp},
\end{equation}

\noindent where $m_1$ is the apparent magnitude of the object, $m_{\rm zp}$ is the magnitude zero-point from \citet{bessell1990}, $F_1$ is the detected flux of the object, and $F_{\rm zp}$ is the flux zero-point from \citet{bessell1998}. %or http://www.astronomy.ohio-state.edu/~martini/usefuldata.html
$m_{\rm zp}$ and $F_{\rm zp}$ for the three filters are listed in Table \ref{table:1}. In our case $m_1$ and $F_1$ are the apparent magnitude and obtained total flux of lightning, respectively.

The optical flux obtained this way will be the one reaching Earth. To get a better estimate, it can be corrected for the telescope observing efficiency in different wavelengths. Here, however I do not consider this effect and only give an upper limit for the fluxes.

%__________________________________________________________________
%__________________________________________________________________
\section{Results and Discussion}  \label{sec:res}

%Table 2 - Cases
\begin{table}  
 \begin{center}
 \small
 \caption{Cases to obtain the optical flux of a single lightning flash (Eq. \ref{eq:dan2}).}
  \begin{tabular}{@{}lll@{}}	
	Case & Power [W] & Filter \\
	\hline
	1 & $10^{15}$ & I \\
	2 & $10^{15}$ & V \\
	3 & $10^{15}$ & U \\
	4 & $3.33 \times 10^{11}$ & I \\
	5 & $3.33 \times 10^{11}$ & V \\
	6 & $3.33 \times 10^{11}$ & U \\
	7 & $1.26 \times 10^9$ & I \\
	8 & $1.26 \times 10^9$ & V \\
	9 & $1.26 \times 10^9$ & U \\
	\hline
   \label{table:case}
  \end{tabular}
 \end{center}
\end{table}
%%Table 

%Table 3 - Results - optical flux of a lightning flash
\begin{table}  
 \begin{center}
 \small
 \caption{Optical flux of a single lightning flash, $I_{\rm opt,fl}$ [Jy], for the nine cases (Table \ref{table:case}) and the three brown dwarfs (Eq. \ref{eq:dan2}).}
  \begin{tabular}{@{}llll@{}}	
	Case & Luhman-16 & SCR 1845-6357 & $\epsilon$ Indi \\
	\hline
	1 & $5.62 \times 10^{-9}$ & $1.51 \times 10^{-9}$ & $1.71 \times 10^{-9}$\\
	2 & $3.84 \times 10^{-9}$ & $1.03 \times 10^{-9}$ & $1.17 \times 10^{-9}$\\
	3 & $2.54 \times 10^{-9}$ & $6.82 \times 10^{-10}$ & $7.74 \times 10^{-10}$\\
	4 & $1.87 \times 10^{-12}$ & $5.02 \times 10^{-13}$ & $5.70 \times 10^{-13}$\\
	5 & $1.28 \times 10^{-12}$ & $3.43 \times 10^{-13}$ & $3.89 \times 10^{-13}$\\
	6 & $8.48 \times 10^{-13}$ & $2.27 \times 10^{-13}$ & $2.58 \times 10^{-13}$\\
	7 & $7.09 \times 10^{-15}$ & $1.90 \times 10^{-15}$ & $2.16 \times 10^{-15}$\\
	8 & $4.85 \times 10^{-15}$ & $1.30 \times 10^{-15}$ & $1.48 \times 10^{-15}$\\
	9 & $3.21 \times 10^{-15}$ & $8.62 \times 10^{-16}$ & $9.77 \times 10^{-16}$\\
	\hline
   \label{table:res1}
  \end{tabular}
 \end{center}
\end{table}
%%Table 

%Table 4a - Results - total optical flux
\begin{table*}  
 \begin{center}
 \small
 \caption{Total optical flux of lightning, $I_{\rm obs}$ [Jy], for $\epsilon$ Indi, SCR 1845-6357, and Luhman-16 (from top to bottom). The nine power-filter cases (Table \ref{table:case}) are listed in the first column, while the flash density (Sect. \ref{ssec:fld}) categories are in the first line of each table. a: $\tau_{\rm fl} = 10^{-4}$ s, b: $\tau_{\rm fl} = 0.3$ s.} 
  \begin{tabular}{lccccc}	
	\multicolumn{6}{c}{$\epsilon$ Indi} \\
	Case/$n_{\rm tot}$ & i & ii & iii & iv & v \\
	\hline
	1 & $3.35 \times 10^{-10}$ & $2.04 \times 10^{-12}$ & $7.42 \times 10^{-12}$ & $1.46 \times 10^{-7}$ & $2.9 \times 10^{-3}$ \\
	2 & $2.29 \times 10^{-10}$ & $1.39 \times 10^{-12}$ & $5.07 \times 10^{-12}$ & $9.96 \times 10^{-8}$ & $1.99 \times 10^{-3}$ \\
	3 & $1.52 \times 10^{-10}$ & $9.24 \times 10^{-13}$ & $3.36 \times 10^{-12}$ & $6.60 \times 10^{-8}$ & $1.32 \times 10^{-3}$ \\
	4 & $3.35 \times 10^{-10}$ & $2.04 \times 10^{-12}$ & $7.42 \times 10^{-12}$ & $1.46 \times 10^{-7}$ & $2.92 \times 10^{-3}$ \\
	5 & $2.29 \times 10^{-10}$ & $1.39 \times 10^{-12}$ & $5.07 \times 10^{-12}$ & $9.96 \times 10^{-8}$ & $1.99 \times 10^{-3}$ \\
	6 & $1.52 \times 10^{-10}$ & $9.24 \times 10^{-13}$ & $3.36 \times 10^{-12}$ & $6.60 \times 10^{-8}$ & $1.32 \times 10^{-3}$ \\
	7a & $4.23 \times 10^{-16}$ & $2.58 \times 10^{-18}$ & $9.37 \times 10^{-18}$ & $1.84 \times 10^{-13}$ & $3.68 \times 10^{-9}$ \\
	7b & $1.27 \times 10^{-12}$ & $7.73 \times 10^{-15}$ & $2.81 \times 10^{-14}$ & $5.52 \times 10^{-10}$ & $1.10 \times 10^{-5}$ \\
	8a & $2.89 \times 10^{-16}$ & $1.76 \times 10^{-18}$ & $6.41 \times 10^{-18}$ & $1.25 \times 10^{-13}$ & $2.52 \times 10^{-9}$ \\
	8b & $8.68 \times 10^{-13}$ & $5.29 \times 10^{-15}$ & $1.92 \times 10^{-14}$ & $3.78 \times 10^{-10}$ & $7.55 \times 10^{-6}$ \\
	9a & $1.92 \times 10^{-16}$ & $1.17 \times 10^{-18}$ & $4.24 \times 10^{-18}$ & $8.34 \times 10^{-14}$ & $1.67 \times 10^{-9}$ \\
	9b & $5.75 \times 10^{-13}$ & $3.50 \times 10^{-15}$ & $1.27 \times 10^{-14}$ & $2.50 \times 10^{-10}$ & $5.00 \times 10^{-6}$ \\
	\hline
   \label{table:eindi1}
  \end{tabular}
  \begin{tabular}{lccccc}	
	\multicolumn{6}{c}{SCR 1845-6357} \\
	Case/$n_{\rm tot}$ & i & ii & iii & iv & v \\
	\hline
	1 & $2.96 \times 10^{-10}$ & $1.80 \times 10^{-12}$ & $6.54 \times 10^{-12}$ & $1.28 \times 10^{-7}$ & $2.57 \times 10^{-3}$ \\
	2 & $2.02 \times 10^{-10}$ & $1.23 \times 10^{-12}$ & $4.47 \times 10^{-12}$ & $8.79 \times 10^{-8}$ & $1.75 \times 10^{-3}$ \\
	3 & $1.34 \times 10^{-10}$ & $8.15 \times 10^{-13}$ & $2.96 \times 10^{-12}$ & $5.82 \times 10^{-8}$ & $1.16 \times 10^{-3}$ \\
	4 & $2.96 \times 10^{-10}$ & $1.80 \times 10^{-12}$ & $6.54 \times 10^{-12}$ & $1.29 \times 10^{-7}$ & $2.57 \times 10^{-3}$ \\
	5 & $2.02 \times 10^{-10}$ & $1.23 \times 10^{-12}$ & $4.47 \times 10^{-12}$ & $8.79 \times 10^{-8}$ & $1.76 \times 10^{-3}$ \\
	6 & $1.34 \times 10^{-10}$ & $8.15 \times 10^{-13}$ & $2.96 \times 10^{-12}$ & $5.82 \times 10^{-8}$ & $1.16 \times 10^{-3}$ \\
	7a & $3.73 \times 10^{-16}$ & $2.27 \times 10^{-18}$ & $8.27 \times 10^{-18}$ & $1.62 \times 10^{-13}$ & $3.25 \times 10^{-9}$ \\
	7b & $1.12 \times 10^{-12}$ & $6.82 \times 10^{-15}$ & $2.48 \times 10^{-14}$ & $4.87 \times 10^{-10}$ & $9.74 \times 10^{-6}$ \\
	8a & $2.55 \times 10^{-16}$ & $1.55 \times 10^{-18}$ & $5.65 \times 10^{-18}$ & $1.11 \times 10^{-13}$ & $2.22 \times 10^{-9}$  \\
	8b & $7.66 \times 10^{-13}$ & $4.66 \times 10^{-15}$ & $1.70 \times 10^{-14}$ & $3.33 \times 10^{-10}$ & $6.66 \times 10^{-6}$ \\
	9a & $1.69 \times 10^{-16}$ & $1.03 \times 10^{-18}$ & $3.74 \times 10^{-18}$ & $7.35 \times 10^{-14}$ & $1.47 \times 10^{-9}$ \\
	9b & $5.07 \times 10^{-13}$ & $3.09 \times 10^{-15}$ & $1.12 \times 10^{-14}$ & $2.21 \times 10^{-10}$ & $4.41 \times 10^{-6}$ \\
	\hline
   \label{table:scr}
  \end{tabular}
  \begin{tabular}{lccccc}	
	\multicolumn{6}{c}{Luhman-16} \\
	Case/$n_{\rm tot}$ & i & ii & iii & iv & v \\
	\hline
	1 & $1.10 \times 10^{-9}$ & $6.71 \times 10^{-12}$ & $2.43 \times 10^{-11}$ & $4.79 \times 10^{-7}$ & $9.58 \times 10^{-3}$ \\
	2 & $7.53 \times 10^{-10}$ & $4.59 \times 10^{-12}$ & $1.67 \times 10^{-11}$  & $3.28 \times 10^{-7}$ & $6.55 \times 10^{-3}$ \\
	3 & $4.99 \times 10^{-10}$ & $3.04 \times 10^{-12}$ & $1.10 \times 10^{-11}$ & $2.17 \times 10^{-7}$ & $4.34 \times 10^{-3}$ \\
	4 & $1.10 \times 10^{-9}$ & $6.71 \times 10^{-12}$ & $2.44 \times 10^{-11}$ & $4.79 \times 10^{-7}$ & $9.58 \times 10^{-3}$ \\
	5 & $7.53 \times 10^{-10}$ & $4.59 \times 10^{-12}$ & $1.67 \times 10^{-11}$ & $3.28 \times 10^{-7}$ & $6.55 \times 10^{-3}$ \\
	6 & $4.99 \times 10^{-10}$ & $3.04 \times 10^{-12}$ & $1.10 \times 10^{-11}$ & $2.17 \times 10^{-7}$ & $4.34 \times 10^{-3}$ \\
	7a & $1.39 \times 10^{-15}$ & $8.47 \times 10^{-18}$ & $3.08 \times 10^{-17}$  & $6.05 \times 10^{-13}$ & $1.21 \times 10^{-8}$ \\
	7b & $4.18 \times 10^{-12}$ & $2.54 \times 10^{-14}$ & $9.24 \times 10^{-14}$ & $1.82 \times 10^{-9}$  & $3.63 \times 10^{-5}$ \\
	8a & $9.52 \times 10^{-16}$ & $5.79 \times 10^{-18}$ & $2.11 \times 10^{-17}$ & $4.14 \times 10^{-13}$ & $8.27 \times 10^{-9}$ \\
	8b & $2.85 \times 10^{-12}$ & $1.74 \times 10^{-14}$ & $6.32 \times 10^{-14}$ & $1.24 \times 10^{-9}$ & $2.48 \times 10^{-5}$ \\
	9a & $6.30 \times 10^{-16}$ & $3.84 \times 10^{-18}$ & $1.39 \times 10^{-17}$ & $2.74 \times 10^{-13}$ & $5.48 \times 10^{-9}$ \\
	9b & $1.89 \times 10^{-12}$ & $1.15 \times 10^{-14}$ & $4.18 \times 10^{-14}$ & $8.22 \times 10^{-10}$ & $1.64 \times 10^{-5}$ \\
	\hline
   \label{table:luh}
  \end{tabular}
 \end{center}
\end{table*}
%%Table 

The parameters introduced in the previous section are combined based on which equations they are used in. This way 9 cases are studied based on the observing filter and lightning optical power combinations. The cases are marked with Arabic numerals and are listed in Table \ref{table:case}. Each case corresponds to a filter-power combination. I obtain optical fluxes for a single lightning flash in each brown dwarf atmosphere using the 9 cases and the distances of the systems (Eq. \ref{eq:dan2}). The results are shown in Table \ref{table:res1}.

As expected, the results show that the larger amount of power is released from a lightning flash, the larger the optical flux reaching us. The results also show that lightning emits most of the flux in the I band (Cases 1, 4, and 7, Table \ref{table:res1}) compared to the other bands assuming the same lightning properties. This is favourable since the Danish telescope's sensitivity in the I band is 90\%. The U band is the least promising in terms of lightning observations, since not only the telescopes sensitivity is only 20\% in this band, but lightning emits the least amount of power in U (Cases 3, 6, and 9, Table \ref{table:res1}) compared to the other bands. However, the difference in flux between the two bands is very small, within an order of magnitude. Finally, the distance will affect the resulting flux as was expected from the inverse-square law. Luhman 16 is the closest brown dwarf binary, therefore, the flux of lightning will be the strongest from its distance (Table \ref{table:res1}). However, the flux from the other two brown dwarfs will be within the same order of magnitude assuming the same physical properties for lightning. 

Table \ref{table:eindi1} lists the total optical flux of lightning that is expected from the three brown dwarf examples, for different parameter combinations. The power-filter cases, again, are shown with Arabic numerals. The roman numerals represent the flash density cases used in the study. Because of the distance dependence, I only analyse further the results for Luhman-16, the closest brown dwarf binary in the sample, which shows the most promising lightning fluxes. The optical fluxes and apparent magnitudes are listed in Tables \ref{table:luh} and \ref{table:res3}, respectively. Depending on what average optical power output we expect from lightning, what duration a discharge has, and statistically how many flashes we can expect to occur (flashes km$^{-2}$ h$^{-1}$), the obtained optical fluxes range between $9.5 \times 10^{-3}$ Jy and $3.8 \times 10^{-18}$ Jy for Luhman 16. These correspond to $\sim 14$ and $\sim 52$ apparent magnitude, respectively. The best case scenario is for an I-band observation, and assumes an optical power output of a single lightning flash $P_{\rm opt,fl} = 10^{15}$ W, a flash duration $\tau_{\rm fl} = 10^{-4}$ s, and a flash density $\rho_{\rm fl} = $ 2000 flashes km$^{-2}$ h$^{-1}$. These values, respectively, are the power and discharge duration estimated by \citet{bailey2014} for a brown dwarf, and the flash density produced by the Mt Redoubt eruption in 2009. The worst case scenario was obtained for a U band observation, and an optical power $P_{\rm opt,fl} = 10^9$ W, a discharge duration $\tau_{\rm fl} = 10^{-4}$ and a flash density  $\rho_{\rm fl} = 10^{-6}$ flashes km$^{-2}$ h$^{-1}$. These values, respectively, correspond to an average optical power of a Saturnian discharge \citep{dyudina2013}, a very fast discharge, and an average global flash density on Jupiter obtained from New Horizons data.

I tested around 140 cases with different lightning optical power output, discharge duration, and flash densities. In $\sim 20$ cases the estimated optical brightness of a lightning storm on the three brown dwarfs is roughly the same as a 13$-$16 magnitude star. In about 30 cases the brightness is between 20$-$30 magnitude, while the rest of the cases show a brightness much fainter than 30 magnitude. Compared to the apparent brightness of the brown dwarfs itself (Table \ref{table:bds}), I find that in a few cases (apparent magnitude of 13$-$16) the expected thunderstorm on the brown dwarf would significantly increase the total brightness of the object. As thunderstorms are generally transient, they could produce a transient brightness change of the brown dwarfs.

%__________________________________________________________________
%__________________________________________________________________
\section{Conclusions} \label{sec:concc}

In this chapter, I conducted a short parameter study to estimate the optical emission of lightning storms on three close-by brown dwarfs, Luhman 16, $\epsilon$ Indi, and SCR 1845-6357. The objects were chosen based on their distances and observability from La Silla, Chile. The fluxes were estimated for three standard filters, I, V, U, and assuming that the observations will be conducted by the Danish 1.54-m telescope.

In summary, I find that the parameters used here are not yet well constrained to give a proper estimate of lightning activity on brown dwarfs. The large variety of them result in a wide range of expected optical fluxes, some of which are in the observable range, with apparent magnitudes of only 13 to 16.
It has been shown that dust clouds form in brown dwarf atmospheres \citep[e.g.][]{helling2008b}. These dust clouds may be very similar to volcanic plumes on Earth, therefore the flash densities produced by volcanic eruptions can be good indicators of lightning activity on brown dwarfs. It is also in favour of the study, that the emitted optical power was obtained from \citet{bailey2014} who estimated the energy release of lightning on brown dwarfs. Nevertheless, this study will be a good guide for future lightning observability studies, which will use better constrained parameters.

%Table 5 - Results - magnitude
\begin{landscape}
\begin{table*}  
 \caption{Apparent magnitude of lightning for $\epsilon$ Indi, SCR 1845-6357, and Luhman-16 (from top to bottom). The nine power-filter cases (Table \ref{table:case}) are listed in the first column, while the flash density categories (Sect. \ref{ssec:fld}) are in the first line of each table. a: $\tau_{\rm fl} = 10^{-4}$ s, b: $\tau_{\rm fl} = 0.3$ s.} 
  \begin{tabular}{cccccc:ccccc:ccccc}	
	 & \multicolumn{5}{c:}{Luhman-16} & \multicolumn{5}{c:}{SCR 1845-6357} & \multicolumn{5}{c}{$\epsilon$ Indi} \\ 
	Case/$n_{\rm tot}$ & i & ii & iii & iv & v & i & ii & iii & iv & v & i & ii & iii & iv & v \\
	\hline
	1 & 31.30 & 36.83 & 35.43 & 24.70 & 13.95 & 	32.72 & 38.26 & 36.86 & 26.13 & 15.37 & 	32.59 & 38.13 & 36.72 & 25.99 & 15.24 \\
	2 & 31.72 & 37.26 & 35.85 & 25.12 & 14.37 & 	33.15 & 38.68 & 37.28 & 26.55 & 15.80 & 	33.01 & 38.55 & 37.15 & 26.41 & 15.66 \\
	3 & 32.18 & 37.72 & 36.31 & 25.58 & 14.83 & 	33.60 & 39.14 & 37.74 & 27.01 & 16.26 & 	33.47 & 39.01 & 37.61 & 26.87 & 16.12 \\
	4 & 31.30 & 36.83 & 35.43 & 24.70 & 13.95 & 	32.72 & 38.26 & 36.86 & 26.13 & 15.38 & 	32.59 & 38.13 & 36.72 & 25.99 & 15.24 \\
	5 & 31.72 & 37.26 & 35.85 & 25.12 & 14.37 & 	33.15 & 38.68 & 37.28 & 26.55 & 15.80 & 	33.01 & 38.55 & 37.15 & 26.41 & 15.66 \\
	6 & 32.18 & 37.72 & 36.31 & 25.58 & 14.83 & 	33.61 & 39.14 & 37.74 & 27.01 & 16.26 & 	33.47 & 39.01 & 37.60 & 26.87 & 16.12 \\
	7a & 46.04 & 51.58 & 50.18 & 39.45 & 28.69 &	47.47 & 53.01 & 51.61 & 40.87 & 30.12 &		47.33 & 52.87 & 51.47 & 40.74 & 29.99 \\
	7b & 37.35 & 42.89 & 41.49 & 30.75 & 20.00 & 	38.78 & 44.32 & 42.91 & 32.18 & 21.43 & 	38.64 & 44.18 & 42.78 & 32.05 & 21.29 \\
	8a & 46.46 & 52.00 & 50.60 & 39.87 & 29.12 &	47.89 & 53.43 & 52.03 & 41.30 & 30.54 & 	47.76 & 53.29 & 51.89 & 41.16 & 30.41 \\
	8b & 37.77 & 43.31 & 41.91 & 31.18 & 20.42 & 	39.20 & 44.74 & 43.34 & 32.60 & 21.85 &		39.06 & 44.60 & 43.20 & 32.47 & 21.71 \\
	9a & 46.92 & 52.46 & 51.06 & 40.33 & 29.57 & 	48.35 & 53.89 & 52.49 & 41.76 & 31.00 & 	48.22 & 53.75 & 52.35 & 41.62 & 30.87 \\
	9b & 38.23 & 43.77 & 42.37 & 31.63 & 20.88 & 	39.66 & 45.20 & 43.80 & 33.06 & 22.31 & 	39.52 & 45.06 & 43.66 & 32.93 & 22.17 \\
	\hline
   \label{table:res3}
  \end{tabular}
\end{table*}
\end{landscape}
%%Table 

%% file: chapters/5_radiomodel.tex
\chapter{Modelling lightning radio energy and testing for Solar System planets} \label{chap:model}

%__________________________________________________________________
%__________________________________________________________________
\section{Introduction} \label{sec:int}

Observations and models suggest that the conditions to develop lightning are present in extrasolar planetary and brown dwarf atmospheres. However, we do not know whether lightning on these objects is similar to or very different from what is known from the Solar System. In this chapter, I estimate the energy radiated from lightning discharges and the total power emitted at radio frequencies in order to study possible differences and similarities of extrasolar and Solar System lightning.
 
The current chapter explores several questions related to lightning properties in exoplanetary and brown dwarf atmospheres, such as: How does lightning radiation on extrasolar bodies compare with what is known from the Solar System? Could lightning be more energetic there than here? Would it produce signatures observable from Earth? What would be the lightning energy deposit into the atmosphere of the extrasolar body? To address these questions, I build a lightning model guided by models developed for Earth lightning. I explore the model's strengths and limitations, and whether this model can be used for making predictions of observations. When modelling extraterrestrial lightning, it is a common practice to utilize parameterization tested for Earth lightning \citep[models partly using terrestrial parameters are, e.g.,][and references therein]{farrell1999,farrell2007,lammer2001,bailey2014,yair2012}. \citet{farrell2007} showed that just by changing one parameter, the duration of the discharge, Saturnian lightning energies may well be in the non-superbolt ($<10^{11}-10^{12}$ J), rather average Earth-like energy range. Although the optical detection of Saturnian lightning \citep{dyudina2010} has confirmed its very high optical energy-release, the results of \citet{farrell2007} show the importance of parameter studies when observations do not constrain physical properties well. Such the case is with exoplanetary lightning studies. The model I present here is based on Earth return stroke models (Chapter \ref{chap:ligform}). It explores the parameter space that affects the outcome, the energy dissipated from lightning discharges and the power radiated at radio frequencies. 

The chapter is organized as follows. Section \ref{sec:model} describes the general dipole model of lightning that I applied for extrasolar atmospheres. In Sect. \ref{sec:param}, I present the computational approach that uses the previously described model, explaining the input and output parameters. The performance test of the model is in Sect. \ref{sec:val}. The results and their discussion are in Sect. \ref{sec:resdis}. I finish the chapter with the conclusions in Sect. \ref{sec:conc}.

%__________________________________________________________________
%__________________________________________________________________
\section{Model Description} \label{sec:model}

In this section, I describe the general modelling ansatz that is applied to explore lightning energetics in exoplanetary and brown dwarf atmospheres. The purpose of the model is to determine the total radiation energy released from a single lightning flash, and explore the properties of the emitted power spectrum. The radiation energy will determine the power emitted at certain frequencies and the radio flux observable from lightning discharges. Once the energy of a single lightning discharge is determined, one can estimate the total energy affecting the atmosphere at a certain time interval in an area or volume, by applying lightning flash densities to the study (e.g. Chapter \ref{chap:stat}). 

A lightning discharge is a complex phenomenon made of several parts (Chapter \ref{chap:ligform}), which are modelled separately \citep[e.g.][]{gordillo2010,ebert2010}. The lightning channel is often tortuous, built up of several segments and branches \citep{levine1978}. \citet{moss2006} used a Monte Carlo model to obtain properties of lightning streamers and found that 10$E_k$\footnote{$E_k$ is the conventional breakdown threshold field \citep{helling2013}.} fields, which are produced at streamer tips, can accelerate part of the low energy electrons emitted from the streamers to energies high enough to function as seed electrons for a thermal runaway electron avalanche. \citet{babich2015} presented a model of the local electric field increase in front of a lightning stepped leader, and showed that the front electrons are capable of initiating relativistic runaway avalanches. They also found that in an inhomogeneous ionization environment, the leader tip enhances the production of runaway electrons compared to homogenous environments. \citet{gordillo2010} modelled the conductivity in a sprite streamer channel (Chapter \ref{chap:ligsig}, Sect. \ref{sec:tle}) and found that the conductivity in the atmosphere lasts for several minutes after the discharge. They suggested that the effects of this long-lasting conductivity have to be taken into account when studying the influence of sprites on the global electric circuit. 

In this chapter, I consider a lightning return stroke model. There are several approaches that can be applied to model the return stroke. My approach is based on a simple dipole radiation model, not taking into account channel tortuosity and branching \citep[e.g.][]{bruce1941, rakov2003}. I use the "engineering" modelling approach \citep[for an extended explanation of the different models see][section 12.2]{rakov2003}. The characteristics of these models include the low number of parameters, and specification of the channel current in order to achieve an agreement between the electromagnetic field predicted by the model and observed at distances ($r$) up to several hundreds of km \citep{rakov2003}. 
My model follows the steps below in order to estimate the radiative energy dissipated from a single lightning discharge: 
	\begin{itemize}
	\item[1.] Consider an electric current, $i(t)$, based on the amount of charges, $Q$, that accumulate in the current channel (Sect. \ref{sec:current}).
	\item[2.] Obtain the electric field, $E(t)$, produced by the dipole moment building up in the current channel with a current of $i(t)$ (Sect. \ref{sec:efield}).
	\item[3.] Calculate the frequency, $E(f)$, and power, $P(f)$, spectra of the electric field, $E(t)$ (Sect. \ref{sec:freqsp}). 
The power spectrum possesses properties important for characterizing lightning radio emission: $f_0$ is the peak frequency, the frequency at which the largest amount of power is released; and $n$ is the negative slope of the power spectrum at high frequencies ($f > f_0$), which carries information on the amount of power released at these frequencies. 
	\item[4.] From the power spectrum one can obtain the radiated power at different frequencies, $P'(f)$, and from $P'(f)$ estimate the total radiated discharge energy, $W_{\rm rad}$ (Sect. \ref{sec:disen}).
	\end{itemize}

Individual parts of the lightning model were used in previously published works \citep[e.g.][]{bruce1941, farrell1999, farrell2007}. For example, \citet{bruce1941} modelled the current in a lightning return stroke channel on Earth. \citet{farrell1999} applied a modified version of the model of \citet{bruce1941} to a Jovian lightning flash, in order to model the radio waveform observed by the \textit{Galileo} probe. \citet{farrell2007} estimated the energy released from Saturnian lightning discharges using the radiated power measured by the \textit{Cassini} spacecraft. I combine these individually tested parts into a modelling approach, which I use to study lightning in exoplanetary and brown dwarf atmospheres. 

%__________________________________________________________________
\subsection{Dipole radiation of lightning} \label{sec:dipole}
 
A common way to model lightning discharges in the Solar System is to assume that lightning radiates as a dipole \citep[e.g.][]{bruce1941}. The parameters that determine the dipole radiation are the length of the dipole, or the characteristic length of charge separation, $h$; the charges that run through the dipole channel, $Q(t)$; the characteristic time of the duration of the discharge, $\tau$; and finally the velocity with which the discharge event occurs, $\v v_0$. In case of return stroke modelling, the velocity, $\v v_0$ will be the velocity of the return stroke itself.

The current, $i(t)$, in the discharge channel is determined by the charges, $Q(t)$, accumulating there:

\begin{equation} \label{eq:charge}
Q(t) = \int_0^t{i(t') dt'}.
\end{equation}

\noindent The charges also determine the electric dipole moment: 

\begin{equation} \label{eq:mom}
M(t) = h Q(t), %2hQ(t), %= 2h\int_0^t{i(t') dt'},
\end{equation}

\noindent where $h$ is the separation of the charged regions \citep{farrell1999} and $Q(t)$ is given by Eq. \ref{eq:charge}. Once the extension of the discharge and the velocity of the event is known, one can calculate the duration of the discharge, $\tau$:  

\begin{equation} \label{eq:tau1}
\tau = \frac{h}{\v v_0}. 
\end{equation}

\noindent At large source-observer distances ($r \geq 50$ km) $\tau$ can be estimated with \citep{volland1984}:

\begin{equation} \label{eq:tau}
\tau = \frac{2 \pi}{\sqrt{\alpha \beta}}, 
\end{equation}

\noindent where $\alpha$ and $\beta$ are frequency-type constants, $\alpha^{-1}$ representing the overall duration of the current flow, while $\beta^{-1}$ representing the rise time of the current wave \citep[$\alpha < \beta$;][]{dubrovin2014}. Because the radio emission of lightning is the result of the acceleration of electrons, the duration of lightning discharge (and consequently the extension of the main lightning channel, with no branches) will determine the frequency (and the wavelength) where the radiated power reaches its peak ($f_0$). Once $\tau$ is determined, I estimate the peak frequency \citep{zarka2004}:

\begin{equation} \label{eq:taufr}
f_0 = \frac{1}{\tau}.
\end{equation}

%__________________________________________________________________
\subsection{Current wave function} \label{sec:current}

%%% Current function %%%

%Table 3 
\begin{table}  
 \centering
 \small
 \caption{Bi-exponential current function parameters, $\alpha$ and $\beta$ (Eq. \ref{eq:1}) as given in the literature for Earth and Jupiter. The last column associates the value-pairs with the plotted current functions in Fig. \ref{fig:compi}.}
  \begin{tabular}{@{}llllc@{}}	
	\hline
	$\alpha$ [$1/s$] & $\beta$ [$1/s$] & Reference & Planet & Fig. \ref{fig:compi} \\ 
	\hline \hline
	$4.4 \times 10^4$ & $4.6 \times 10^5$ & \citet{bruce1941} & Earth & $[1]$ \\
	$2 \times 10^4$ & $2 \times 10^5$ & \citet{levine1978} & Earth & $[2]$ \\
	$1.5 \times 10^3$ & $1.75 \times 10^3$ & \citet{farrell1999} & Jupiter & $[3]$ \\
	\hline
  \label{table:3}
  \end{tabular}
\end{table}
%Table 

\begin{figure}
  \centering
  \includegraphics[scale=0.7]{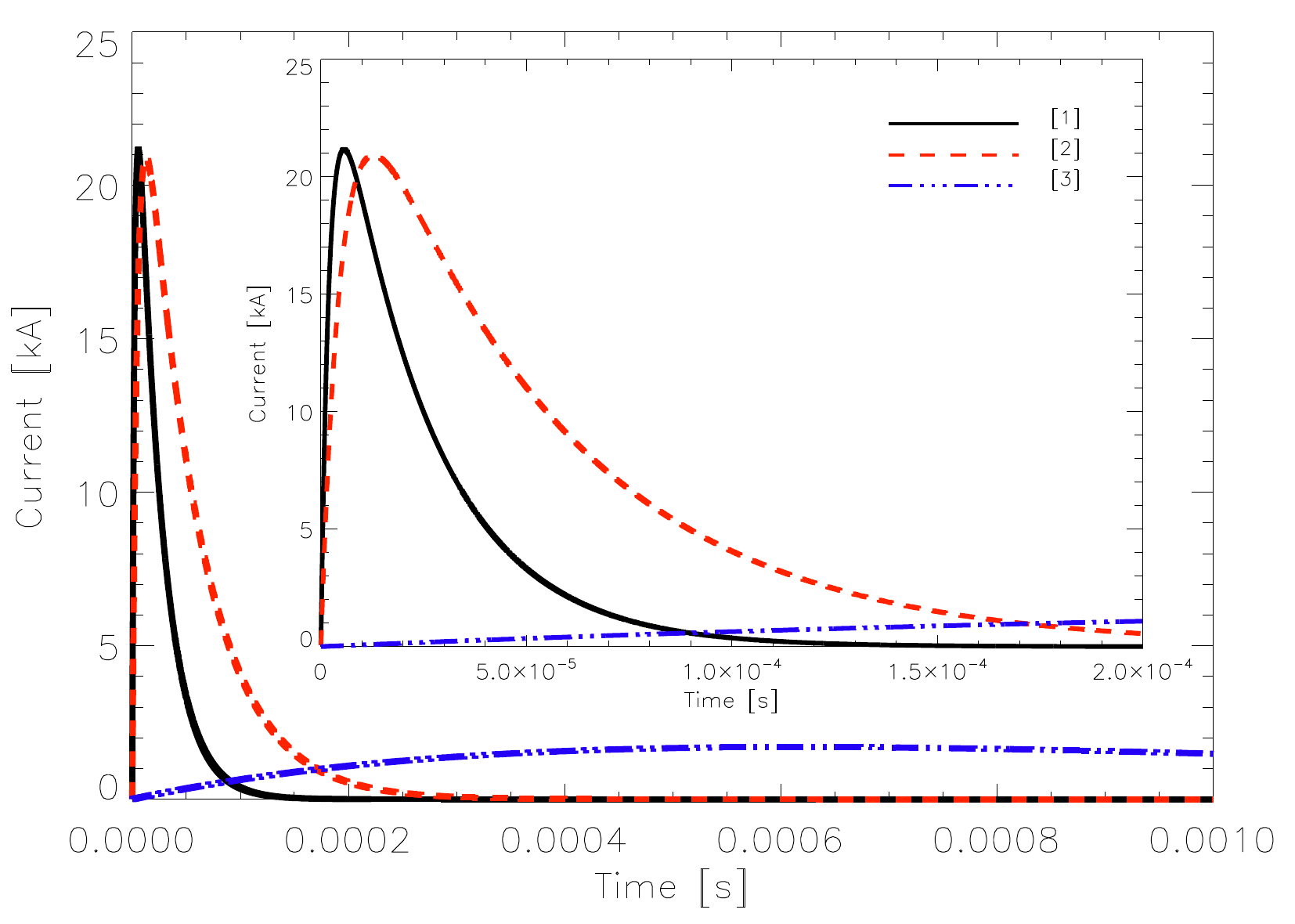}
  \caption{Bi-exponential current function, $i(t)$ (Eq. \ref{eq:1}), for different combinations of $\alpha$ and $\beta$ parameters, [1], [2], and [3], as listed in Table \ref{table:3}. The larger $\alpha$ the longer the discharge event is, while the larger $\beta$ the slower the current reaching its peak. The current peak, $i_0$, was set to 30 kA in all cases.}
  \label{fig:compi}
	\vspace{0.8cm}
\end{figure}

%Table - compare H and bi-exp 
\begin{table*}  
\resizebox{\columnwidth}{!}{
\begin{threeparttable}
 \centering
 \small
 \caption{Parameters for the bi-exponential (Eq. \ref{eq:1}) and the Heidler (Eq. \ref{eq:3}) current functions shown in Fig. \ref{fig:ibh}. The individual cases are annotated by $[1]-[6]$.}
  \begin{tabular}{@{}lllllll@{}}
	\hline
	\multicolumn{7}{c}{Bi-exponential function} \\
	\hline
	Figure \ref{fig:ibh} & \multicolumn{2}{l}{$i_0$ [kA]} & $\alpha$ [$1/s$] & \multicolumn{2}{l}{$\beta$ [$1/s$]} & Comments \\
	\hline \hline
	$[1]$ & \multicolumn{2}{l}{\vtop{\hbox{\strut 30.0}\hbox{\strut \citep{farrell1999}}}} & \vtop{\hbox{\strut $4.4 \times 10^4$}\hbox{\strut \citep{bruce1941}}} & \multicolumn{2}{l}{\vtop{\hbox{\strut $4.6 \times 10^5$}\hbox{\strut \citep{bruce1941}}}} & \vtop{\hbox{\strut \citet{bruce1941} used $i_0=20$kA}\hbox{\strut (see paragraph above their sect. (3.3))}} \\
	$[2]$ & \multicolumn{2}{l}{30.0} & $4.4 \times 10^4$ & \multicolumn{2}{l}{$4.6 \times 10^5$} & \dittoclosing \\
	$[3]$ & \multicolumn{2}{l}{30.0} & $4.4 \times 10^4$ & \multicolumn{2}{l}{$4.6 \times 10^5$} & \dittoclosing \\
	$[4]$ & \multicolumn{2}{l}{30.0} & $4.4 \times 10^4$ & \multicolumn{2}{l}{$4.6 \times 10^5$} & \dittoclosing \\
	$[5]$ & \multicolumn{2}{l}{30.0} & 6970.88 & \multicolumn{2}{l}{2202643.3} & $\alpha$ and $\beta$ are given by Eq. \ref{eq:ab} \\
	$[6]$ & \multicolumn{2}{l}{30.0} & 6970.88 & \multicolumn{2}{l}{2202643.3} & $\alpha$ and $\beta$ are given by Eq. \ref{eq:ab} \\

	\hline	\hline
	\multicolumn{7}{c}{Heidler function}	\\
	\hline
	Figure \ref{fig:ibh} & $i_0$ [kA] & $\eta$ & $\tau_1$ [$\mu$s] & $\tau_2$ [$\mu$s] & $m$ & Comments \\ 
	\hline \hline
	$[1]$ & 30.0 & 0.73 & 0.3 & 0.6 & 2 & \citet[][table 1]{diendorfer1990} \\ 
	$[2]$ & 30.0 & 0.37 & 0.3 & 0.6 & 2 & $\eta$ calculated by Eq. \ref{eq:eta} \\
	$[3]$ & 30.0 & 0.92 & 0.454 & 143.0 & 2 & \vtop{\hbox{\strut $\eta$ calculated by Eq. \ref{eq:eta};}\hbox{\strut $\tau_1$ and $\tau_2$ are values for subsequent stroke;}\hbox{\strut \citet[][table 1]{heidler2002}}} \\
	$[4]$ & 50.0 & 0.92 & 0.454 & 143.0 & 2 & \vtop{\hbox{\strut $\eta$ calculated by Eq. \ref{eq:eta}; $i_0$, $\tau_1$ and $\tau_2$}\hbox{\strut from \citet[][table 1 and 2]{heidler2002}}} \\
	$[5]$ & 30.0 & 0.92 & 0.454 & 143.0 & 2 & \vtop{\hbox{\strut $\eta$ calculated by Eq. \ref{eq:eta};}\hbox{\strut $\tau_1$ and $\tau_2$ from  \citet{heidler2002}}} \\
	$[6]$ & 30.0 & 1.0 & 0.454 & 143.0 & 2 & \vtop{\hbox{\strut no correction for $i_0$ ($\eta = 1$);}\hbox{\strut $\tau_1$ and $\tau_2$ from \citet{heidler2002}}}\\
	\hline
  \label{table:ibh}
  \end{tabular}
\end{threeparttable}
}
\end{table*}
%Table

\begin{figure*}
\advance\leftskip-1.0cm
\advance\rightskip-1.0cm
 % \centering
\begin{subfigure}[b]{0.57\textwidth}
  \includegraphics[width=\columnwidth]{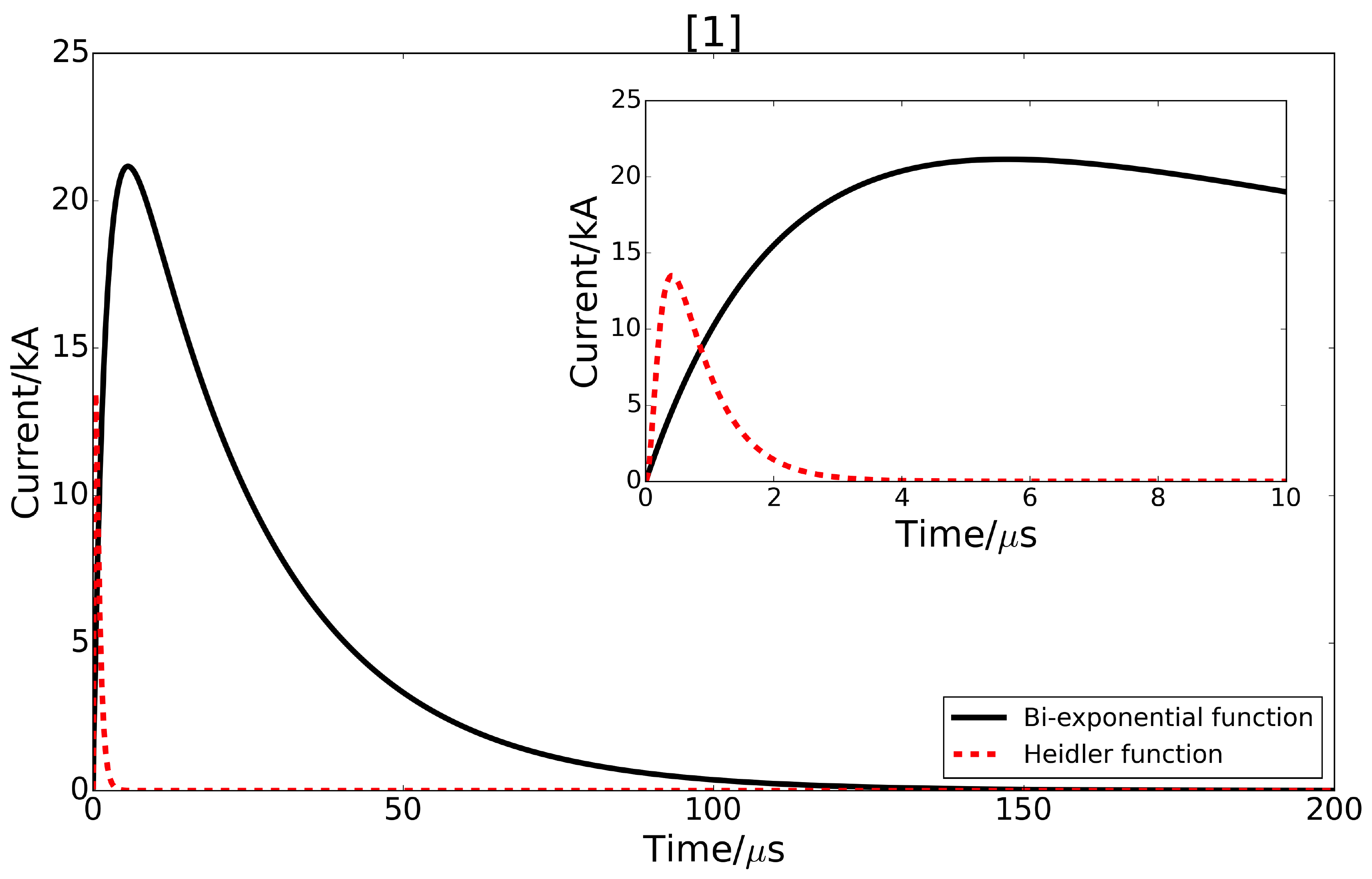}
  \includegraphics[width=\columnwidth]{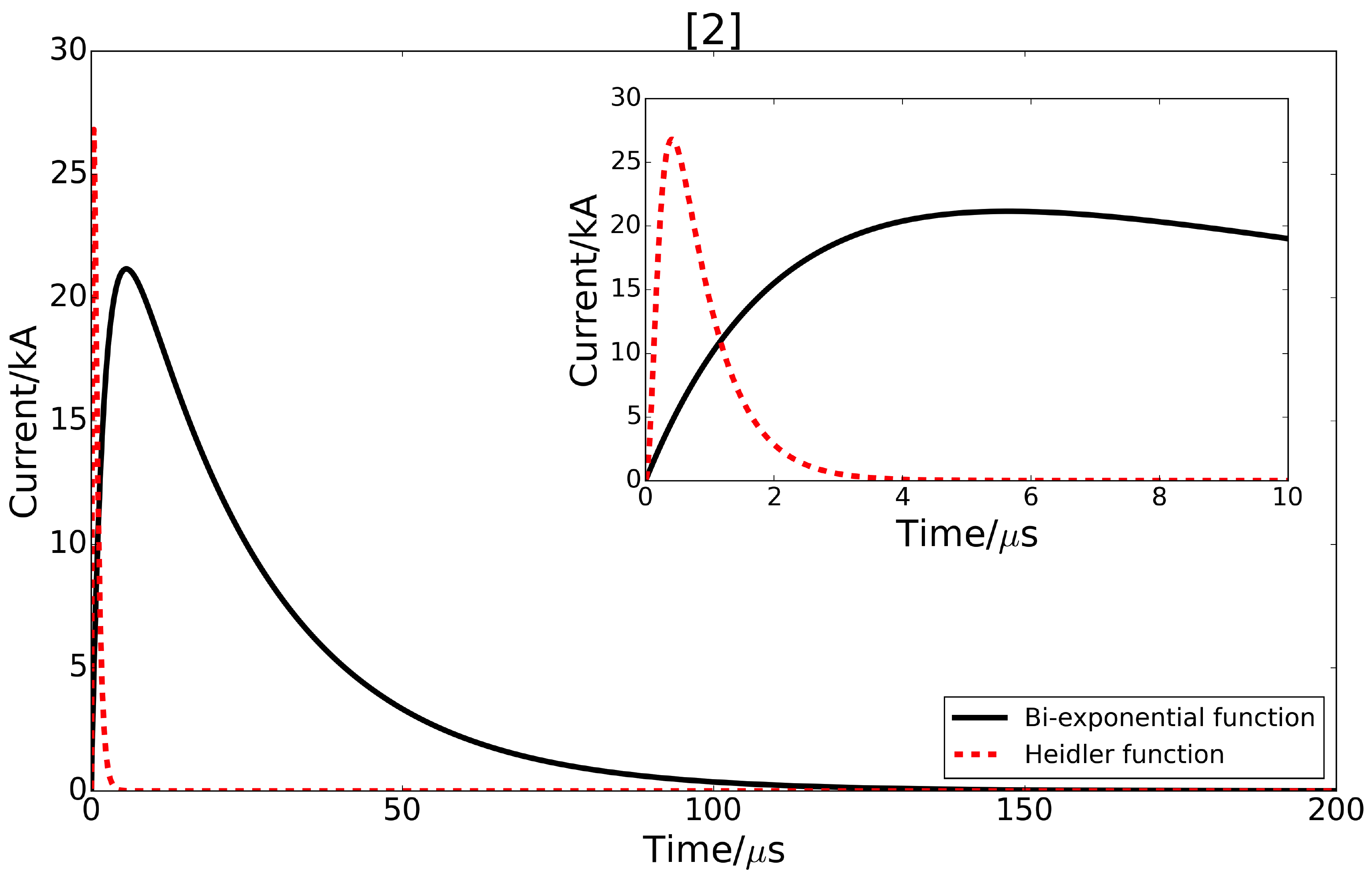}
  \includegraphics[width=\columnwidth]{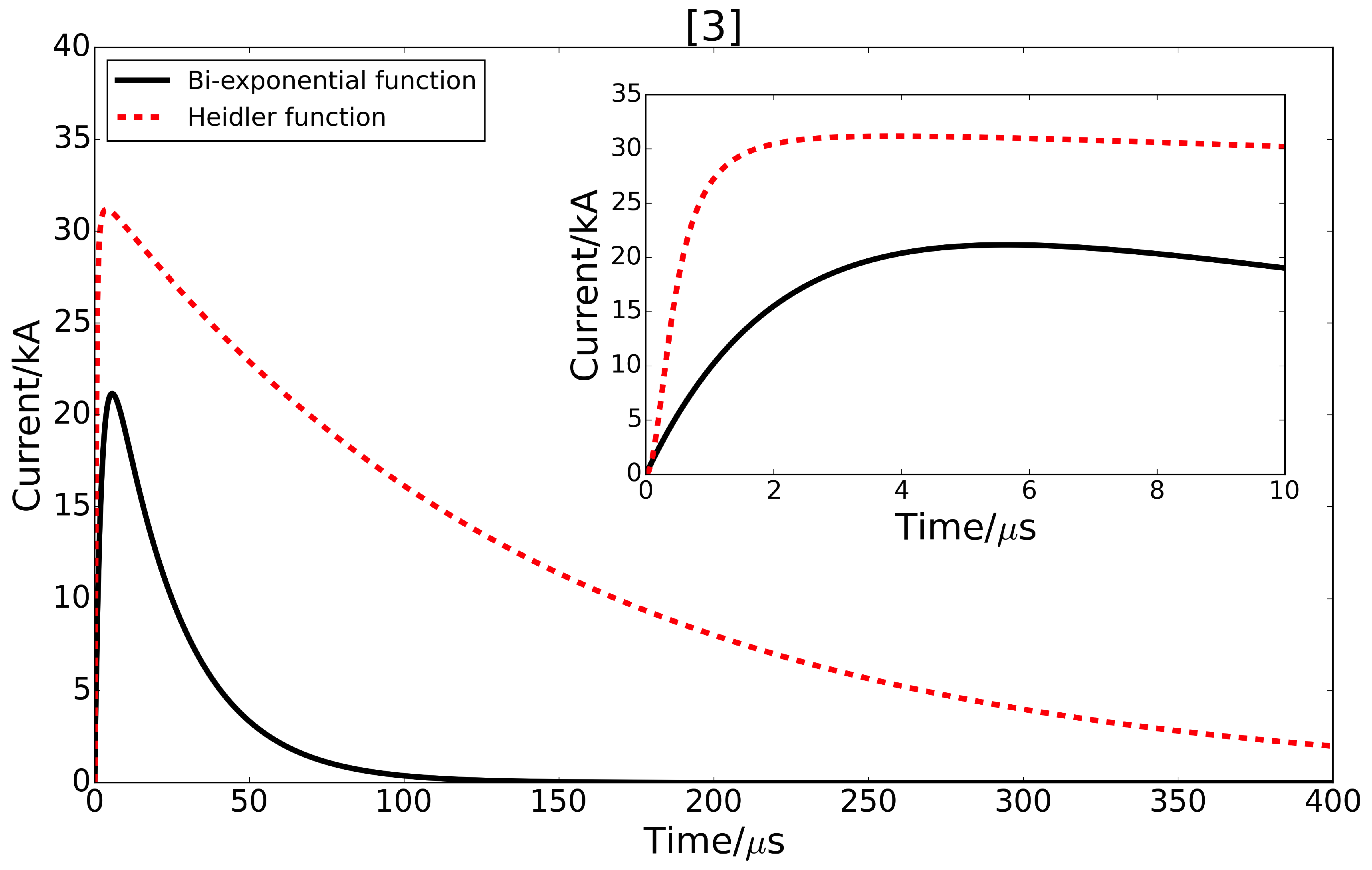}
\end{subfigure} 
\begin{subfigure}[b]{0.57\textwidth}
  \includegraphics[width=\columnwidth]{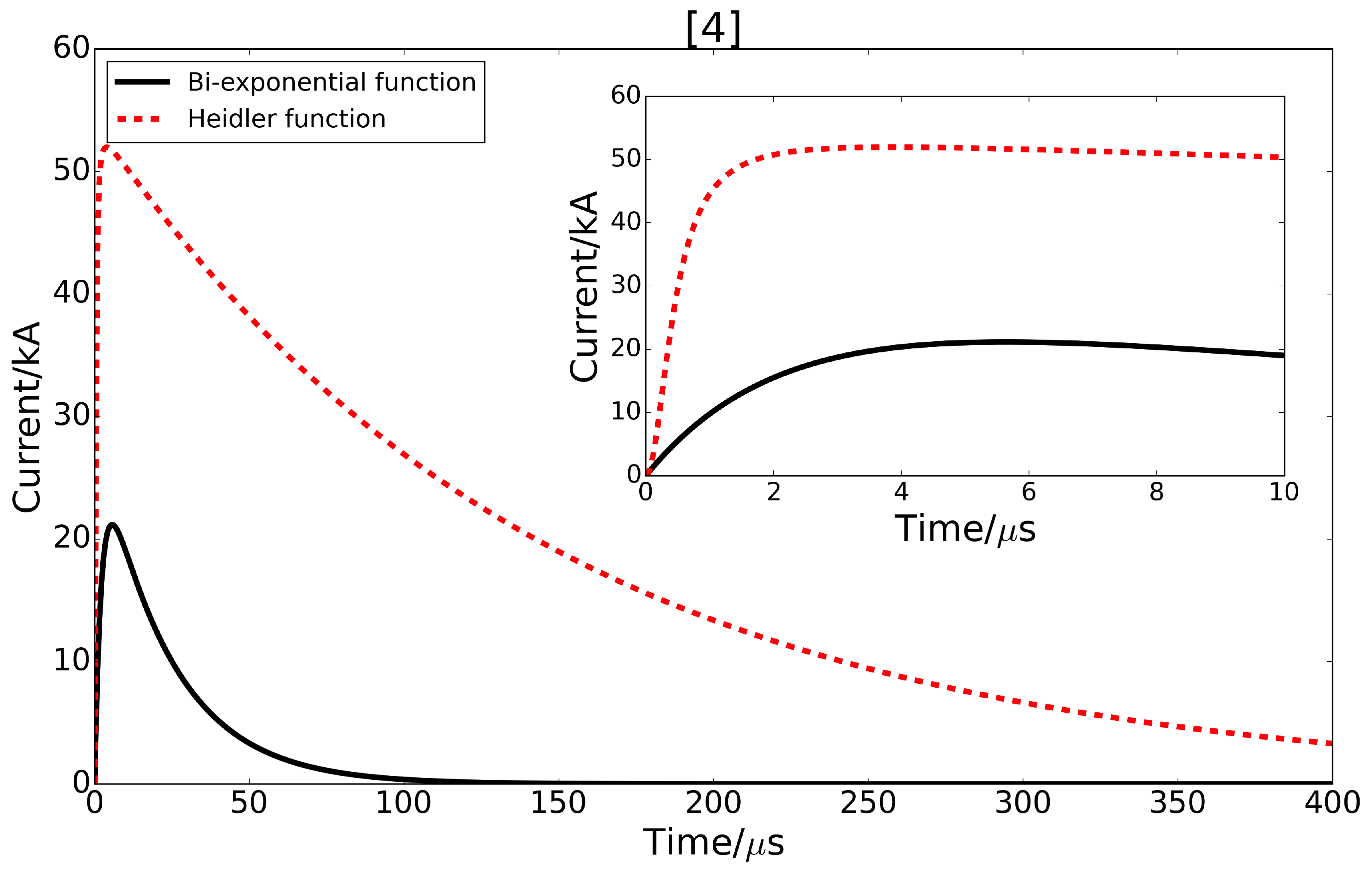}
  \includegraphics[width=\columnwidth]{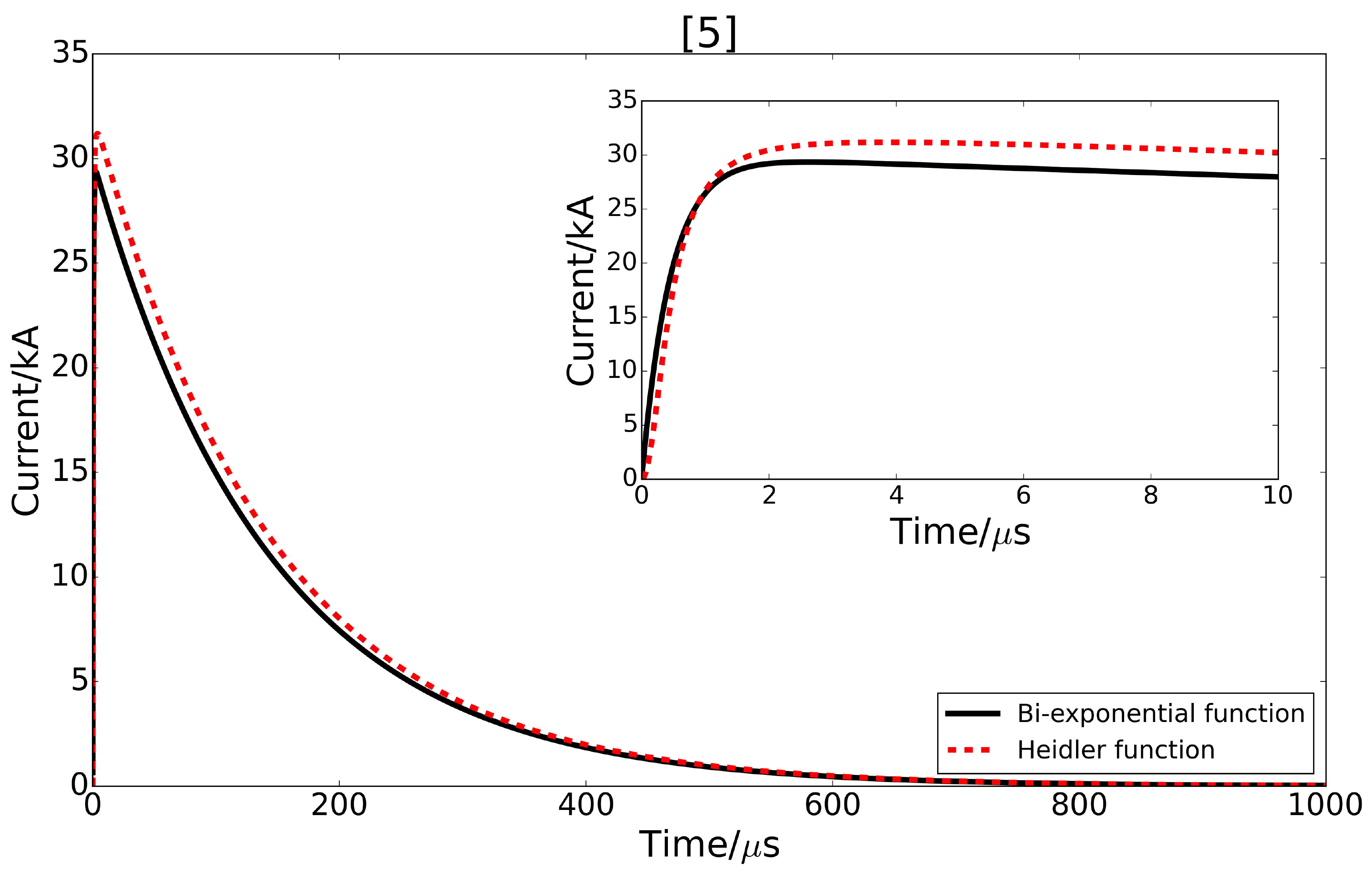}
  \includegraphics[width=\columnwidth]{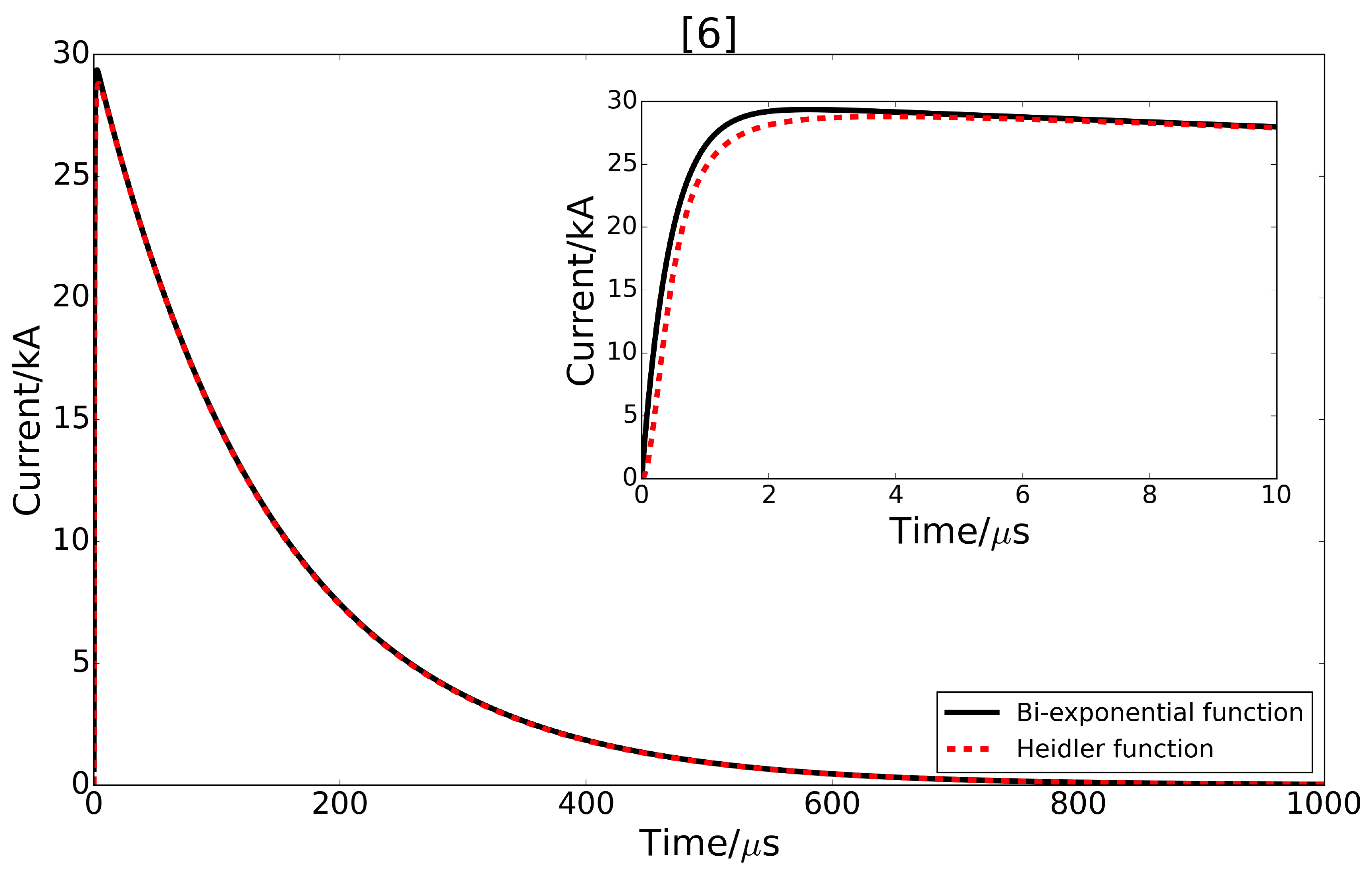}
\end{subfigure} 
  \caption{Comparison of the bi-exponential (black, solid line) and the Heidler (red, dashed line) current functions, with parameters listed in Table \ref{table:ibh}. Both current functions represent the same current when $\alpha$ and $\beta$ are expressed by $\tau_1$ and $\tau_2$ (Eq. \ref{eq:ab}) and $\eta$ is set to 1 (see panel $[6]$ of this figure and rows $[6]$ of Table \ref{table:ibh}).}
  \label{fig:ibh}
	\vspace{0.8cm}
\end{figure*}

The current at the channel base ($z = 0$) produced by electrons moving from one charged region of a cloud to another, has been modelled by various current functions in the literature. The most used ones are the double- (or bi-) exponential function (Eq. \ref{eq:1}) introduced by \citet{bruce1941}, and the Heidler function (Eq. \ref{eq:3}) first used by \citet{heidler1985}. It is common to include a combination of multiple Heidler functions or bi-exponential and Heidler functions to the model when describing the current in the lightning channel, so that they would reproduce the observed current shape better \citep[for references see][Section 4.6.4]{rakov2003}. I discuss these two current functions in order to study how the way they represent the current in the lightning channel may affect the resulting electric field, frequency and power spectra. I investigate their sensitivity against the parameters derived from measurements for Solar System planets.

The bi-exponential current function: 
\begin{equation} \label{eq:1}
i(z=0,t) = i(t) = i_0(e^{-\alpha t}-e^{-\beta t}),
\end{equation}

\noindent where $\alpha$ and $\beta$ are the same parameters as in Eq. \ref{eq:tau}, and $i_0$ is the current peak, the global maximum of the current function, $z=0$ represents the channel base. For simplicity I use $i(t)$ when referring to $i(0,t)$. Table \ref{table:3} summarizes different values for $\alpha$ and $\beta$ suggested by several authors. \citet{bruce1941}, \citet{levine1978} and \citet{farrell1999} all used the same current peak ($i_0=30$ kA) in their work, independent of the object considered. The difference between the used $\alpha$ and $\beta$ parameters results from different assumptions of the current channel: \citet{levine1978} considered the tortuosity of the channel, while \citet{bruce1941} did not. Unlike the first two authors, \citet{farrell1999} modelled lightning on Jupiter trying to reproduce a current waveform less steep than the one for Earth lightning, resulting in lower $\alpha$ and $\beta$ parameters. \citet{levine1978} slightly modified the bi-exponential current function by adding an intermediate current (with a current peak of 2.5 kA) to the formula, and making it continuous in $t=0$. Hence the lower $\alpha$ and $\beta$ parameters that describe the main current pulse in their work. Figure \ref{fig:compi} illustrates the effect of changing the parameters $\alpha$ and $\beta$ on the shape of the current function. It shows that as $\alpha$ decreases, the duration of the discharge event becomes longer, while as $\beta$ decreases, the rise time of the current (the time between $t=0$ s and the peak) becomes longer.

The Heidler function \citep{heidler1985}: 
\begin{equation} \label{eq:3}
i(t) = \frac{i_0}{\eta}\frac{\left(\frac{t}{\tau_1}\right)^m}{\left(\frac{t}{\tau_1}\right)^m+1}e^{-\frac{t}{\tau_2}},
\end{equation}
\noindent where $m \in \mathbb{N}$, $i_0$ [kA] is the current peak, $\eta$ is the correction factor for the current peak given by Eq. \ref{eq:eta} 
\citep[][p. 8]{paolone2001},  
$\tau_1$ [s] is the time constant determining the current-rise time and $\tau_2$ [s] is the time constant determining the current-decay time \citep{diendorfer1990},

\begin{equation} \label{eq:eta}
\eta = e^{-\frac{\tau_1}{\tau_2}\left(m\frac{\tau_2}{\tau_1}\right)^{1/m}}.
\end{equation}

The Heidler function is preferred to the bi-exponential because its time-derivative is zero at $t=0$ (unlike the bi-exponential function, which shows a discontinuity at $t=0$), which is consistent with the measured return-stroke current wave shape \citep{paolone2001, heidler2002}. 

Figure \ref{fig:ibh} shows how the bi-exponential and the Heidler function behave when changing the parameters in Equations \ref{eq:1} and \ref{eq:3}. The used parameters are listed in Table \ref{table:ibh}. $\alpha$ and $\beta$ can be related to $\tau_1$ and $\tau_2$ through: 

\begin{equation} \label{eq:ab}
\alpha = \frac{1}{\tau_1+\tau_2}, \\
\beta = \frac{1}{\tau_1}
\end{equation}

\noindent based on the definitions of $\alpha$, $\beta$, $\tau_1$ and $\tau_2$ given above. The panels of Fig. \ref{fig:ibh} show that the two types of current functions represent the same current if $\eta$ is set to 1 and $\alpha$ and $\beta$ are expressed by $\tau_1$ and $\tau_2$ as in Eq. \ref{eq:ab} (Fig. \ref{fig:ibh}, [6]; Table \ref{table:ibh}, [6]). The Heidler function has the possibility of fine-tuning the shape of the curves by introducing a more complex function form. Later in the study I use the bi-exponential current function for the model for its simplicity, as I want to derive the principal effect of the current running through the channel. 

%__________________________________________________________________
\subsection{Electric field} \label{sec:efield}

\begin{figure*}
\advance\leftskip-1.0cm
\advance\rightskip-1.0cm
\begin{subfigure}[b]{0.57\textwidth}
  \includegraphics[width=\columnwidth, trim=0cm 0cm 0cm 0cm]{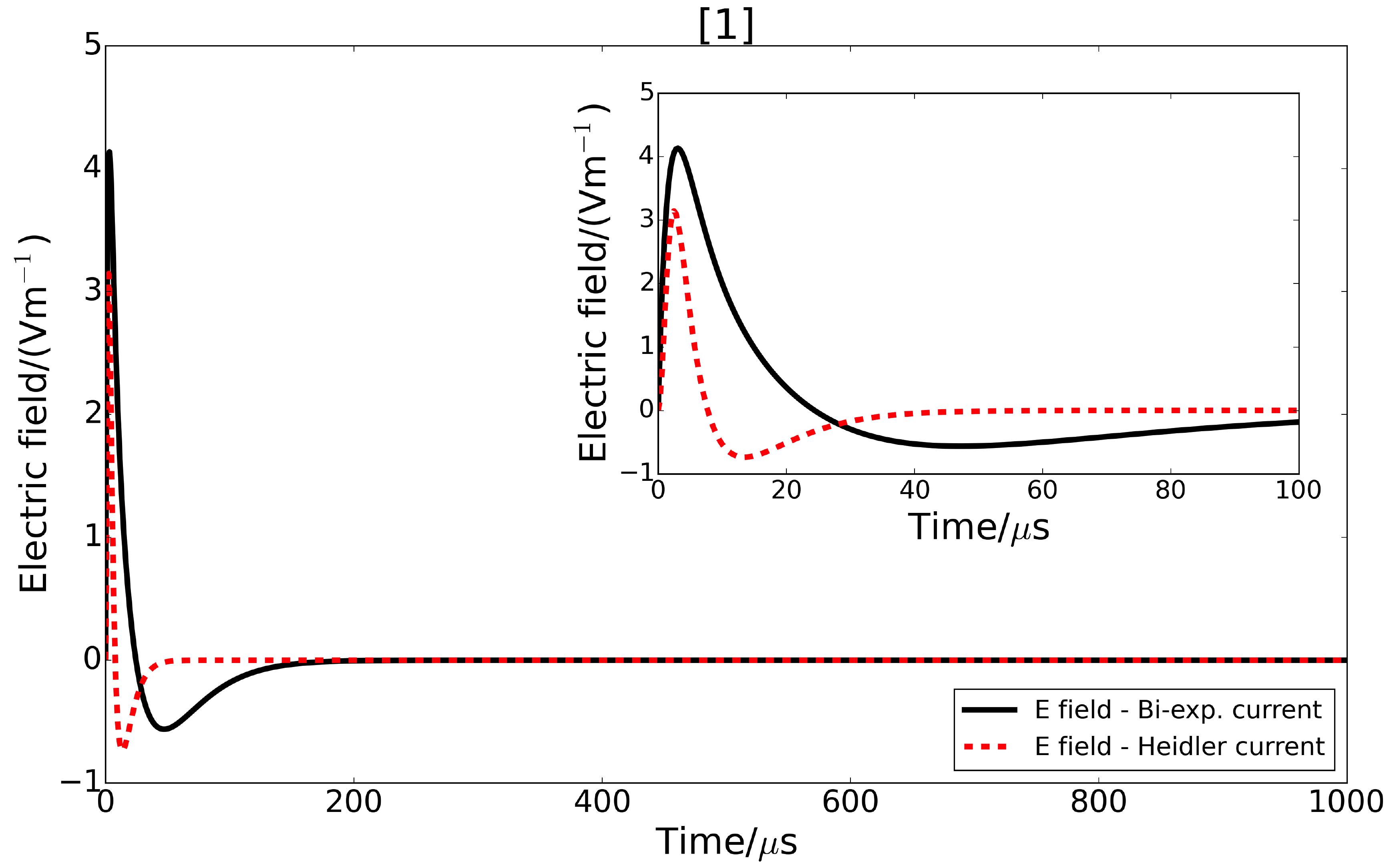}
  \includegraphics[width=\columnwidth, trim=0cm 0cm 0cm 0cm]{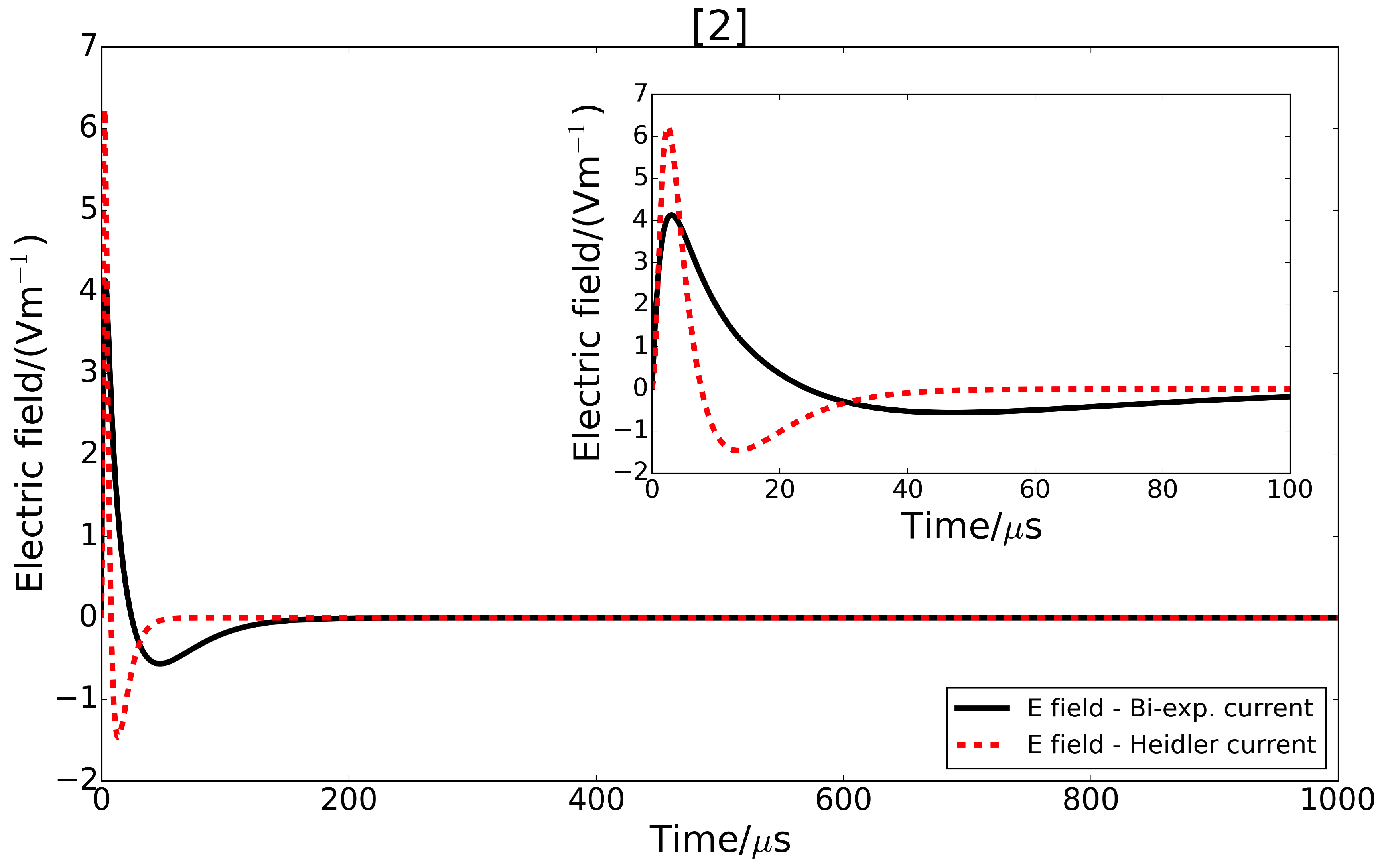}
  \includegraphics[width=\columnwidth, trim=0cm 0cm 0cm 0cm]{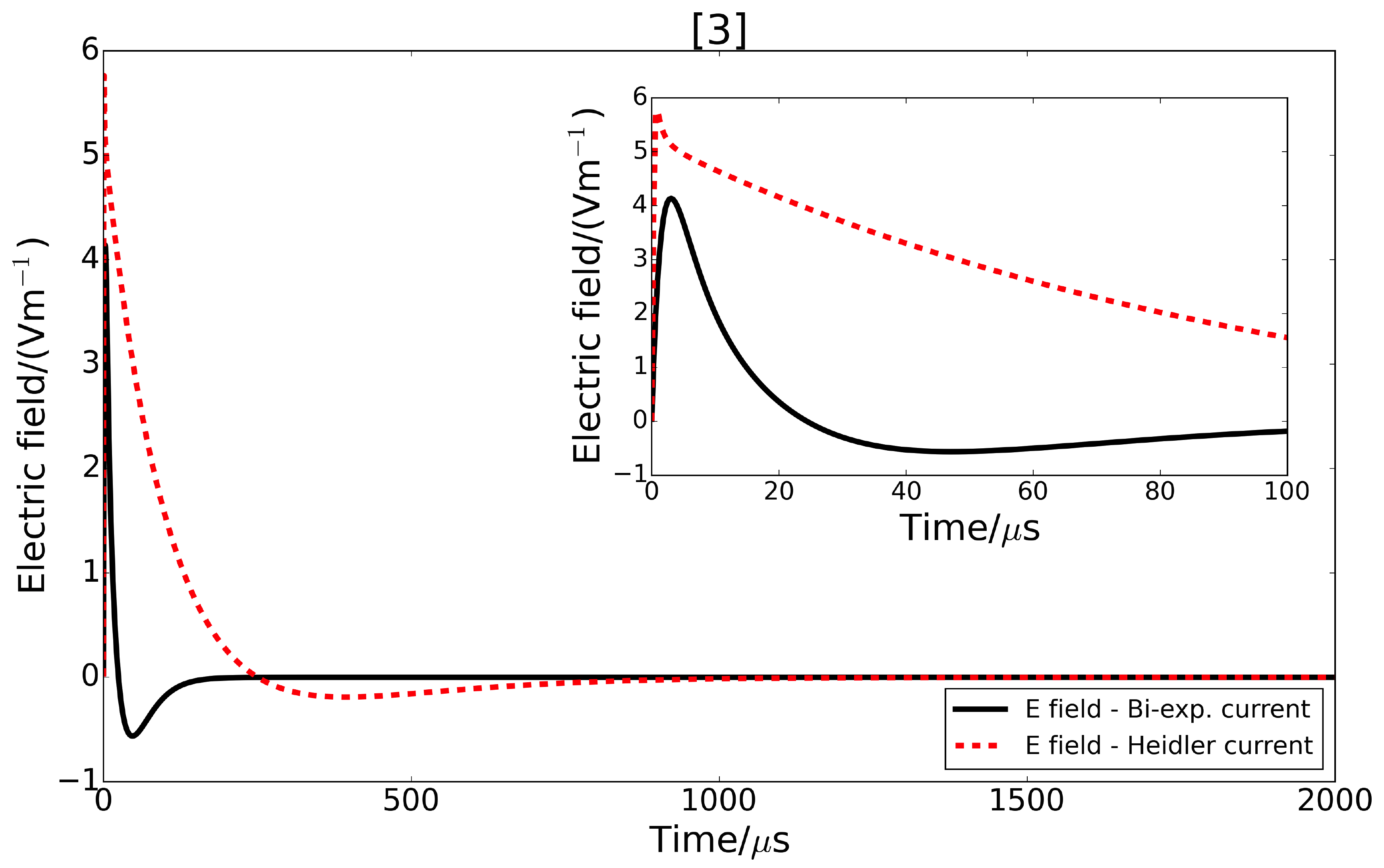}
\end{subfigure} 
\begin{subfigure}[b]{0.57\textwidth}
  \includegraphics[width=\columnwidth, trim=0cm 0cm 0cm 0cm]{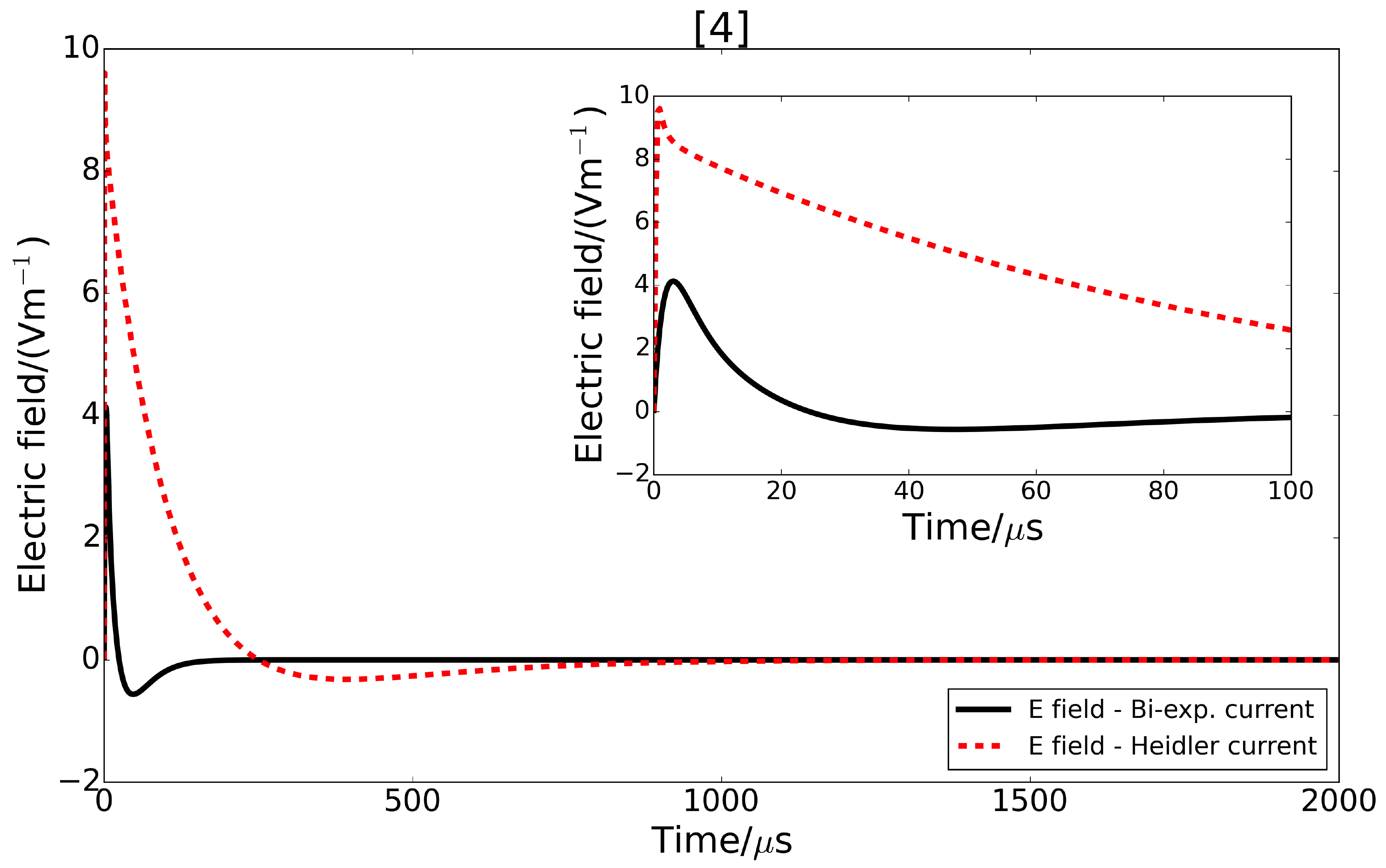}
  \includegraphics[width=\columnwidth, trim=0cm 0cm 0cm 0cm]{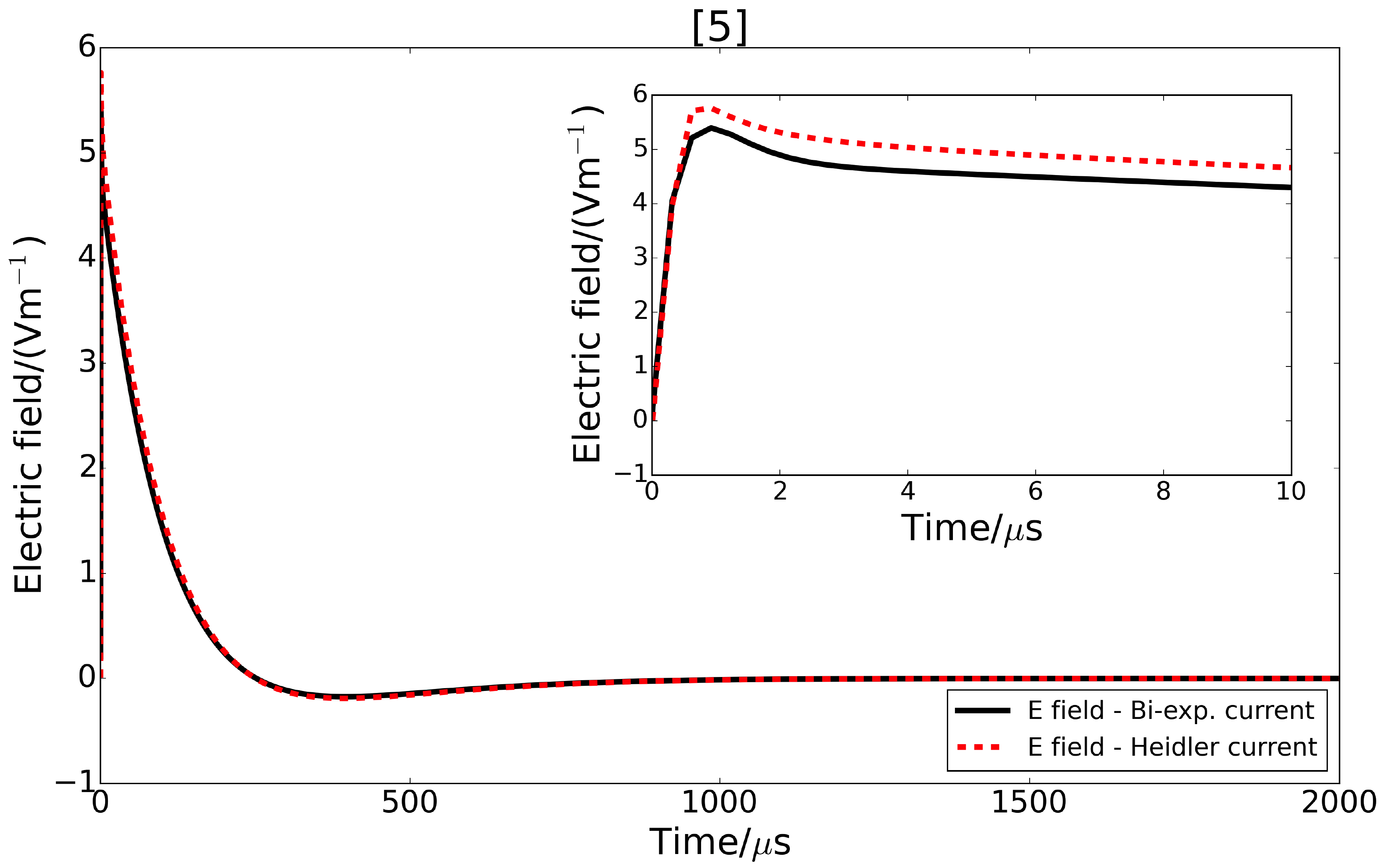}
  \includegraphics[width=\columnwidth, trim=0cm 0cm 0cm 0cm]{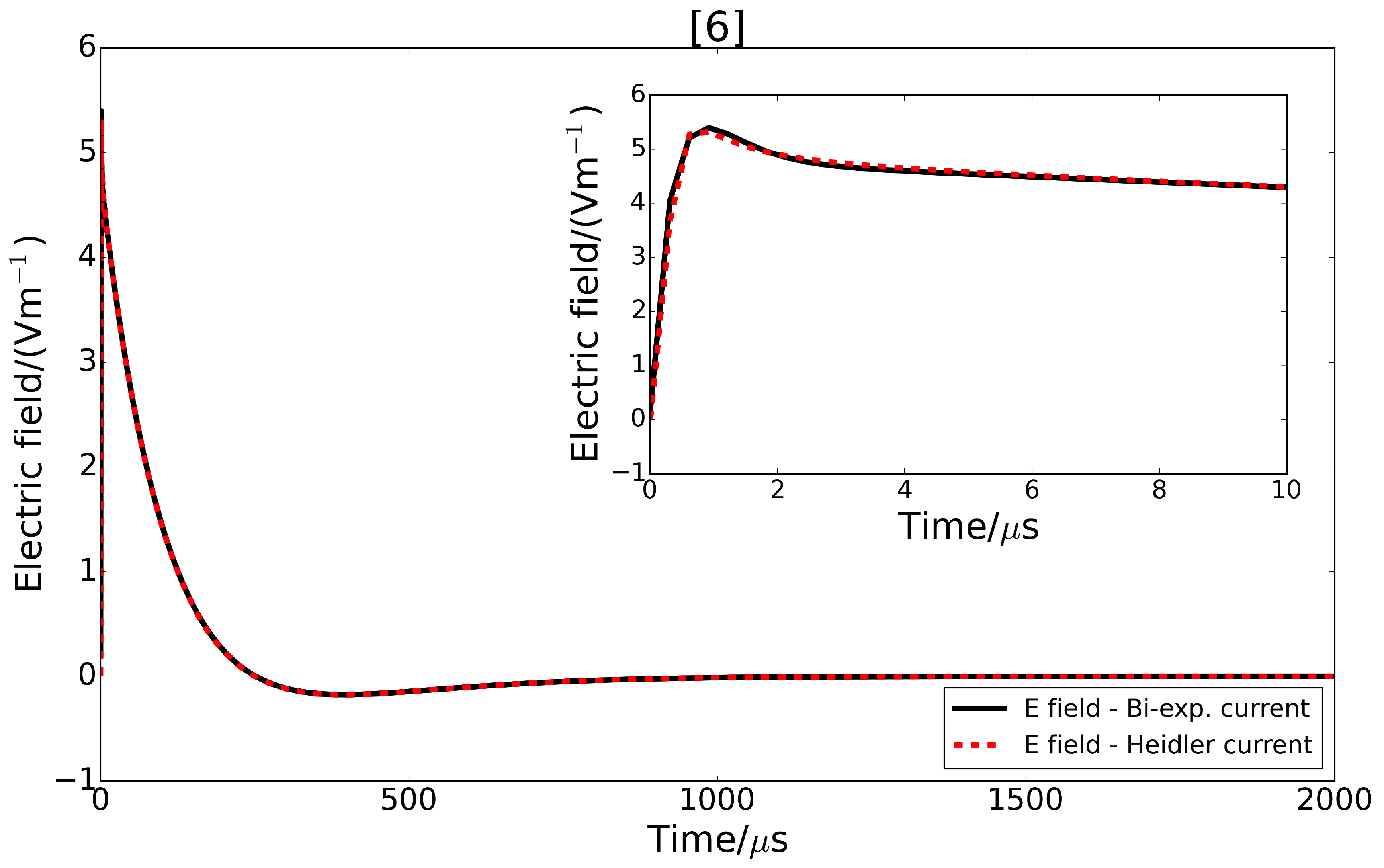}
\end{subfigure} 
  \caption{Electric fields calculated from the bi-exponential (black, solid line) and the Heidler (red, dashed line) current functions represented in Fig. \ref{fig:ibh}. The interchangeability of the two current functions as shown on Fig. \ref{fig:ibh} carries on to the electric fields obtained from them.}
  \label{fig:ebh}
	\vspace{0.8cm}
\end{figure*}

%%% Electric field %%%

The electric field which results from the electric current as described by Eq. \ref{eq:1} (or Eq. \ref{eq:3}) is given by the following general expression for a linear dipole model of a lightning discharge \citep{bruce1941}:
\begin{equation} \label{eq:4}
E(t) = \frac{1}{4 \pi \epsilon_0}\left[\frac{M(t)}{r^3}+\frac{1}{c r^2}\frac{dM(t)}{dt}+\frac{1}{c^2 r}\frac{d^2M(t)}{dt^2}\right],
\end{equation}

\noindent where $M(t)$ is the electric dipole moment given by Eq. \ref{eq:mom}, and $\epsilon_0$ is the permittivity of the vacuum in F$/$m (Farads per meter), and all physical quantities have units in SI. The first term in Eq. \ref{eq:4} is the electrostatic field, the second term is the magnetic induction field and the third term is the radiation field \citep{bruce1941}.

The time derivative of the dipole moment is
\begin{equation} \label{eq:5}
\frac{dM(t)}{dt} = 2i(t)\int_0^t{\v v(t') dt'},
\end{equation}

\noindent where $i(t)$ is the current in the lightning channel at time $t$ and $\v v(t)$ is the velocity of the return stroke \citep{bruce1941}. Eq. \ref{eq:5} can be derived from Eq. \ref{eq:mom}, if $\int_0^t{\v v(t') dt'} \equiv \Delta v \Delta t \equiv h$, and $\frac{d}{dt}\left[\int_0^t{i(t') dt'}\right] \equiv i(t)$. \citet{bruce1941} found that this velocity decreases as the lightning stroke propagates upwards, since they conducted observations from the ground, which results in the following expression:

\begin{equation} \label{eq:6}
\v v(t) = \v v_0 e^{-\gamma t},
\end{equation}
\noindent where the values for Earth are: $\v v_0 \sim 8 \times 10^7$ m s$^{-1} = 0.3 c$, with $c$ the speed of light, and $\gamma \sim 3 \times 10^4$ s$^{-1}$ \citep{bruce1941}. However, the drop of the velocity may be due to the loss of energy, in which case it would decrease not only upward-propagating but downward-propagating as well.

\subsubsection{Electric field from the bi-exponential current function}

First, I consider the bi-exponential function (Eq. \ref{eq:1}) as the current function at the channel base ($i(0,t) = i(t)$). Two approaches can be followed when calculating the electric field: the velocity of the return stroke can be considered either constant ($\v v(t) = \v v_0$) or varying in time (Eq. \ref{eq:6}).
Combining Eqs \ref{eq:1}, \ref{eq:5} and \ref{eq:6} results in:
\begin{equation} \label{eq:7}
\frac{dM(t)}{dt} = 2\frac{i_0\v v_0}{\gamma}(e^{-\alpha t}- e^{-\beta t})(1 - e^{-\gamma t}),
\end{equation}
which can be used to evaluate Eq. \ref{eq:4}.
 
Eq. \ref{eq:7} simplifies slightly when $\v v$ does not vary in time, resulting in

\begin{equation} \label{eq:9}
\frac{dM(t)}{dt} = 2 i_0 \v v_0 t(e^{-\alpha t}- e^{-\beta t}).
\end{equation}

This approach was used by \citet{farrell1999} when deriving the electric field of Jovian lightning discharges. Combining Eqs \ref{eq:4} and \ref{eq:9} with the derivative of Eq. \ref{eq:9} result in %Differentiating Eq. \ref{eq:9} and combining that with Eqs \ref{eq:4} and \ref{eq:9} result in

\begin{dmath} \label{eq:10}
E(t) =\frac{2 i_0 \v v_0}{4\pi \epsilon_0}\left[\frac{1}{c r^2}t(e^{-\alpha t}- e^{-\beta t})+\frac{1}{c^2 r}(e^{-\alpha t}- e^{-\beta t}-t\alpha e^{-\alpha t}+t\beta e^{-\beta t})\right].
\end{dmath}

\noindent 
In Eq. \ref{eq:10} the first term of Eq. \ref{eq:4} is neglected due to the large distance between the source and the observer ($r >> h$; Sect. \ref{ssec:elcom}). 

\begin{figure*}
\advance\leftskip-1.0cm
\advance\rightskip-1.0cm
 % \centering
\begin{subfigure}[b]{0.58\textwidth}
  \includegraphics[width=\columnwidth, trim=0cm 0cm 0cm 0cm]{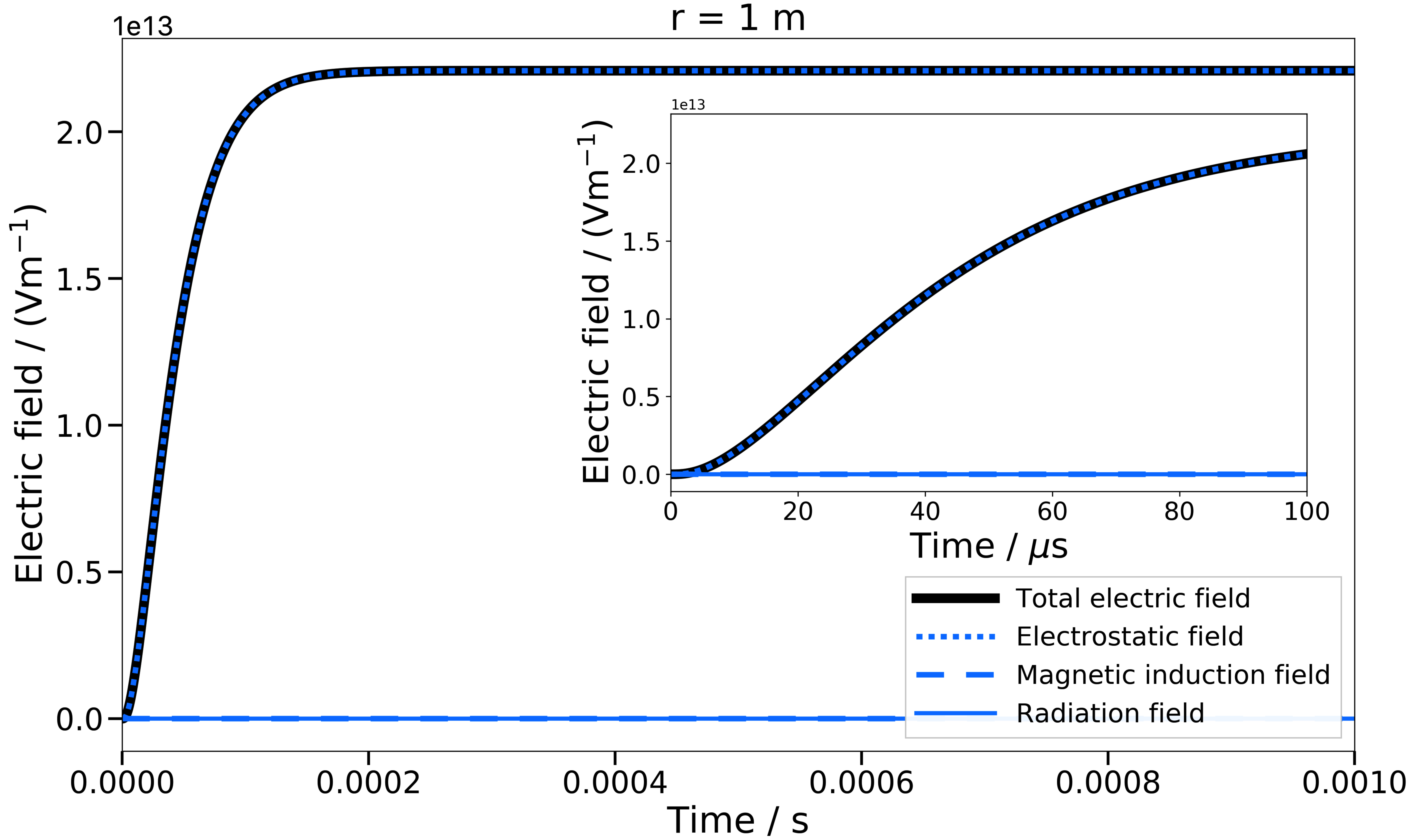}
  \includegraphics[width=\columnwidth, trim=0cm 0cm 0cm 0cm]{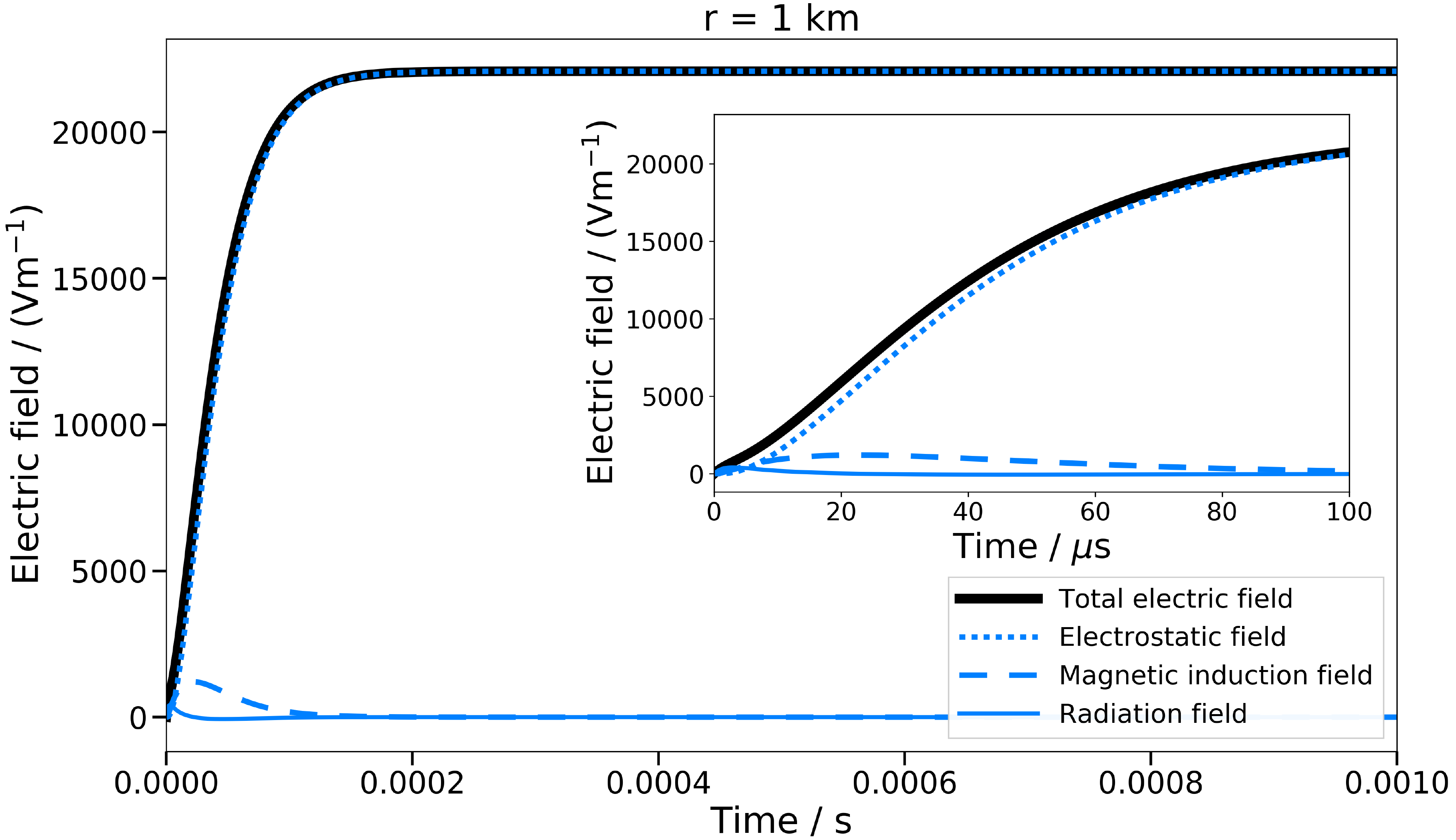}
  \includegraphics[width=\columnwidth, trim=0cm 0cm 0cm 0cm]{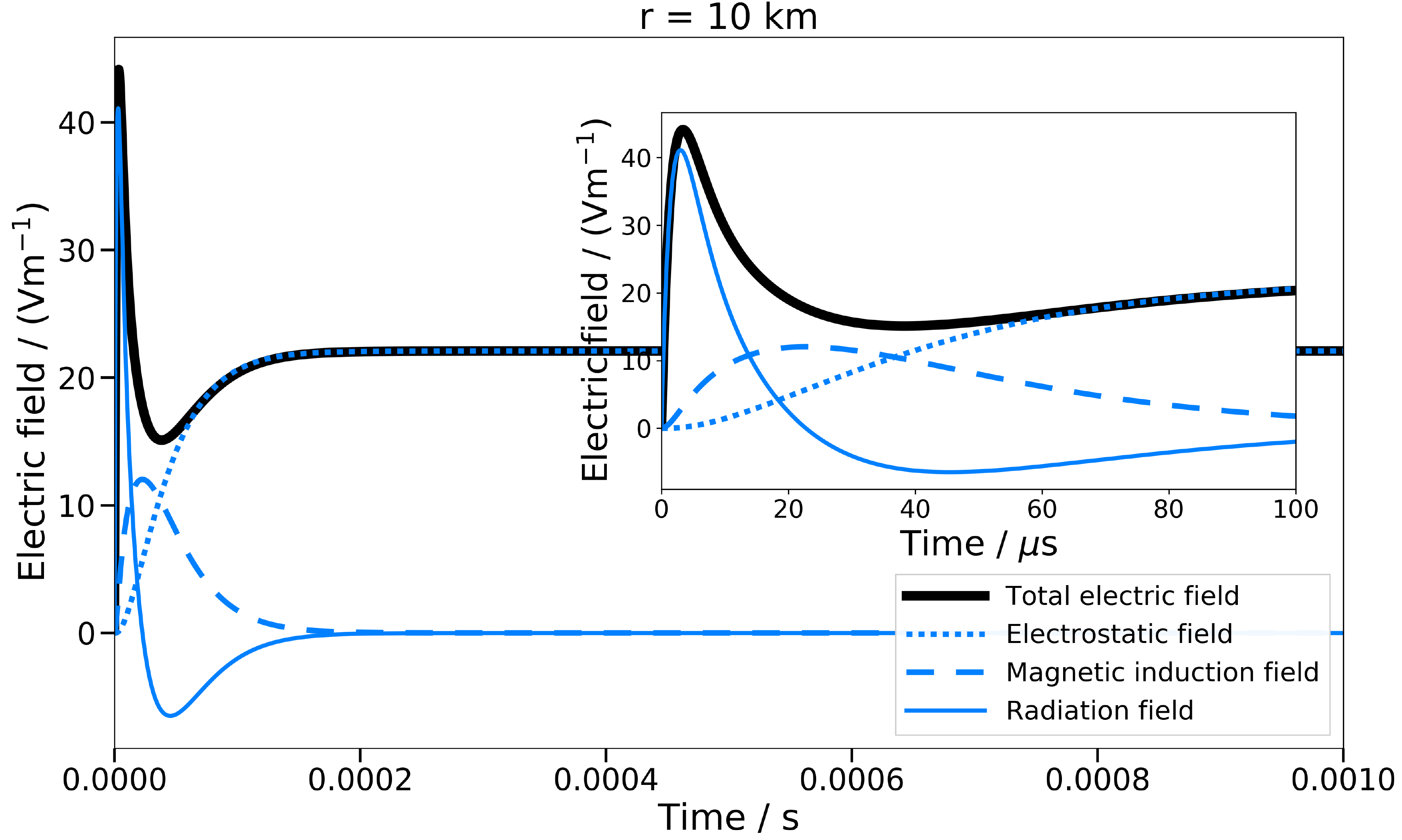}
\end{subfigure} 
\begin{subfigure}[b]{0.57\textwidth}
  \includegraphics[width=\columnwidth, trim=0cm 0cm 0cm 0cm]{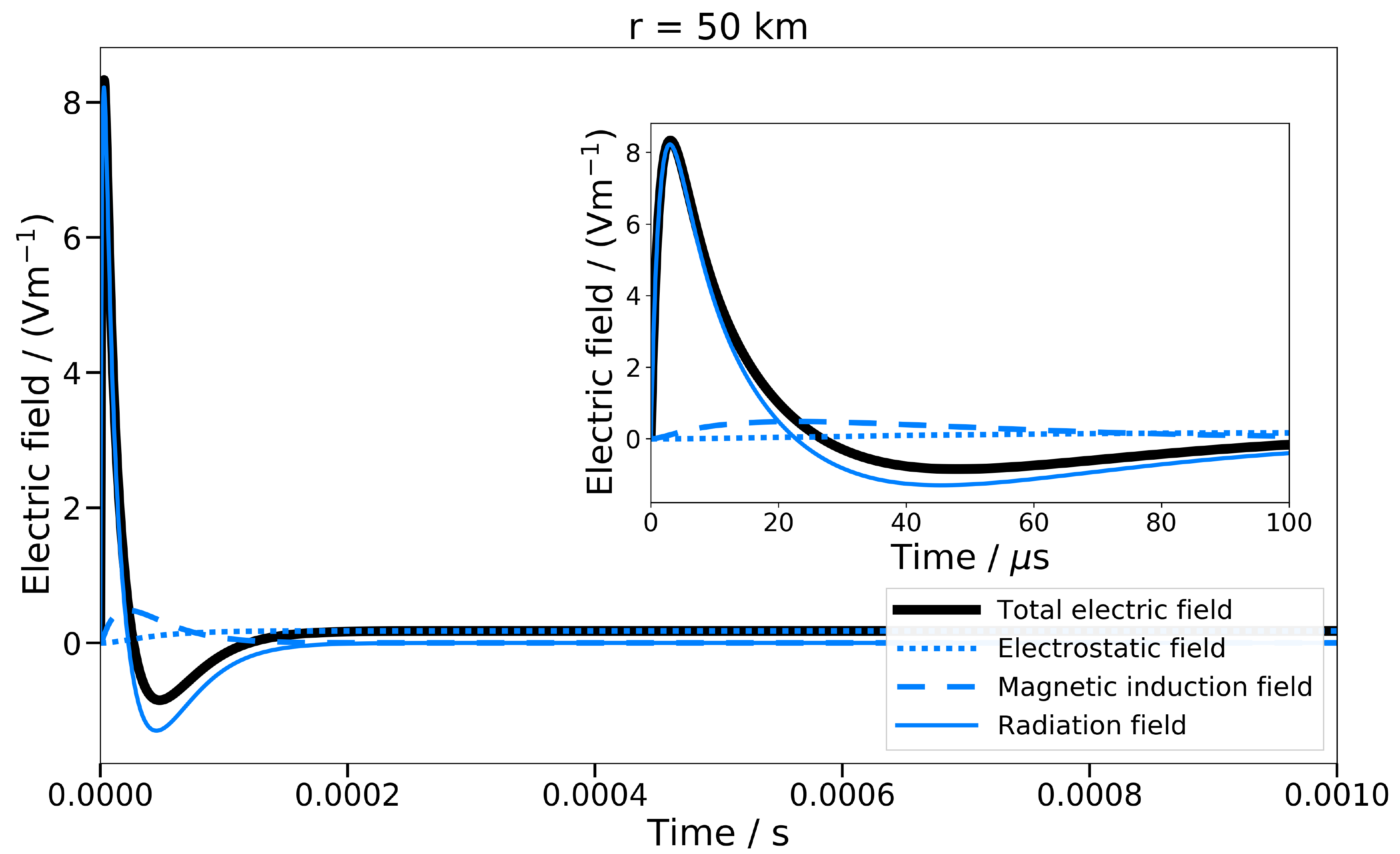}
  \includegraphics[width=\columnwidth, trim=0cm 0cm 0cm 0cm]{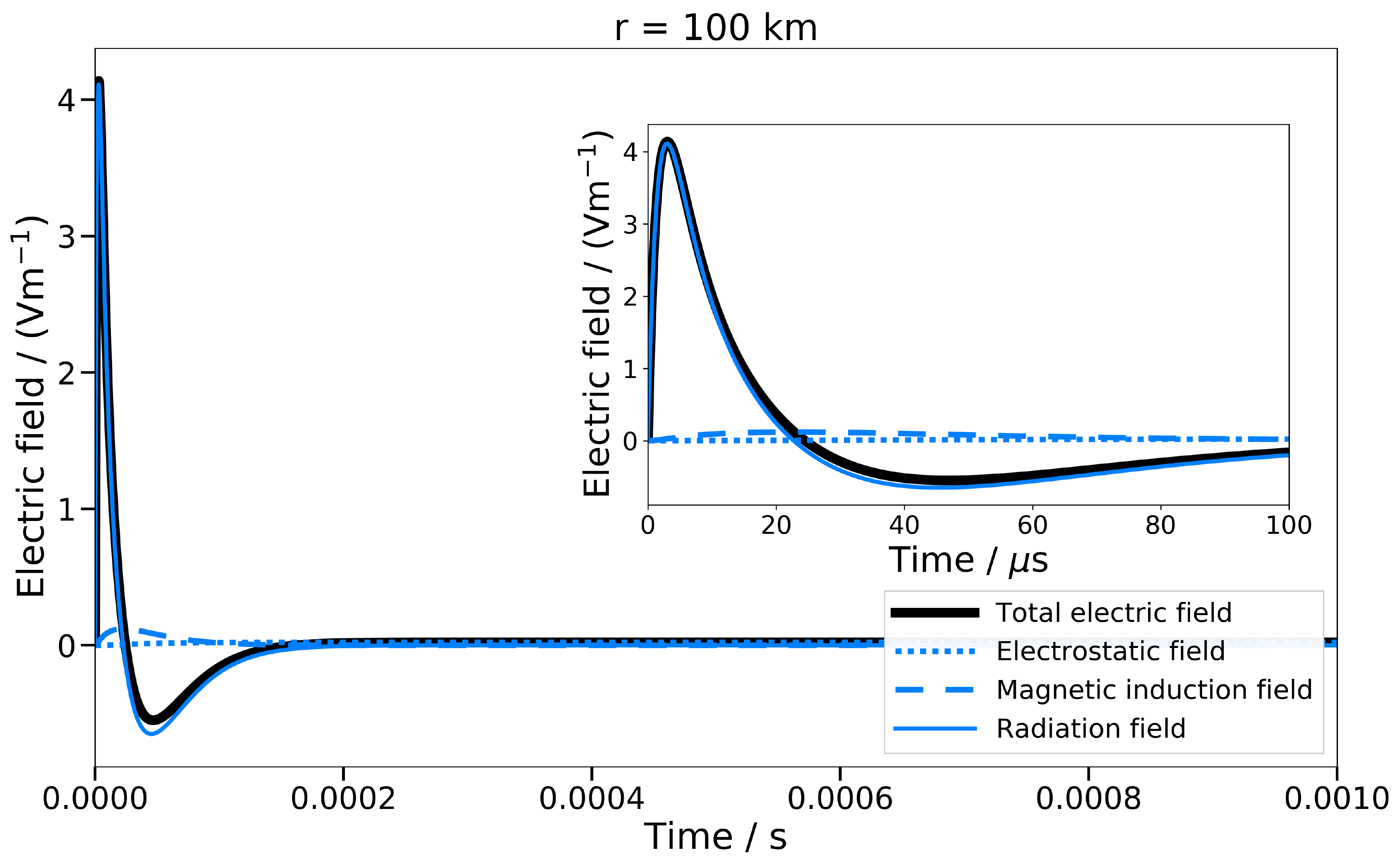}
  \includegraphics[width=\columnwidth, trim=0cm 0cm 0cm 0cm]{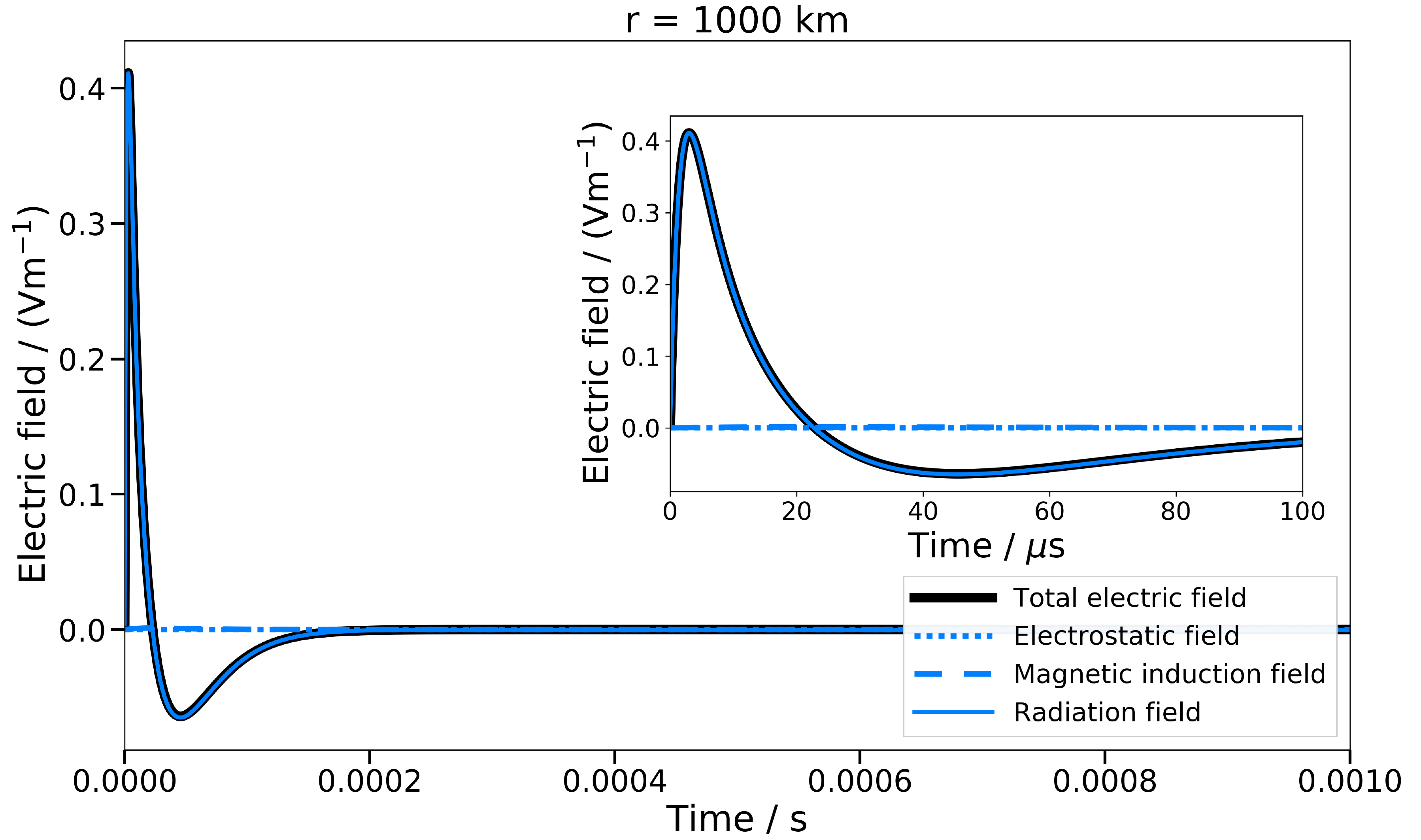}
\end{subfigure}
  \caption{Comparing the three components of the electric field (Eq. \ref{eq:4}). Total electric field: solid black line; Electrostatic field: dotted blue line; Magnetic induction field: dashed blue line; Radiation field: solid blue line. The different panels demonstrate how the electric field components change with lightning-observer distance, $r$ (Sect. \ref{ssec:elcom}).}
  \label{fig:ecomp}
\vspace{0.8cm}
\end{figure*}

\subsubsection{Electric field from the Heidler current function} \label{ssec:heid}

Now I consider a current given by the Heidler function (Eq. \ref{eq:3}). For simplicity the velocity is considered to be constant. This way Eq. \ref{eq:9} changes to

\begin{equation} \label{eq:hdm}	%hdm - heidler moment 1st derivative
\frac{dM(t)}{dt} = 2 \v v_0 t \frac{i_0}{\eta}\frac{\left(\frac{t}{\tau_1}\right)^m}{\left(\frac{t}{\tau_1}\right)^m+1}e^{-\frac{t}{\tau_2}},
\end{equation}
and the second derivative of the dipole moment is
\begin{equation} \label{eq:hddm}	%hddm - heidler moment 2nd derivative
\frac{d^2M(t)}{dt^2} = \frac{2 \v v_0 i_0}{\eta} \left[\frac{\left(\frac{t}{\tau_1}\right)^m}{\left(\frac{t}{\tau_1}\right)^m+1}e^{-\frac{t}{\tau_2}}\left(1-t \frac{1}{\tau_2}\right) + t e^{-\frac{t}{\tau_2}} \frac{\frac{m}{\tau_1} (\frac{t}{\tau1})^{m-1}}{((\frac{t}{\tau_1})^m + 1)^2}\right].
\end{equation}

By combining Eqs \ref{eq:hdm} and \ref{eq:hddm} with Eq. \ref{eq:4}, I calculate the electric field at large distances from the source ($r >> h$). Figure \ref{fig:ebh} demonstrates how the electric field changes when derived from different current functions (Fig. \ref{fig:ibh}). The distance between the source and observer for all figures is $r = 100$ km, and the propagation velocity is constant, $v_0 = 8.0 \times 10^7$ m s$^{-1}$. Due to the large distance, $r$, the first term in Eq. \ref{eq:4} was neglected in Fig. \ref{fig:ebh} (I explain why some of the terms of the electric field can be neglected in Sect. \ref{ssec:elcom}). To derive the electric field, SI units are used. The curves of the electric fields on Fig. \ref{fig:ebh} show the same forms relative to each other that are shown by the current functions in Fig. \ref{fig:ibh}. When the current function has a larger peak, the corresponding electric field will show a larger peak compared to the field calculated from the other current function. When the two current functions are calculated with parameters $\eta=1$ and $\alpha$ and $\beta$ being expressed by $\tau_1$ and $\tau_2$ as in Eq. \ref{eq:ab}, then the electric fields resulting from these current functions are the same (Figs \ref{fig:ibh} and \ref{fig:ebh}, [6]; and Table \ref{table:ibh}, [6]).

\subsubsection{Effect of the three electric field components} \label{ssec:elcom}

The electric field of a lightning discharge has three main components (Eq. \ref{eq:4}): electrostatic, magnetic induction and radiation fields. The three components depend on $r$, the distance between the lightning channel and the observer (Eq. \ref{eq:4}). Figure \ref{fig:ecomp} illustrates this dependence. The six figures show six distances, increasing from top to bottom, left to right, between 1 m and 1000 km. All other parameters are the same for all figures ($i_0=30 \ {\rm kA;} \ \v v_0=8\times10^{7} \ {\rm m s}^{-1}$), and all electric fields were calculated from the bi-exponential current function ($\alpha=4.4\times10^4 \ {\rm s}^{-1}; \ \beta=4.6\times10^6 \ {\rm s}^{-1}$). It is clearly seen, that the electrostatic field (blue dotted line) has an effect on the total electric field (black solid line) very close to the discharge event. This effect rapidly decreases as we get further away from the source and at a few tens of km it becomes negligible. At the meantime the induction field (blue dashed line) becomes stronger and still affects the overall shape at $\sim 50$ km, slightly increasing the total electric field. Kilometres away from the source the effect of the radiation field increases, while the induction field decreases. From a few tens$-$hundreds of km, it is the dominant component of the electric field. The same effect of the source-observer distance on the electric field components can be seen in figure 1 of \citet{dubrovin2014}, who analysed the electric field of a Saturnian lightning discharge.

The purpose of showing how the different components dominate the total electric field is to demonstrate, that at large distances ($r>50$ km) the radiation field will be the dominant component of the electric field, with a slight contribution from the induction field. Further in this study I use only these parts of the electric field to calculate lightning frequency spectrum and radio energy of lightning originating from exoplanets parsecs away from our Solar System.

%__________________________________________________________________
\subsection{Frequency and power spectra} \label{sec:freqsp}

\begin{figure*}
\advance\leftskip-1.0cm
\advance\rightskip-1.0cm
 % \centering
\begin{subfigure}[b]{0.57\textwidth}
  \includegraphics[width=\columnwidth, trim=0cm 0cm 0cm 0cm]{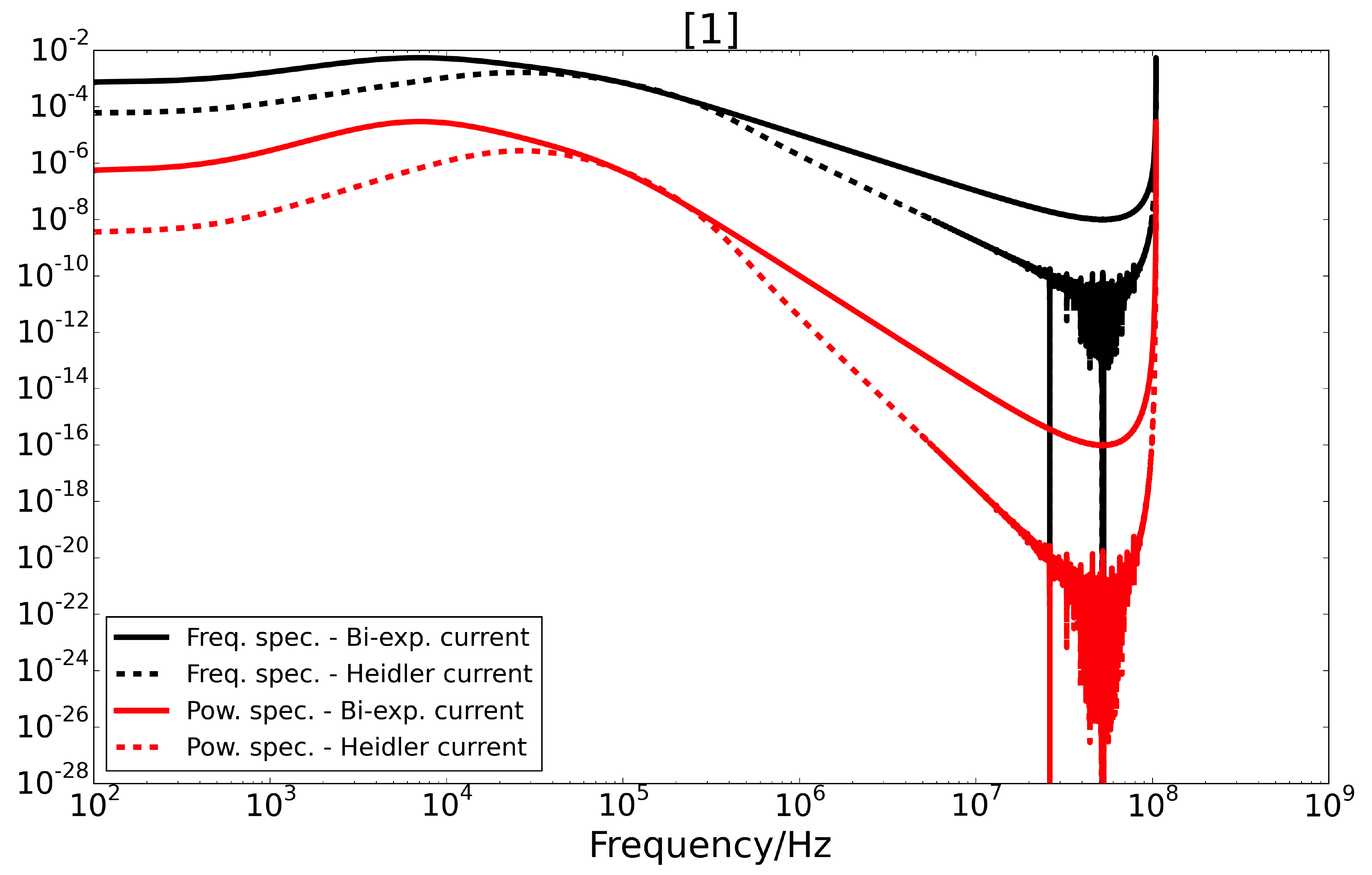}
  \includegraphics[width=\columnwidth, trim=0cm 0cm 0cm 0cm]{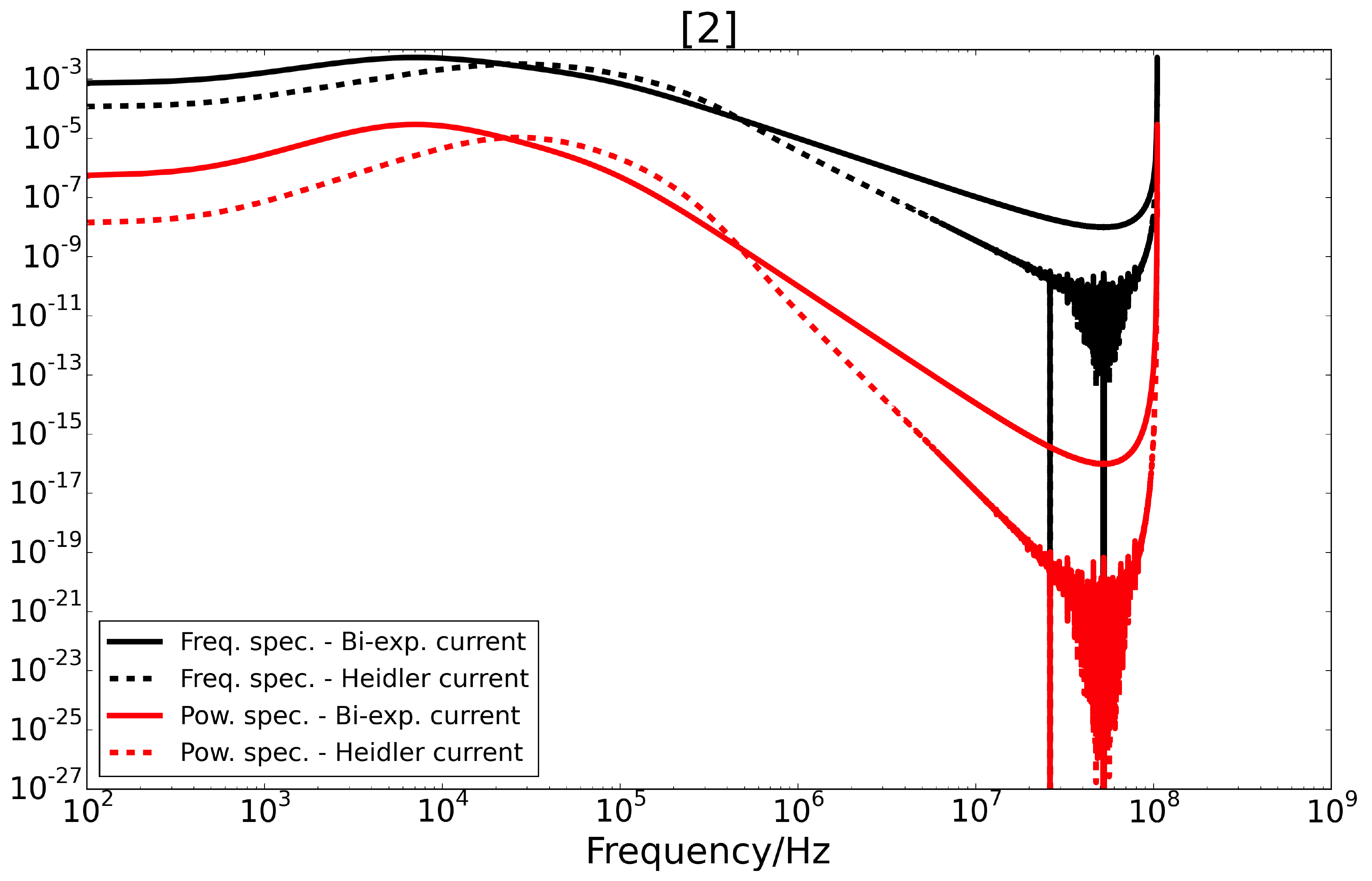}
  \includegraphics[width=\columnwidth, trim=0cm 0cm 0cm 0cm]{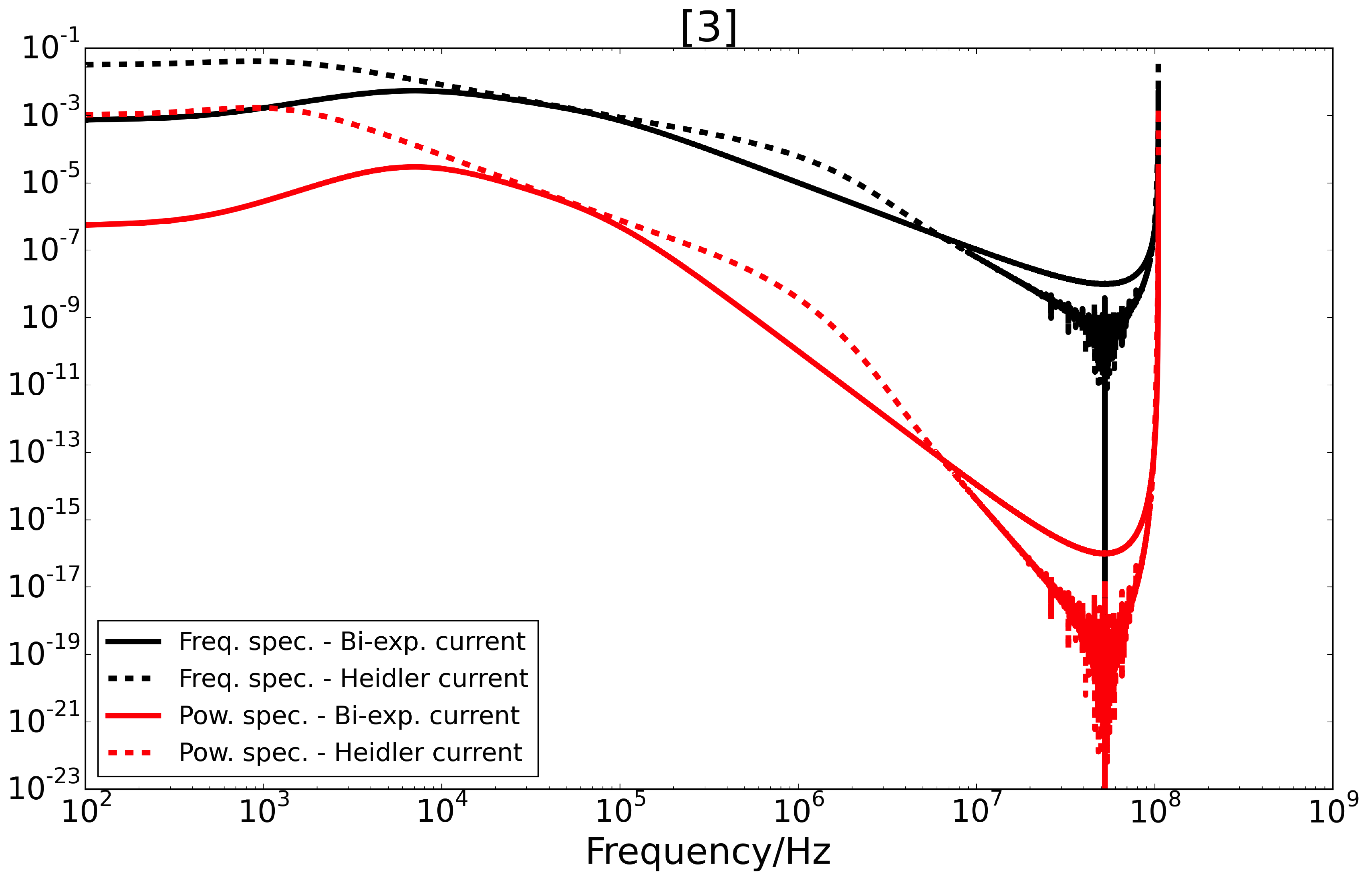}
\end{subfigure} 
\begin{subfigure}[b]{0.57\textwidth}
  \includegraphics[width=\columnwidth, trim=0cm 0cm 0cm 0cm]{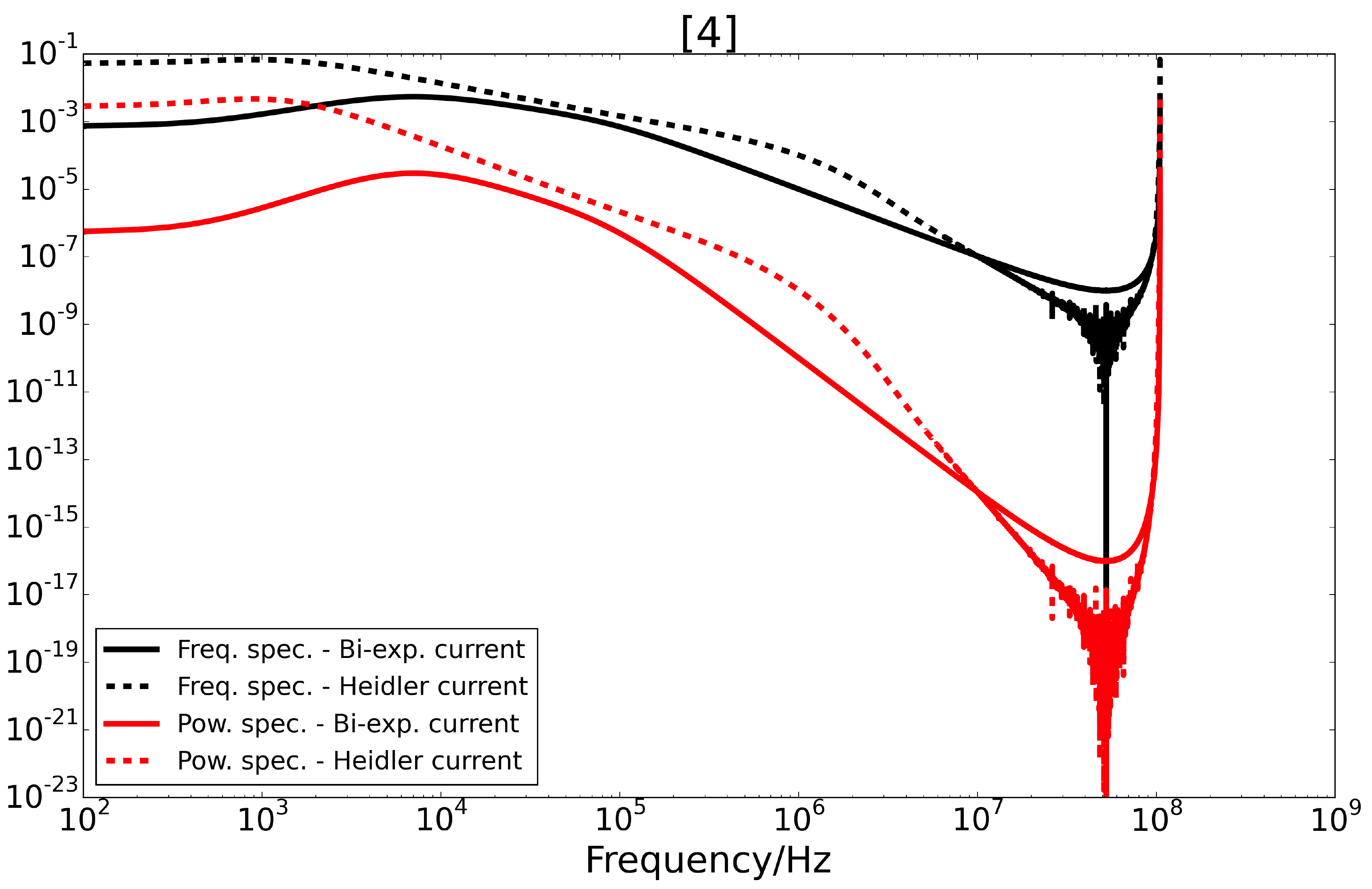}
  \includegraphics[width=\columnwidth, trim=0cm 0cm 0cm 0cm]{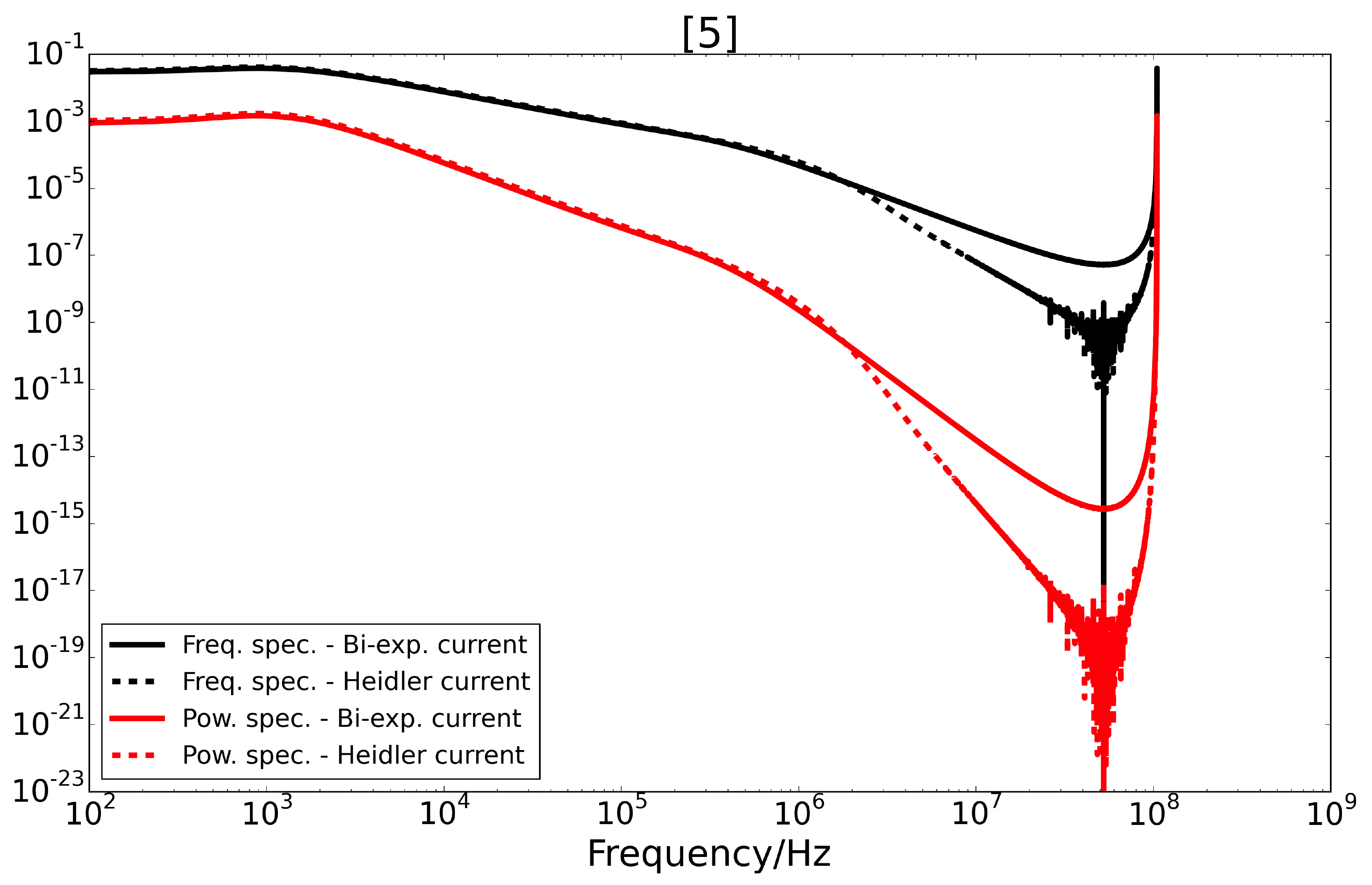}
  \includegraphics[width=\columnwidth, trim=0cm 0cm 0cm 0cm]{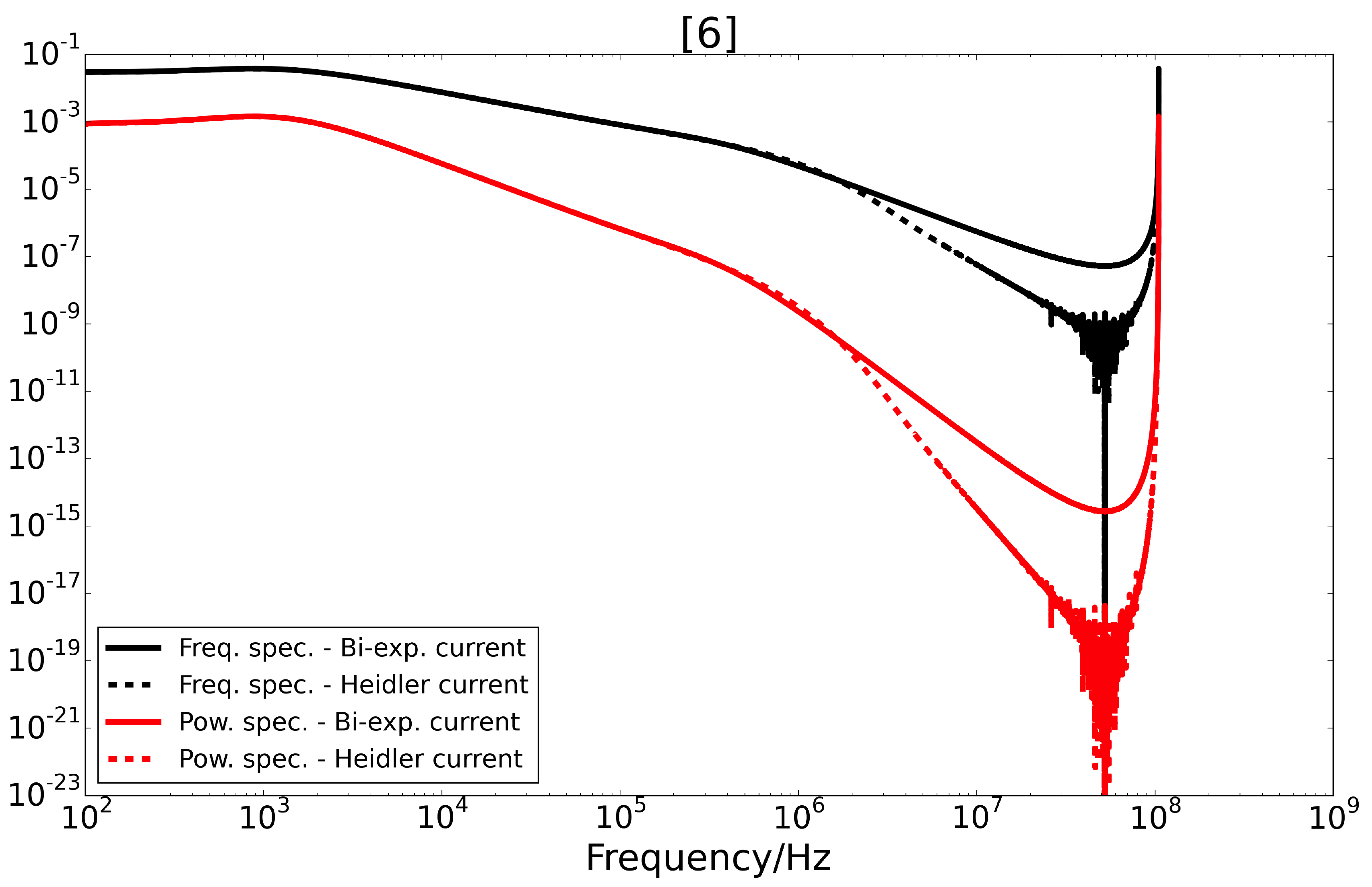}
\end{subfigure}
  \caption{Comparing frequency (black) and power (red) spectra calculated from the electric fields of the bi-exponential (solid line) and the Heidler (dashed line) current functions (Fig. \ref{fig:ebh}), with parameters listed in Table \ref{table:ibh}. The y-axis show relative units.}
  \label{fig:fpbh}
	\vspace{0.8cm}
\end{figure*}

The power spectrum of the electric field carries information on the amount of power released at certain frequencies. It quickly reaches its peak then slowly decreases with a power law. The characteristics of the power spectrum (peak frequency, $f_0$, and spectral roll-off, $f^n$) help predict the amount of power released by lightning in a frequency band, i.e. at a band that is used for observations. Because of the ionosphere and other limitation factors, such as the large distance between the observer and the source, which triggers the contamination of the radio signal by background and foreground sources (Chapter \ref{chap:ligsig}, Sect. \ref{ssec:radsig}), only part of the radio lightning spectrum can be observed, and the peak of the emission is often unknown. In my model, I generate a radio electric field frequency spectrum and from that the power spectrum in order to determine total lightning power released at radio frequencies and the lightning radiated energy at these frequencies. I also estimate $f_0$, and $n$, where possible, which will be important for predictions of future lightning radio observations.

The frequency spectrum, $E(f)$, is the Fourier transform (FT) of the electric field, $E(t)$. Eq. \ref{eq:8}. defines the relation between the electric field in the time domain and the frequency domain,

\begin{equation} \label{eq:8}
E(f) = \int_{0}^{\infty}{e^{-i 2 \pi f t}E(t)dt},
\end{equation}
where $i = \sqrt{-1}$ and $f$ is the frequency in [Hz].

I derive the following analytical form of the FT of the electric field, $E(t)$ given in Eq. \ref{eq:10}:
\begin{dmath}\label{eq:88}
E(f) = \frac{2 i_0 \v v_0}{4\pi \epsilon_0}\left[\frac{1}{c r^2}\left(\frac{1}{(\alpha+i 2 \pi f)^2}-\frac{1}{(\beta+ i 2 \pi f)^2}\right)+\frac{1}{c^2 r}\left(\frac{\beta}{(\beta+ i 2 \pi f)^2}-\frac{\alpha}{(\alpha + i 2 \pi f)^2}-\frac{1}{\beta + i 2 \pi f}+\frac{1}{\alpha + i 2 \pi f}\right)\right],
\end{dmath}

\noindent In practice the Fourier Transform of a function is usually calculated using the numerical Fast Fourier Transform (FFT) method. For basic functions such as Eq. \ref{eq:4}, it is easy to determine the analytical form of the frequency spectrum, however for more complicated ones, such as the electric field derived from the Heidler-function (Sect. \ref{ssec:heid}), it is easier to use the FFT, which I am going to use in the following calculations\footnote{IDL's inbuilt \textit{fft} function}. Generally, the FT of a function is calculated when the function contains a periodic signal. The challenge in using numerical FFT on the non-periodic electric field of a lightning discharge is, firstly, numerical integrations cannot be made to infinity, which means a part of the function has to be cut out, and secondly, by cutting the function a sharp edge will appear at the point where it was cut out which results in a badly calculated FT. To solve this problem the Hann Window function was used which is a technique for signal processing and it smooths the edges of the curve so that the FFT could be calculated properly. Furthermore, a numerical FT code does not know the time step with which the data is sampled, it assumes that the sample has a step-size of 1 s. This is not true in the case in this Chapter, and to correct for it I multiply the result of the FFT my own time-step, which depends on the duration of the discharge.

The power spectrum, $P(f)$, is the FT of the square of the electric field, $E(t)^2$. By squaring the electric field one obtains the power radiated by the field. To express the power in the frequency domain, I calculate the Fourier Transform:

\begin{equation} \label{eq:pow}
P(f) = \int_{0}^{\infty}{e^{-i 2 \pi f t}E(t)^2dt}, 
\end{equation}

\noindent I obtain the exact form of Eq. \ref{eq:pow} with the help of \textbf{WolframAlpha}\footnote{https://www.wolframalpha.com/}, and with $E(t)$ as in Eq. \ref{eq:10}:

{\small
\begin{dmath} \label{eq:pow_an}
P(f) = \left(\frac{2 i_0 v_0}{4 \pi \epsilon_0}\right)^2 \left[\frac{1}{c^2 r^4} \left( \frac{1}{4(\alpha+i \pi f)^3} + \frac{1}{4(\beta+i \pi f)^3} - \frac{2}{(\alpha+\beta+2 i \pi f)^3}\right) + \frac{1}{c^4 r^2} \left(\frac{1}{2(\alpha+i \pi f)} + \frac{1}{2(\beta+i \pi f)} + \frac{\alpha^2}{4(\alpha+i \pi f)^3} + \frac{\beta^2}{4(\beta+i \pi f)^3} + \frac{2(\alpha+\beta)}{(\alpha+\beta+2 i \pi f)^2} 
- \frac{4 \alpha \beta}{(\alpha+\beta+2 i \pi f)^3} - \frac{2}{\alpha+\beta+2 i \pi f} - \frac{\alpha}{2(\alpha+i \pi f)^2} - \frac{\beta}{2(\beta+i \pi f)^2} \right) + 
\frac{1}{c^3 r^3} \left(\frac{4 (\alpha+\beta)}{(\alpha+\beta+2 i \pi f)^3}  
- \frac{4}{(\alpha+\beta+2 i \pi f)^2} - \frac{\alpha}{2(\alpha+i \pi f)^3} - \frac{\beta}{2(\beta+i \pi f)^3} + \frac{1}{2(\alpha+i \pi f)^2} + \frac{1}{2(\beta+i \pi f)^2}\right)\right].
\end{dmath}
}

Then, the power spectral density, $P'(f)$ [W Hz$^{-1}$], at the source of the emission can be expressed by:

\begin{equation} \label{eq:pow2}
P'(f) = P(f) 2 c \pi \epsilon_0 r^2.
\end{equation} 

\noindent Equation \ref{eq:pow2} is the time-averaged power, assuming sinusoidal wave functions\footnote{The time-averaged power, in general, is the intensity, $I$, times the area of a sphere, where the source radiates to: $<P> = I 4 \pi r^2$. The time average of a sinusoidal wave function is $1/2$, resulting in $I = (1/2) c \epsilon_0 E^2$, where $E$ is the radiating electric field. Hence, $P(f)$ in Eq. \ref{eq:pow2} is multiplied by 2 instead of 4.}. $P'(f)$ will depend on the frequency at which I measure it. To obtain the total released power, one has to integrate $P'(f)$ over the frequency range it has been released at.

\begin{figure}
  \centering
  \includegraphics[scale=0.4]{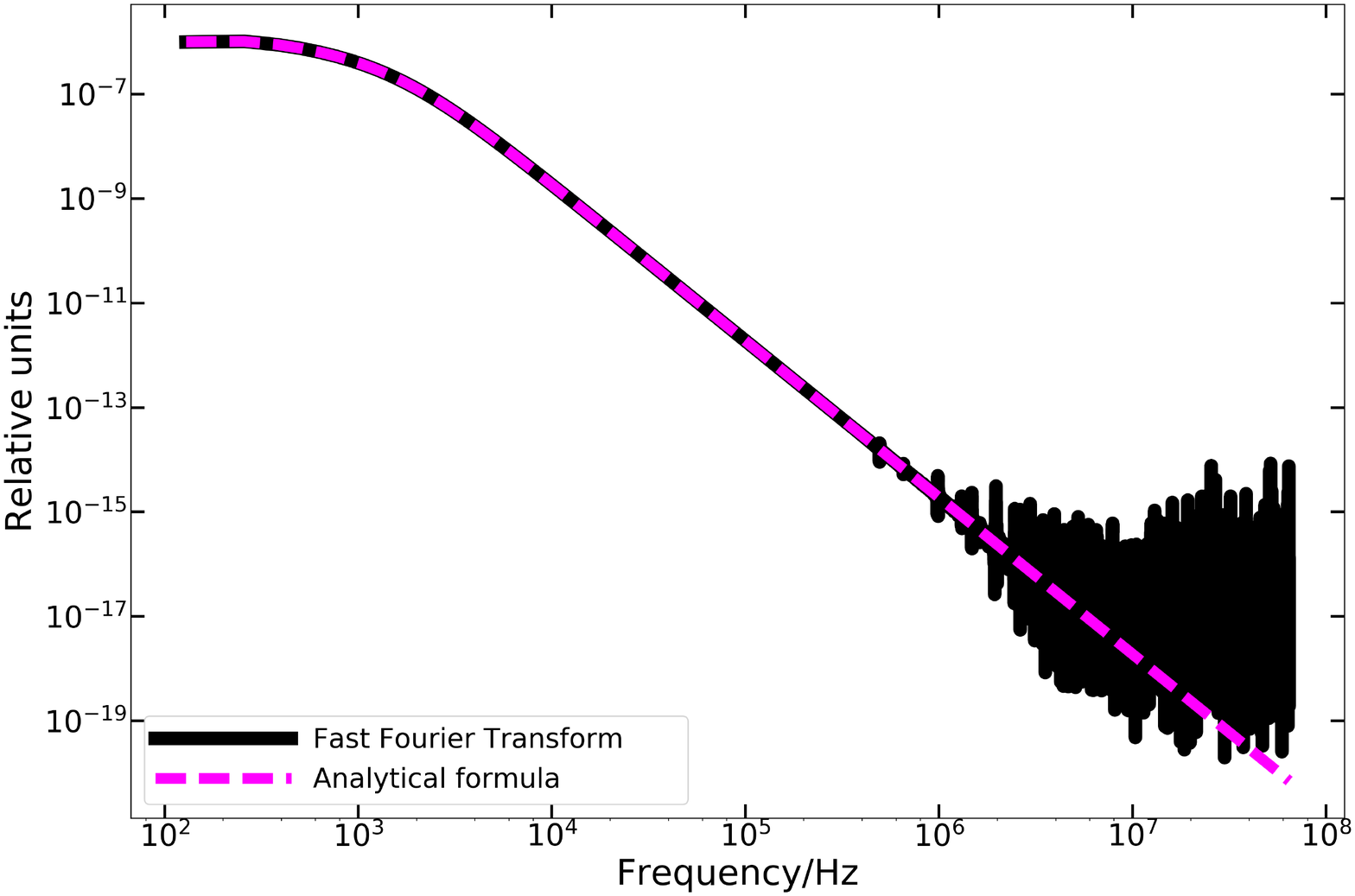}
  \caption{Numerical Fourier transform (black, solid) compared to the analytical Fourier transform (magenta, dashed) of the same electric field function. Used parameters: $\alpha = 4.4 \times 10^4$ s$^{-1}$, $\beta = 4.6 \times 10^5$ s$^{-1}$, $i_0 = 30$ kA, $\v v_0 = 0.3$ c, $r = 1000$ km. The numerical version (calculated by IDL's \textit{fft} function) carries a forest-like noise at very high frequencies, which is eliminated from the model by using the analytical formula (Eq. \ref{eq:pow_an}).}
  \label{fig:facom}
	\vspace{0.5cm}
\end{figure}

Figure \ref{fig:fpbh} shows the frequency and power spectra resulting from the electric fields shown in Fig. \ref{fig:ebh}. It is seen from Fig. \ref{fig:fpbh} that the same trend of matching the spectra as a result of the bi-exponential and the Heidler functions that has been seen in Figs \ref{fig:ibh}-\ref{fig:ebh} is also present. For example, when $\eta=1$ in the Heidler function, and $\alpha$ and $\beta$, parameters of the bi-exponential current function, are expressed by $\tau_1$ and $\tau_2$ of the Heidler function (Eq. \ref{eq:ab}) as in panel [6] of Figs \ref{fig:ibh}, \ref{fig:ebh} and \ref{fig:fpbh}, then the frequency and power spectra will represent the same spectra resulting from the electric fields of the two current functions. At very high frequencies a forest-like noise is introduced with the numerical FFT. This noise is eliminated from the model by using the analytical form of the FT (Fig. \ref{fig:facom}).

To calculate the discharge energy, I need to know how the power spectrum varies with frequency (what is its slope at higher frequencies, $n$) and where its peak ($f_0$) is. To obtain the slope, I fit a linear function to the part of the spectrum that is at frequencies larger than the peak frequency. The peak is obtained from the duration of the discharge (Eq. \ref{eq:taufr}). In the further sections, I use the analytical form of the Fourier Transform obtained from the bi-exponential function (Eqs \ref{eq:88} and \ref{eq:pow_an}), as it does not introduce a numerical noise at large frequencies (Fig. \ref{fig:facom}).

%__________________________________________________________________
\subsection{Radiated discharge energy and radiated power density} \label{sec:disen}

The radiated discharge energy, $W_{\rm rad}$, and the discharge dissipation energy, $W_d$, are calculated in order to estimate how energetic lightning discharges can be, occurring on extrasolar objects with different atmospheric conditions (T$_{\rm gas}$, p$_{\rm gas}$, chemical composition, etc.). By knowing the energy dissipated form lightning, I can estimate the changes in the local chemical composition of the atmosphere and determine whether observable signatures can be produced as a result of the production of non-equilibrium species (Chapter \ref{chap:ligsig}). I obtain the total radiated discharge energy, $W_{\rm rad}$ [J], from the total radiated power, $P_{\rm rad}$ [W], released during the discharge: 
\begin{equation} \label{eq:nwd}	%nwd == new wd
W_{\rm rad} = P_{\rm rad} \tau,
\end{equation}

\noindent where $\tau$ [s] is the discharge duration.
The total radiated power is given by 
\begin{equation} \label{eq:ptot}
P_{\rm rad} = \int_{f_{\rm min}}^{f_{\rm max}}{P'(f') df'}.
\end{equation}

\noindent The integral boundaries are calculated when the time sample of the model is converted into frequencies. $f_{\rm min} = t_{\rm max}^{-1}$ and $f_{\rm max} = t_{\rm min}^{-1}$, where $t_{\rm min}+t_{\rm max} = \tau$.

I mention that the total dissipation energy of lightning, $W_d$, is obtained from $W_{\rm rad}$, assuming a radio efficiency, $k$. $k$ represents the amount of energy radiated into the radio frequencies from the total dissipated energy, and is between 0 and 1:

\begin{equation} \label{eq:wd2}
W_d = \frac{1}{k} W_{\rm rad}
\end{equation}

%__________________________________________________________________
%__________________________________________________________________
\section{Computational approach} \label{sec:param}

\begin{figure}
  \centering
  \includegraphics[scale=0.2]{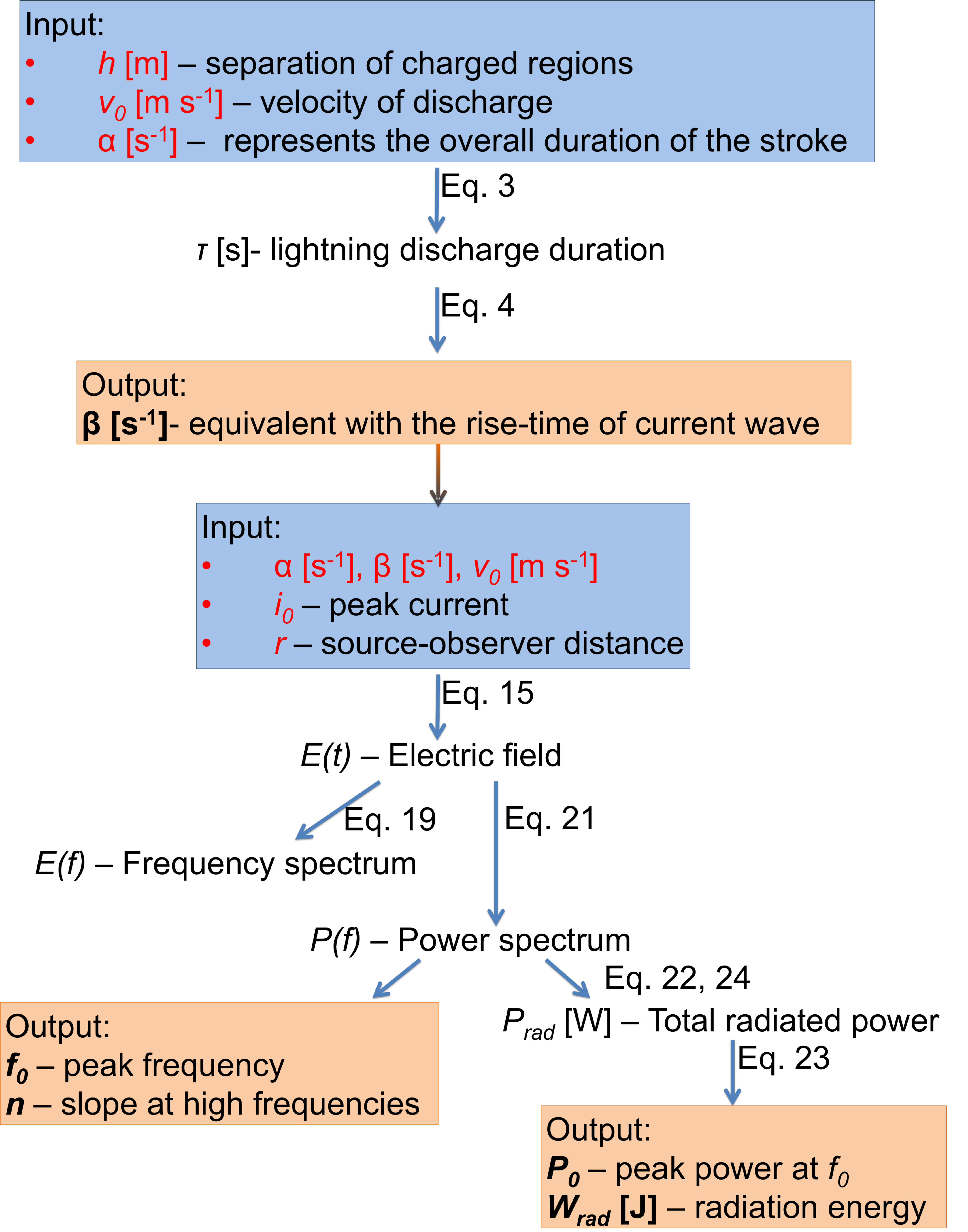}
  \caption{Flow chart of the lightning radio model and its input parameters and outputs described in Sects \ref{sec:model} and \ref{sec:param}. The chart works as a guide of the equations used in the computational approach. 
	}
  \label{fig:chart2}
	\vspace{0.8cm}
\end{figure}

In this section, I summarize my computational approach (Fig. \ref{fig:chart2}): Starting with a bi-exponential current function (Eq. \ref{eq:1}), I calculate the electric field and its frequency and power spectra. Next, I obtain properties of the power emitted at high frequencies: $f_0$ is the frequency at which the peak of the power is emitted; $n$ is the spectral roll-off of the power spectrum at frequencies $f>f_0$. These properties of the power spectrum are necessary when making predictions of the lightning radio power emitted at chosen observed frequencies. Finally, I estimate the lightning energy radiated into the radio frequencies. The free parameters in the model are:

\vspace{0.2cm}

\noindent $r$: The distance between the source and the observer. This would be the distance between the extrasolar object and Earth.

\vspace{0.2cm}

\noindent $h$: The characteristic length of the charge separation, or length of the discharge. $h$ is taken from \citet[][their fig. 9, right]{bailey2014}, who calculated it for several atmosphere models. The length, $h$ and the properties of the atmospheres (effective temperature T$_{\rm eff}$, surface gravity log($g$), metallicity [M/H]) are listed in Table \ref{table:h}.

\vspace{0.2cm}

\noindent $\v v_0$: The velocity of the return stroke was taken from \citet{bruce1941}, who estimated $\v v_0 = 0.3 {\rm c} = 8.0 \times 10^7$ m s$^{-1}$ for Earth lightning. 

\vspace{0.2cm}

\noindent $i_0$: The current peak is based on Earth values as well. I used $30$ kA, which is the average current in a terrestrial negative return stroke, as the reference value and tested how the output changes when I only change the current. 

\vspace{0.2cm}

My method is summarized on a flow chart in Fig. \ref{fig:chart2}: From the two variables $h$ and $\v v_0$, one can determine $\tau$, the discharge duration using Eq. \ref{eq:tau1}. After that the following calculations are made: $\alpha$, parameter of the bi-exponential current function (Eq. \ref{eq:1}), is randomly picked from a Gaussian distribution, which has a mean and standard deviation that ensure that $\alpha$ is $\sim 1$ orders of magnitude smaller than $\beta$. This is an empirical choice based on $\alpha$ and $\beta$ listed in the literature (Table \ref{table:3}). From $\tau$ and $\alpha$ one can determine $\beta$, the other parameter of the current function, using Eq. \ref{eq:tau}. The next step is to determine the electric field produced by the current. For that I use Eq. \ref{eq:10}, with physical quantities having SI units. I obtain the frequency and power spectra from the electric field as explained in Sect. \ref{sec:freqsp}. The power spectrum represents the distribution of radiated power in frequency space. I also estimate the total power of lightning radiated at radio frequencies (Eqs \ref{eq:pow2} and \ref{eq:ptot}), which will determine the energy radiated at radio from lightning discharges (Eq. \ref{eq:nwd}). 

%Table h + atmospheres
\begin{table}
 \begin{center}
 \scriptsize
 \caption{Total length of the discharge ($h$) adapted from \citet[][their fig. 9, right]{bailey2014}, the minimum number of charges ($Q_{\rm min}$) needed to initiate a discharge adapted from \citet[][their fig. 7]{bailey2014}, and the properties of the atmospheres where the discharges lengths are reached. The subscript '1' indicates an atmosphere of solar metallicity ([M/H] = 0.0), while subscript '2' indicates an atmosphere of sub-solar metallicity ([M/H] = -3.0). The peak current, $i_0$, is calculated by Eq. \ref{eq:iq} from $Q_{\rm min}$, and the discharge duration, $\tau$, is obtained from $h$ through Eq. \ref{eq:tau1} with $\v v_0=0.3$c (Fig. \ref{fig:exotau}).}
  \begin{tabular}{@{}lllllllllll@{}}	
	\hline
	 & T$_{\rm eff}$ [K] & $h_1$ [m] & $\tau_1$ [s] & $Q_{\rm min,1}$ [C] & $i_{0,1}$ [A] & $h_2$ [m] & $\tau_2$ [s] & $Q_{\rm min,2}$ [C] & $i_{0,2}$ [A] \\ 
	\hline
	\multirow{4}{*}{\vtop{\hbox{\strut Brown dwarf}\hbox{\strut log($g$) = 5.0}}} & 1500 & 168 & $2.1 \times 10^{-6}$ & 70 & $3.3 \times 10^7$ & 890 & $1.1 \times 10^{-5}$ & 312 & $2.8 \times 10^7$ \\
	 & 1600 & 66 & $8.2 \times 10^{-7}$ & 33 & $4.0 \times 10^7$ & 753 & $9.4 \times 10^{-6}$ & 237 & $2.5 \times 10^7$ \\	
	 & 1800 & 58 & $7.3 \times 10^{-7}$ & 22 & $3.0 \times 10^7$ & 623 & $7.8 \times 10^{-6}$ & 216 & $2.8 \times 10^7$ \\
	 & 2000 & 27 & $3.3 \times 10^{-7}$ & 12 & $3.6 \times 10^7$ & 286 & $3.6 \times 10^{-6}$ & 80 & $2.2 \times 10^7$ \\	
	\hline
  \end{tabular}
	
	\vspace{0.5cm}
	
  \begin{tabular}{@{}llllllllll@{}}	
	\hline
	 & T$_{\rm eff}$ [K] & $h_1$ [m] & $\tau_1$ [s] & $Q_{\rm min,1}$ [C] & $i_{0,1}$ [A] & $h_2$ [m] & $\tau_2$ [s] & $Q_{\rm min,2}$ [C] & $i_{0,2}$ [A] \\ 
	\hline
	\multirow{4}{*}{\vtop{\hbox{\strut Giant gas planet}\hbox{\strut log($g$) = 3.0}}} & 1500 & 69 & $8.6 \times 10^{-7}$ & $3.8 \times 10^3$ & $4.5 \times 10^9$ & 2494 & $3.1 \times 10^{-5}$ & $1.0 \times 10^5$ & $3.2 \times 10^9$ \\
	 & 1600 & 20 & $2.5 \times 10^{-7}$ & $1.8 \times 10^3$ & $7.5 \times 10^9$ & 2370 & $3.0 \times 10^{-5}$ & $8.2 \times 10^4$ & $2.8 \times 10^9$ \\	
	 & 1800 & 19 & $2.3 \times 10^{-7}$ & $1.7 \times 10^3$ & $7.2 \times 10^9$ & 1844 & $2.3 \times 10^{-5}$ & $6.2 \times 10^4$ & $2.7 \times 10^9$ \\
	 & 2000 & 15 & $1.9 \times 10^{-7}$ & $1.1 \times 10^3$ & $5.6 \times 10^9$ & 942 & $1.2 \times 10^{-5}$ & $3.4 \times 10^4$ & $2.9 \times 10^9$ \\	
	\hline
  \end{tabular}
	\label{table:h}
 \end{center}
\vspace{0.5cm}
\end{table}
%Table 

I also consider $i_0$ obtained from the minimum number of charges, $Q_{\rm min}$, necessary to overcome the electrostatic breakdown field in extrasolar atmospheres according to \citet[their fig. 7]{bailey2014}. \citet{bailey2014} used the classical breakdown field to determine $Q_{\rm min}$, which does not include the idea of runaway breakdown \citep{rousseldupre2008}, therefore overestimates the critical field strength necessary to initiate a breakdown \citep{bailey2014}. This suggests that the obtained $Q_{\rm min}$ for each atmosphere will be an upper limit necessary for breakdown and the actual values in nature may be lower. 
Here, $Q_{\rm min}$ replaces $i_0$ as an input parameter in the model.
The used values are listed in Table \ref{table:h}. Based on Eq. \ref{eq:charge}, I obtained $i_0$ from $Q_{\rm min}$:
\begin{equation} \label{eq:iq}
i_0 = \frac{\Delta Q}{\Delta t},
\end{equation}

\noindent where $\Delta Q \equiv Q_{\rm min}$ and $\Delta t \equiv \tau$ (discharge duration).

The results of the calculations are the properties of the power spectrum, $f_0$ and $n$, the duration of the discharge, $\tau$, the peak current, $i_0$, the total radio power, $P_{\rm rad}$, the peak radio power density, $\frac{P_0}{\Delta f}$, and the radio energy of lightning, $W_{\rm rad}$. Some of the outputs, $f_0$, $n$, $\frac{P_0}{\Delta f}$, will be used to estimate the radiated power density, and hence the observability of lightning radio emission, at a given frequency, where observations are planned (Chapter \ref{chap:concl}).

%__________________________________________________________________
%__________________________________________________________________
\section{Model performance} \label{sec:val}

In this section, I test the model presented in Sects \ref{sec:model} and \ref{sec:param} with observed and modelled terrestrial, Jovian and Saturnian lightning properties and model parameters.
I can use these tests, since my model only becomes dependent on the atmosphere of planets, exoplanets, and brown dwarfs, once, in the final step, I include the extension of the discharge channel and the charges accumulated in that channel. For this final step, values specific for the Solar System planets can be used. Therefore, the model is universal and can be used for Solar System as well as extrasolar lightning modelling. 

%__________________________________________________________________
\subsection{Earth} \label{sec:earval}

\begin{figure}
  \centering
  \includegraphics[scale=0.089]{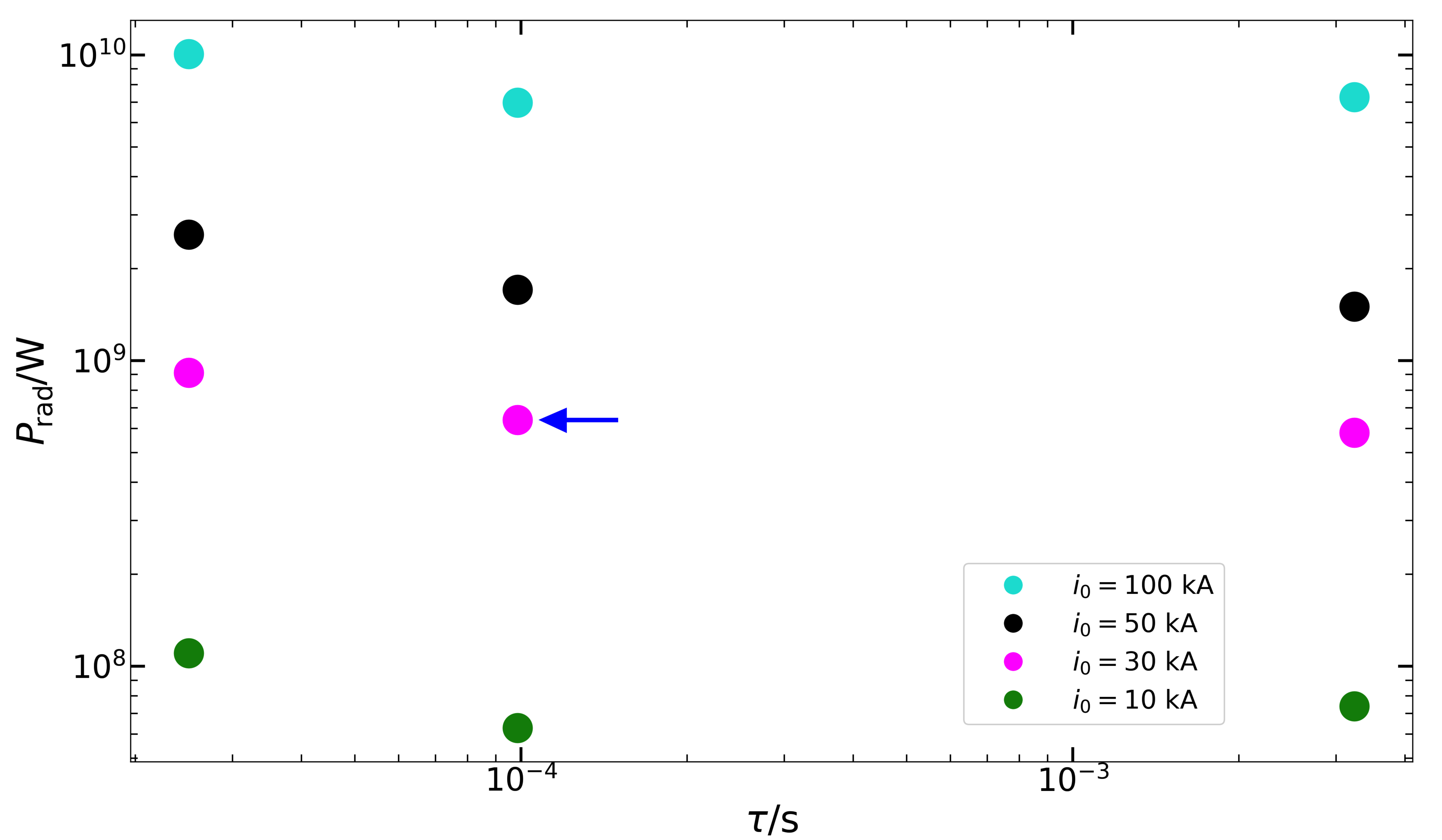}
	\includegraphics[scale=0.37]{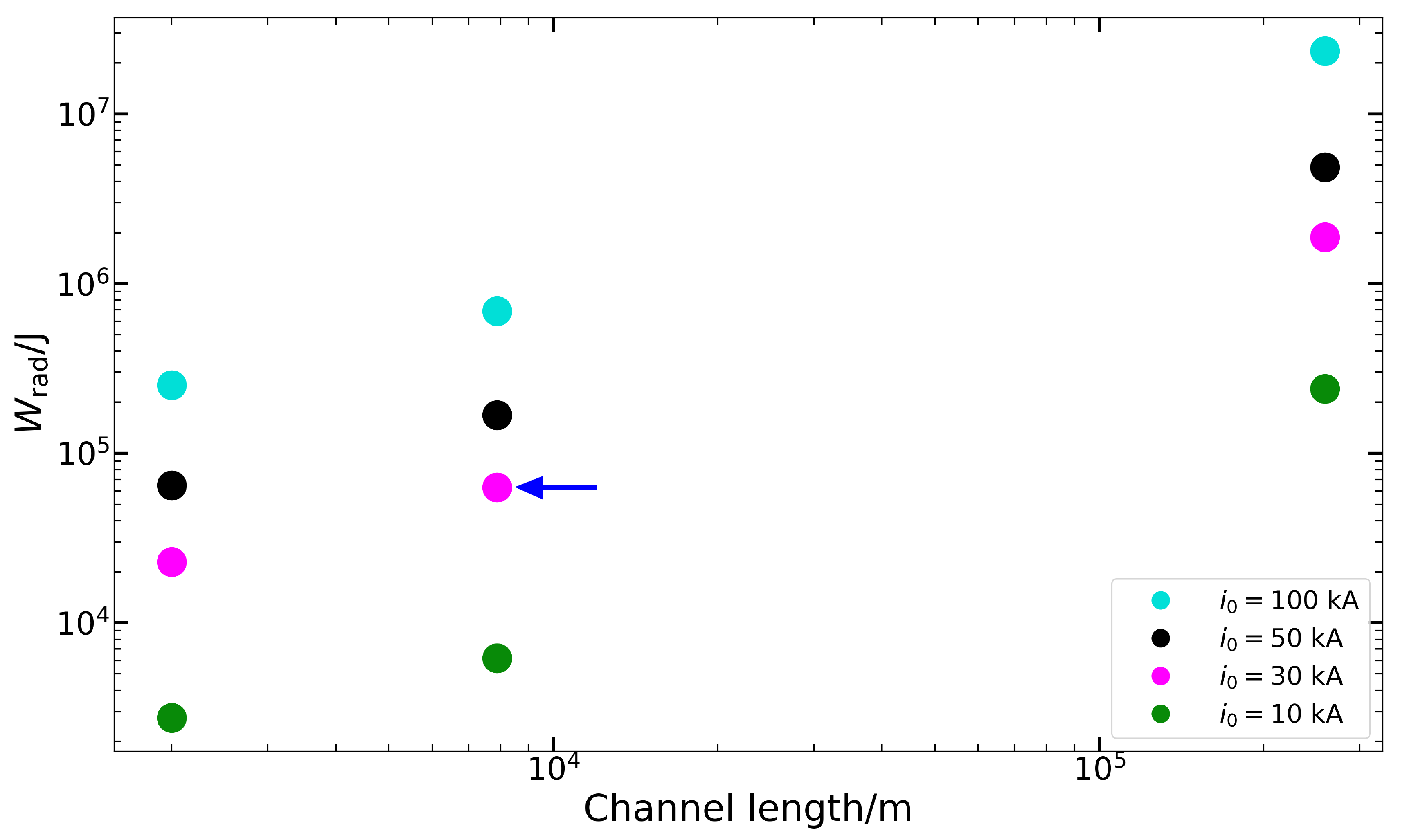}
  \caption{Total radio power of lightning, $P_{\rm rad}$, vs discharge duration, $\tau$ (top), and radiated energy, $W_{\rm rad}$, vs discharge length, $h$ (bottom). The four colours indicate four different peak currents used for the calculations. The two panels depend on each other through $h$, as given by Eq. \ref{eq:tau1}. The input parameters, $h$ and resulting $\tau$, and $i_0$ are listed in Table \ref{table:yv}. Even though $P_{\rm rad}$ does not change significantly with $\tau$ and hence with $h$, $W_{\rm rad}$ will be higher if $h$, and therefore $\tau$ is larger, because W$_{\rm rad}$ is determined through Eq. \ref{eq:nwd}. The blue arrow points to a typical value of an Earth lightning stroke, with $\tau = 100$ $\mu$s, and $W_{\rm rad} \sim 6 \times 10^4$ J \citep[][table 6.2]{volland1984}.}
  \label{fig:yv}
\end{figure}

\begin{table}[!h] 
	\small
 \begin{center}
 \caption{Test parameters for Fig. \ref{fig:yv}. These values were selected in order to represent discharge lengths, $h$, and peak currents, $i_0$, observed and modelled on Earth, therefore suggesting the most realistic parameter combinations for an Earth-like discharge. $\tau$ depends on $h$ through Eq. \ref{eq:tau1}, where $\v v_0 = 0.3 {\rm c}$ is constant for each case.}
  \begin{tabular}{@{}lll:ll@{}} 
	\hline
	$h$ [m] & Reference ($h$) & $\tau$ [s] & $i_0$ [A] & Reference ($i_0$) \\
	\hline
	$2 \times 10^3$ & \citet{baba2007} & $2.5 \times 10^{-5}$ & $10^4$ & arbitrary example \\
	$7.89 \times 10^3$ & \citet[][p. 124]{rakov2003} & $10^{-4}$ & $3 \times 10^4$ & \citet{farrell1999} \\
	$2.59 \times 10^5$ & \citet{bruning2015} & $3 \times 10^{-3}$ & \vtop{\hbox{\strut $5 \times 10^4$}\hbox{\strut $10^5$}} & \vtop{\hbox{\strut \citet{heidler2002}}\hbox{\strut arbitrary example}} \\
	\hline
  \end{tabular}
	\label{table:yv}
 \end{center}
 \vspace{0.5cm}
\end{table}

%Table Validation
\begin{table*}
\resizebox{\columnwidth}{!}{
\begin{threeparttable}  
\footnotesize
\setlength{\tabcolsep}{5.5pt}
 \caption{Testing the model for Earth discharges based on \citet{volland1984}, table 6.1, first row ("$G-R_1$"). The first row serves as reference for input values and comparison for the results of this model. $k$ is the radio energy efficiency used in Eq. \ref{eq:wd2}. Steps of calculation: from $\alpha$ and $\beta$ obtain $\tau$ with Eq. \ref{eq:tau}; from $\tau$ calculate $h$ (Eq. \ref{eq:tau1} with $\v v_0 = 0.3$c), $f_0$ (Eq. \ref{eq:taufr}), and $Q$ (Eq. \ref{eq:iq}). Rest as described in Sect. \ref{sec:param}.}
  \begin{tabular}{@{}lllllllllll@{}} 
	\hline
	 & $i_0$ [kA] & $\alpha$ [s$^{-1}$] & $\beta$ [s$^{-1}$] & $\tau$ [s] & $h$ [m] & $|Q|$ [C] & $f_0$ [kHz] & $W_d$ [J] & $W_{\rm rad}$ [J] & $k$ \\
	\hline
	\citet[][table 6.2]{volland1984} & 30 & $2 \times 10^4$ & $2 \times 10^5$ & $9.9 \times 10^{-5}$\tnote{(1)} & 7890 & 1.35 & 10.1 & $6.97 \times 10^7$ & $1.46 \times 10^6$\tnote{(2)} & 0.021 \\
	\hdashline
	Model input & 30 & $2 \times 10^4$ & $2 \times 10^5$ & - & - & - & - & - & - & 0.021 \\
	Model output (this work) & - & - & - & $9.93 \times 10^{-5}$ & 7948 & 2.98 & 10.06 & $6.96 \times 10^6$\tnote{(3)} & $1.46 \times 10^5$ & - \\
	\hline
  \end{tabular}
  \begin{tablenotes}
	\item[(1)] $\tau$ is not listed in \citet[][table 6.2]{volland1984}. Applying Eqs \ref{eq:tau1} and \ref{eq:tau} on the values in the first row, the obtained $\tau$ is 98.5 and 99.35 $\mu$s, respectively.
	\item[(2)] $W_{\rm rad}$ obtained from $W_d$ and $k$, which are listed in \citet[][table 6.2]{volland1984}.
	\item[(3)] Calculated from $W_{\rm rad}$ (model output) with Eq. \ref{eq:wd2}.
  \end{tablenotes}
	\label{table:val}
 \vspace{0.5cm}
\end{threeparttable}
}
\end{table*}

Earth lightning is the most well-studied form of lightning discharges in the Solar System. The modelling of Earth lightning goes back to the first half of $20^{\rm th}$ century \citep[e.g.][]{bruce1941, drabkina1951}, and then was further developed in the 1970s-80s \citep[e.g.][and references therein]{uman1969, heidler1985, rakov2003}. Therefore, I apply measured parameters of Earth lightning discharges as a template to test the model.  First, I use various discharge extension, $h$, and peak current, $i_0$, combinations to obtain the total radio power, $P_{\rm rad}$, and radiated energy, $W_{\rm rad}$. The used values are listed in Table \ref{table:yv} and the results are shown in Fig. \ref{fig:yv}. The top panel of Fig. \ref{fig:yv} presents the obtained $P_{\rm rad}$ vs the discharge duration, $\tau$. The different colours indicate different $i_0$-s used in the model. The bottom panel of the same figure shows how the obtained $W_{\rm rad}$ varies with $h$. $h$ and $\tau$ are directly proportional to each other as in Eq. \ref{eq:tau1}, with $\v v_0 = 0.3$c. The figure shows that the larger the peak current the more power and energy is released from a lightning stroke. It also demonstrates that the slower the discharge the more energy is radiated from lightning, even though the power released rather decreases with larger $\tau$. The blue arrow indicates a lightning return stroke with $\tau = 100$ $\mu$s, which corresponds to a peak frequency, $f_0 = 10$ kHz, and $i_0 = 30$ kA. The radio energy of such a stroke is $\sim 6 \times 10^4$ J, while the radio power is $\sim 6 \times 10^8$ W. \citet{borovsky1998} estimated the energy dissipated from a return stroke from the electrostatic energy density stored around the lightning channel. They found the dissipation energy per unit length to be $2 \times 10^2 - 10^4$ J m$^{-1}$. They note that their result is in agreement with previous studies considering hydrodynamic models for lightning channel expansion \citep[e.g.][]{plooster1971}. \citet{plooster1971} found the total energy of lightning to be $4 \times 10^2 - 9 \times 10^2$ J m$^{-1}$ for $i_0=20$ kA, and $1.7 \times 10^3$ J m$^{-1}$ for $i_0 = 40$ kA. \citet{borovsky1998}, however, also mentions that their results are 1-2 orders of magnitude lower than calculated by e.g. \citet{krider1968}, who estimated the total dissipation energy from optical measurements assuming an optical efficiency of 0.38 to be $2.3 \times 10^5$ J m$^{-1}$. Assuming a channel length of $h = 7890$ m, as for the data point marked by a blue arrow in Fig. \ref{fig:yv}, the dissipated energy of lightning, $W_d$, according to the above authors is between $1.6 \times 10^6$ J and $1.8 \times 10^9$ J. Applying Eq. \ref{eq:wd2} with a radio efficiency $k=0.01$, the radiated energy in the radio band is between $1.6 \times 10^4$ J and $1.8 \times 10^7$ J. My value of $\sim 6 \times 10^4$ J is within this range, however, closer to the values obtained by \citet{borovsky1998} and \citet{plooster1971}.

For the second test, I set the model up to reproduce values found in table 6.2 of \citet{volland1984}. I chose the first row of the table ("$G-R_1$"), which was derived from parameters used in \citet{bruce1941}. It represents a first return stroke of a lightning discharge. I list the parameters and the results of my model in Table \ref{table:val}. The table suggest that my results of discharge energy are approximately one order of magnitude lower than the ones obtained by \citet{volland1984}.

%__________________________________________________________________
\subsection{Saturn} \label{sec:satval}

%Table Validation - Saturn
\begin{table*} 
\resizebox{\columnwidth}{!}{
\begin{threeparttable} 
\setlength{\tabcolsep}{5.5pt}	
\footnotesize
 \caption{Testing the model with \textbf{Saturnian values}. Two main approaches were followed: (1) The model was set up for a quick discharge, $\tau_{\rm stroke} = 1$ $\mu$s, as in \citet{farrell2007}. (2) The model was set up to reproduce the measured energy of SEDs and shape of their power spectra, with $\tau_{\rm stroke} = 100$ $\mu$s as in \citet{mylostna2013}. The extension of the discharge, $h$, was obtained from Eq. \ref{eq:tau1} with $\v v=0.3$ c. For a detailed description of the various cases, including references for input parameters, see Sect. \ref{sec:satval}. Input values are marked with italics.}
  \begin{tabular}{@{}llllllllllll@{}}
	\hline
	 & $\tau_{\rm stroke}$ [$\mu$s] & $\tau_{\rm SED}$ [s] & $i_0$ [kA] & $|Q|$ [C] & $P_{\rm rad}$ [W]\tnote{(1)} & $W_{\rm rad}$ [J]\tnote{(1)} & $W_{\rm SED}$ [J] & $W_d$ [J]\tnote{(2)} & $k$ & \vtop{\hbox{\strut stroke/}\hbox{\strut SED}} & \vtop{\hbox{\strut SED/}\hbox{\strut flash}} \\
	\hline
	(1a) & \textit{1} & $\mathit{10^{-6}}$ & \textit{30} & 0.03 & $6.4 \times 10^8$ & $6.4 \times 10^2$ & $6.4 \times 10^2$ & $6.4 \times 10^5$ & \textit{0.001} & \textit{1} & \textit{1} \\
	(1b) & \textit{1} & $\mathit{10^{-6}}$ & 35000 & 35 & $10^{15}$ & $10^9$ & $10^9$ & $\mathit{10^{12}}$ & \textit{0.001} & \textit{1} & \textit{1} \\
	 & \textit{1} & \textit{0.23} & 75 & 0.075 & $4.5 \times 10^9$ & $4.5 \times 10^3$ & $10^9$ & $\mathit{10^{12}}$ & \textit{0.001} & $2.3 \times 10^5$ & \textit{1} \\
	\hdashline
	(2) & \textit{100} & \textit{0.23} & 75 & 7.5 & $4.5 \times 10^9$ & $4.5 \times 10^5$ & $10^9$ & $\mathit{10^{12}}$ & \textit{0.001} & $2.3 \times 10^3$ & \textit{1} \\
	 & \textit{100} & \textit{0.035} & 135 & 13.5 & $1.6 \times 10^{10}$ & $1.6 \times 10^6$ & $5.5 \times 10^8$ & $\mathit{1.1 \times 10^{12}}$ & \textit{0.001} & 350 & \textit{2} \\
	\hline
  \label{table:val3}
  \end{tabular}
  \begin{tablenotes}
	\item[(1)] Here $P_{\rm rad}$ represents stroke power, and $W_{\rm rad}$ represents the stroke energy.
	\item[(2)] $W_d$ is the total dissipation energy of a lightning flash.
  \end{tablenotes}
 \end{threeparttable}
 }
 \vspace{0.5cm}
\end{table*}

\begin{table}
	\small 
	\begin{center}
	\caption{Used stroke durations for testing the model on Saturnian discharges, and the resulting parameters that depend on $\tau$ through Equations \ref{eq:tau1}, \ref{eq:tau}, \ref{eq:taufr}. These values are outputs from the model as described in Sect. \ref{sec:satval} and Table \ref{table:val3}.}
  \begin{tabular}{@{}llllll@{}}
  \hline
   & $\tau_{\rm stroke}$ [$\mu$s] & $\alpha$ [s$^{-1}$] & $\beta$ [s$^{-1}$] & $h$ [m] & $f_0$ [kHz] \\
	 \hline
  (1) & \textit{1} & $3.5 \times 10^6$ & $1.1 \times 10^7$ & 90 & 1000 \\
	(2) & \textit{100} & $3.3 \times 10^4$ & $1.2 \times 10^5$ & 9000 & 10 \\
  \hline
  \end{tabular}
	\label{table:satval_2}
	\end{center}
	\vspace{0.5cm}
\end{table}

I discussed Saturn Electrostatic Discharges (SED) in Chapter \ref{chap:liginout}, Sect. \ref{sec:int_sat}, however, for the sake of completeness, here I repeat and discuss in more detail the relevant information for my work. SEDs have been observed since \textit{Voyager 1} and \textit{2} passed by the planet \citep{warwick1981, zarka1983}. The measured SED spectrum shows a relatively flat part below ~10 MHz, and it becomes a bit steeper till 40 MHz (Voyager PRA cut-off limit) with a slope of $f^{-1}-f^{-2}$ \citep{zarka1983,zarka2004}. \citet{warwick1981} deduced the shortest time structure of SEDs to be 140 $\mu$s, while \citet{zarka1983} measured a burst duration of 30 to 450 ms from Voyager data. \textit{Cassini} data showed a slight roll-off of $f^{-0.5}$ of the spectrum at the range of 4$-$16 MHz, with power spectral density of 40 to 220 W Hz$^{-1}$, and bust duration of 15 to 450 ms \citep{fischer2006}. The peak frequency of SED emission cannot be determined from the data, which means it is below the ionospheric cut-off. Assuming an Earth-like discharge, with peak frequencies around 10 kHz, to reproduce the measured power densities \citep[on average $\sim 60$ W Hz$^{-1}$, ][]{zarka2004}, a very strong discharge is needed, with energies of the order of $10^{13}$ J. Therefore, \citet{farrell2007} suggested a much faster discharge, which would result in a peak frequency at $\sim 1$ MHz, and would shift the whole power spectrum to higher frequencies. The result of this would be that discharges less energetic than previously estimated \citep[$\sim 10^6$ J,][]{farrell2007} could produce the observed power density. However, this theory was excluded by the first optical detections of lightning on Saturn \citep{dyudina2010}. \citet{dyudina2010} measured an optical energy of $10^9$ J of a single lightning flash, while \citet{dyudina2013} obtained a total flash power of $10^{13}$ W and optical energies between $10^8-10^9$ J, both suggesting that the total energy of a lightning flash is of the order of $10^{12}$ J, assuming that 0.1\% of the total energy of lightning is radiated in the optical \citep{borucki1987}. Furthermore, \citet{mylostna2013} observed SEDs with the Ukrainian T-shaped Radio telescope (UTR-2), and mapped their temporal structure. They found that the finest structure observable was 100 $\mu$s short. They also measured a spectral roll off of $f^{-2}$ between 20 kHz and 200 kHz, and a peak frequency $f_0 = 17$ kHz. Their findings further support the super-bolt scenario of Saturnian lightning flashes.

In order to test the model, I set it up for various cases: 
(1) The model was set up for a quick discharge, $\tau_{\rm stroke} = 1$ $\mu$s, as in \citet{farrell2007}. (2) The model was set up to reproduce the measured total energy of a flash \citep[e.g.]{dyudina2010,dyudina2013,mylostna2013} and shape of SED power spectra \citep[e.g.][]{zarka1986,fischer2006}, with $\tau_{\rm stroke} = 100$ $\mu$s as in \citet{mylostna2013}. Part (1a) is a test of the model with an Earth-like current peak \citep[$i_0 = 30$ kA,][]{volland1984}, while Part (1b) takes into account the obtained total energy from optical measurements \citep[$W_d = 10^{12}$ J][]{dyudina2013}, and iterates the corresponding peak current (increasing $i_0$ with 5000 kA during each step). (1b) also considers an SED built up of several strokes (second row of part (1b) in Table \ref{table:val3}). Part (2) consists of various cases depending on how many strokes/SED and SED/flash are used, and each time it iterates the current peak (with 15 kA during each step) to match $W_d = 10^{12}$ J. In each case, the distance, $r$, was taken to be the distance between the \textit{Cassini} spacecraft and Saturn during the measurements, $r = 46.5$ $R_{\rm Sat}$ \citep{fischer2007}. I assume that the radio efficiency, $k$, is the same as the optical one determined for Jupiter by \citet{borucki1987}. Each time the steps described in Sect. \ref{sec:param} were followed (Fig. \ref{fig:chart2}), except the starting input parameter was $\tau$ and not $h$. Furthermore, I consider cases when the duration of an SED burst is not equal to the duration of a stroke or a flash. $\tau_{\rm SED} = 0.23$ s is the average value in \citet{fischer2006} for an SED burst duration, while $\tau_{\rm SED} = 0.035$ s is the duration \citet{dyudina2013} considered for their energy estimates. In this case, \citet{dyudina2013} also assumed that one flash has a duration of 70 ms, therefore, one flash consisting of two SEDs. The "stroke/SED" and "SED/flash" values listed in Table \ref{table:val3} are a direct result of the time considerations.

Table \ref{table:val3} lists the input parameters and the results of the tests. Case (1) shows the what a quick discharge would look like on Saturn. First, I assume that nothing else is know about SEDs than what was considered in \citet{farrell2007}, I also assume that one flash consists of one SED, which consists of one stroke. I find that such a discharge would release $6 \times 10^2$ J energy in the radio band, and dissipates $6 \times 10^5$ energy in total. Next, I apply the total dissipation energy previously obtained and confirmed by optical measurements, $W_d = 10^{12}$ J, to the quick discharge theory, and find that, an incredibly large, 35000 kA current is necessary to produce such and energy from one flash consisting of one SED made up of one stroke. Finally, it is known that SED bursts are longer in duration than 1 $\mu$s. Here, I apply the average SED duration found in \citet{fischer2006}, $\tau_{\rm SED} = 0.23$ s. I find that 75 kA current has to run through one stroke, to produce $4.5 \times 10^3$ J radio energy, and $2.3 \times 10^5$ strokes are needed to produce $W_d = 10^{12}$ J energy of a flash consisting of one SED with the duration of $\tau_{\rm SED} = 0.23$ s.

\begin{figure}
  \centering
  \includegraphics[scale=0.6]{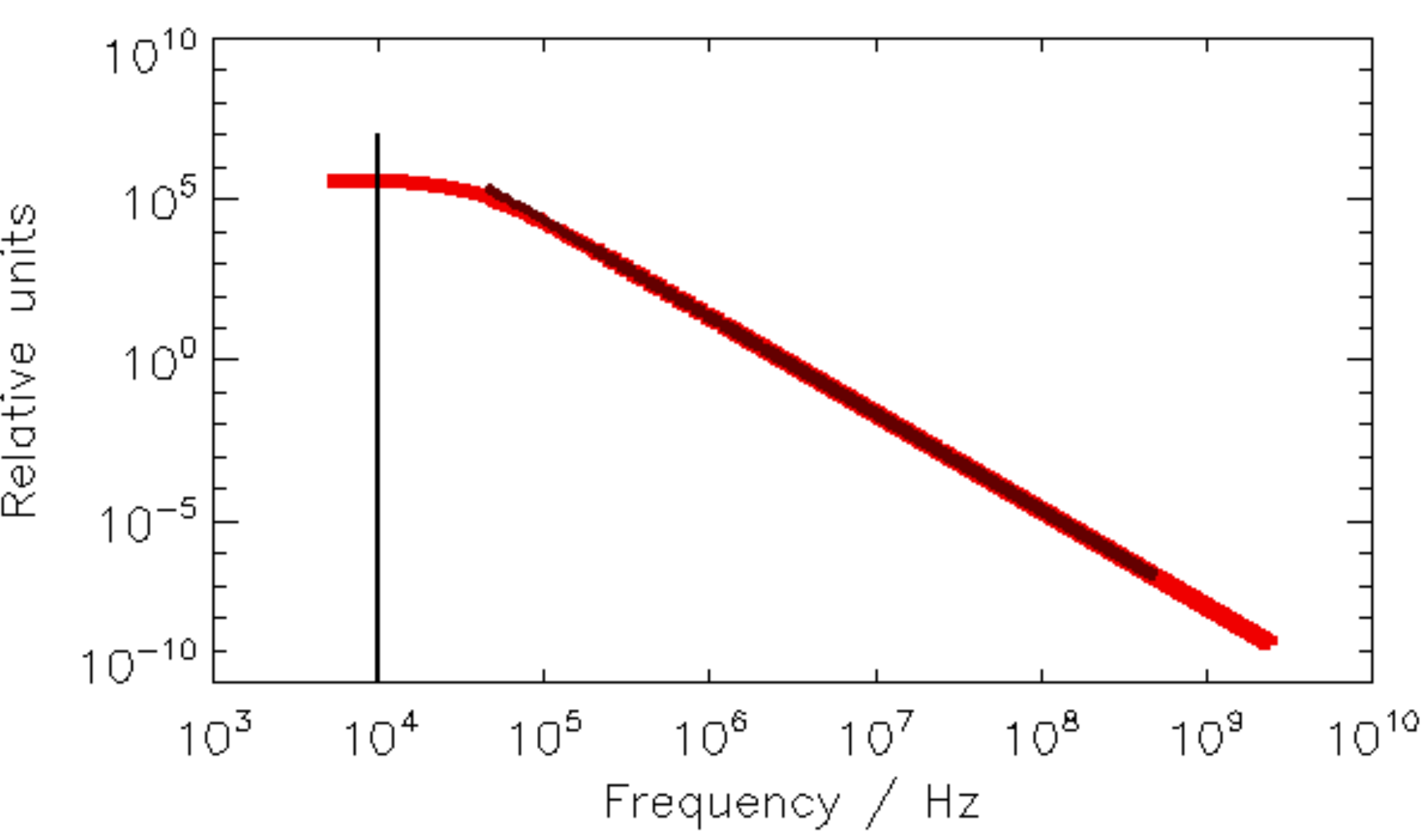} 
  \caption{Saturnian lightning electric field power spectrum (red). The fitted black line varies with frequency as $f^{-3}$, which is the slope of the power spectrum. The vertical black line indicates the peak frequency, $f_0 = 10$ kHz. The figure was produced with the parameters listed in the second row of approach (2) in Tables \ref{table:val3} and \ref{table:satval_2}.}
  \label{fig:satshape}
	\vspace{0.8cm}
\end{figure}

Case (2) uses more realistic values for SED and stroke duration, based on previous measurements. The stroke duration was assumed to be $\tau_{\rm stroke} = 100$ $\mu$s in each case, after \citet{mylostna2013}. First, I consider an SED of $\tau_{\rm SED} = 0.23$ s \citep[average in][]{fischer2006}, and one SED in one flash. The results suggest that 75 kA peak current would be enough to produce a dissipation energy of $10^{12}$ J, with 2300 strokes/SED with $W_{\rm rad} = 4.5 \times 10^5$ J/stroke. Second, I assume that $\tau_{\rm SED} = 0.035$ s, and one flash consists of two SED bursts. I find that 350 strokes/SED would produce enough energy, with $i_0 = 135$ kA to account for the total dissipation energy of a flash, $W_d = 10^{12}$ J. In Table \ref{table:satval_2} we further see that a quick, $\tau = 1$ $\mu$s, discharge would propagate to an unrealistic, 90 m, while a slower discharge of $\tau = 100$ $\mu$s, would result in an extension, $h = 9$ km, similar to Earth discharges.

Finally, I also tested the model against the shape of the observed power spectrum. Figure \ref{fig:satshape} shows a representative curve. All other curves show the same shape, with a shifted peak depending on the discharge duration. This suggests that my model, which is based on a simple, vertical dipole radiation model, with no branches and tortuosity of the channel, does not reproduce the shape of the power spectrum well. The resulting power spectrum shows a slope of $f^{-3}$ ($n = 3$) at high frequencies, no matter what the input parameters are. Since for Saturn, a flatter spectrum has been observed, my results may underestimate the power released at these frequencies, and therefore, overestimate the peak current obtained for each case in this section.

%__________________________________________________________________
\subsection{Jupiter} \label{sec:jupval}

%Table Validation - Jupiter
\begin{table*}  
 \resizebox{\columnwidth}{!}{
 \begin{threeparttable}
 \caption{Testing the model with \textbf{Jovian values}. Approaches (i) and (ii). (i) The model is set up to reproduce values obtained from the data of the Galileo probe by \citet{rinnert1998}. For the sake of comparison, in this approach I use $\v v=0.1$c instead of $\v v = 0.3$c, which is used in all other approaches. (ii) The discharge duration, $\tau$, is the average of the duration interval (260$-$520 $\mu$s) given by \citet{rinnert1998}, and the radio efficiency is taken to be the same as the optical efficiency determined by \citet{borucki1987}. The difference between (ii,a) and (ii,b) is the used radio efficiency. In each case, the peak current is iterated so that the final dissipation energy would reach the pre-set value of $10^{12}$ J. For further information, see Sect. \ref{sec:jupval}.}
  \begin{tabular}{@{}llllllllll@{}}
	\hline
   & $\tau$ [$\mu$s] & $i_0$ [kA] & $|Q|$ [C] & $M(t)$ [C m] & $h$ [m] & $P_{\rm rad}$ [W] & $W_{\rm rad}$ [J] & $W_d$ [J] & $k$ \\
	\hline
	(i), \citet{rinnert1998} & 240 & $6 \times 10^3$ & 1500 & $10^7$ & 7000 & $10^{14}$ & $2.5 \times 10^{10}$ & $10^{12}$ & 0.025 \\
	Model input & 240 & - & - & - & - & - & - & $10^{12}$ & 0.025 \\
	Model output (this work) & - & $2.5 \times 10^4$ & 5930 & $4.3 \times 10^7$ & 7190 & $10^{14}$ & $2.6 \times 10^{10}$ & - & - \\
	\hdashline
	(ii,a), Model input & 390\tnote{(1)} & - & - & - & - & - & - & $10^{12}$ & 0.001\tnote{(2)} \\
	Model output (this work) & - & $2.8 \times 10^3$ & 1100 & $3.87 \times 10^7$ & $3.5 \times 10^4$ & $3 \times 10^{12}$ & $1.1 \times 10^9$ & - & -  \\
	(ii,b), Model input & 390\tnote{(1)} & - & - & - & - & - & - & $10^{12}$ & 0.025 \\
	Model output (this work) & - & $1.24 \times 10^4$ & 4850 & $1.7 \times 10^8$ & $3.5 \times 10^4$ & $6.6 \times 10^{13}$ & $2.6 \times 10^{10}$ & - & -  \\
	\hline
  \end{tabular}
	\label{table:val2}
  \begin{tablenotes}
 	\item[(1)] Average duration of 260 $\mu$s and 520 $\mu$s from \citet{rinnert1998}.
	\item[(2)] Optical efficiency from \citet{borucki1987}. I assume that the radio efficiency is the same.
  \end{tablenotes}
 \end{threeparttable}
 }
 \vspace{0.5cm}
\end{table*}

%Table Validation - Jupiter
\begin{table*}
 \setlength{\tabcolsep}{5.5pt}
 \footnotesize
 \begin{center}
 \caption{Approach (iii): Testing the model with \textbf{Jovian values}, based on the modelling approach of \citet{farrell1999}. The input parameters for my model are: $\alpha = 1.5 \times 10^3$ s$^{-1}$, $\beta = 1.75 \times 10^3$ s$^{-1}$ \citep{farrell1999}; $W_d = 10^{12}$ \citep[e.g.][]{rinnert1998}; $k = 0.001$. \citet{farrell1999} modelled Jovian discharges with peak frequency, $f_0 = 500$ Hz, and duration of 1 to 2 ms. For further information, see Sect. \ref{sec:jupval}.}
  \begin{tabular}{@{}lllllllllll@{}}
	\hline
   & $\tau$ [$\mu$s] & $i_0$ [kA] & $|Q|$ [C] & $M(t)$ [C m] & $h$ [m] & $P_{\rm rad}$ [W] & $W_{\rm rad}$ [J] & $W_d$ [J] & $k$ & $f_0$ [Hz] \\
	\hline
  (iii) & 3800 & $3.2 \times 10^3$ & $1.25 \times 10^4$ & $4.37 \times 10^9$ & $3.5 \times 10^5$ & $2.6 \times 10^{11}$ & $10^9$ & $10^{12}$ & 0.001 & 257 \\
	\hline
  \end{tabular}
	\label{table:val2_2}
\end{center}
\vspace{0.5cm}
\end{table*} 

I discussed Jovian lightning in Chapter \ref{chap:liginout}, Sect. \ref{sec:int_jup}, however, just like for Saturn, here I repeat and discuss the relevant information for my work in more detail. Jovian lightning was observed by several spacecraft both in the optical and radio bands \citep[e.g.][]{cook1979,borucki1982,borucki1992,rinnert1998,little1999,baines2007}. The \textit{Voyagers}, \textit{Galileo}, \textit{Cassini}, and \textit{New Horizons} all measured the average optical power of lightning on Jupiter to be $\sim 10^9$ J, with values between $3.4 \times 10^8$ J \citep{baines2007} and $2.5 \times 10^{10}$ J \citep{rinnert1998}. \citet{borucki1987} determined from laboratory experiments that the optical efficiency of lightning on Jupiter is 0.001. \citet{rinnert1998} estimated both the radio energy and total dissipation energy from data gathered by the Galileo probe during its descent into Jupiter's atmosphere. Their results suggest that the radio efficiency is 0.025, with $W_{\rm rad} = 2.5 \times 10^{10}$ J, and $W_d = 10^{12}$ J. The data of the probe provide us with valuable information on the radio spectrum of lightning on the gas giant. \citet{rinnert1998} obtained pulse durations between 266 and 522 $\mu$s, with inter-pulse gaps between 680 $\mu$s and 1 s. Such slow discharges have their peak power radiated at $\sim 500$ Hz \citep{farrell1999}. \citet{rinnert1998} estimated several properties of discharges on Jupiter, which I use as comparison for the model results, and list them in Table \ref{table:val2}. I follow two approaches, (i) first, I test the model against the example given in \citet{rinnert1998}, who deduced lightning parameters from Galileo probe data. For the sake of comparison, in this approach I use $\v v=0.1$c, like \citet{rinnert1998}, instead of $\v v = 0.3$c, which is used in all other approaches. (ii), next, I apply information of the duration of the discharge measured by the Galileo probe \citep{rinnert1998}, and experimental results from \citet{borucki1987}, who estimated the optical efficiency of lightning on Jupiter. Here, I assume this efficiency is the same for radio emission. Finally, Table \ref{table:val2_2} lists my results for the thirds approach, (iii) when I ran the model with input parameters from \citet{farrell1999}. Each time the peak current was iterated to match the dissipation energy with the pre-set limit of $W_d = 10^{12}$ J. $r = 1000$ km in each case \citep[like in][]{farrell1999}.

The results of Approach (i), in Table \ref{table:val2}, show that to reach the desired dissipation energy, $W_d = 10^{12}$ J, my model requires $\sim 4$ times more charges and more peak current in the channel, than what was estimated by \citet{rinnert1998}. This suggests that the model underestimates the released power, and the shape of the power spectrum is flatter than what I obtain. Approach (ii,a) and (ii,b), in Table \ref{table:val2} illustrates the importance of radio efficiency, $k$, in the model. When $k$ is lower, a lower amount of radio energy is necessary to obtain the required total dissipation energy, which results in lower number of necessary charges and amount of peak current in the channel. Table \ref{table:val2_2} lists the results when the \citet{farrell1999} set-up is applied to the model. In this case a ten times slower discharge is considered than before. Though the necessary peak current to obtain $W_d = 10^{12}$ J is not much higher than, e.g., in Approach (ii,a), the resulting charges and charge moment are orders of magnitude larger. This is because a ten times slower discharge will create a ten times longer discharge channel with the same velocity, resulting in very large $Q$ and $M(t)$ values.

%__________________________________________________________________
\subsection{Evaluation} \label{sec:5_eval}

In summary, I find that the model described in this chapter works relatively well considering that it is a single, dipole model, which does not include the effects of channel tortuosity and branching. With the help of the three Solar System planet data, I can evaluate the performance of my model. I find that the shape of the power spectrum is not well reproduced. This is the main cause of the following inconsistencies between the model and observed values of lightning parameters:

\begin{itemize}
	\vspace{-0.5cm}
		\item The tests for Earth lightning show that I underestimate the energy by one order of magnitude. This can be because the produced power spectrum has a slope of $f^{-3}$, instead of the observed $-2$ and $-3$. I suggest that the overall shape of the electric field power spectrum of lightning is relatively steeper than the observed one, and therefore, result in lower amount of calculated power and energy.
		\item The tests for Saturn suggest that Saturnian discharges are indeed super-bolt-like discharges, with peak currents around 70$-$130 kA. However, the shape of the spectrum in the model is not as flat as the observed ones, which likely overestimates the necessary current in the channel to produce an observed discharge dissipation energy of $W_d = 10^{12}$ J.
		\item Finally, the tests for Jupiter help us give a constrain on how different the model is from observed Jovian and Saturnian like discharges. The measurements of the Galileo probe \citep{rinnert1998,lanzerotti1996} provide us with valuable information on the behaviour of lightning radio emission on Jupiter. It seems, the electric field power spectrum of these discharges \citep[$f^{-1.5} - f^{2}$,][]{farrell1999} is much flatter than what is known from Earth \citep[$f^{-2}-f^{-4}$,][]{rakov2003}, but not as flat as Saturnian spectra \citep[$f^{-0.5}-f^{-2}$,][]{fischer2006,mylostna2013}. My results suggest that the model overestimates the necessary peak current to reach $W_d = 10^{12}$ J by a factor of 4. This means the produced power is underestimated, due to a modelled power spectrum that is steeper ($f^{-3}$) than observations suggest.  
		\item In conclusion, my model seems to underestimate the released power and energy by a factor of four to ten. I will consider this as source of uncertainty when I discuss the results of exoplanetary lightning modelling.
\end{itemize}

%__________________________________________________________________
%__________________________________________________________________
\section{Results and Discussion} \label{sec:resdis}

In Sect. \ref{sec:current}, I analysed the differences between the two most commonly used current functions, the bi-exponential (Eq. \ref{eq:1}) and the Heidler function (Eq. \ref{eq:3}). I showed that the bi-exponential function represent the same current shape as the Heidler function, if the input parameters are well chosen (see Table \ref{table:ibh} and Fig. \ref{fig:ibh}). I used the bi-exponential function in my model, since it is easier to implement and it describes the current well. In Sects \ref{sec:efield} and \ref{ssec:elcom}, I analysed the components of the electric field resulting from the current running in the channel. I showed that when observing a lightning discharge from large distances ($r>50$ km), most of the information is carried in the induction and radiation fields, and the electrostatic part is negligible (Fig. \ref{fig:ecomp}).

I tested the model in Sect. \ref{sec:val} against data from three Solar System planets, Earth, Jupiter, and Saturn. The results suggest that the model underestimates the released power and energy by a factor of four to ten. This is because the shape of the electric field power spectrum does not change during our modelling approach, as I do not include the effects of channel tortuosity and branching. As I change $\tau$, the discharge duration, the resulting power spectrum shifts, and its peak will closely follow the expression in Eq. \ref{eq:taufr}.

I analysed the effects of the different input parameters on the energy release and radio power output of lightning discharges (Fig. \ref{fig:yv}, Table \ref{table:yv}). These parameters are $h$ the extension of the discharge, $\v v_0$ the velocity of the return stroke, and $i_0$ the peak current. $h$ and $\v v_0$ determine $\tau$, the duration of the discharge, through Eq. \ref{eq:tau1}, therefore, I only discuss the effects of $\tau$. From Fig. \ref{fig:yv} I determine that the larger the peak current the more power and energy are released from a lightning stroke. I also find that the slower the discharge the more energy is radiated from lightning, even though the power released rather slightly decreases with larger $\tau$.

%__________________________________________________________________
\subsection{Energy-release of lightning on exoplanets and brown dwarfs} \label{sec:exoen}

\begin{figure}
  \centering
  \includegraphics[width=0.85\columnwidth]{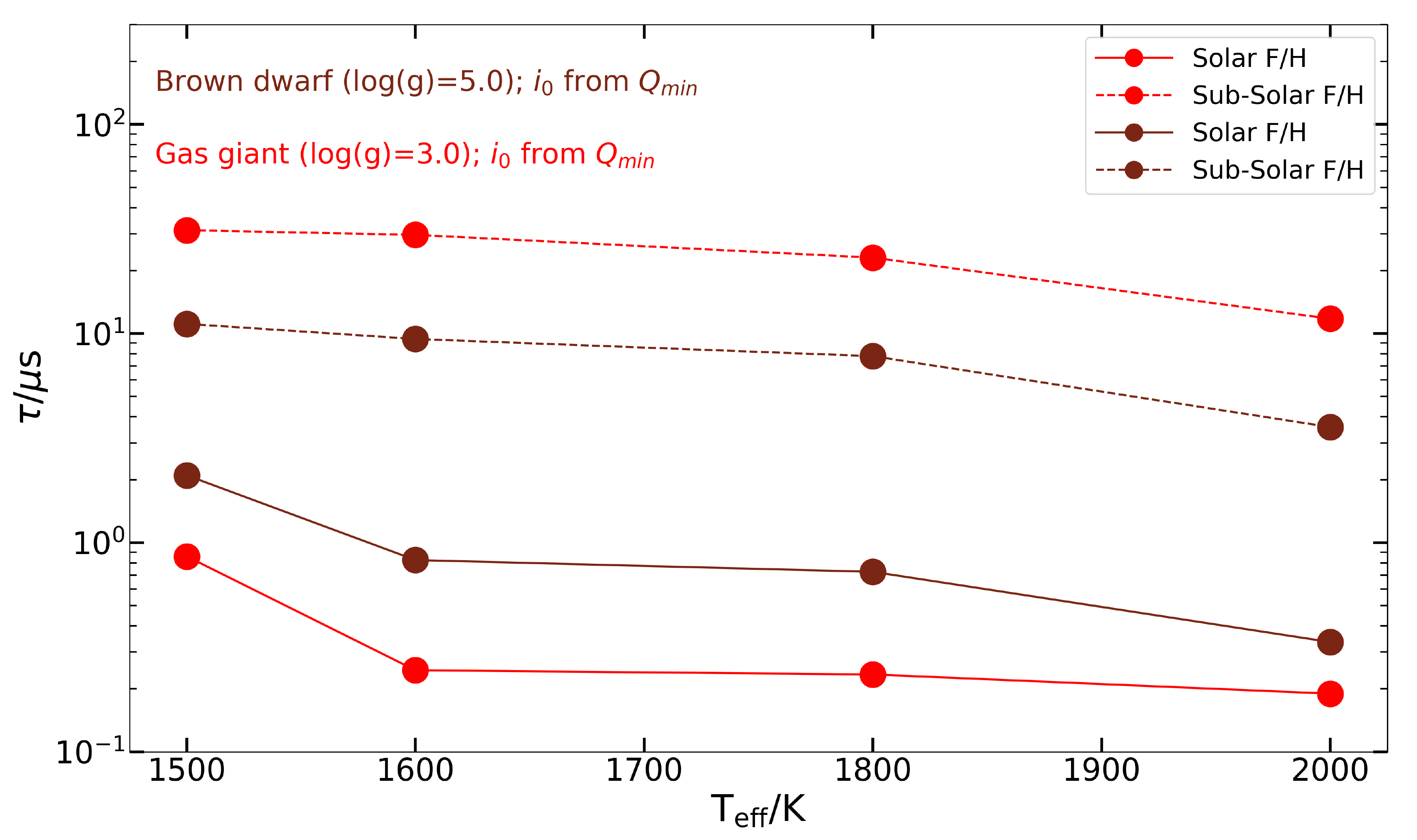}
  \caption{Discharge duration, $\tau$, calculated for different types of extrasolar bodies based on discharge lengths obtained by \citet{bailey2014} and listed in Table \ref{table:h}. $\tau$ was calculated from Eq. \ref{eq:tau1} using a current velocity of $\v v_0 = 0.3$c.}
  \label{fig:exotau}
	\vspace{0.8cm}
\end{figure}

\begin{figure*}
  \centering
  \includegraphics[width=0.85\columnwidth]{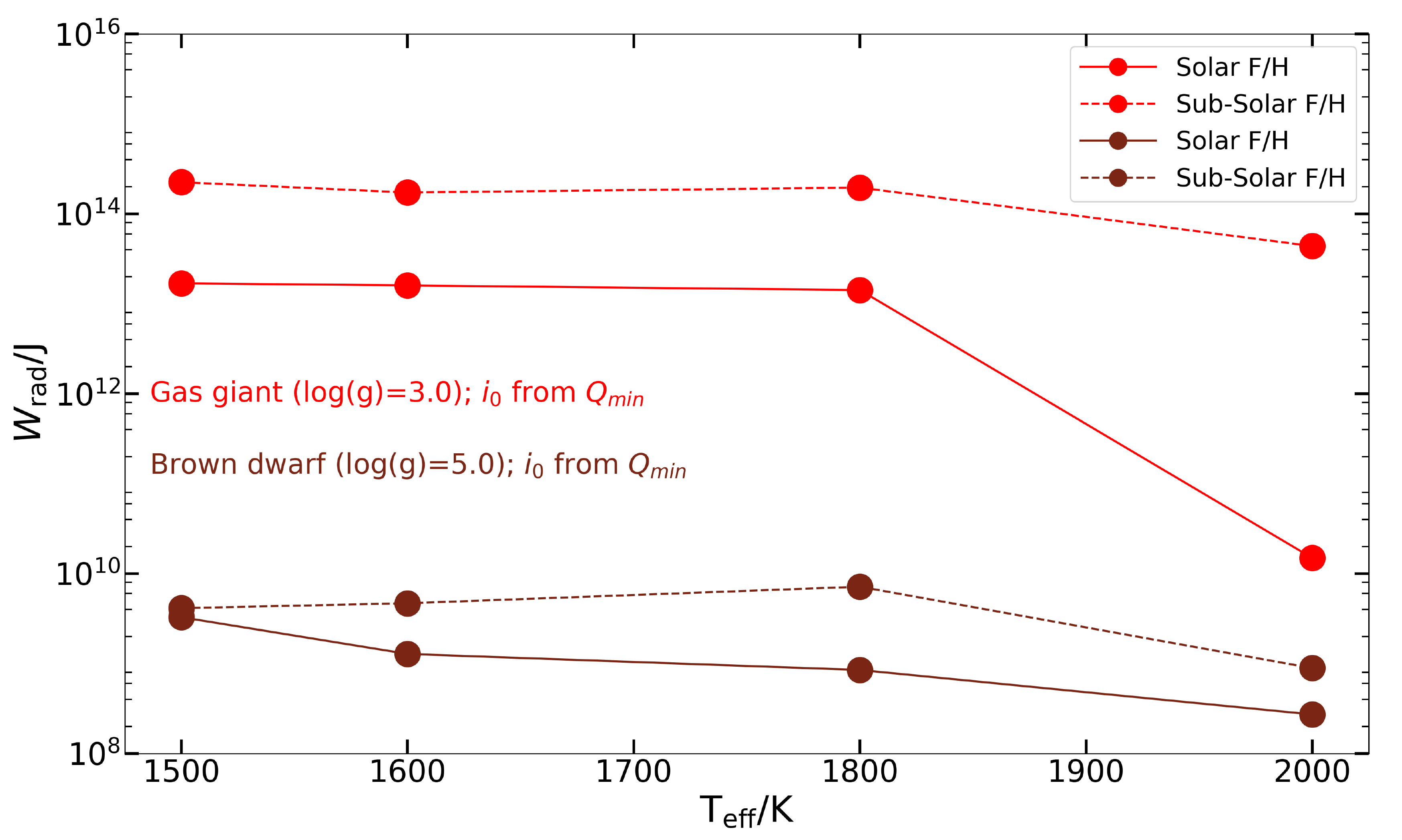}
  \includegraphics[width=0.85\columnwidth]{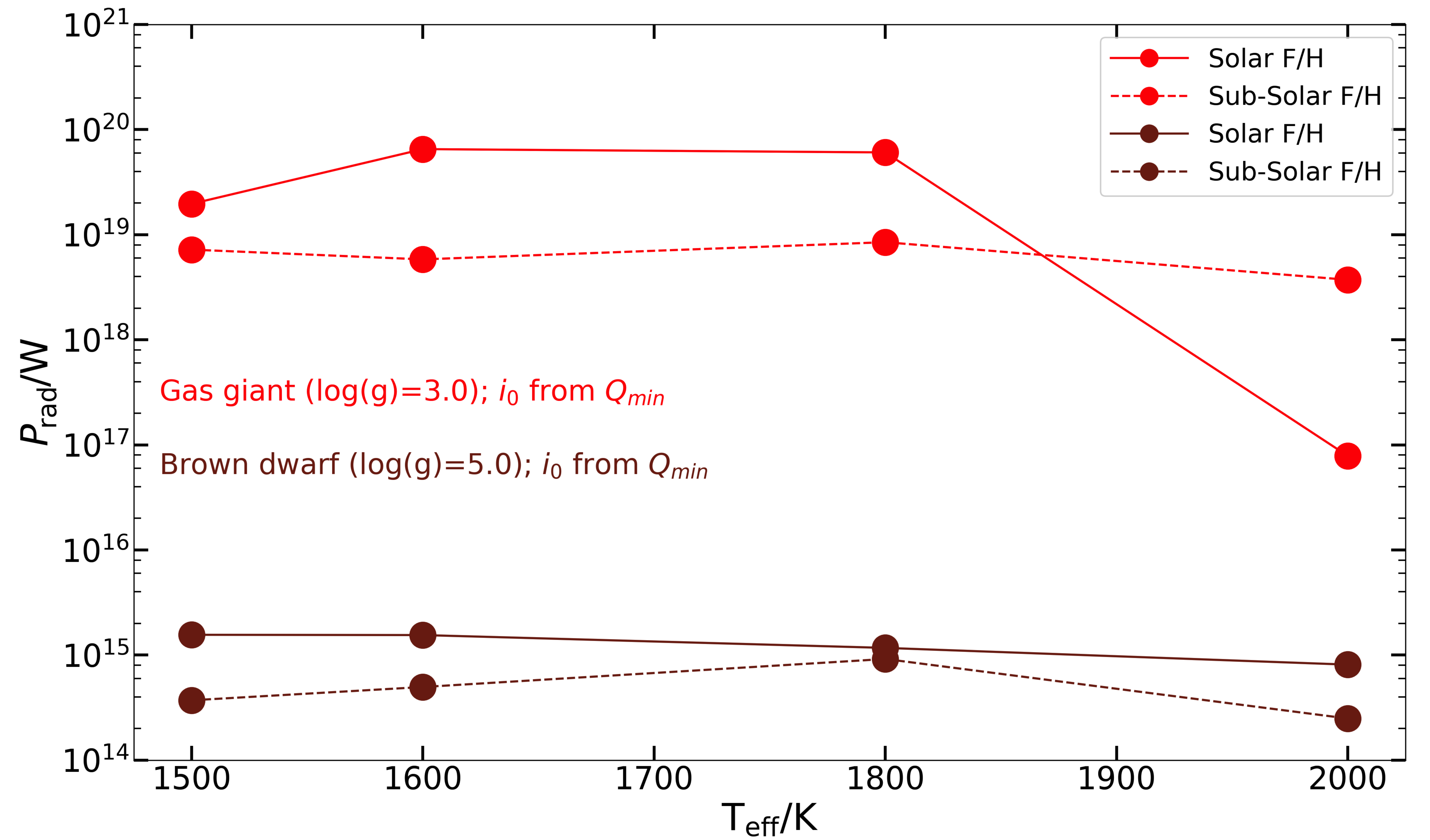}
  \caption{Total radiated energy (top) and total radio power (bottom) released from a lightning discharge estimated using peak currents, $i_0$, obtained from the minimum charges necessary to initiate a discharge (Table \ref{table:h}), based on \citet[][their fig. 7]{bailey2014}, using Eq. \ref{eq:iq}.}
  \label{fig:exoiq}
	\vspace{0.8cm}
\end{figure*}

\begin{figure*}
  \centering
  \includegraphics[width=0.85\columnwidth]{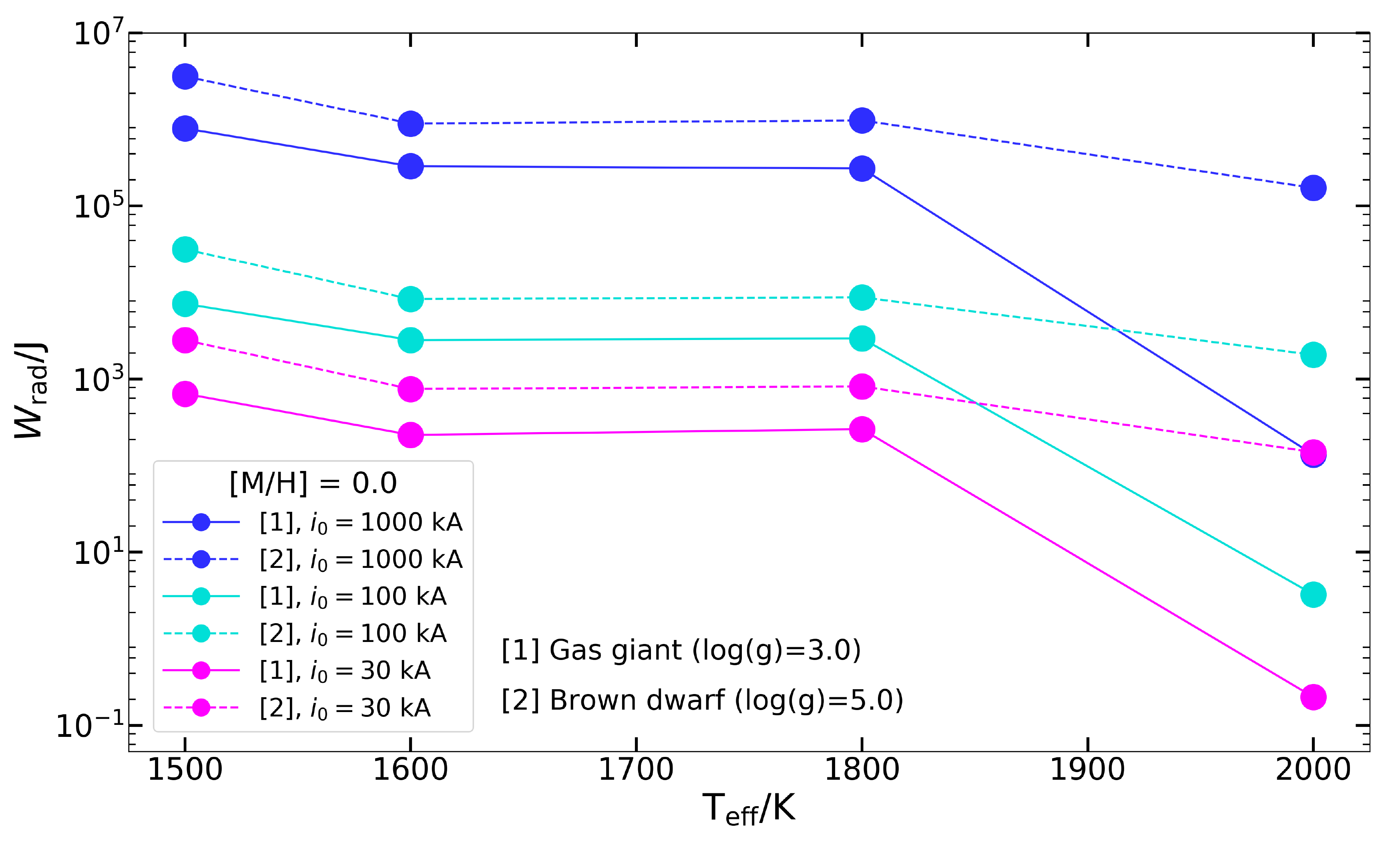} 
  \includegraphics[width=0.85\columnwidth]{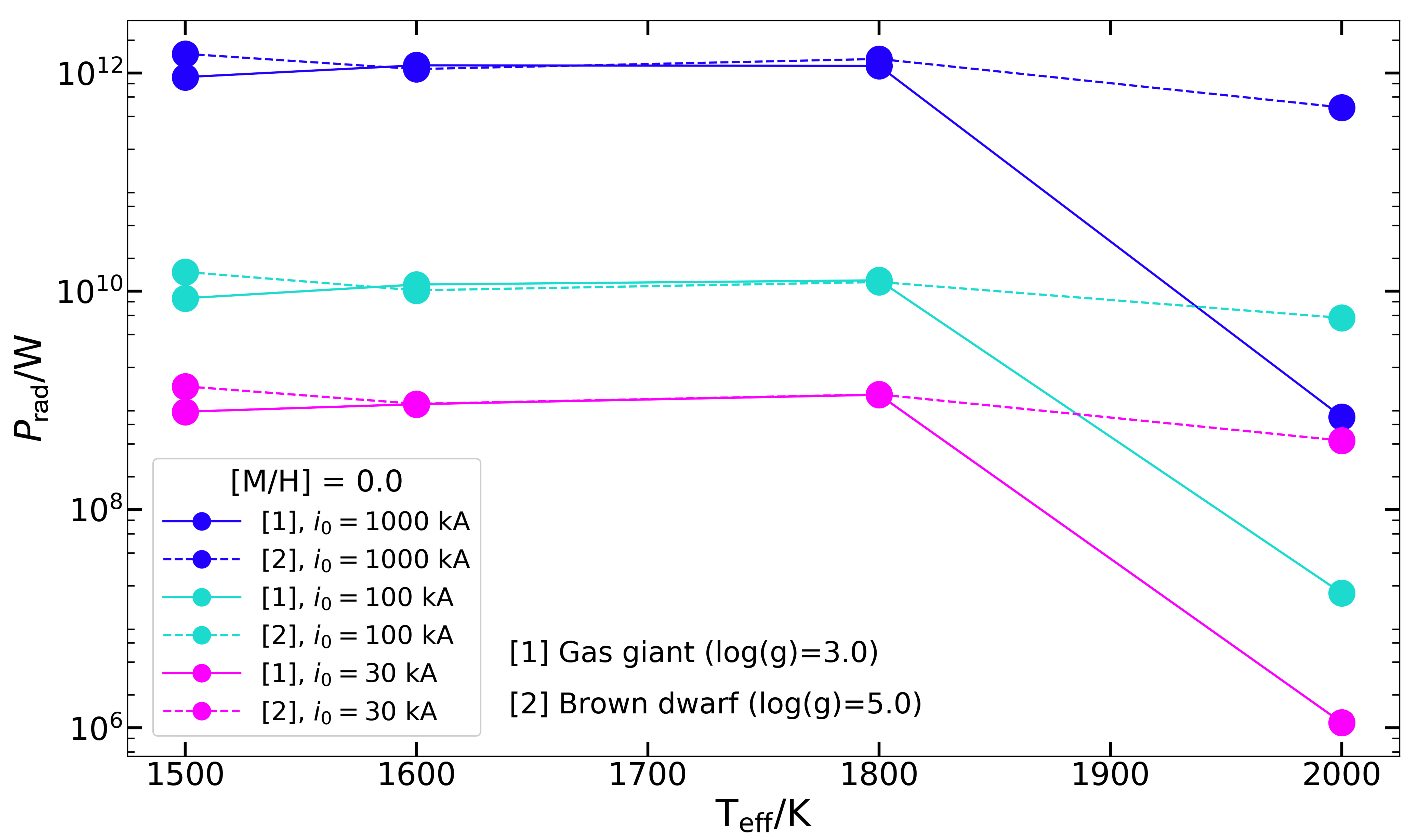} 
  \caption{Total radiated energy (top) and total radio power (bottom) released from lightning in different extrasolar atmospheres with solar metallicity ([M/H]=0.0), for peak currents $i_0 =$ 30, 100, 1000 kA (magenta, cyan, and blue colours, respectively). The different atmospheres are represented in my model by the extension of the discharge, $h$, as in \citet[][their fig. 9, right]{bailey2014} and Table \ref{table:h}.}
  \label{fig:exo11}
	\vspace{0.8cm}
\end{figure*}

\begin{figure*}
  \centering
  \includegraphics[width=0.85\columnwidth]{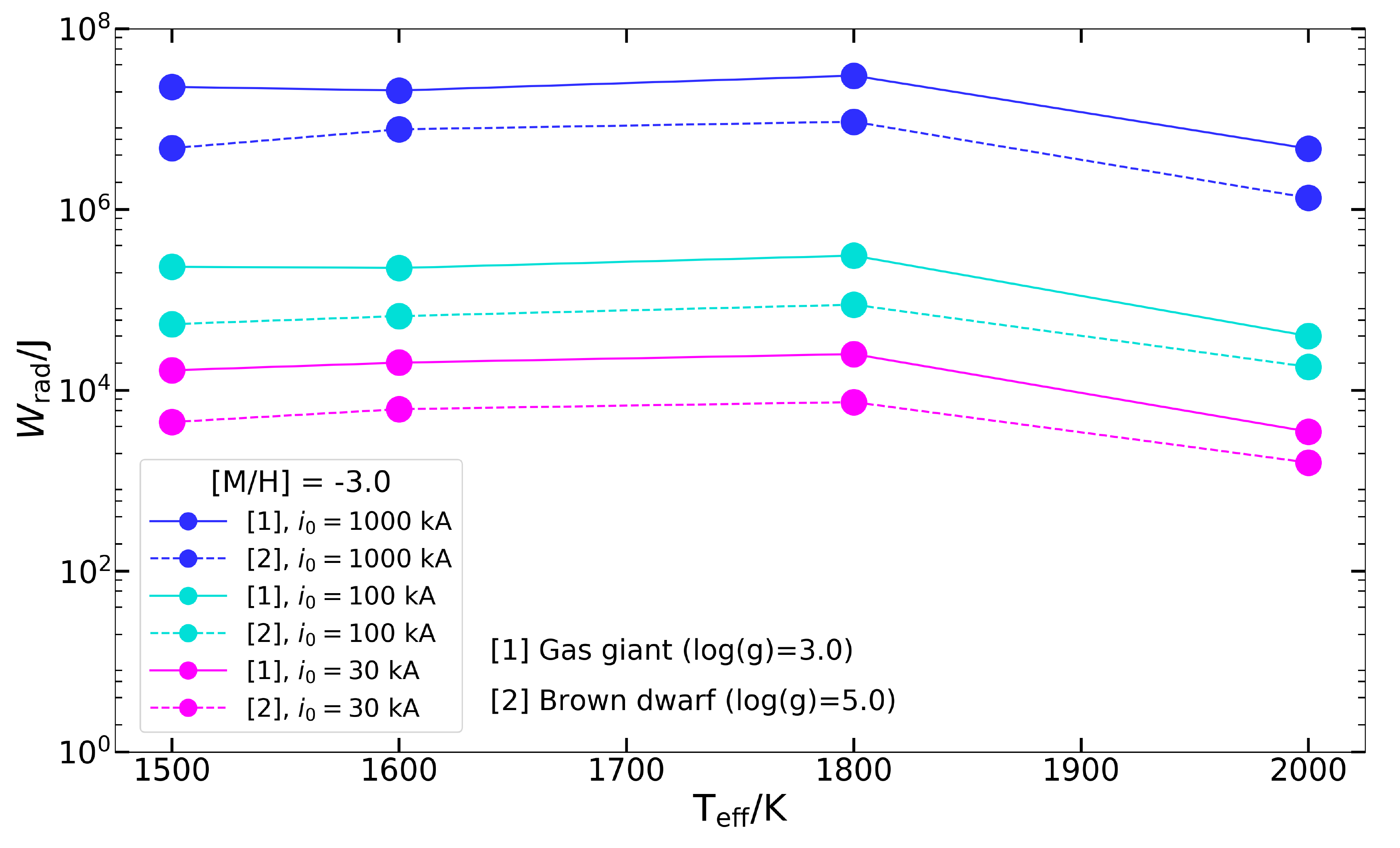} 
  \includegraphics[width=0.85\columnwidth]{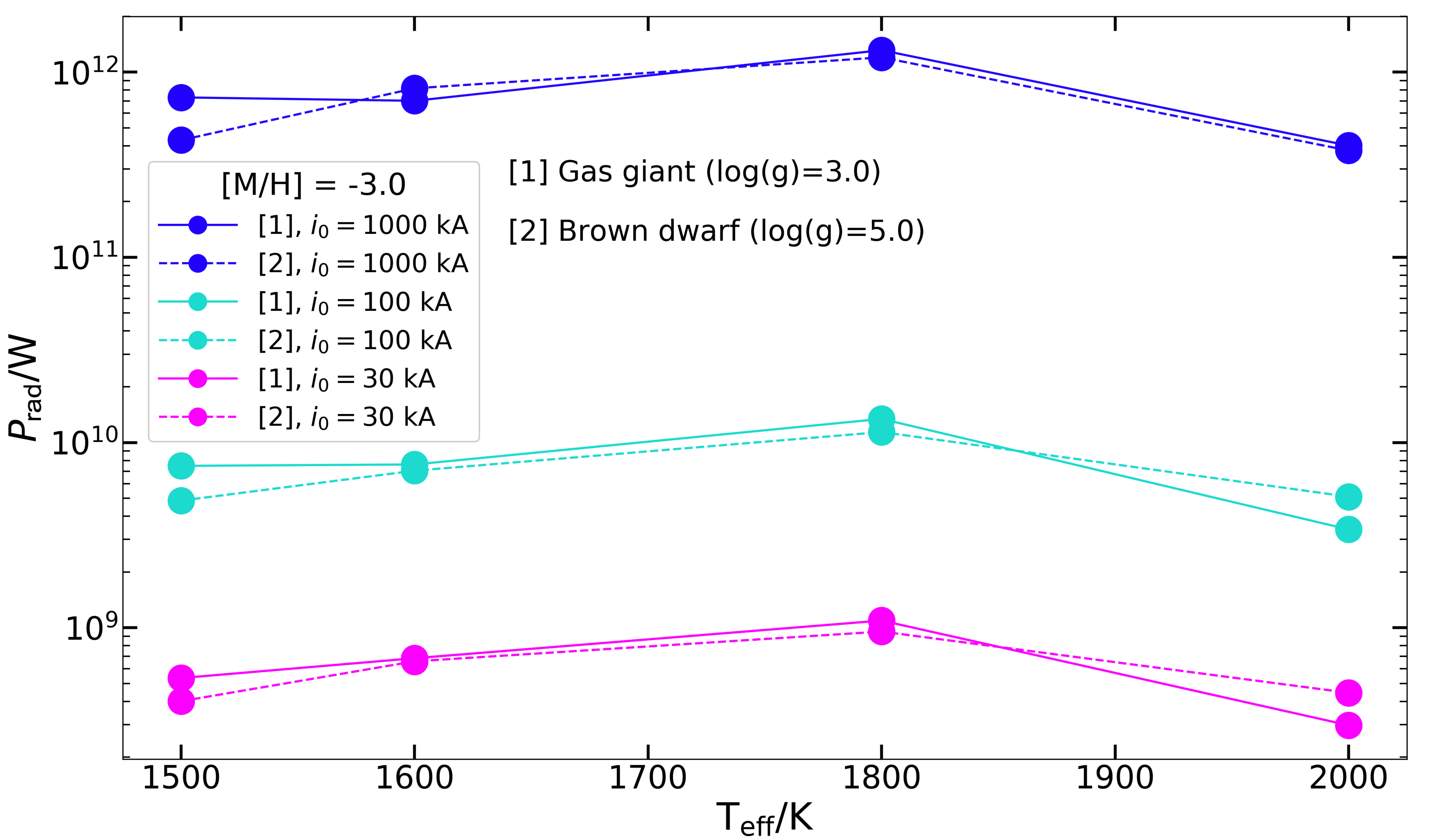}
  \caption{Total radiated energy (top) and total radio power (bottom) released from lightning in different extrasolar atmospheres with sub-solar metallicity ([M/H]=-3.0), for peak currents $i_0 =$ 30, 100, 1000 kA (magenta, cyan, and blue colours, respectively). The different atmospheres are represented in my model by the extension of the discharge, $h$, as in \citet[][their fig. 9, right]{bailey2014} and Table \ref{table:h}.}
  \label{fig:exo12}
	\vspace{0.8cm}
\end{figure*}

The radio energy radiated by lightning discharges depends on the peak current, $i_0$, and the duration of the discharge, $\tau$. $\tau$ also determines the frequency at which the peak power is released, while the strength of the electric field is determined by $i_0$. In my model, $\tau$ depends on the extrasolar object's properties through the extension of the discharge, $h$, while $i_0$ through the minimum number of charges, $Q_{\rm min}$, necessary to initiate a breakdown \citep[For comparison, the average amount of charges in a lightning channel on Earth is 30 C,][]{bruce1941}. The number of charges ($Q$) in the lightning channel is unknown, therefore \citet{bailey2014} considered two cases: case i) assume that $Q$ is constant in the channel then evaluate the properties of the local electric field, and case ii) derive $Q_{\rm min}$ such that $Q_{\rm min}$ = $Q(E_{\rm b}(p))$, where $E_{\rm b}(p)$ is the pressure-dependent breakdown field.
I used the results of their second approach, as that is connected to the properties of the atmosphere through the breakdown electric field. The effective temperature, $T_{\rm eff}$, metallicity, [M/H], and surface gravity, log($g$), of the object determine the local temperature and pressure profile of the extrasolar atmosphere, as given by the \textsc{Drift-Phoenix} model atmospheres \citep{helling2008c,helling2008d,witte2011,bailey2014}. \citet{bailey2014} found that the extension of the discharge will be larger in high-pressure atmospheres where the surface gravity is large or the metallicity is low. They also found that $h$ will decrease with increasing effective temperature. High-pressure atmospheres require a larger $Q_{\rm min}$ for lightning to be initiated. However, \citet{bailey2014} showed that in brown dwarfs, where clouds form at higher pressures, the minimum number of charges necessary for breakdown is smaller than in giant gas planets. They reason this with the extension of the cloud deck in brown dwarfs being shorter than in giant gas planets, resulting in a larger electric field throughout the cloud, which means that a lower number of charges is sufficient to initiate the breakdown in those atmospheres \citep{bailey2014}.

To explore lightning energy and power release on exoplanets and brown dwarfs, I followed three approaches: (I) Both $Q_{\rm min}$ (and directly $i_0$) and $h$ are taken from \citet{bailey2014}. This way I accept their results, and assume that both values are valid for the studied extrasolar objects. (II) $h$ is from \citet{bailey2014}, and $i_0$ is chosen from values observed in the Solar System. These are: $i_0 = 1000, 100, 30$ kA \citep[arbitrary example, arbitrary example][respectively]{farrell1999}. This approach assumes that charge accumulation in the channel on extrasolar objects is similar to their Solar System counterparts. (III) $Q_{\rm min}$ is from \citet{bailey2014}, but $h$ is chosen from Solar System values: $h = 259, 7.89, 2$ km \citep[][respectively]{bruning2015,rakov2003,baba2007}. Such discharges would have an "extrasolar-like" current, but Solar System-like discharge channel length. In each case, I used Eq. \ref{eq:tau1} to calculate the discharge duration from $h$ and the current velocity in the channel, $\v v_0 = 0.3$c.

\textbf{(I), $h$ and $Q_{\rm min}$ in Table \ref{table:h}:}
The discharge durations and peak currents for the different extrasolar bodies are listed in Table \ref{table:h}. Fig. \ref{fig:exotau} further illustrates the obtained $\tau$ values. My results of power and energy release in the investigated exoplanet and brown dwarf atmospheres are shown in Fig. \ref{fig:exoiq}. The top panel shows the radiated energy of lightning, and the bottom panel depicts the total radio power of the discharge. Each figure shows different case studies based on the three planetary parameters: $T_{\rm eff}$, [M/H], and log($g$). To calculate the power and energy released from lightning, I used the discharge durations presented in Fig. \ref{fig:exotau}, and the peak current, $i_0$, calculated using Eq. \ref{eq:iq} from the minimum number of charges, $Q_{\rm min}$, necessary to initiate a breakdown according to \citet[][their fig. 7]{bailey2014}. These figures indicate that lightning in giant gas planets, or low-gravity, young, brown dwarfs, with log($g$)$=3.0$, reaches higher energies than in brown dwarfs with log($g$)$=5.0$. \citet{bailey2014} found $h$ to increase with decreasing metallicity, therefore I found $\tau$ to increase with decreasing metallicity as well. Though the radiated power of lightning is higher in solar composition atmospheres (Fig. \ref{fig:exoiq}, right panel), the energy is higher in sub-solar compositions, because of the way $\tau$ acts in these atmospheres. Finally, the figures also show that the released lightning energy and power are less dependent on the bodies' effective temperature, than on the surface gravity or the chemical composition. The breakdown field that determines whether a lightning discharge will develop or not, does not strongly depend on the chemical composition (i.e. ionisation energy) of the gas. However, it depends on the local pressure, which is determined by the opacity in an atmosphere \citep{helling2013}. The very high currents resulting from high $Q_{\rm min}$ (Table \ref{table:h}) produce an electric field that will release very high energy and power in the radio bands: $\sim 10^{8}-10^{10}$ J and $\sim 10^{13}-10^{15}$ W in brown dwarf atmospheres, and $10^{13}-10^{14}$ J and $10^{19}-10^{20}$ W in giant gas planet atmospheres. Applying a $k = 0.01$ radio efficiency \citep[$\sim 1$\% on Earth][]{volland1984}, the total dissipation energy for these objects is $W_d \sim 10^{15}-10^{16}$ J for gas giants and $W_d \sim 10^{10}-10^{12}$ J for brown dwarfs, latter one being comparable to lightning on Jupiter and Saturn. If I assume that the radio efficiency is 0.001, as was experimentally suggested for Jupiter by \citet{borucki1987}, the dissipation energy becomes even higher: $W_d \sim 10^{16}-10^{17}$ J for gas giants and $W_d \sim 10^{11}-10^{13}$ J for brown dwarfs. If I further consider the factor of 4 to 10 underestimate of the energy, as is suggested by our tests in Sect. \ref{sec:val}, the resulting energies will further increase.

\textbf{(II), $h$ as in Table \ref{table:h}, $i_0 = 1000, 100, 30$ kA:}
I tested how the lightning radio power and energy release behaves assuming a fix peak current for each atmosphere. I note that this way I disregard the findings of \citet{bailey2014} about $Q_{\rm min}$, however, this study gives us valuable information on how, $W_{\rm rad}$ and $P_{\rm rad}$ changes with $h$ and therefore with $\tau$. Figures \ref{fig:exo11} and \ref{fig:exo12} show the results for the three example current peaks. The top panels show the radiated energy, while the bottom panels demonstrate the total radio power. Figures \ref{fig:exo11} presents the results for solar metallicity atmospheres, Fig. \ref{fig:exo12} shows results for sub-solar metallicity atmospheres. As expected, the larger the peak current, the more power and energy is released from lightning. A more interesting result is that while in solar metallicity atmospheres lightning is more energetic in giant gas planets, in sub-solar compositions lightning releases more energy in higher gravity environments (brown dwarfs). With the highest peak current, $i_0 = 1000$ kA, the released radio energy is $W_{\rm rad} \sim 10^6-5 \times 10^7$ J for both gas giants and brown dwarfs, with slightly lower energy release from the latter type of objects.

\begin{sidewaysfigure}
  \centering
  \includegraphics[scale = 0.09]{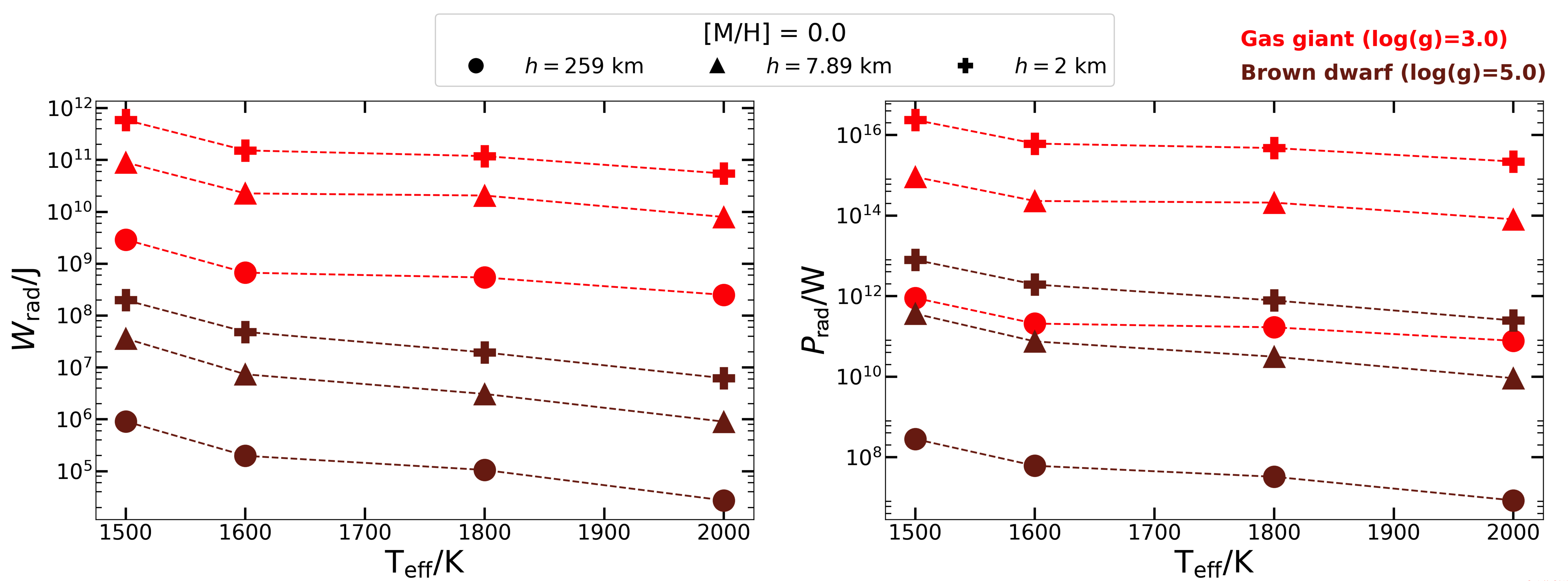} \\
  \includegraphics[scale = 0.09]{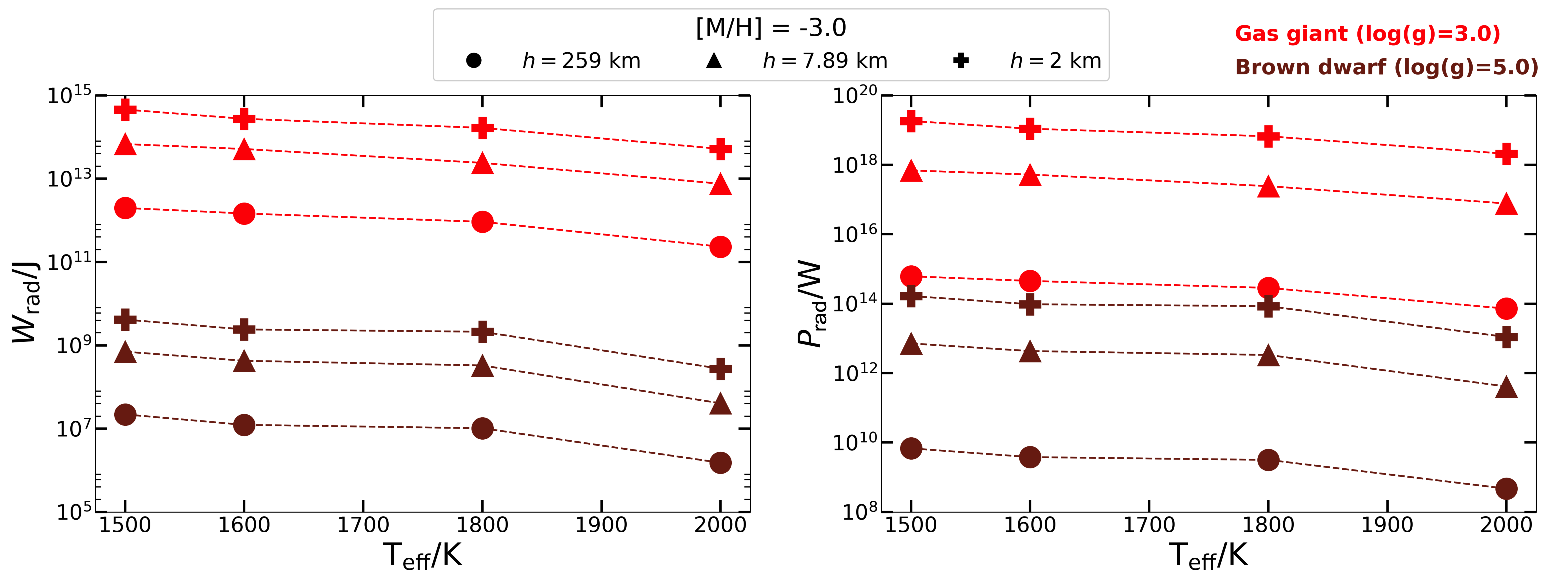}
  \caption{Total radiated energy (left) and total radio power (right) released from lightning in different gas giant (red) and brown dwarf (brown) atmospheres, for discharge extensions $h =$ 2, 7.89, 259 km (cross, triangle, circle symbols, respectively). The durations of discharge for each extension are $\tau = 2.5 \times 10^{-5}, 9.9 \times 10^{-5}, 3.2 \times 10^{-3}$ s, respectively, calculated with Eq. \ref{eq:tau1} and $\v v = 0.3$c. The top panels show solar metallicity atmospheres ([M/H]=0.0), while the bottom panels show sub-solar metallicity atmospheres ([M/H]=-3.0). The different atmospheres are represented in my model by the minimum charges necessary to initiate a discharge, $Q_{\rm min}$, as in \citet[][their fig. 9, right]{bailey2014} and Table \ref{table:h}.}
  \label{fig:exo2}
	\vspace{0.8cm}
\end{sidewaysfigure}

\textbf{(III), $Q_{\rm min}$ as in Table \ref{table:h}, $h = 2, 7.89, 259$ km:}
Next, I fixed $h$ for each atmosphere, and use $Q_{\rm min}$ from \citet{bailey2014}. Again, this way I disregard their findings about the discharge channel, however, we gain clear information about energy and power behaviour with varying current peak, $i_0$ (Table \ref{table:h}) in the example atmospheres. Figure \ref{fig:exo2} shows our results. The top panels present the results for solar metallicity, while the bottom ones show lightning radio energies (left) and powers (right) in sub-solar metallicity atmospheres. It is clear from the figure, that sub-solar metallicity atmospheres produce lightning with more radio energy and power release than solar compositions. For the same $h$, in higher surface gravity environments (i.e. brown dwarfs) lightning releases less energy than in lower surface gravity objects. The released power and energy slightly decreases with effective temperature. Also, the shorter the discharge channel the larger $P_{\rm rad}$ and $W_{\rm rad}$ J. In this case, giant gas planetary lightning produces higher energy every time. The released radio energy for gas giants is $W_{\rm rad} \sim 10^8-10^{15}$ J, and for brown dwarfs $W_{\rm rad} \sim 5 \times 10^4 - 10^{10}$ J. Applying various radio efficiencies, the total dissipation energy can be 2 to 3 orders of magnitude larger than $W_{\rm rad}$, and further applying the uncertainty factor from our tests in Sect. \ref{sec:val}, the results can be even higher by an order of magnitude.

My results suggest that the discharge energy will strongly depend on the process through which the cloud particles are charged and the processes that cause the electrostatic potential to build up. A discussion of processes for brown dwarfs and giant gas planets can be found in \citet{helling2016b} and in comparison to the Solar System in \citet{helling2016}.
However, no consistent description of large-scale lightning discharges is available for the atmospheres discussed here, nor for any of the Solar System planets.

%__________________________________________________________________
\subsection{The effect of parameter uncertainties} \label{subs:uncert}

%Table Alpha test
\begin{table}  
\small
 \begin{center}
 \caption{Statistical analysis of the changes in the total discharge energy, $W_d$, and the total radio power of lightning, $P_{\rm rad}$ due to changes in the $\alpha$ parameter. $\alpha$ is randomly picked for 100 times, for each extrasolar case study in Table \ref{table:h} (in total 16 cases, as in Approach (I)). The statistical values here are the minimum, maximum, average and median of the results of the 16 cases. The maximum value suggest that there is a 200\% change in the results, however, that is only valid for one data point, which corresponds to the outlier data points in Figs \ref{fig:exoiq}, \ref{fig:exo11}, and \ref{fig:exo12} ([M/H]=0.0, log($g$)=3.0, $T_{\rm eff} = 2000$ K). After removing this outlier, we get a more informative result.}
  \begin{tabular}{@{}lllll@{}}	
	\hline
	 & minimum & maximum & average & median \\
	\hline
	$\alpha$ & 8.7\% & 9.5\% & 9.3\% & 9.3\% \\
	$W_{\rm rad}$ & 12.5\% & 200.0\% & 32.9\% & 21.4\%  \\
	$P_{\rm rad}$ & 12.5\% & 200.0\% & 32.9\% & 21.4\% \\
	\hdashline
	\multicolumn{5}{c}{After removing the 200\% outlier} \\
	$\alpha$ & 8.7\% & 9.5\% & 9.3\% & 9.3\% \\
	$W_{\rm rad}$ & 12.5\% & 36.2\% & 21.8\% & 19.8\%  \\
	$P_{\rm rad}$ & 12.5\% & 36.2\% & 21.8\% & 19.8\% \\
	\hline
  \label{table:test}
  \end{tabular}
 \end{center}
\end{table}
%Table 

%Table tau, i0 test
\begin{table}  
\small
 \begin{center}
 \caption{Effects of changing discharge duration, $\tau$ (left), and current peak, $i_0$ (right), compared to a base value, on the total radio power, $P_{\rm rad}$, and the radiated discharge energy, $W_{\rm rad}$. All other parameters remain unchanged. The base value for $\tau$ was 100 $\mu$s, while for $i_0$ it was 30 kA. $\Delta$ sign represents the change or uncertainty in the value.}
  \begin{tabular}{@{}lllllllll@{}}	
	\cline{1-3}\cline{7-9}
  $\Delta \tau$ [s] & $\Delta P_{\rm rad}$ & $\Delta W_{\rm rad}$ & & & & $\Delta i_0$ [A] & $\Delta P_{\rm rad}$ & $\Delta W_{\rm rad}$ \\ \cline{1-3}\cline{7-9}
	$10^{-6}$ & 0.13 \% & 1.13 \% & & & & $10^3$ & 6.8\% & 6.8\% \\
	$10^{-5}$ & 1.2 \% & 11.3 \% & & & & $10^4$ & 77.8\% & 77.8\% \\ \cline{7-9}
	$10^{-4}$ & 6.6 \% & 113.3 \% & & & & & & \\ \cline{1-3}
  \label{table:test2}
  \end{tabular}
 \end{center}
\end{table}
%Table 

\begin{enumerate}

\item[a)] \textbf{$\alpha$ $[{\rm s}^{-1}]$:}
I tested our results against the uncertainty in the only (semi-)randomly chosen parameter, $\alpha$, a frequency type constant introduced in the current function (Eq. \ref{eq:1}). I randomly choose $\alpha$ from a normal distribution with mean and standard deviation that ensures that $\alpha$ is $\sim 1$ order of magnitude lower than $\beta$, the other frequency type parameter of the bi-exponential current function (Eq. \ref{eq:1}). This choice is based on the commonly used $\alpha$ and $\beta$ pairs in the literature (Table \ref{table:3}). I carried out the numerical experiment for a hundred runs for each exoplanet and brown dwarf types (Table \ref{table:h}). The results are listed in Table \ref{table:test}, and show that a roughly 9.3\% change in $\alpha$ as a result of the random pick throughout the 100 runs for each case, results in a 21.8\% average and 19.8\% median variation in both the total power ($P_{\rm rad}$) and the total discharge energy ($W_d$). The variation in both $W_d$ and $P_{\rm rad}$ is between 12.5\% and 36.2\% depending on the object, and with that on the extension of the discharge, $h$, and the number of charges in the channel, $Q_{\rm min}$. However, the tests did not include the testing of the effects of $h$ and $Q_{\rm min}$. These values are calculated after removing the outlier of the data set, appearing in both Figs \ref{fig:exoiq}, \ref{fig:exo11}, and \ref{fig:exo12} ([M/H]=0.0, log($g$)=3.0, $T_{\rm eff} = 2000$ K). This atmosphere alone suggests that the variations caused by $\alpha$ can be up to 200\%. However, this data point causes errors in the calculations, most probably due to numerical effects caused by the combination of a very short channel and a large $Q_{\rm min}$.

\vspace{0.2cm}

\item[b)] \textbf{$\tau$ [s] and $i_0$ [A]:}
The discharge duration, $\tau$, and the peak current, $i_0$, are the two values that will affect the results the most (Sect. \ref{sec:resdis}). Therefore, I tested how much changing these values compared to a base value will affect the resulting radiated discharge energy, $W_{\rm rad}$, and total radio power, $P_{\rm rad}$. I gradually increased $\tau$ and $i_0$ separately, starting from a base or comparison value, while all the rest of the input parameters were fixed (i.e $\alpha$, $\beta$, see Sect. \ref{sec:current}). The base value for $\tau$ was 100 $\mu$s, while for $i_0$ it was 30 kA. The results are shown in Table \ref{table:test2}. The tests showed, that the energy is fairly sensitive to the changes in the discharge duration, while the power seems to be less sensitive. A 100 $\mu$s change in $\tau$ causes only about a percent change in the energy, and 0.1\% change in the power. Two orders of magnitude change in $\tau$ changes the power only by 6.6\%, but changes the energy by more than a 100\%. I also point out that increasing $\tau$ results in a increasing $P_{\rm rad}$ and $W_{\rm rad}$. Similarly, I tested the sensitivity of the end results to $i_0$. The results show that a 10 kA change in the peak current results in a $\sim 77.8$\% change in both the total radio power, and the radiation energy. This is twice as much as the uncertainty caused by the $\alpha$ parameter. The duration of the return stroke is not well determined for different planets. For Earth it is around 50-100 $\mu$s \citep{rakov2003}, while for Saturn the duration of the SEDs (Saturn Electrostatic Discharges) were measured to be between 30 ms and more than 0.3 s \citep{zarka2004}. This suggest that the duration of exoplanetary lightning will span a fairly large interval, resulting in large variations in the released energy and radio power. In this study we calculated the peak current based on the obtained charges in \citet[][case ii)]{bailey2014}. These currents are of the order of $10^7-10^9$ A, therefore the changes due to the current in the resulting energies and radio powers are significant.
\end{enumerate}

%__________________________________________________________________
%__________________________________________________________________
\section{Conclusions} \label{sec:conc}

It has been shown before that lightning may occur in extrasolar planetary and brown dwarf atmospheres \citep[e.g.][]{helling2013,helling2013b}.
However, due to the lack of actual "exo-lightning" observations, we do not know how similar or different lightning is in extrasolar planetary atmospheres compared to what is known from the Solar System. In this chapter, I estimated the energy radiated at radio frequencies and dissipated from lightning discharges and the total power emitted at radio frequencies in order to study the differences and similarities of extrasolar and Solar System lightning. 

In general, the radiated power and emitted energy of a lightning discharge depends on two properties: the discharge duration, $\tau$, and the peak current, $i_0$. The quicker the lightning discharge, the larger the power and energy released from it when the peak current is constant. Short lightning channels are the result of high local pressure in the atmosphere \citep[e.g. for large surface gravity and low metallicity;][]{bailey2014}. The larger the peak current, the larger the power and energy released from the lightning discharge, hence, atmospheres that are exposed or intrinsically produce high-efficient cloud charging may produce a large current flow, and therefore a large radio signal. I also found that for a shorter discharge, a smaller peak current is necessary to obtain the same power density measured at a given radio frequency. Therefore, quick discharges do not require a large amount of current in order to produce observable radio fluxes at high frequencies. However, what "quick discharge" and "large amount of current" mean, will depend on the distance of the observed planet, and the surrounding radio noise. \citet{xue2015} measured the spectrum of natural cloud-to-ground discharges, and arrived to a similar conclusion that the structure of the spectrum is mostly affected by the current magnitude and duration. 

My results suggest that lightning on extrasolar, planetary objects can be very different from what we know from the Solar System. I related the model to extrasolar atmospheres through the extension of the discharge, $h$, and the charges in the current channel, $Q_{\rm min}$, as described in Sects \ref{sec:param} and \ref{sec:exoen}. I note that the actual number of charges resulting in the potential difference of a cloud that precedes a lightning discharge, as well as the extension of the discharge channel are unknown, therefore I followed three approaches to determine the energy and radio power of lightning in different extrasolar atmospheres: 

\begin{itemize}
\setlength\itemsep{1.0em}
\item \textbf{(I), $h$ and $Q_{\rm min}$ as in Table \ref{table:h} \citep{bailey2014}}: 
$Q_{\rm min}$ is obtained for each atmosphere type so that it is the minimum number of charges necessary to initiate a pressure-dependent breakdown field in the individual atmospheres \citep{bailey2014}. $h$ is also determined based on propagation effects in an extrasolar atmosphere \citep{bailey2014}. The results indicate that due to the very short channels, and the large amount of charges in the channel, very quick discharges occur with very large peak currents in giant gas planets. Moreover, Figure \ref{fig:exoiq} suggests that total dissipated energies can reach as high as $10^{11}-10^{13}$ J in brown dwarfs (log($g$)=5.0) and $10^{16}-10^{17}$ J in giant gas planets (log($g$)=3.0).
\item \textbf{(II), $h$ as in Table \ref{table:h}, $i_0 = 1000, 100, 30$ kA:} 
$Q_{\rm min}$ is obtained from the same peak current, $i_0$, for each atmosphere type, which are represented by $h$ from \citet{bailey2014}. I find that lightning in giant gas planets, or low-gravity, young, brown dwarfs, with log($g$)= 3.0, reaches higher energies than in brown dwarfs with log($g$)= 5.0, if the metallicity is sub-solar, however higher surface gravity objects with solar metallicity host more energetic lightning flashes. In general, atmospheres with sub-solar metallicity host stronger flashes than in solar compositions. The released lightning energy and power are less dependent on the bodies' effective temperature, than on the surface gravity or the chemical composition of the object (Figs \ref{fig:exo11} and \ref{fig:exo12}). 
\item \textbf{(III), $Q_{\rm min}$ as in Table \ref{table:h}, $h = 2, 7.89, 259$ km:}
I apply the results from \citet{bailey2014} for $Q_{\rm min}$, and an $h$ that is measured for discharges in the Solar System. I estimate the energy and power for three different extensions. The results (Fig. \ref{fig:exo2}) suggest that sub-solar metallicity atmospheres host more energetic, more powerful lightning flashes. For the same $h$, higher surface gravity objects host less powerful and energetic flashes. I also find, that the shorter the discharge channel the higher the released energy and power are. $W_{\rm rad} \sim 10^8-10^{15}$ J for gas giant planets, and $W_{\rm rad} \sim 5 \times 10^4 - 10^{10}$ J for brown dwarfs.
\end{itemize}

\noindent I further note that our results may underestimate the actual energy-release by a factor of four to ten, as it is suggested by our tests with Solar System lightning (Sect. \ref{sec:val}). Uncertainty in the results is also introduced by the random pick of the $\alpha$ parameter (Sect. \ref{subs:uncert}), which can be on average 20\%.

The results suggest that lightning releases more energy and radio power in certain exoplanetary and brown dwarf atmospheres than in Solar System planetary atmospheres. The considered objects are much more different from Earth, Jupiter, and Saturn, where lightning has been observed, therefore such energy release may not be unreasonable.

%% file: chapters/6_conclusions.tex
\chapter{Summary and Future Research} \label{chap:concl}

The presently known ensemble of exoplanets is extremely diverse, including Earth-like planets and giant gas planets some of which resemble brown dwarfs. A large number of these objects have atmospheres where clouds form. Our current knowledge of cloud formation on exoplanets and brown dwarf \citep[e.g.][]{sing2015,helling2008b,helling2011,helling2011b}, and lightning activity in the Solar System \citep[e.g][]{rakov2003,yair2008,yair2012,helling2016b} suggest that lightning occurs in extrasolar atmospheres. 
The electrostatic field breakdown that is associated with lightning is relatively independent of the chemical composition of the atmospheric gas. The local population of thermal electrons provides the seed electrons to initiate such a field breakdown also in extrasolar atmospheres \citep{helling2013}. Cosmic rays will enhance this population of seed electrons in particular in the upper part of the atmosphere. Therefore, large-scale discharges in the form of lightning should be expected to occur in the upper, electron-dominated part of atmospheric clouds \citep{helling2013}.

Studying lightning in extrasolar atmospheres, as well as on Solar System planet, is important to understand atmospheric electrification, convection and cloud dynamics. Lightning discharges largely affect the chemistry of the local atmosphere, and it has been shown that they produce prebiotic molecules \citet{miller1953,miller1959}, species important for the formation of life. Since "exo-lightning" has not been observed yet, we can only model the properties of discharges on exoplanets and brown dwarfs based on knowledge obtained from Solar System lightning. The models, however, can be extended and applied for extrasolar bodies. In this thesis I presented a first, in-depth study of exoplanetary and brown dwarf lightning properties and observability. Here, I summarize this work and will present an outlook on future research prospects.

\section{Properties of "exo-lightning"}

Lightning has not been detected outside the Solar System, therefore, in order to address lightning activity on extrasolar objects, we have to look around us and see what we already know about lightning on Earth and on other Solar System planets. In this thesis I studied two major characteristics of lightning: its spatial and temporal distribution, and its energy and power emission. 

In Chapter \ref{chap:stat} I presented a statistical study of lightning activity on Earth, Jupiter, Saturn, and Venus, based on optical and/or radio measurements. The obtained information in Sections \ref{sec:earth} and \ref{sec:solsys} was used to estimate lightning occurrence on example extrasolar planetary objects, including transiting and directly imaged planets, and brown dwarfs. The results suggest that volcanically very active planets, and objects with clouds of similar compositions as volcano plumes, would show the largest lightning flash densities if lightning occurred at the same rate on these planets as it does in volcano plumes on Earth. 

In Chapter \ref{chap:hatp11b} I gave a first estimate of lightning activity on the exoplanet HAT-P-11b based on observational data. I found that the tentative radio emission detected from the direction of the exoplanet, could be produced by extremely large lightning activity, if lightning has the same energetic properties that we know from Saturn. The included parameter study suggests that under certain conditions lower flash density-storms could produce the observed radio signal, with flash densities of the order of a few tens of flashes km$^{-2}$ h$^{-1}$. This is only ten times larger than what the average largest storms show in the USA, and is not unprecedented in the Solar System. In Chapter \ref{chap:danish}, I studied what the optical emission of "exo-lightning" would be, if it had the same statistical and radiating characteristics that we know from the Solar System. I estimated optical fluxes and apparent magnitudes of lightning storms on the three closest brown dwarfs, Luhman-16, $\epsilon$ Indi, and SCR 1845-6357, in standard I, V, and U bands. The results suggest that lightning will occur in a large variety of extrasolar objects, with various flash densities, emitted powers and discharge durations. Some of the parameter combinations with the largest optical power output and largest flash densities could favour lightning observations in the optical band, since lightning storms on the investigated brown dwarfs could be as bright us a 13 to 16 magnitude star.

After studying lightning in the Solar System and applying that knowledge to extrasolar planets, in Chapter \ref{chap:model} I asked how different lightning can be from this picture on exoplanets and brown dwarfs. Is it possible that discharges release more energy and power outside the Solar System than on Solar System planets? The energy released from lightning will determine the strength of the electromagnetic waves emitted from the discharge, and the amount of non-equilibrium species produced in the atmosphere. Ultimately, it will determine whether "exo-lightning" signatures can be observed from Earth or not. In Chapter \ref{chap:model}, I presented a model built from previously tested lightning models, and extended it so that exoplanetary lightning can be studied as well. The simple dipole model calculates the energy and total radio power released by lightning, and connects these estimates to the atmospheric properties through the extension of the discharge and the peak of the current flowing in the discharge channel. I found that in objects quite different from Solar System planets (log($g$)=3.0 and 5.0; T$_{\rm eff}$ = 1500 \dots 2000; [M/H] = 0.0 and -0.3), lightning may be $2-8$ orders of magnitude more energetic than Earth lightning, and up to $5$ orders of magnitude more energetic than Saturnian and Jovian lightning.

The results of this PhD project are essential in understanding how atmospheric electricity may work in exoplanetary and brown dwarf atmospheres, as it provides the community with a first estimate of "exo-lightning" occurrence rate and energy release. This work shows a new way of interpreting observational data. It suggests that the source of observed signals may be very different from the usually accepted interpretations, e.g. what one might think is a radio signal resulting from magnetosphere-stellar wind interactions, it may actually be lightning induced radio emission.

\section{Future research possibilities}

There are several open questions related to the scientific field of exoplanetary atmospheric electricity. 
A project directly following the study presented in Chapter \ref{chap:model} has already been started. The results I discussed in the chapter focus on the energy and total radio power released from lightning on the investigated extrasolar objects. However, it does not provide observability estimates. This new study will apply the results of Chapter \ref{chap:model}, and will further analyse the obtained lightning power spectra in order to estimate the radio flux of lightning at certain frequency bands from certain exoplanets and brown dwarfs using the methods presented in Chapters \ref{chap:hatp11b} and \ref{chap:danish}. Combining the obtained radio flux of one lightning flash with lightning climatology studies presented in Chapter \ref{chap:stat}, I will be able to estimate the total radio flux of lightning discharges from extrasolar objects. The study will further explore the implications of possible radio signal detection, e.g. it will provide estimates of ionospheric plasma frequency.

The already existing predictions can be tested by actual observations. As the results of this thesis suggest, a multi-wavelength observational campaign could lead to the first detection of lightning induced signatures on an extrasolar object, most probably a close-by faint brown dwarf. Radio and optical observations simultaneously could produce yet un-seen and un-explained signals, which may be the result of lightning activity. Such observations could be followed up by infrared telescopes, which could detect spectral signatures of enhanced non-equilibrium species produced by lightning in the atmosphere of the observed object. Low-frequency radio arrays like LOFAR or LWA (Long Wavelength Array), high-precision optical and NIR telescopes, like the ones of the Gemini Observatory, and future space missions such as JWST (James Webb Space Telescope) could all contribute to the future study of extrasolar lightning. 

Such project would involve the addressing of the following questions: which objects are the best candidates for lightning detection with current technology, and what sensitivity would be needed to observe lightning on an Earth-like planet in the habitable zone? With current radio facilities, what depths can we hope to reach? What radio bands are the most promising for observations? How could the observations be improved by potential space radio-arrays? The data obtained by ground-based telescopes could further provide with information on Earth lightning as a noise-source in the data of extra-terrestrial observations. The outcome of such observational research project would be beneficial and significant no matter whether lightning is detected on an exoplanet or brown dwarf, or not. The detection of "exo-lightning", combined with planetary lightning observations with, e.g. the \textit{Juno} spacecraft, would support theoretical works regarding cloud formation, convection, ionization, and electricity in extrasolar atmospheres, or would encourage the further development of such work. Non-detection of lightning would suggest that a better understanding of exo-climates, cloud patterns and theoretical lightning formation is needed, carving the path for future studies.

The modelling work presented in Chapter \ref{chap:model} largely focuses on hot Jupiters and brown dwarfs. However, there is a large population of extrasolar objects that it does not address, such as super-Earths, Earth-size planets, directly imaged young planets, etc. Therefore, further modelling work can improve our current understanding of lightning on exoplanets. The work I presented in Chapter \ref{chap:model} based on input parameters from \citet{bailey2014}, who estimated several properties of large-scale discharges in hot exoplanets and brown dwarfs. A coupling between their code and my energy-estimate code would lead to lightning energy estimates in several exoplanet types other than what has already been studied.